\documentclass[prb,aps,nofootinbib,amssymb,twocolumn,superscriptaddress,10pt,floatfix]{revtex4-2}
\usepackage[T1]{fontenc}
\usepackage{graphicx}
\usepackage{xcolor}
\usepackage{dcolumn}
\usepackage{comment}
\usepackage{amsmath,amssymb,bbold,bm}
\usepackage{hyperref}
\usepackage{amsfonts}
\usepackage{float,latexsym}
\usepackage{multirow}
\usepackage{color}
\usepackage{cleveref}
\usepackage{physics}
\usepackage{tabularx}
\usepackage{array}
\usepackage[caption=false]{subfig}
\usepackage{siunitx}
\usepackage{gensymb}
\usepackage{pifont}
\usepackage{placeins}
\usepackage{mathtools}
\usepackage{longtable}
\usepackage{multirow}
\usepackage{nicematrix}
\usepackage{tikz}
\usepackage{xstring}
\usepackage{makecell}
\usepackage{xr}
\usepackage{ifthen}
\usetikzlibrary{decorations.markings}
\usetikzlibrary{positioning}
\usetikzlibrary{arrows.meta}
\usepackage{tikz-feynman}
\usetikzlibrary{calc}
\tikzfeynmanset{warn luatex=false}
\let\vec\mathbf

\usepackage{chemformula}
\usepackage{pdfpages}
\usepackage{cancel}

\makeatletter
\def\maketag@@@#1{\hbox{\m@th\normalfont\normalsize#1}}
\makeatother

\crefname{appendix}{Appendix}{Appendices}
\crefname{equation}{Eq.}{Eqs.}
\crefname{figure}{Fig.}{Figs.}
\crefname{table}{Table}{Tables}
\crefname{section}{Section}{Sections}
\crefname{enumi}{Point}{Points}
\creflabelformat{appendix}{[#2#1#3]}
\crefmultiformat{figure}{Figs.~#2#1\xdef\mycreffirstarg{#1}#3}
{ and~#2#1#3}{, #2#1#3}{ and~#2#1#3}

\makeatletter     
\renewcommand\onecolumngrid{
	\do@columngrid{one}{\@ne}%
	\def\set@footnotewidth{\onecolumngrid}
	\def\footnoterule{\kern-6pt\hrule width 1.5in\kern6pt}%
}

\makeatletter
\AtBeginDocument{\let\LS@rot\@undefined}
\makeatother

\RequirePackage[normalem]{ulem} 
\RequirePackage{color}\definecolor{RED}{rgb}{1,0,0}\definecolor{BLUE}{rgb}{0,0,1} 
\RequirePackage{listings} 
\RequirePackage{color} 
\lstdefinelanguage{DIFcode}{ 
	moredelim=[il][\color{red}\sout]{\%DIF\ <\ }, 
	moredelim=[il][\color{blue}\uwave]{\%DIF\ >\ } 
} 
\lstdefinestyle{DIFverbatimstyle}{ 
	language=DIFcode, 
	basicstyle=\ttfamily, 
	columns=fullflexible, 
	keepspaces=true 
} 
\lstnewenvironment{DIFverbatim}{\lstset{style=DIFverbatimstyle}}{} 
\lstnewenvironment{DIFverbatim*}{\lstset{style=DIFverbatimstyle,showspaces=true}}{} 

\newcommand{\cre}[2]{\hat{#1}^\dagger_{#2}}
\newcommand{\des}[2]{\hat{#1}_{#2}}

\newcommand{\ie}{{\it i.e.}}
\newcommand{\eg}{{\it e.g.}}

\usepackage{enumitem}

\newcommand{\ba}{\begin{equation}\begin{aligned}}
		\newcommand{\ea}{\end{aligned}\end{equation}}

\def\qq{\mathbf{q}}
\def\kk{\mathbf{k}}
\def\KK{\mathbf{K}}

\def\pp{\mathbf{p}}

\def\RR{\mathbf{R}}
\def\rr{\mathbf{r}}
\def\GG{\mathbf{G}}
\def\QQ{\mathbf{Q}}
\def\uu{\mathbf{u}}
\def\dd{\mathbf{d}}
\def\tt{\mathbf{t}}

\newcommand{\titlePaper}{
	Quantum geometry and critical temperature enhancement in \ch{MgB2} superconductivity
}

\newcommand{\paperAuthors}{
	\author{Yi Jiang}
	\thanks{These authors contributed equally to this work.}
	\affiliation{Donostia International Physics Center (DIPC), Paseo Manuel de Lardizábal. 20018, San Sebastián, Spain}
	\author{Haoyu Hu}
	\thanks{These authors contributed equally to this work.}
	\affiliation{Donostia International Physics Center (DIPC), Paseo Manuel de Lardizábal. 20018, San Sebastián, Spain}
	\affiliation{Department of Physics, Princeton University, Princeton, NJ 08544, USA}
	\affiliation{Department of Physics, University of Science and Technology of China, Hefei, Anhui 230026, China}
	
	\author{Dumitru C\u{a}lug\u{a}ru}
	\affiliation{Department of Physics, Princeton University, Princeton, NJ 08544, USA}
	\affiliation{Rudolf Peierls Centre for Theoretical Physics, University of Oxford, Oxford OX1 3PU, United Kingdom}
	
	\author{Kaja H.~Hiorth}
	\affiliation{Department of Applied Physics, Aalto University School of Science, FI-00076 Aalto, Finland}
	\author{Martin Gutierrez-Amigo}
	\affiliation{Department of Applied Physics, Aalto University School of Science, FI-00076 Aalto, Finland}
	
	\author{Junze Deng}
	\affiliation{Department of Applied Physics, Aalto University School of Science, FI-00076 Aalto, Finland}
	
	\author{Hanqi Pi}
	\affiliation{Donostia International Physics Center (DIPC), Paseo Manuel de Lardizábal. 20018, San Sebastián, Spain}
	
	\author{Handong Chen}
	\affiliation{Department of Physics, Princeton University, Princeton, New Jersey 08544, USA}
	
	\author{Maia G.~Vergniory}
	\affiliation{Département de Physique et Institut Quantique, Université de Sherbrooke, Sherbrooke, J1K 2R1 Québec, Canada}
	\affiliation{Donostia International Physics Center (DIPC), Paseo Manuel de Lardizábal. 20018, San Sebastián, Spain}
	
	\author{Ion Errea}
	\affiliation{University of the Basque Country (UPV/EHU), Europa Plaza 1, 20018 Donostia/San Sebastián, Spain}
	\affiliation{Centro de
		Física de Materiales (CSIC-UPV/EHU), Manuel de Lardizabal Pasealekua 5, 20018 Donostia/San Sebastián, Spain}
	\affiliation{Donostia International Physics Center (DIPC), Paseo Manuel de Lardizábal. 20018, San Sebastián, Spain}
	
	\author{Emilia Morosan}
	\affiliation{Department of Physics and Astronomy, Rice University, Houston, TX 77005, USA}
	\affiliation{Rice Center for Quantum Materials (RCQM), Rice University, Houston, TX 77005, USA}
	\affiliation{Smalley-Curl Institute, Rice University, Houston, TX 77005, USA}
	
	\author{Leslie M.~Schoop}
	\affiliation{Department of Chemistry, Princeton University, Princeton, New Jersey 08544, USA}
	
	\author{Claudia Felser}
	\affiliation{Max Planck Institute for Chemical Physics of Solids, N\"{o}thnitzer Str. 40, Dresden 01187, Germany}
	
	\author{Miguel A.L. Marques}
	\affiliation{
		Research Center Future Energy Materials and Systems of the University Alliance Ruhr and Interdisciplinary Centre for Advanced Materials Simulation$,$ Ruhr University Bochum$,$ Universitätsstraße 150$,$ D-44801 Bochum$,$ Germany}
	
	\author{P\"{a}ivi T\"{o}rm\"{a}}
	\affiliation{Department of Applied Physics, Aalto University School of Science, FI-00076 Aalto, Finland} 
	
	\author{Daniel Agterberg}
	\affiliation{Department of Physics, University of Wisconsin–Milwaukee, Milwaukee, Wisconsin 53201, USA}
	
	\author{B.~Andrei Bernevig}
	\email{bernevig@princeton.edu}
	\affiliation{Department of Physics, Princeton University, Princeton, NJ 08544, USA}
	\affiliation{Donostia International Physics Center (DIPC), Paseo Manuel de Lardizábal. 20018, San Sebastián, Spain}
	\affiliation{IKERBASQUE, Basque Foundation for Science, Bilbao, Spain}
}

\crefname{appendix}{Appendix}{Appendices}
\crefname{equation}{Eq.}{Eqs.}
\crefname{figure}{Fig.}{Figs.}
\crefname{table}{Table}{Tables}
\crefname{section}{Section}{Sections}
\creflabelformat{appendix}{[#2#1#3]}

\makeatletter
\AtBeginDocument{\let\LS@rot\@undefined}
\makeatother

\makeatletter
\renewcommand\onecolumngrid{
\do@columngrid{one}{\@ne}%
\def\set@footnotewidth{\onecolumngrid}
\def\footnoterule{\kern-6pt\hrule width 1.5in\kern6pt}%
}

\allowdisplaybreaks

\begin{document}
\title{\titlePaper}
\paperAuthors
\let\oldaddcontentsline\addcontentsline

\begin{abstract}
\ch{MgB2}, a phonon-mediated superconductor with record-high critical temperature $T_c\simeq 39$~K, is revisited to obtain a comprehensive theory of electrons, phonons, and their coupling with minimal \textit{ab initio} input. We construct compact analytic models for the electronic structure, phonons, and electron-phonon coupling (EPC) of \ch{MgB2}. We show that strong in-plane B $sp^2$ bonding realizes an \emph{obstructed} band structure whose natural description is a bond-centered kagome lattice, yielding small quasi-2D $\sigma$-band Fermi-surface cylinders and pronounced quantum-geometric effects. The phonon spectrum is found to closely track that of a graphene-like boron layer, but the heavy intercalated Mg atoms dominate the three acoustic branches and rigidly lift the boron modes into the optical sector, while the in-plane B--B bond-stretching mode exhibits a pronounced softening along $\Gamma$--A. By symmetry, this $\Gamma$-point bond-stretching mode is the only $\Gamma$ phonon that can couple to the $\sigma$ Fermi surface, explaining its dominant contribution to the EPC. Upon electron doping toward the doubly degenerate band edge of the $\sigma$ sheets, we find that a reduced density of states competes with enhanced EPC matrix elements. At light electron doping, \textit{ab initio} calculations show that the EPC enhancement dominates, leading to an increase in $T_c$ (within the clean doping limit without disorder effects). Using the Gaussian approximation for the EPC tensor, we further show that this enhancement is overwhelmingly quantum geometric in origin, arising from a geometric EPC contribution of the small $\sigma$ Fermi surface peaked at $\Gamma$. Overall, our results provide a transparent, symmetry-based account of superconductivity in \ch{MgB2} and suggest that quantum-geometric effects can be essential for shaping doping trends in phonon-mediated superconductors. 
\end{abstract}

\maketitle

\paragraph{Introduction.}

Since the discovery of superconductivity at $T_c\simeq 39\,\mathrm{K}$ in \ch{MgB2} in 2001~\cite{Nagamatsu2001_MgB2, larbalestier2001strongly, bud2001boron, yildirim2001giant, osborn2001phonon, lorenz2001high, buzea2001review}, this simple binary compound has served as a benchmark for phonon-mediated superconductivity at unusually high temperature among ambient-pressure materials. Early first-principles studies rapidly established that the pairing is driven by strong coupling between the in-plane boron $sp^2$ carriers and the B--B bond-stretching phonons~\cite{AnPickett2001_MgB2, Kortus2001_MgB2, liu2001beyond, kong2001electron}. Subsequent anisotropic calculations and experiments clarified its multiband nature, with distinct superconducting gaps on the quasi-2D $\sigma$ sheets and the 3D $\pi$ sheets~\cite{szabo2001evidence, bouquet2001specific, giubileo2001two, angst2002temperature, souma2003origin, Choi2002_MgB2_twoGap, choi2002first, mazin2002superconductivity, gurevich2003enhancement, braccini2005high, choi2009anisotropic}. 
Tuning \ch{MgB2} by chemical substitution has been extensively explored, but in most cases $T_c$ is reduced~\cite{dou2002enhancement}: for example, electron doping via Al substitution on the Mg site suppresses superconductivity~\cite{Slusky2001_AlDoping}, and carbon substitution on the B site likewise degrades $T_c$~\cite{lee2003carbon, kazakov2005carbon}. An important exception is provided by epitaxial thin films under in-plane tensile strain~\cite{pogrebnyakov2004enhancement, xi2007mgb2}, where an enhanced $T_c$ up to $\sim 41.8\,\mathrm{K}$ was reported together with a further strain-induced softening of the bond-stretching mode~\cite{zheng2006searching, bekaert2017evolution, zhang2011effect, johansson2022effect}.

\begin{figure*}[htbp]
\centering
\includegraphics[width=\linewidth]{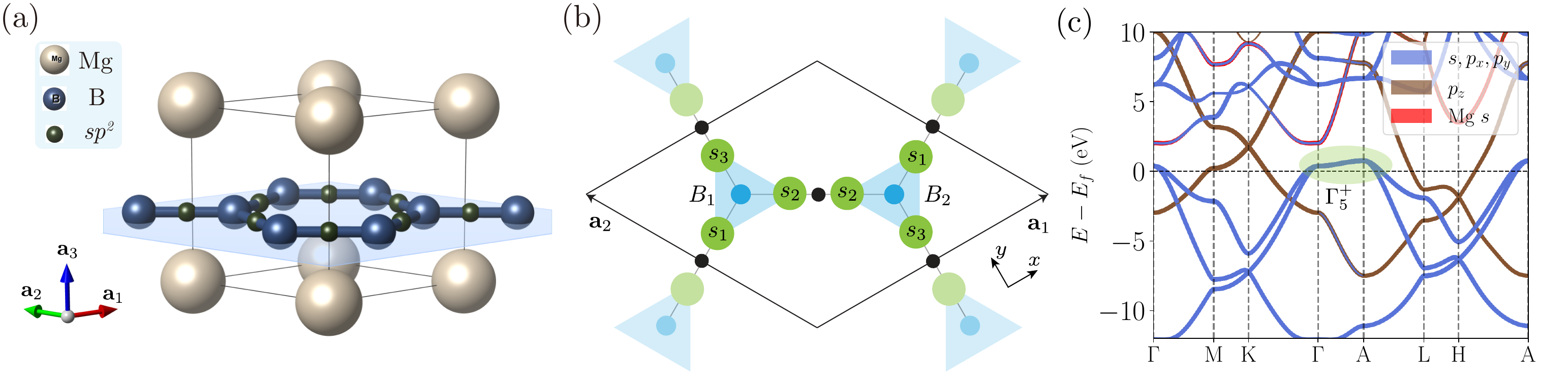}
\caption{Crystal structure and electronic properties of \ch{MgB2}. 
(a) Side view of the \ch{MgB2} structure: boron atoms (blue) form honeycomb layers, and Mg atoms (gray) occupy interlayer sites. Black spheres mark the obstructed kagome sites at B--B bond centers, which host the $sp^2$ bonding states. 
(b) Top view of a boron honeycomb layer. On each B site (blue), the $(s,p_x,p_y)$ orbitals hybridize into three bond-directed, $s$-like orbitals (green) located along the $sp^2$ bonds. The two bond-directed orbitals on a given B--B bond further combine into bonding and antibonding states centered at the bond midpoint (black), forming a kagome lattice. 
(c) Electronic band structure of \ch{MgB2}, with color and thickness indicating orbital weight from boron $(s,p_x,p_y)$ (blue), $p_z$ (brown), and Mg $s$ orbitals. The $p_z$ Dirac bands have a relatively strong $k_z$-dispersion and form the $\pi$ Fermi surface. The three lowest blue bands, originating from the bond-centered $sp^2$ bonding manifold on a kagome lattice, are quasi-2D and form the $\sigma$ Fermi surface along $\Gamma$--A with $\Gamma_5^+$ representation. }
\label{fig:structure-band-FS}
\end{figure*}

Beyond the conventional emphasis on the density of states (DOS), recent work has highlighted that the \emph{quantum geometry}~\cite{resta2011insulating, provost1980riemannian} of Bloch wavefunctions encoded, \eg, in the quantum metric, and related interband coherence effects, can directly control superconducting observables such as the superfluid weight, particularly in flat-band and multiband settings~\cite{PeottaTorma2015_superfluidWeight,Julku2016_quantumMetric, peotta2308quantum, torma2022superconductivity, torma2023essay, yu2025quantum, liu2025quantum, verma2025quantum}. In parallel, quantum geometric structures can also enter pairing interactions: in multiorbital systems, symmetry-enforced degeneracies and rapidly varying wavefunctions can enhance effective couplings through geometric terms, including in electron-phonon problems where the electron-phonon coupling (EPC) can be decomposed into energetic and geometric contributions within controlled approximations~\cite{yu2024non}. These insights provide a natural conceptual framework for \ch{MgB2}, where the $sp^2$ states are bond-centered and exhibit strong wavefunction variation near the doubly degenerate point at the Brillouin zone center.

In this work, we revisit \ch{MgB2} by combining \textit{ab initio} calculations with the development of analytical models that capture the symmetry and quantum-geometry properties essential to its superconductivity. We show that the electronic structure, phonons, and EPC in \ch{MgB2} can be captured with minimal \emph{ab initio} input. Strong covalent $sp^2$ bonding produces an \emph{obstructed}~\cite{xu2021three, xu2024filling} effective kagome lattice centered on B–B bonds, yielding a small $\sigma$ Fermi surface. The appearance of this $\sigma$ Fermi surface can be understood as modeled in second-order perturbation theory regarding the Mg atom. The resulting obstructed orbitals naturally endow the $\sigma$ bands with large quantum-geometric weight and strong EPC. The phonon spectrum closely mirrors that of graphene, but with an overall lifting of boron phonons due to the intercalated, heavy Mg atoms. 
A pronounced softening of the in-plane bond-stretching mode along $\Gamma$--A is also observed. By symmetry, this $\Gamma$-point bond-stretching ($\Gamma_5^+$) mode is the only $\Gamma$ mode that couples to the $\sigma$ Fermi surface in \ch{MgB2}. Finally, when electrons are doped toward the twofold degenerate band edge of the $\sigma$ Fermi surface, our calculation reveals a competition between a decreasing DOS and an increasing EPC matrix element. In the small-doping regime, the EPC enhancement dominates and raises $T_c$. Following Ref.~\cite{yu2024non}, we adopt the Gaussian approximation to trace this enhancement primarily to the quantum-geometric contribution to the EPC. 
While \ch{MgB2} has been extensively studied, our work introduces two key advances. First, we develop a symmetry-based theoretical framework for quantum geometric EPC, which provides a powerful strategy for identifying new superconducting materials. Second, our theory enables predictions of $T_c$ as a function of doping. In particular, we theoretically predict an enhancement in $T_c$ under light electron doping. These findings establish new methodological routes for the discovery of novel superconductors.

\paragraph{Electronic properties.}
\ch{MgB2} crystallizes in the \ch{AlB2} structure (space group 191, $P6/mmm$), in which boron atoms form graphene-like honeycomb layers and Mg ions occupy interlayer triangular sites, as shown in \cref{fig:structure-band-FS} (a, b). The in-plane B–B network is strongly covalent from the $sp^2$ bonding, while the intercalated Mg atoms primarily act as electron donors. The total number of valence electrons per unit cell in \ch{MgB2} is the same as in graphene, but their physical properties are puzzlingly distinct, as we detail in the following. 

The electronic bands of \ch{MgB2} are shown in \cref{fig:structure-band-FS} (c). Naively, one might expect the band structure to closely mimic that of graphene: out-of-plane B $p_z$ orbitals form $\pi$ bands with the Dirac crossings at K and H, while in-plane $sp^2$ states form deep $\sigma$ bands~\cite{yao2007spin, abergel2010properties} (see \cref{app:sec:DFT_electron}). However, the presence of the Mg layers qualitatively reshapes the filling. Although the Mg orbitals couple relatively weakly to the in-plane $sp^2$ network, they hybridize more strongly with the B $p_z$ orbitals, effectively pushing down the $\pi$ bands and increasing their filling (see analytical description in \cref{app:sec:emergence_sigma_FS}). Charge neutrality then forces a reduced filling of the in-plane B $sp^2$ bands, driving them across the Fermi level and contributing to the $\sigma$ Fermi surface, with irreducible representation (IRREP) $\Gamma_5^+$ at $\Gamma$ [the notation follows \textit{Bilbao Crystallographic Server}~\cite{aroyo2011crystallography, aroyo2006bilbao1, aroyo2006bilbao2}]. The resultant Fermi surface consists of quasi-two-dimensional, hole-like $\sigma$ cylinders around the $\Gamma$–A line originating from B $sp^2$ bands, and compensated by electron-like $\pi$ sheets associated with the Dirac bands near K.

To analyze the $sp^2$ sector in more detail, we construct a minimal tight-binding (TB) model that makes explicit the \textit{obstructed} nature of the $\sigma$ states (more details are given in \cref{app:sec:DFT_electron} and \cref{app:sec:electron_analytic}). Starting from the $(s,p_x,p_y)$ orbitals from the two boron atoms at honeycomb sites $2d$ in the unit cell, we first build a 6-orbital TB Hamiltonian. We then perform a symmetry-guided change of basis to three $s$-like orbitals located along the B–B $sp^2$ bonds (non-maximal Wyckoff position $6m$, following the \textit{Bilbao Crystallographic Server} convention), as shown in \cref{fig:structure-band-FS} (b). On each B–B bond, the two $s$-like orbitals from two B atoms can be further recombined into bonding and antibonding states, \ie, 
$\psi^{\pm}_{sp^2} = \frac{1}{\sqrt{2}}(\phi_{1} \pm \phi_{2})$, where $\psi^+_{sp^2}$ and $\psi^-_{sp^2}$ denotes the bonding and antibonding combination of the two $s$ orbitals $\phi_{1,2}$, respectively. The three bonding combinations per unit cell form an effective kagome lattice at Wyckoff position $3g$.  
In this “bonding-only” description, the $\sigma$ bands near $E_f$ are well captured by a kagome tight-binding model defined on the bond centers. This is the hallmark of an \emph{obstructed atomic limit}~\cite{xu2021three,xu2024filling, gao2022unconventional, yang2024superconductivity, wang2022two, cualuguaru2025probing, holbrook2024real}: the symmetry-allowed Wannier centers of the occupied bands are fixed at the bond centers, away from the physical boron atoms at the honeycomb sites, which typically leads to non-trivial quantum geometry. 
In \cref{app:sec:hopping_integral}, we show that the hopping amplitudes of our analytical tight-binding model based on DFT are accurately reproduced by hydrogen-like orbital-overlap integrals, without explicit \textit{ab initio} input. This indicates that the small $\sigma$ Fermi surface and its low-energy electronic Hamiltonian can already be captured, to good accuracy, within this simple overlap-based construction.

This construction of effective orbitals leads to an ``onsite–hopping duality'' characteristic of obstructed band structures, where the roles of “onsite” and “nearest-neighbor” hopping terms in the atomic $(s,p_x,p_y)$ basis are effectively exchanged in the kagome bonding basis. Detailed derivations are left in \cref{app:sec:simple_Hel_model}. The latter description makes the \textit{obstructed}, kagome-like character of the $\sigma$ bands manifest and, as we show below, it is precisely this bond-centered, obstructed structure that underlies the strong quantum-geometry effects and the unusually large EPC in \ch{MgB2}.

\begin{figure}[htbp]
\centering
\includegraphics[width=\linewidth]{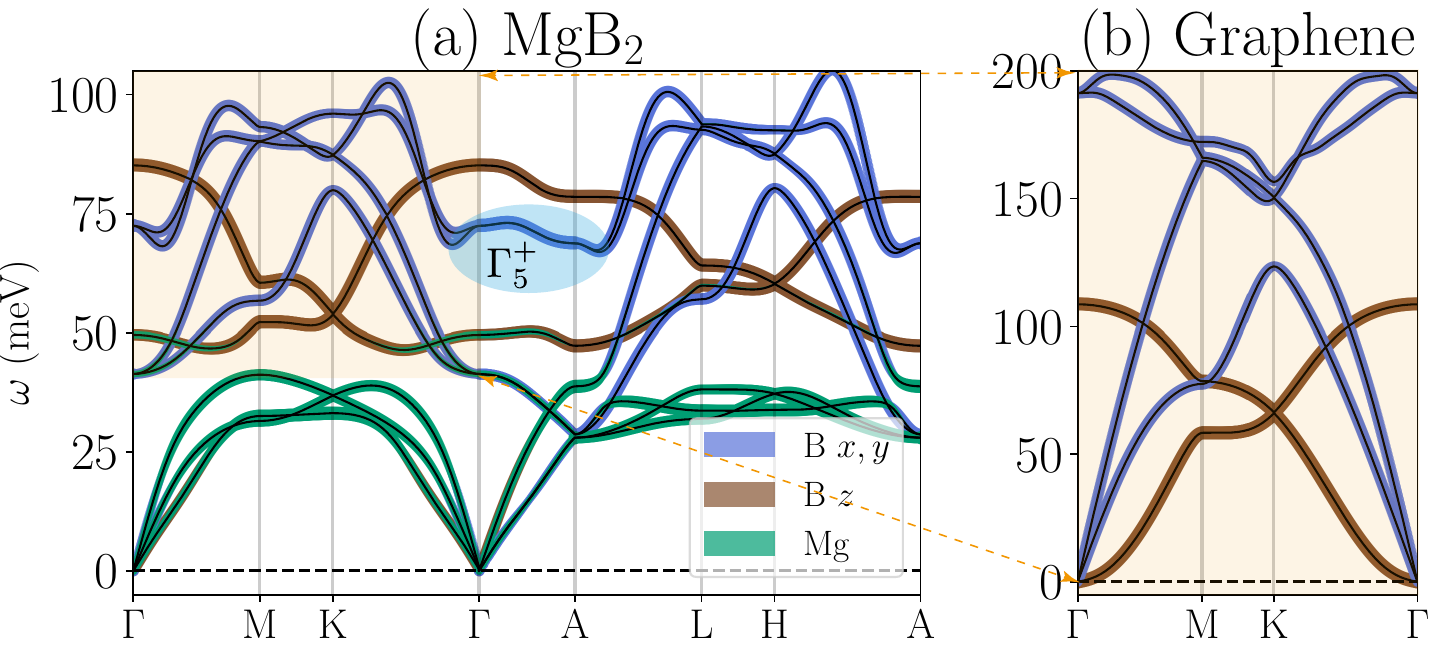}
\caption{Comparison of the phonon spectrum of (a) \ch{MgB2} and (b) graphene. In both plots, the blue (brown) bands denote the weight from the $x,y$ ($z$) phonon from boron in (a) and carbon in (b). In (a), the B–B bond-stretching modes with $\Gamma_5^+$ IRREP are marked with a blue circle, which give a strong softening and the dominant contribution to the electron-phonon coupling (EPC) in \ch{MgB2}. The yellow-shaded regions serve as a visual guide, indicating that the graphene phonon is close to the \ch{MgB2} optical phonon, primarily originating from the boron atoms on the $k_z=0$ plane, with an overall frequency shift induced by the heavy magnesium atoms. 
}
\label{fig:main:phonon}
\end{figure}

\paragraph{Phonon properties.}

The phonon dispersion of \ch{MgB2}, shown in \cref{fig:main:phonon}(a), separates broadly into three low-frequency branches with predominantly Mg character and six higher-frequency branches dominated by B vibrations. Its overall structure closely resembles the graphene phonon spectrum shown in \cref{fig:main:phonon}(b), but with two notable differences. First, the graphene-like boron branches are shifted to higher energies relative to the Mg-dominated acoustic modes and appear as optical branches in \ch{MgB2}. Second, the in-plane B--B bond-stretching branch undergoes pronounced softening along $\Gamma$--A. At $\Gamma$, this mode transforms as the 2D IRREP $\Gamma_5^+$ and provides the main EPC channel to the $\sigma$ bands. The maximum phonon frequency of graphene is approximately twice that of \ch{MgB2}, reflecting the shorter C--C bond length and the correspondingly larger in-plane bond-stretching force constants in graphene.

A simple consequence of the large mass separation between the light B layers and the heavy intercalated Mg layers in \ch{MgB2} is that the two parts of the phonon spectrum play very different roles. The heavy Mg atoms mainly form the acoustic branches, while the B-derived modes retain the structure of an isolated graphene-like boron layer but are pushed upward almost rigidly into the optical sector. Thus, in \ch{MgB2}, the phonon spectrum can be understood as a graphene-like B spectrum lifted by the coupling to the heavy Mg subsystem, together with low-frequency acoustic modes dominated by Mg motion.

\begin{figure*}[htbp]
\centering
\includegraphics[width=\linewidth]{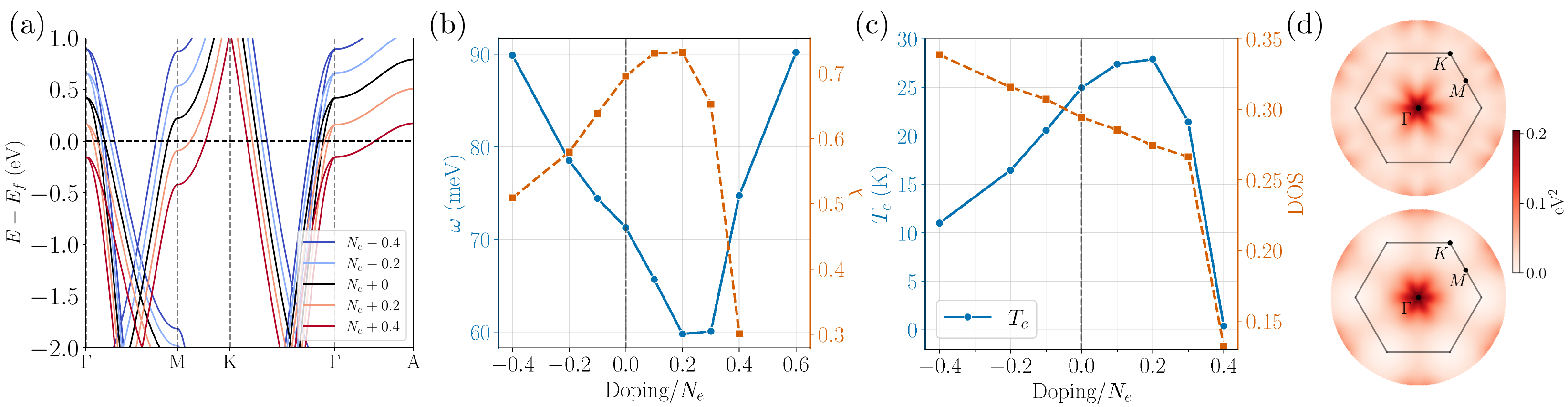}
\caption{\textit{Ab initio} doping dependence of the superconducting properties of \ch{MgB2}. 
(a) Electronic band structures at representative doping levels, where $N_e+n$ denotes $n$ electrons are added. 
(b) The blue curve represents the average frequency of the in-plane B--B bond-stretching mode along $\Gamma$--A as a function of doping, showing a pronounced minimum at light electron doping regime. The orange curve shows the doping-dependence of total EPC strength $\lambda$. 
(c) DOS at the Fermi level $N(E_f)$ and \textit{ab initio} superconducting transition temperature $T_c$ versus doping: while $N(E_f)$ decreases monotonically from hole to electron doping, $T_c$ is non-monotonic, increasing at light electron doping and decreasing at larger electron doping.  
(d) Band-resolved EPC $G_{\kk}^m$ in the band basis on the $k_z=0$ plane (defined in \cref{eq:EPC_band_basis_GM5+}) for the two $sp^2$ bands contributing to the $\sigma$ Fermi surface, with the top (bottom) plot from the upper (lower) $sp^2$ band, exhibiting a strong peak near $\Gamma$. }
\label{fig:main:doping}
\end{figure*}

Among the boron-derived phonons, the in-plane B--B bond-stretching mode at $\Gamma$ transforms as the two-dimensional irrep $\Gamma_5^+$ (the analogue of the $E_{2g}$ mode in graphene) and exhibits a pronounced softening in \ch{MgB2}. By symmetry, this $\Gamma_5^+$ mode is the only $\Gamma$-point phonon that can couple to the $\sigma$ Fermi surface~\cite{liu2001beyond}. Generally, let a phonon with momentum $\mathbf{Q}$ transform under its little group $\mathcal G_{\mathbf Q}$ according to the representation $D^{p}(\mathbf Q)$. For an electron on the Fermi surface at momentum $\mathbf K$ such that $\KK+\QQ$ also lies on the Fermi surface, we first collect the (electron) momenta generated by $\mathcal G_{\mathbf Q}$, \ie, $S_{\mathbf Q,\mathbf K}=\{\,g\mathbf K \mid g\in \mathcal G_{\mathbf Q}\,\}$. 
Momentum conservation then restricts the particle-hole operators that can couple to the phonon to
$O_{\mathbf K,\mathbf Q}=\{\,\gamma^\dagger_{\mathbf k+\mathbf Q}\gamma_{\mathbf k}\mid \mathbf k\in S_{\mathbf Q,\mathbf K}\,\}$,
where we suppress band indices for simplicity. Under $\mathcal G_{\mathbf Q}$, this operator set transforms as the tensor product $D(\mathbf K+\mathbf Q)\otimes D^*(\mathbf K)$. The EPC is symmetry-allowed only if the product representation contains the trivial irrep, \ie,
\begin{equation}
D(\mathbf K+\mathbf Q)\otimes D^*(\mathbf K)\otimes D^{p, *}(\mathbf Q) \supset \mathbf 1,
\end{equation}
or equivalently, $D(\mathbf K+\mathbf Q)\otimes D^*(\mathbf K) \supset D^{p}(\mathbf Q)$. 
For \ch{MgB2}, we focus on the small $\sigma$ Fermi surface around $\Gamma$, whose low-energy states transform as $\Gamma_5^+$. At $\mathbf Q=\Gamma$, the relevant decomposition is
\begin{equation}
\Gamma_5^+\otimes \Gamma_5^+ = \Gamma_1^+ \oplus \Gamma_2^+ \oplus \Gamma_5^+ .
\label{eq:epc_selection_rule}
\end{equation}
Therefore, only $\Gamma$-point phonons transforming as $\Gamma_1^+$, $\Gamma_2^+$, or $\Gamma_5^+$ can couple to the $\sigma$ states. In the phonon spectrum of \ch{MgB2}, however, $\Gamma_{1,2}^+$ is absent, and the only mode present is precisely the bond-stretching $\Gamma_5^+$ mode. A similar argument can be applied to the $\Gamma$--A line. This selection rule thus explains why the softened $\Gamma_5^+$ phonon provides the unique dominant $\Gamma$-point EPC channel for the small $\sigma$ Fermi surface in \ch{MgB2}. The large softening in this mode is due to the strong EPC of this bond-stretching mode.

\paragraph{Electron-phonon coupling and doping effect on $T_c$.}
The mode- and momentum-resolved EPC strength $\lambda_{\mathbf q\nu}$ in \ch{MgB2} is highly concentrated in the vicinity of the in-plane B--B bond-stretching branch along $\Gamma$--A, contributing over 70\% of the total $\lambda$, whereas all other phonons contribute comparatively weakly (see \cref{app:sec:DFT_EPC_Tc} for details). This selectivity is a direct consequence of the \textit{obstructed}, bond-centered nature of the $sp^2$ electrons: the $\sigma$ Fermi surface originates from B $sp^2$ bonding states that form an effective kagome lattice centered on B--B bonds. Consequently, phonon distortions that predominantly stretch these bonds couple most efficiently to the $\sigma$ carriers, while modes corresponding to essentially unidirectional displacements of the B atoms couple much more weakly~\cite{AnPickett2001_MgB2, Kortus2001_MgB2}. The symmetry analysis in \cref{eq:epc_selection_rule} further shows that at $\Gamma$ the bond-stretching $\Gamma_5^+$ mode is the only $\Gamma$-point phonon symmetry-allowed to couple to the $\sigma$ Fermi surface, fully consistent with the \textit{ab initio} $\lambda_{\mathbf{q}\nu}$ distribution.

We next examine how this dominant EPC channel evolves under carrier doping in \textit{ab initio} (details in \cref{app:sec:DFT_doping_results}). The relaxed lattice constants expand monotonically under electron doping. The DFT band structures at representative doping levels are shown in \cref{fig:main:doping} (a). Electron doping monotonically shrinks the $\sigma$ Fermi surface and pushes the Fermi level toward the two-dimensional $\Gamma_5^+$ band edge, decreasing the DOS at the Fermi level [\cref{fig:main:doping} (c)]. 
Interestingly, we find a competition between a decreasing DOS and an increasing EPC strength. In the light electron-doping regime, the EPC enhancement dominates, leading to an initial increase of $T_c$ despite the reduced DOS. At larger electron doping, the DOS depletion eventually prevails and $T_c$ drops to zero, as shown in \cref{fig:main:doping} (c). The phonon response tracks the behavior of EPC strength: the averaged frequency of the B--B bond-stretching mode along $\Gamma$--A exhibits a pronounced minimum at light electron doping [\cref{fig:main:doping} (b)], indicating the strongest softening precisely where the EPC is maximal. 
We note that this prediction contradicts several experiments in which electrons are introduced via substitutional dopants and $T_c$ decreases~\cite{dou2002enhancement, Slusky2001_AlDoping, lee2003carbon, kazakov2005carbon}. However, such doping may induce additional effects, including disorder, which have not been explicitly accounted for in our DFT calculations. 
Remarkably, this \textit{ab initio} doping result is qualitatively consistent with the experiments in Ref.~\cite{pogrebnyakov2004enhancement, xi2007mgb2}, where \ch{MgB2} thin films with in-plane tensile strain are reported to have a higher $T_c$ of 41.8 K. The strain expands the in-plane lattice constants, shrinks the $\sigma$ Fermi surface, and increases the filling of $sp^2$ bands, leading to a stronger softening of the bond-stretching modes, and enhanced $T_c$ (see \cref{app:sec:strained_Tc} for more details). This trend is in qualitative agreement with our prediction for light electron doping. 
We remark that our \textit{ab initio} SC calculation is restricted to the $sp^2$ sector, neglecting the $\pi$ Fermi surface. This gives an underestimated $T_c$ at zero doping, but allows us to isolate the competition between DOS and EPC. A full anisotropic Eliashberg treatment is left for future study.

The rise of $T_c$ at light electron doping implies that the increase in EPC strength outweighs the decrease in the DOS. To explicitly show that the EPC is enhanced as we dope towards the doubly-degenerate electron $\Gamma_5^+$ node, we evaluate the band-basis EPC on the $k_z=0$ plane at phonon momentum $\mathbf{q}=\mathbf{0}$ and sum over the two bond-stretching components of the $\Gamma_5^+$ mode with
$u_1=\frac{1}{\sqrt{2}}(u_{B_{1x}}-u_{B_{2x}})$ and
$u_2=\frac{1}{\sqrt{2}}(u_{B_{1y}}-u_{B_{2y}})$, \ie, 
\begin{equation}
G^m_{\mathbf k} = \big|G^{m,m,u_1}_{\mathbf k,\mathbf 0}\big|^2 + \big|G^{m,m,u_2}_{\mathbf k,\mathbf 0}\big|^2,
\label{eq:EPC_band_basis_GM5+}
\end{equation}
with $m=1,2$ labeling the two $sp^2$ bands forming the $\sigma$ Fermi surface, $u_{B_{ix/y}}$ is the $x/y$ directional movement of $B_i$ atom, and $G_{\kk,\qq}^{mu\nu}$ is the EPC tensor in the band basis. 
As shown in \cref{fig:main:doping}(d), $G^m_{\mathbf k}$ is sharply peaked at $\Gamma$ and decays away from it. It also remains comparatively large along the $\Gamma$-K direction, reflecting the structure of the $sp^2$ bonding wavefunctions (as captured by the analytic model in \cref{app:sec:EPC_sp2_basis_analytic_model}). 

Performing an angular average of $G^m_{\mathbf k}$ over circles of radius $k$ around $\Gamma$, we find that the resulting Fermi-surface-averaged EPC decays approximately as $k^2$ away from $\Gamma$. In \cref{app:sec:EPC_hamiltonian}, we reproduce this $k^2$ scaling with a minimal analytic EPC Hamiltonian formulated in the $sp^2$ bonding basis, including onsite and bond EPC terms. In \cref{app:sec:analytic_estimation_lambda}, we then combine the Fermi surface–averaged EPC strength with an analytic $\kk\cdot\pp$ DOS model for the $\Gamma_5^+$ node, showing that the EPC enhancement as electron doping approaches the degeneracy point of the two bands can outweigh the (approximately linear) reduction of the DOS, and thus produce the observed increase of $\lambda$ and $T_c$ at light electron doping. This demonstrates that the doping trend is governed primarily by wavefunction effects, even as the DOS decreases.

\begin{figure}[htbp]
\centering
\includegraphics[width=\linewidth]{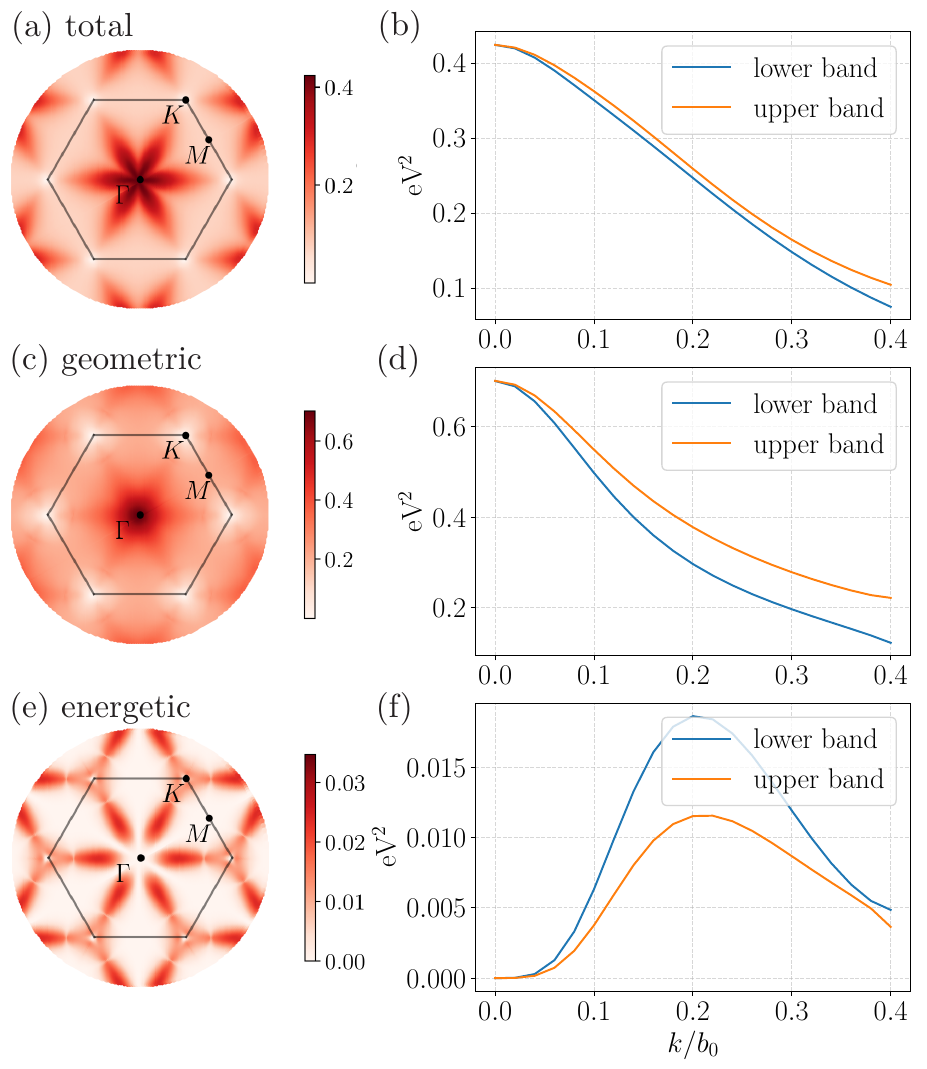}
\caption{EPC in the band basis obtained from the analytical Gaussian approximation (GA), where the total EPC is separated into the geometric and energetic contributions. In each row, the first panel shows the band-resolved EPC $G^m_{\mathbf{k}}$ (defined in \cref{eq:EPC_band_basis_GM5+}) on the $k_z=0$ plane for phonon momentum $\mathbf{q}=0$ and the upper $sp^2$ bands. The second panel presents the radial average of $G^m_{\mathbf{k}}$ versus the distance $k$ from $\Gamma$ ($b_0$ being the length of the in-plane reciprocal lattice vector). The total GA EPC reproduces the \textit{ab initio} trend in \cref{fig:main:doping}: $G_{\mathbf{k}}$ peaks at $\Gamma$, remains elevated along the $\Gamma$–K line, and decays approximately as $k^{2}$ away from $\Gamma$. The geometric part shows an even sharper maximum at $\Gamma$ compared with the total EPC, whereas the energetic part is small in magnitude and vanishes at $\Gamma$. For simplicity, the geometry–energy cross term is omitted, which is negative near $\Gamma$ and lowers the total EPC compared with the geometric part.}
\label{fig:main:GA}
\end{figure}

\paragraph{Quantum geometric origin of the increase in $T_c$.} 
To disentangle the microscopic origin of the enhanced EPC near the $\Gamma_5^+$ node, we evaluate the EPC within the Gaussian approximation (GA) and then separate it into the geometric and energetic parts~\cite{yu2024non} (see \cref{app:sec:EPC_from_GA} for more details).  GA assumes two-center \textit{direct} hopping and models each hopping amplitude by a Gaussian in the bond displacement,
$t_{ij}(\mathbf R)=t^0_{ij}\exp\Big[-\frac{\gamma_{i,j}}{2}\big|\mathbf R+\mathbf r_j-\mathbf r_i\big|^2\Big]$, 
where $\rr_i$ is the position of $i$-th atom, and $\RR$ denotes a lattice translation. 
Under GA, the linear response of the hoppings to atomic displacements, which is nothing but the EPC, can be obtained analytically. In momentum space, this yields a compact  Gaussian form of EPC:
\begin{equation}
f^{i j}_{\mu}(\mathbf k)=
i \gamma_{ij}\,\partial_{k_\mu} t_{ij}(\mathbf k)
\qquad (\mu=x,y,z),
\label{eq:GA_f_def_main}
\end{equation}
and, assuming a common decay factor $\gamma_{ij}\equiv\gamma$, we have
$f_{\mu}(\mathbf k)=i\gamma\,\partial_{k_\mu} h(\mathbf k)$. 
Writing the electronic Hamiltonian in its spectral decomposition
$h(\mathbf k)=\sum_{n}\epsilon_{n\mathbf k}P^{n}(\mathbf k)$, the Gaussian form of EPC separates naturally into an \emph{energetic} part and a \emph{geometric} part, \ie, $f_{\mu}(\mathbf k) = f_{\mu}^{\mathrm{E}}(\mathbf k)+f_{\mu}^{\mathrm{geo}}(\mathbf k)$, where
\begin{equation}\begin{aligned}
f_{n\mu}^{\mathrm{E}}(\mathbf k) &= i\gamma\,(\partial_{k_\mu}\epsilon_{n\mathbf k})\,P^{n}(\mathbf k),\\
f_{n\mu}^{\mathrm{geo}}(\mathbf k) &= i\gamma\,\epsilon_{n\mathbf k}\,\partial_{k_\mu}P^{n}(\mathbf k),
\label{eq:GA_E_geo_main}
\end{aligned}
\end{equation}
The geometric term is controlled by momentum derivatives of the projectors and hence by the quantum geometry of the Bloch wavefunctions, which will diverge near the symmetry-protected degenerate points. 

The resulting EPC tensor in momentum space takes a simple difference form,
\begin{equation}
g^{ij,l\mu}_{\mathbf k,\mathbf q}
=
f^{ij}_{\mu}(\mathbf k+\mathbf q)\,\delta_{jl}
-
f^{ij}_{\mu}(\mathbf k)\,\delta_{il},
\label{eq:GA_gkq_main}
\end{equation}
This framework therefore yields a controlled decomposition of the band-basis EPC, and hence of the mode-resolved coupling $\lambda_{\mathbf q\nu}$ and total EPC strength $\lambda$ into three contributions:
\begin{equation}
    \lambda = \lambda^{\mathrm{geo}} + \lambda^{\mathrm{E}} + \lambda^{\mathrm{geo\text{-}E}},
    \label{eq:lambda_GA_decomposition}
\end{equation}
\ie, geometric, energetic, and their cross-term (see \cref{app:sec:EPC_from_GA} for more details). 
While the GA expressions are initially proposed for $s$-orbital with isotropic hopping~\cite{yu2024non}, in \cref{app:sec:EPC_from_GA_MgB2} we generalize the construction to the $(s,p_x,p_y)$ manifold in \ch{MgB2} using the Slater-Koster parameterization with angular dependence, with 
\begin{equation}
\begin{aligned}
    g_{\kk,\qq}^{i j,l\mu} &= \left(f_\mu^{ij}(\kk+\qq) \delta_{jl} - f_\mu^{ij}(\kk) \delta_{il} \right)- \\
    & \sum_{\nu} \frac{\epsilon_{\nu\mu}}{i\gamma_{ij} r_{ij}^2} \left\{ \left[\tilde{\sigma}_y, f_{\nu}(\kk+\qq)\right]_{ij} \delta_{jl} - \left[\tilde{\sigma}_y, f_{\nu}(\kk)\right]_{ij} \delta_{il} \right\},
    \label{eq:GA_EPC_spxpy}
\end{aligned}
\end{equation}
where $\tilde{\sigma}_y=\text{Diag}[0,\sigma_y]$, and $r_{ij}$ is the distance between $ij$ atoms. In \cref{eq:GA_EPC_spxpy}, the summation term comes from the angular dependence of the $p$ orbitals, which is absent in the $s$-orbital GA in \cref{eq:GA_gkq_main}. 

Using the GA parametrization, we further evaluate the doping dependence of the total EPC strength $\lambda$ and superconducting properties (see more details in \cref{app:sec:EPC_from_GA_MgB2}). Remarkably, the GA reproduces the same qualitative doping trend as the fully \textit{ab initio} calculations: upon light electron doping, $\lambda$ (and hence $T_c$) increases even though the DOS decreases, and then drops to zero at larger electron doping when the DOS depletes. Decomposing $\lambda$ into the geometric, energetic, and cross contributions as in \cref{eq:lambda_GA_decomposition}, 
we find that this non-monotonic behavior is driven by the geometric component: $\lambda^{\mathrm{geo}}$ follows the rise-and-fall trend and dominates the total coupling, whereas $\lambda^{\mathrm{E}}$ decreases monotonically and largely tracks the DOS. This provides an independent confirmation that the enhancement of $\lambda$ and $T_c$ at light electron doping is controlled primarily by wavefunction (quantum-geometric) effects rather than by the DOS.

We further consider the GA EPC in the band basis, as shown in \cref{fig:main:GA}. The GA band-basis EPC $G^m_{\mathbf k}$ (defined in \cref{eq:EPC_band_basis_GM5+}) reproduces all salient features of the \textit{ab initio} result in \cref{fig:main:doping} (d): $G^m_{\mathbf k}$ is sharply peaked at $\Gamma$, remains enhanced along the $\Gamma$-K direction (due to the form of the $sp^2$ band wavefunctions), and decays approximately as $k^{2}$ with the in-plane distance from $\Gamma$. Crucially, the peak structure is overwhelmingly geometric in origin. The geometric contribution exhibits an even sharper maximum at $\Gamma$ (compared with the total EPC) and dominates the total signal, while the energetic part is much smaller and vanishes at $\Gamma$. Therefore, the pronounced enhancement of the EPC near $\Gamma$, and the resulting rise of $\lambda$ and $T_c$ when the Fermi level approaches this region under light electron doping, can be traced directly to the geometric contribution. 

The maximum of the quantum geometric contribution can be seen also in the superfluid weight of \ch{MgB2}. We calculate it from \textit{ab initio} band dispersions and Bloch states~\cite{hiorth2026ab, Liang2017, Huhtinen2022}. We find that the superfluid weight is dominated by the conventional contribution, reflecting the highly dispersive nature of the Fermi-level $\sigma$ and $\pi$ bands. The smaller geometric contribution is mainly carried by the $\sigma$ bands. It shows a pronounced peak near the twofold-degenerate 2D states at $\Gamma$. More details can be found in \cref{app:sec:superfluid_weight}.

\paragraph{Discussion.}\label{sec:discussion} 
In this work, we revisited \ch{MgB2} from a symmetry-based perspective and developed compact analytical models for its electronic structure, lattice dynamics, and EPC, using limited input from \textit{ab initio} calculations. The strong in-plane B $sp^2$ bonding leads to an \textit{obstructed}, bond-centered description of the $\sigma$ bands, while Mg hybridizes primarily with the B $p_z$ orbitals. This separation naturally accounts for the small quasi-2D $\sigma$ Fermi surfaces along $\Gamma$--A. The phonon spectrum can likewise be understood starting from a graphene-like boron layer: coupling to the heavier Mg sublattice shifts the predominantly boron-derived branches to higher frequencies, reflecting the strong mass separation between the two sublattices. Within this framework, symmetry singles out the in-plane B--B bond-stretching $\Gamma_5^+$ mode as the principal linear coupling channel to the $\sigma$ Fermi surfaces.

The \textit{obstructed} nature of the $\sigma$ bands also provides a unified explanation for the magnitude, selectivity, and doping dependence of the EPC. The relevant low-energy orbitals reside primarily on B--B bonds and form an effective kagome-like network. Consequently, bond-stretching distortions couple efficiently to the carriers, whereas distortions that do not directly modulate these bonds have much weaker effects. Upon electron doping, the calculated $T_c$ varies non-monotonically because the reduction of the density of states competes with an increase in the EPC as the Fermi level approaches the doubly degenerate $\Gamma_5^+$ electronic states near $\Gamma$. Using the GA decomposition, we find that this enhancement originates predominantly from the quantum-geometric contribution to the band-basis EPC, associated with the rapid momentum-space variation of the $\sigma$-band wavefunctions near $\Gamma$, rather than from the purely energetic contribution. This mechanism leads to an enhanced $T_c$ in the lightly electron-doped regime. The trend is qualitatively consistent with experiments on tensile-strained \ch{MgB2} thin films, in which changes in the effective filling of the $sp^2$ bands were accompanied by critical temperatures approaching $42$~K~\cite{pogrebnyakov2004enhancement,xi2007mgb2}. Electron-side chemical doping may therefore provide another route for tuning $T_c$, provided that disorder and impurity scattering remain sufficiently weak.

\ch{MgB2} is also an instructive example of superconductivity emerging from a low carrier density sector: the $\sigma$ bands contribute only a modest DOS, while the $\pi$ Dirac bands, despite a large Fermi surface, play a minor role in the pairing channel relevant to high $T_c$. If the $\pi$ Fermi surface is neglected, \ch{Mgb2} would resemble a lightly doped semiconductor (it is interesting to note that a high-throughput study of hydrates found many promising superconductor candidates to resemble degenerate semiconductors~\cite{Pires2026}). 
This suggests a broader design principle: high $T_c$ need not require a large DOS if the EPC matrix elements are strongly enhanced by wavefunction (quantum-geometric) effects and couple to high-frequency phonons. In particular, semiconductors or semimetals with band edges hosting rapidly varying Bloch wavefunctions, for instance, near symmetry-protected degeneracies or near band inversions, could exhibit similarly enhanced geometric EPC when doped into the relevant bands. If such systems can be effectively doped while maintaining strong coupling to light-atom bond-stretching phonons and avoiding strong carrier localization, they may offer a promising route toward phonon-mediated superconductivity with elevated $T_c$ in a low-DOS regime.

\section*{Acknowledgments}
\paragraph*{Funding:} 
We thank Jiabin Yu for the helpful discussion. 
We thank the technical support provided by Donostia International Physics Center Supercomputing Center. 
The simulations presented in this article were partially performed on computational resources managed and supported by Princeton Research Computing, a consortium of groups including the Princeton Institute for Computational Science and Engineering (PICSciE) and the Office of Information Technology's High Performance Computing Center and Visualization Laboratory at Princeton University. We acknowledge the computational resources provided by the Aalto Science-IT project. 
Y.J. was supported by the European Research Council (ERC) under the European Union’s Horizon 2020 research and innovation program (Grant Agreement No. 101020833), as well as by the IKUR Strategy under the collaboration agreement between Ikerbasque Foundation and DIPC on behalf of the Department of Education of the Basque Government. 
M.G.V and H.P. were supported by the Ministry for Digital Transformation and of Civil Service of the Spanish Government through the QUANTUM ENIA project call - Quantum Spain project, and by the European Union through the Recovery, Transformation and Resilience Plan - NextGenerationEU within the framework of the Digital Spain 2026 Agenda. 
M.G.V. thanks support to the Deutsche Forschungsgemeinschaft (DFG, German Research Foundation) GA 3314/1-1 – FOR 5249 (QUAST), the Spanish Ministerio de Ciencia e Innovacion (PID2022-142008NB-I00) and the Canada Excellence Research Chairs Program for Topological Quantum Matter. 
D.C. acknowledges support from the UKRI Horizon Europe Guarantee Grant No. EP/Z002419/1, and the support provided by the Leverhulme Trust.  
B.A.B. and H.H. were supported by the Gordon and Betty Moore Foundation through Grant No. GBMF8685 towards the Princeton theory program, the Gordon and Betty Moore Foundation’s EPiQS Initiative (Grant No. GBMF11070), the Global Collaborative Network Grant at Princeton University, the Simons Investigator Grant No. 404513, the NSF-MERSEC (Grant No. MERSEC DMR 2011750), the Simons Collaboration on New Frontiers in Superconductivity (Grant No. SFI-MPS-NFS-00006741-01), Princeton Catalysis Initiative (PCI), the Schmidt Foundation at the Princeton University and the National Science Foundation through the AI Research Institutes program Award No. DMR-2433348. 
This work was supported by a collaboration between The Kavli Foundation, Klaus Tschira Stiftung, and Kevin Wells, and by the Jane and Aatos Erkko Foundation, the Keele Foundation and the Magnus Ehrnrooth Foundation, as part of the SuperC collaboration. 
B.A.B., P.T., D.F.A., I.E., and M.A.L.M. were supported by a grant from the Simons Foundation (SFI-MPS-NFS-00006741-02, D.F.A.; SFI-MPS-NFS-00006741-01, B.A.B.; SFI-MPS-NFS-00006741-10, I.E.; SFI-MPS-NFS-00006741-13 (M.A.L.M.); SFI-MPS-NFS-00006741-12, P.T.) in the Simons Collaboration on New Frontiers in Superconductivity. This work is part of the Finnish Centre of Excellence in Quantum Materials (QMAT).

\renewcommand{\addcontentsline}[3]{}
%

\let\addcontentsline\oldaddcontentsline

\renewcommand{\thetable}{S\arabic{table}}
\renewcommand{\thefigure}{S\arabic{figure}}
\renewcommand{\theequation}{S\arabic{section}.\arabic{equation}}

\onecolumngrid
\pagebreak
\thispagestyle{empty}

\clearpage
\begin{center}
	\textbf{\large Supplementary Information for ''\titlePaper{}``}\\[.2cm]
\end{center}

\appendix
\renewcommand{\thesection}{\Roman{section}}
\tableofcontents
\let\oldaddcontentsline\addcontentsline

\newpage

\section{Review of electron-phonon coupling Hamiltonian}
In this section, we review the formalism of electron-phonon coupled systems with the electron-phonon coupling (EPC) Hamiltonian, including the symmetry properties and superconducting properties.  We also review the EPC from Gaussian approximation (GA), an analytic approximation of the matrix elements of the EPC introduced in \cite{yu2024non}, and its application.

\subsection{Electron Hamiltonian}
We start from the electron Hamiltonian. 
Define the real-space electron operator as $\cre{c}{\RR, i}$, which creates an electron at $\rr_i$ position in the unit cell labeled by $\RR$. The corresponding momentum-space operator $\cre{c}{\kk, i}$ is defined by the Fourier transformation (FT):
\begin{equation}\begin{aligned}
\cre{c}{\RR, i} &=\frac{1}{\sqrt{N_e}}\sum_{\kk} \cre{c}{\kk, i} e^{-i\kk(\RR+\rr_i)}, \\
\cre{c}{\kk, i} &=\frac{1}{\sqrt{N_e}}\sum_{\RR} \cre{c}{\RR, i} e^{i\kk(\RR+\rr_i)}, \\
\label{app:eq:FT_electron_op}
\end{aligned}\end{equation}
where $N_e=N_e^1\times N_e^2\times N_e^3$ denotes the number of unit cells. A generic single-particle electron tight-binding Hamiltonian has the form
\begin{equation}\begin{aligned}
\hat{H}_{el} &= \sum_{\RR,\Delta\RR,ij} t_{ij}(\Delta\RR) \cre{c}{\RR,i} \des{c}{\RR+\Delta\RR,j} \\
&= \sum_{\kk, ij} h_{ij}(\kk) \cre{c}{\kk,i} \des{c}{\kk,j},
\label{app:eq:H_el_def}
\end{aligned}\end{equation}
where $h_{ij}(\kk)=\sum_{\Delta\RR} t_{ij}(\Delta\RR) e^{i\kk\cdot(\Delta\RR+\rr_j-\rr_i)}$. 
We the FT gauge in \cref{app:eq:FT_electron_op} throughout the manuscript, where the symmetry operators will acquire phase factors from sublattice shifts $\rr_i$. 
Note in \cref{app:eq:H_el_def}, we implicitly assume the two-center approximation of the hopping (see discussion in \cref{app:sec:two-center-approx}), such that the hopping $t_{ij}(\Delta\RR)$ is a function of the \textit{relative position} of two orbitals, \ie, $\Delta\RR+\rr_j-\rr_i$, but does not depend on the \textit{specific} positions of orbitals. 

We then introduce the electron band basis $\cre{\gamma}{\kk,n}$ for $\hat{H}_{el}$, \ie, 
\begin{equation}\begin{aligned}
\sum_{j} h_{ij}(\kk) U_{j,n}(\kk) &= \epsilon_{\kk,n} U_{i,n},\\
\cre{\gamma}{\kk,n} &= \sum_{i} U_{i,n}(\kk) \cre{c}{\kk,i},\\ 
\cre{c}{\kk,i} &= \sum_n U^*_{i,n}(\kk) \cre{\gamma}{\kk, n}, \\
\Rightarrow H_{el} &= \sum_{\kk,n}\epsilon_{\kk,n} \cre{\gamma}{\kk,n}\des{\gamma}{\kk,n}.
\end{aligned}\end{equation}
The eigenvectors are orthonormal: 
\begin{equation}\begin{aligned}
\sum_{i} U^*_{i,n}(\kk) U_{i,m}(\kk)=\delta_{mn},\quad
\sum_{n} U^*_{i,n}(\kk) U_{j,n}(\kk)=\delta_{ij}.
\end{aligned}\end{equation}
The Hamiltonian and eigenvectors have an additional embedding matrix when shifted by a reciprocal lattice vector $\GG$
\begin{equation}\begin{aligned}
U_{i,n}(\kk+\GG) &= U_{i,n}(\kk) e^{-i\GG\cdot\rr_i}, \\
h_{ij}(\kk+\GG) &= h_{ij}(\kk) e^{i\GG\cdot(\rr_j-\rr_i)}.
\end{aligned}\end{equation}
Remark that one can also use the lattice gauge, \ie, $\cre{\bar{c}}{\kk, i} =\frac{1}{\sqrt{N_e}}\sum_{\RR} \cre{c}{\RR, i} e^{i\kk\cdot\RR}$ without the sublattice embedding matrix, and $\bar{h}_{ij}(\kk)=\sum_{\Delta\RR} h_{ij}(\Delta\RR) e^{i\kk\cdot \Delta\RR} = h_{ij}(\kk)e^{-i\kk\cdot(\rr_j-\rr_i)}$. The eigenvectors $\bar{U}_{i,n}$ in the lattice gauge has the relation $\bar{U}_{i,n}=e^{i\kk\cdot\rr_i} U_{i,n}$.

\subsection{Phonon Hamiltonian}
Let $\hat{u}_{\RR,i\nu}$ be the phonon displacement operator for the $\nu \in\{x,y,z\}$-directional movement of the $i$-th atom in unit cell $\RR$. The corresponding momentum operator is $\hat{P}_{\RR,i\mu}$. 
Then the phonon Hamiltonian has the form
\begin{equation}\begin{aligned}
\hat{H}_{ph} &= \sum_{\RR,i,\mu} \frac{1}{2M_i} \hat{P}_{\RR,i\mu}^2 + \sum_{\RR,\RR', i\mu,j\nu} \frac{1}{2} \Phi_{i\mu,j\nu}(\RR' - \RR) \hat{u}_{\RR,i\mu} \hat{u}_{\RR',j\nu},
\end{aligned}\end{equation} 
where $\Phi_{i\mu,j\nu}(\RR)$ is the real-space force constant matrix, and $M_i$ is the mass of the $i$-th atom. 

Define the FT of the phonon operators 
\begin{equation}\begin{aligned}
\des{u}{\RR,i\mu} &= \frac{1}{\sqrt{N_p}}\sum_{\qq} \des{u}{\qq,i\mu} e^{i\qq(\RR+\rr_i)}, \\
\des{P}{\RR,i\mu} &= \frac{1}{\sqrt{N_p}}\sum_{\qq} \des{P}{\qq,i\mu} e^{-i\qq(\RR+\rr_i)}, 
\end{aligned}\end{equation}
where $N_p=N_p^1\times N_p^2\times N_p^3$ denotes the number of unit cells for phonon. Note that $\des{u}{\RR,i\mu}$ has the same FT convention as $\des{c}{\RR,i}$ instead of $\cre{c}{\RR,i}$. As $\cre{u}{\RR,i\mu}=\des{u}{\RR,i\mu}$, we have $\des{u}{\qq,i\mu}=\des{u}{-\qq,i\mu}^*$. As the real-space force constant matrix is real, we have the Hermitian condition
\begin{equation}
    \Phi_{i\mu,j\nu}(\RR-\RR')=
\Phi_{j\nu,i\mu}(\RR'-\RR),\quad 
\Phi_{i\mu,j\nu}(\qq)=
\Phi_{j\nu,i\mu}(-\qq).
\end{equation}

The phonon Hamiltonian in momentum space is
\begin{equation}\begin{aligned}
\hat{H}_{ph} &= \sum_{\qq,i,\mu} \frac{1}{2M_i} \hat{P}_{\qq,i\mu}\hat{P}_{-\qq,i\mu} + \sum_{\qq, i\mu,j\nu} \frac{1}{2} \Phi_{i\mu,j\nu}(\qq) \hat{u}_{-\qq,i\mu} \hat{u}_{\qq, j\nu},
\end{aligned}\end{equation} 
where
\begin{equation}\begin{aligned}
\Phi_{i\mu,j\nu}(\qq) &= \sum_{\RR} \Phi_{i\mu,j\nu}(\RR) e^{i\qq(\RR+\rr_j-\rr_i)}.
\end{aligned}\end{equation} 

The dynamic matrix $D_{i\mu,j\nu}(\qq)$ is defined as the mass-scaled momentum-space force constants:
\begin{equation}\begin{aligned}
D_{i\mu,j\nu}(\qq)= \Phi_{i\mu,j\nu}(\qq) \frac{1}{\sqrt{M_i M_j}}.
\end{aligned}\end{equation}
The Hermicity of the dynamical matrix gives 
$D_{i\mu,j\nu}(\qq)=
D_{j\nu,i\mu}(-\qq)$. 
The eigen equation for phonon modes is
\begin{equation}\begin{aligned}
\sum_{j\nu} \Phi_{i\mu,j\nu}(\qq) u_{j\nu,n}(\qq) &= M_i \omega_{\qq,n}^2 u_{i\mu,n}(\qq),\\
\sum_{j\nu} D_{i\mu,j\nu}(\qq) U^{p}_{j\nu,n}(\qq) &= \omega_{\qq,n}^2 U^p_{i\mu,n}(\qq),
\label{app:eq:phonon_eig_eq}
\end{aligned}\end{equation}
where $u_{i\mu,n}(\qq)=\frac{1}{\sqrt{M_i}} U^{p}_{i\mu,n}(\qq)$, with $u_{i\mu,n}(\qq)$ being the atomic displacements, and $U^{p}_{i\mu,n}(\qq)$ the eigenvectors of dynamical matrix, satisfying $\sum_{i,\mu} U^{p*}_{i\mu,m}(\qq) U^{p}_{i\mu,n}(\qq)=\delta_{mn}$. 

We define the phonon operator in the band basis 
\begin{equation}\begin{aligned}
\hat{w}_{\qq,n} &= \sum_{i\mu} U^{p,*}_{i\mu,n}(\qq) \tilde{\hat{u}}_{\qq,i\mu},\quad
\hat{p}_{\qq,n} = \sum_{i\mu} U^p_{i\mu,n}(\qq) \tilde{\hat{P}}_{\qq,i\mu}, \\
\hat{b}_{\qq,n} &= \sqrt{\frac{\omega_{\qq,n}}{2\hbar}} \hat{w}_{\qq,n} + \frac{i}{\sqrt{2\hbar\omega_{\qq,n}}} \hat{p}_{-\qq,n},\quad
\hat{b}^\dagger_{\qq,n} = \sqrt{\frac{\omega_{\qq,n}}{2\hbar}} \hat{w}_{-\qq,n} - \frac{i}{\sqrt{2\hbar \omega_{\qq,n}}} \hat{p}_{\qq,n},
\end{aligned}\end{equation}
where $\tilde{\hat{u}}_{\RR,i\mu} = \sqrt{M_i} \hat{u}_{\RR, i\mu},  \tilde{\hat{P}}_{\RR,i\mu} = \frac{1}{\sqrt{M_i}} \hat{P}_{\RR, i\mu}$ are the mass-scaled phonon operators. 
The phonon Hamiltonian in the band basis has the form
\begin{equation}\begin{aligned}
H_{ph} &= \sum_{\qq,n}\frac{1}{2}\left(
\des{p}{\qq,n}\des{p}{-\qq,n} + \omega_{\qq,n}^2 \des{w}{\qq,n}\des{w}{-\qq,n} \right) \\
&= \sum_{\qq,n} \hbar\omega_{\qq,n} (\cre{b}{\qq,n}\des{b}{\qq,n}+\frac{1}{2}).
\label{app:eq:free_phonon_ham}
\end{aligned}\end{equation}

\paragraph{\textbf{Acoustic sum rule.}} 
The force constant matrix $\Phi_{i\mu,j\nu}(\RR-\RR')$ satisfies the following sum rules due to the translation-invariance: 
\begin{equation}\begin{aligned}
\sum_{\RR',j} \Phi_{i\mu,j\nu}(\RR-\RR')=0.
\label{Eq: sum_rule_realspace}
\end{aligned}\end{equation}
This constraint can be understood as follows.  The harmonic energy of a phonon system with displacement $u_{\RR,i\mu}$ is 
$U=\sum_{\RR,\RR', i\mu,j\nu} \frac{1}{2} \Phi_{i\mu,j\nu}(\RR' - \RR) u_{\RR,i\mu} u_{\RR',j\nu}$, and the force on atom $(i,\RR)$ is 
$F_{i\mu}(\RR) = -\frac{\partial U}{\partial u_{\RR,i\mu}}$. 
Now consider a global translation by $a_\nu$  applied to all atoms. Such a uniform displacement should not change the energy of the system. Consequently, the corresponding force must vanish:
$F_{i\mu}(\RR)=-\sum_{\RR',j\nu} \Phi_{i\mu,j\nu}(\RR' - \RR) a_{\nu}=0$. 
Since this must hold for any $a_\nu$, we obtain the real-space sum rule given in \cref{Eq: sum_rule_realspace}. 
\cref{Eq: sum_rule_realspace} gives $N\times3\times3$ constraint equations ($N$ is the number of atoms), which can be rewritten as 
\begin{equation}
    \Phi_{i\mu,i\nu}(\RR) = -\sum_{\RR'\neq \bm{0} \text{ or } j\ne i} \Phi_{i\mu,j\nu}(\RR-\RR')
\end{equation}
\ie, the diagonal blocks $\Phi_{i\mu,i\nu}(\RR)$ are determined by off-diagonal blocks if the acoustic sum rule is enforced. 

The momentum space constraints are
\begin{equation}\begin{aligned}
\sum_{j} \Phi_{i\mu,j\nu}(\kk=0)=0.
\end{aligned}\end{equation}
or in terms of the dynamical matrix
\begin{equation}\boxed{ \begin{aligned}
\sum_{j} D_{i\mu,j\nu}(\kk=0)\sqrt{M_j}=0.
\end{aligned}}\end{equation}

\subsection{EPC Hamiltonian}\label{app:sec:intro-EPC-ham}
Consider a generic one-body electron Hamiltonian $\hat{H}_e=\hat{T}_e+V_{KS}(\rr)$, where $\hat{T}_e$ is the electron kinetic term, and $V_{KS}$ is the self-consistent one-body Kohn-Sham potential obtained from DFT, with the form
$V_{KS}=V_{ext}+V_H[n]+V_{xc}[n]$, where the three terms are the electron-ion potential, the Hartree term, and the exchange-correlation term, respectively. 
The real-space electron-phonon coupling (EPC) matrix element is defined as the first derivative of $\hat{H}_e$ with respect to the atomic displacement:
\begin{equation}\begin{aligned}
g_{\RR_e,\RR_{p}}^{ij,l\mu} &= \langle W_{i \RR} | \partial_{l\mu (\RR+\RR_{p})} \hat{H}_e | W_{j (\RR+\RR_e)} \rangle 
= \langle W_{i\bm{0}_e} | \partial_{l\mu \RR_{p}} \hat{H}_e| W_{j\RR_e} \rangle \\
&= \langle W_{i\bm{0}_e} | \partial_{l\mu \RR_{p}} V^{KS}| W_{j\RR_e} \rangle, 
\label{app:eq:def_epc_realspace}
\end{aligned}\end{equation}
where $|W_{i\RR} \rangle$ is the $i$-th Wannier orbital in cell $\RR$, and $\partial_{l\mu\RR_p}$ is the partial derivative with respective to the $\mu$ movement of $l$ atom in the $\RR_{p}$ cell. 
We notice that the electron kinetic operator is independent of atomic displacement, thus it does not contribute to the derivative.  
Remark that the generic EPC does not satisfy the two-center approximation in general (see discussion in \cref{app:sec:two-center-approx}). 
In practical DFPT calculations \cite{giustino2017electron}, the electron–phonon coupling (EPC) is first obtained in the band basis, $G^{mn\nu}_{\kk,\qq}$ (see \cref{app:eq:EPC_band_basis})), and only then transformed to the Wannier basis~\cite{EPCpaper}. For clarity of presentation, however, we introduce the EPC directly in the Wannier basis first.

The EPC Hamiltonian has the form
\begin{equation}\begin{aligned} 
\hat{H}_{epc} &= \sum_{\RR,\RR_e,\RR_{p},i,j,l,\mu} 
g_{\RR_e,\RR_{p}}^{ij,l\mu} \cre{c}{\RR,i} \des{c}{\RR+\RR_e,j} \des{u}{\RR+\RR_{p},l\mu} \\
&=\frac{1}{\sqrt{N_p}} \sum_{\kk,\qq,i,j,l\mu} g_{\kk,\qq}^{ij,l\mu}  \cre{c}{\kk+\qq,i} \des{c}{\kk,j}\des{u}{\qq,l\mu} 
\end{aligned}\end{equation} 
where the momentum-space EPC matrix element is 
\begin{equation}\begin{aligned} 
g_{\kk,\qq}^{ij,l\mu}= \sum_{\RR_{e},\RR_{p}} g_{\RR_e,\RR_{p}}^{ij,l\mu} e^{i\kk\cdot(\RR_e+\rr_j-\rr_i)+i\qq\cdot(\RR_{p}+\rr_l-\rr_i)}.
\label{app:eq:epc_FT_formula}
\end{aligned}\end{equation} 
\cref{app:eq:epc_FT_formula} holds when $N_e\leq N_p$, where $N_e$ ($N_p$) is the number of unit cells for electron (phonon). Note that $N_e$ cannot be smaller than $N_p$, otherwise $\kk+\qq$ does not fit on the grid of $N_e$. 
Note that the momentum-space EPC matrix has embedding matrix $g_{\kk+\GG,\qq+\GG'}^{ij,l\mu} = g_{\kk,\qq}^{ij,l\mu} e^{i\GG\cdot(\rr_j-\rr_i)+i\GG'\cdot(\rr_l-\rr_i)}$. 
The Hermitian condition of the EPC Hamiltonian leads to 
\begin{equation}\begin{aligned}
g^{ij,l\mu}_{\RR_e,\RR_p} &= (g^{ji,l\mu}_{-\RR_e,\RR_p-\RR_e})^*, \\
g^{ij,l\mu}_{\kk,\qq} &= (g^{ji,l\mu}_{\kk+\qq,-\qq})^*.
\end{aligned}\end{equation} 

We then transform the EPC Hamiltonian into the band basis. To do so, we define the electrons and phonon operators in the band basis, i.e., $\cre{\gamma}{\kk,n}$ and phonons $\des{w}{\qq,s}$:
\begin{equation}\begin{aligned} 
\cre{c}{\kk,i} &= \sum_n U^*_{i,n}(\kk) \cre{\gamma}{\kk, n}, \\
\des{u}{\qq,i\mu} &= \sum_{s} \frac{U_{i\mu, s}^p(\qq)}{\sqrt{M_i}}  \des{w}{\qq,s} \\
&= \sum_{s} \sqrt{\frac{\hbar}{2M_i \omega_{\qq,s}}} U_{i\mu, s}^p(\qq) (\cre{b}{-\qq,s} + \des{b}{\qq,s}).
\end{aligned}\end{equation} 
Then the EPC Hamiltonian in the band basis has the form
\begin{equation}\begin{aligned} 
\hat{H}_{epc} = & \frac{1}{\sqrt{N_p}} 
\sum_{\kk,\qq,mns} 
\tilde{G}^{mns}_{\kk,\qq} \cre{\gamma}{\kk+\qq,m} \des{\gamma}{\kk,n} w_{\qq,s} \\ 
&= \frac{1}{\sqrt{N_p}} \sum_{\kk,\qq,mns} G^{mns}_{\kk,\qq}
\cre{\gamma}{\kk+\qq,m} \des{\gamma}{\kk,n} (\cre{b}{-\qq,s} + \des{b}{\qq,s}) \\
&= \frac{1}{\sqrt{N_p}} \sum_{\kk,\qq,mns} G^{mns}_{\kk,\qq}
\cre{\gamma}{\kk+\qq,m} \des{\gamma}{\kk,n} \des{b}{\qq,s} + h.c.,
\label{app:eq:epc_ham_band_basis}
\end{aligned}\end{equation}
where
\begin{equation}\begin{aligned}
\tilde{G}_{\kk,\qq}^{mns} &= \sum_{ij,l\mu} g_{\kk,\qq}^{ij,l\mu} U_{i,m}^*(\kk+\qq) U_{j,n}(\kk) \frac{1}{\sqrt{M_l}} U^p_{l\mu,s}(\qq), \\
G^{mns}_{\kk,\qq} &= \sqrt{\frac{\hbar}{2\omega_{\qq,s}}} \tilde{G}_{\kk,\qq}^{mns}. 
\label{app:eq:EPC_band_basis}
\end{aligned}\end{equation} 
Note that $G^{mns}_{\kk,\qq}$ enters the expressions of superconducting-related quantities defined in \cref{app:sec:SC_properties_formalism}. The real-space EPC matrix element $g^{ij,l\mu}_{\RR_e,\RR_p}$ has unit eV/\AA, while the band basis EPC $G^{mns}_{\kk,\qq}$ has unit eV, where the phonon frequency $\omega_{\qq,s}$ takes the unit of Hz.

The EPC tensor also satisfies the acoustic sum rule. This is because under an arbitrary uniform translation $\des{u}{\RR+\RR_{p},l\mu}\rightarrow \des{u}{\RR+\RR_{p},l\mu}+\bm{t}_{\mu}$, the system should be invariant. Thus
\begin{equation}\begin{aligned}
\Delta\hat{H}_{epc} &= 
\sum_{\RR,\RR_e,i,j,\mu} \left(\sum_{\RR_{p},l}
g_{\RR_e,\RR_{p}}^{ij,l\mu}\right) \cre{c}{\RR,i} \des{c}{\RR+\RR_e,j} \bm{t}_{\mu} =0 \\
\Rightarrow
&\sum_{\RR_{p},l} 
g_{\RR_e,\RR_{p}}^{ij,l\mu} =0,\quad \forall i,j, \RR_e, \mu. 
\label{app:eq:EPC_sum_rul}
\end{aligned}\end{equation}
The momentum-space EPC satisfies the following form of the acoustic sum rule
\begin{equation}\boxed{ \begin{aligned}
\sum_{l} g^{ij,l\mu}_{\kk,\qq=0} &=
\sum_{\RR_e} e^{i\kk\cdot (\RR_e+\rr_j-\rr_i)} \sum_{\RR_p,l} g_{\RR_e,\RR_{p}}^{ij,l\mu} =0.
\label{app:eq:EPC_sum_rule}
\end{aligned}}\end{equation}
This is equivalent to requiring the EPC of three acoustic modes at $\qq=0$ to be zero. A more detailed discussion of the acoustic sum rule and its dependence on the atomic wavefunctions is deferred to Ref.~\cite{EPCpaper}.

We remark that the EPC tensor in the band basis $G^{mns}_{\kk,\qq}$ has the same form under the atomic (with the sublattice embedding matrix) and lattice gauge (without the sublattice embedding matrix), since the $e^{i\kk\cdot\rr_i}$ phases cancel in the FT to momentum space step (\cref{app:eq:epc_FT_formula}) and the transformation to the band basis (\cref{app:eq:EPC_band_basis}). However, $G^{mns}_{\kk, \qq}$ is not gauge-invariant, as the electron and phonon band bases are defined up to a $U(1)$ phase (or a unitary matrix that mixes degenerate states). The quantities $\sum_{mn}|G^{mn, s}_{\kk, \qq}|^2$ and $\sum_{ms}|G^{mn, s}_{\kk, \qq}|^2$ are both gauge-invariant when the summations contain the degenerate subspaces. 

The real-space EPC tensor $g_{\RR_e,\RR_p}^{ij,l\mu}$ can be obtained from first-principles calculations~\cite{giustino2007electron,lee2023electron,ponce2016epw,margine2013anisotropic, zhou2021perturbo}. 
In practice, this tensor can be truncated to finite electron and phonon real-space ranges, because its short-range component decays rapidly, typically exponentially, with respect to both $\RR_e$ and $\RR_p$. 
For insulators and semiconductors, the analytic long-range electrostatic contribution is first separated out, so that only the short-range EPC component is Wannier-interpolated in real space~\cite{giustino2007electron, zhou2021perturbo}.

Ref.~\cite{luo2024data} further proposed a data-compression scheme based on singular value decomposition (SVD) of the real-space EPC tensor. 
For each fixed composite channel $a\equiv(ij,l\mu)$, the short-range tensor is viewed as a matrix $g^{a}_{\RR_e,\RR_p}$, 
where $\RR_e$ labels the electron real-space coordinate and $\RR_p$ labels the phonon real-space coordinate. 
One then performs the SVD 
\[
    g^{a}_{\RR_e,\RR_p}
    =
    \sum_{\lambda}
    s^{a}_{\lambda}\,
    u^{a}_{\lambda}(\RR_e)\,
    \left[v^{a}_{\lambda}(\RR_p)\right]^* ,
\]
where $s^{a}_{\lambda}$ are singular values, while $u^{a}_{\lambda}(\RR_e)$ and $v^{a}_{\lambda}(\RR_p)$ can be interpreted as effective electronic and vibrational real-space patterns. 
Keeping only the dominant singular components gives the low-rank approximation
\[
    g^{a}_{\RR_e,\RR_p}
    \approx
    \sum_{\lambda=1}^{N_{\rm sv}}
    s^{a}_{\lambda}\,
    u^{a}_{\lambda}(\RR_e)\,
    \left[v^{a}_{\lambda}(\RR_p)\right]^* ,
\]
which substantially reduces storage and interpolation cost. 
Remarkably, retaining only a small fraction of the singular values, typically of order 1-2\%, is sufficient to reproduce EPC-driven material properties with high accuracy~\cite{luo2024data}.

\subsection{Symmetry constraints}\label{app:sec:symmetry_constraints}
In this part, we consider the symmetry constraints on the EPC Hamiltonian. For a given symmetry operation $g=\{R_g|\mathbf{t}_g\}$, it acts on the real-space vector $\rr$ as $g\rr=R_g\rr+\mathbf{t}_g$. 

\paragraph{Electron.} 
We consider the real-space electron operator $\cre{c}{\RR,i\alpha}$, where $i$ is the atom index and $\alpha$ the orbital index. It has the transformation 
\begin{equation}\begin{aligned}
g \cre{c}{\RR,i\alpha} g^{-1} &= \sum_{j\beta} \cre{c}{g(\RR+\rr_i)-\rr_j,j\beta} D_{j\beta,i\alpha}(g)\\
&= \sum_{j\beta} \cre{c}{\RR',j\beta} D_{j\beta,i\alpha}(g) \delta_{\RR', g(\RR+\rr_i)-\rr_j},
\label{app:eq:el_rep_definition}
\end{aligned}\end{equation}
where $D(g)$ is the unitary representation matrix of $g$. Note that $D_{j\beta,i\alpha}(g)$ is nonzero only if $g(\RR+\rr_i)-\rr_j$ is a lattice vector, \ie, atom $i$ and $j$ are related by $g$. 
Similarly, the momentum space operator has the transformation
\begin{equation}\begin{aligned}
g\cre{c}{\kk,i\alpha} g^{-1} = \sum_{j\beta}\cre{c}{g\kk,j\beta} D_{j\beta,i\alpha}(g,\kk),\quad 
D_{j\beta,i\alpha}(g,\kk)= D_{j\beta,i\alpha}(g) e^{-i\kk\cdot R^{-1}_g\mathbf{t}_g}.
\end{aligned}\end{equation}
Note that an extra phase factor in the representation matrix arises from the translational part of $g$. The operator in the band basis $\cre{\gamma}{\kk,n}$ transforms as
\begin{equation}\begin{aligned}
g \cre{\gamma}{\kk,n} g^{-1} &= \sum_{i\alpha} U_{i\alpha,n}(\kk) \sum_{j\beta} \cre{c}{g\kk,j\beta} D_{j\beta,i\alpha}(g,\kk) \\
&= \sum_{i\alpha,j\beta} \left(\sum_{n'}\cre{\gamma}{g\kk,n'} U^*_{j\beta,n'}(g\kk)\right)
D_{j\beta,i\alpha}(g,\kk) U_{i\alpha,n}(\kk) 
\\
&= \sum_{n'} \cre{\gamma}{g\kk,n'} \tilde{D}_{n'n}(g,\kk),\\
\tilde{D}_{n'n}(g,\kk) &= \sum_{i\alpha,j\beta}  U_{j\beta,n'}^*(g\kk) D_{j\beta,i\alpha}(g,\kk) U_{i\alpha,n}(\kk) 
\label{app:eq:rep_mat_el_band_basis}
\end{aligned}\end{equation}
The representation matrix $\tilde{D}_{n'n}(g,\kk)$ in the band basis can be simplified into the block-diagonal form of irreducible representations (IRREPs). Assume the band index $n$ is decomposed into two indices $s\gamma$, where $s$ denotes sets of degenerate bands belonging to the same IRREP, and $\gamma$ is the band inside the $s$-th set. Then $\tilde{D}_{n'n}(g,\kk)=\tilde{D}^{s}_{\gamma'\gamma}(g,\kk)$, and 
\begin{equation}\begin{aligned}
\sum_{\gamma} \tilde{D}^{s}_{\gamma'\gamma}(g,\kk) \tilde{D}^{s,*}_{\gamma''\gamma}(g,\kk)= 
\sum_{\gamma} \tilde{D}^{s}_{\gamma\gamma'}(g,\kk) \tilde{D}^{s,*}_{\gamma\gamma''}(g,\kk)= 
\delta_{\gamma',\gamma''},
\label{app:eq:rep_band_basis_ortho_relation}
\end{aligned}\end{equation}
where $\gamma,\gamma',\gamma''$ are band indices belonging to the $s$-th degenerate set. 

The electron Hamiltonian transforms as 
\begin{equation}\begin{aligned}
g \hat{H}_{el} g^{-1} =& \sum_{\RR,\Delta\RR,i\alpha j\beta} t_{i\alpha,j\beta}(\Delta\RR) g \cre{c}{\RR,i\alpha} g^{-1} g \des{c}{\RR+\Delta\RR,j\beta} g^{-1} \\
=& \sum_{\RR',\Delta\RR',i'\alpha',j'\beta'}
\left(
\sum_{i\alpha,j\beta} t_{i\alpha,j\beta}(\Delta\RR') D_{i'\alpha',i\alpha}(g) D^*_{j'\beta',j\beta}(g)
\right) 
\cre{c}{\RR',i'\alpha'} \des{c}{\RR'+\Delta\RR',j'\beta'} 
\\
& \quad \delta_{\RR',g(\RR+\rr_i)-\rr_{i'}}\delta_{\Delta\RR', R_g(\Delta\RR+\rr_j-\rr_i)-(\rr_{j'}-\rr_{i'})} \\
=& \sum_{\RR', \Delta\RR',i'\alpha',j'\beta'} t_{i'\alpha',j'\beta'}(\Delta\RR') \cre{c}{\RR',i'\alpha'} \des{c}{\RR'+\Delta\RR',j'\beta'} =\hat{H}_{el},\\
\Rightarrow 
t_{i'\alpha',j'\beta'}(\Delta\RR') =& \sum_{i\alpha,j\beta} D_{i'\alpha',i\alpha}(g) t_{i\alpha,j\beta}(\Delta\RR) D_{j'\beta',j\beta}^*(g) \delta_{\Delta\RR', R_g(\Delta\RR+\rr_j-\rr_i)-(\rr_{j'}-\rr_{i'})}.
\end{aligned}\end{equation}
Similarly, $\hat{H}_{el}$ in momentum space operators transform as 
\begin{equation}\begin{aligned}
g \hat{H}_{el} g^{-1} &= \sum_{\kk, i\alpha,j\beta} h_{i\alpha,j\beta}(\kk) g \cre{c}{\kk,i\alpha} g^{-1} g \des{c}{\kk,j\beta} g^{-1}, \\
&= \sum_{\kk,i'\alpha',j'\beta'} \left(
\sum_{i\alpha,j\beta} h_{i\alpha,j\beta}(\kk) D_{i'\alpha',i\alpha}(g,\kk) D^*_{j'\beta',j\beta}(g,\kk)
\right)
\cre{c}{g\kk,i'\alpha'} \des{c}{g\kk,j'\beta'} 
=\hat{H}_{el}, \\
\Rightarrow h_{i'\alpha',j'\beta'}(g\kk) &= \sum_{i\alpha,j\beta} D_{i'\alpha',i\alpha}(g, \kk) h_{i\alpha,j\beta}(\kk) D^*_{j'\beta',j\beta}(g, \kk)  \\
&= \sum_{i\alpha,j\beta} D_{i'\alpha',i\alpha}(g) h_{i\alpha,j\beta}(\kk) D^*_{j'\beta',j\beta}(g). 
\label{app:eq:Hk_sym_transform}
\end{aligned}\end{equation} 
Note that in the last equation of \cref{app:eq:Hk_sym_transform}, the $\kk$-dependence in $D_{i'\alpha',i\alpha}(g, \kk)$ is eliminated. 
$\hat{H}_{el}$ in band basis operators transform as 
\begin{equation}\begin{aligned}
g \hat{H}_{el} g^{-1} &= \sum_{\kk,n} \epsilon_{\kk,n} 
\sum_{n'} \cre{\gamma}{g\kk,n'} \tilde{D}_{n'n}(g,\kk)
\sum_{n''} \des{\gamma}{g\kk,n''} \tilde{D}^*_{n''n}(g,\kk)\\
&= \sum_{\kk,n',n''} \cre{\gamma}{g\kk,n'} \des{\gamma}{g\kk,n''} \left(\sum_{n} \epsilon_{\kk,n}  \tilde{D}_{n'n}(g,\kk) \tilde{D}^*_{n''n}(g,\kk)
\right) \\
&= \sum_{\kk, s,\gamma',\gamma''} \cre{\gamma}{g\kk,s\gamma'} \des{\gamma}{g\kk,s\gamma''} \epsilon_{\kk,s} \sum_{\gamma}  \tilde{D}^s_{\gamma'\gamma}(g,\kk) \tilde{D}^{s,*}_{\gamma''\gamma}(g,\kk)
\\
&= \sum_{\kk,s,\gamma} \epsilon_{\kk,s} \cre{\gamma}{g\kk,s,\gamma} \des{\gamma}{g\kk,s,\gamma},
\end{aligned}\end{equation}
where $s$ denotes the degenerate sets of bands and $\gamma$ is the band within the degenerate set. We have used the orthogonal relation of bands of different IRREPs in the band basis defined in \cref{app:eq:rep_band_basis_ortho_relation}.

\paragraph{Phonon.} 
Similar to the electron operator, the real-space, momentum-space, and band basis phonon operators transform under symmetry $g$ as
\begin{equation}\begin{aligned}
g\des{u}{\RR,i\mu}g^{-1} &= \sum_{j\nu} \des{u}{g(\RR+\rr_i)-\rr_j,j\nu} D^p_{j\nu,i\mu}(g), \\
g\des{u}{\qq,i\mu}g^{-1} &= \sum_{j\nu} \des{u}{g\qq,j\nu} D^p_{j\nu,i\mu}(g,\qq),\quad
D_{j\nu,i\mu}^p(g,\qq)= D^p_{j\nu,i\mu}(g) e^{i\qq\cdot R^{-1}_g\mathbf{t}_g}, \\
g\des{w}{\qq, n}g^{-1} &= \sum_{n'} \des{w}{g\qq, n'} \tilde{D}^p_{n'n}(g, \qq),\quad
\tilde{D}_{n'n}^p(g,\qq)= \sum_{i\mu j\nu} U^{p}_{j\nu,n'}(g\qq) D^{p}_{j\nu,i\mu}(g,\qq) U^{p,*}_{i\mu,n}(\qq),
\label{app:eq:rep_mat_ph_band_basis}
\end{aligned}\end{equation}
where $D^p_{j\nu,i\mu}(g)=D^{perm}_{ji}(g)R_{\nu\mu}(g)$, with $D^{perm}(g)$ being the permutation matrix of atomic positions under $g$, and $R(g)$ the $O(3)$ rotation matrix of $g$. $\tilde{D}^p_{n'n}(g, \qq)$ is the representation matrix in the phonon band basis. Note that $\tilde{D}^p(g,\qq)$ is obtained using the eigenvectors $U^p(\qq)$ of the dynamical matrix (not scaled by the atomic masses, see \cref{app:eq:phonon_eig_eq}), which are orthonormal. To derive the last line in \cref{app:eq:rep_mat_ph_band_basis}, recall that 
$\hat{w}_{\qq,n} = \sum_{i\mu} U^{p,*}_{i\mu,n}(\qq) \sqrt{M_i} \hat{u}_{\qq,i\mu}$, thus we have
\begin{equation}\begin{aligned}
g\hat{w}_{\qq,n} g^{-1} &= \sum_{i\mu} U^{p,*}_{i\mu,n}(\qq) \sqrt{M_i} g \hat{u}_{\qq,i\mu} g^{-1}  =  \sum_{i\mu} U^{p,*}_{i\mu,n}(\qq) \sqrt{M_i}   \sum_{j\nu} D^p_{j\nu,i\mu}(g,\qq) \des{u}{g\qq,j\nu} \\
&= \sum_{i\mu} U^{p,*}_{i\mu,n}(\qq) \sqrt{M_i} \sum_{j\nu} D^p_{j\nu,i\mu}(g,\qq) \left( \frac{1}{\sqrt{M_j}} \sum_{n'} U^p_{j\nu,n'}\des{w}{g\qq,n'}\right)
= \sum_{n'} \des{w}{g\qq, n'} \tilde{D}^p_{n'n}(g, \qq),
\end{aligned}\end{equation}
where we have used the fact that $M_i=M_j$ holds for symmetry-related atoms. 

The real-space force constants $\Phi(\RR)$ transform as
\begin{equation}\begin{aligned}
\Phi_{i'\mu',j'\nu'}(\Delta\RR') =& \sum_{i\mu,j\nu} D^p_{i'\mu',i\mu}(g) \Phi_{i\mu,j\nu}(\Delta\RR) D_{j'\nu',j\nu}^{p*}(g) \delta_{\Delta\RR', R_g(\Delta\RR+\rr_j-\rr_i)-(\rr_{j'}-\rr_{i'})}.
\end{aligned}\end{equation}
The momentum-space force constants transform as 
\begin{equation}\begin{aligned}
\Phi_{i'\mu',j'\nu'}(g\qq) &= \sum_{i\mu,j\nu} D^p_{i'\mu',i\mu}(g) \Phi_{i\mu,j\nu}(\qq) D^{p,*}_{j'\nu',j\nu}(g).
\end{aligned}\end{equation}
Note that the $\qq$-dependence in $D^{p}_{j'\nu',j\nu}(g,\qq)$ is eliminated. The dynamical matrix transforms in the same way. Note that symmetry operations can only connect atoms with the same mass.

\paragraph{Electron-phonon coupling.} 
The EPC Hamiltonian in real-space operators transforms as
\begin{equation}\begin{aligned} 
g\hat{H}_{epc} g^{-1} =& \sum_{\RR,\RR_e,\RR_{p},i\alpha,j\beta,l,\mu} 
g_{\RR_e,\RR_{p}}^{i\alpha,j\beta,l\mu} 
\sum_{i'\alpha'} \cre{c}{g(\RR+\rr_i)-\rr_{i'},i'} D_{i'\alpha',i\alpha}(g) 
\sum_{j'\beta'} \des{c}{g(\RR+\RR_e+\rr_j)-\rr'_j,j'} D^*_{j'\beta',j\beta}(g) \\
& \sum_{l'\mu'} \des{u}{g(\RR+\RR_{p}+\rr_l)-\rr_l, l'\mu'} D^p_{l'\mu',l\mu}(g) \\
=& \sum_{\RR',\RR'_e,\RR'_p,i'\alpha',j'\beta',l'\mu'}
\left( \sum_{i\alpha,j\beta,l\mu} 
g_{\RR_e,\RR_p}^{i\alpha,j\beta,l\mu} D_{i'\alpha',i\alpha}(g) D_{j'\beta',j\beta}^*(g) D^p_{l'\mu',l\mu}(g) \right) 
\cre{c}{\RR',i'\alpha'} \des{c}{\RR'+\RR'_e,j'\beta'} \des{u}{\RR'+\RR'_p,l'\mu'} \\
& \delta_{\RR',g(\RR+\rr_i)-\rr_{i'}} 
\delta_{\RR'_e, R_g(\RR_e+\rr_j-\rr_i)-(\rr_{j'}-\rr_{i'})}
\delta_{\RR'_p, R_g(\RR_p+\rr_l-\rr_i)-(\rr_{l'}-\rr_{i'})} \\
=& \sum_{\RR',\RR'_e,\RR'_p,i'\alpha',j'\beta,l'\mu'} 
g_{\RR'_e,\RR'_p}^{i'\alpha',j'\beta',l'\mu'} 
\cre{c}{\RR',i'\alpha'} \des{c}{\RR'+\RR'_e,j'\beta'} \des{u}{\RR'+\RR'_p,l'\mu'}
= \hat{H}_{epc} \\
\end{aligned}\end{equation} 
Thus the real-space EPC tensor transforms as
\begin{equation}\begin{aligned}
g_{\RR'_e,\RR'_p}^{i'\alpha',j'\beta',l'\mu'}  =&
\sum_{i\alpha,j\beta,l\mu} g_{\RR_e,\RR_p}^{i\alpha,j\beta,l\mu} D_{i'\alpha',i\alpha}(g) D_{j'\beta',j\beta}^*(g) D^p_{l'\mu',l\mu}(g)  
\delta_{\RR'_e, R_g(\RR_e+\rr_j-\rr_i)-(\rr_{j'}-\rr_{i'})}
\delta_{\RR'_p, R_g(\RR_p+\rr_l-\rr_i)-(\rr_{l'}-\rr_{i'})}
\end{aligned}\end{equation}
The momentum-space EPC tensor transforms as 
\begin{equation}\begin{aligned}
g\hat{H}_{epc} g^{-1} =& 
\sum_{\kk,\qq,i'\alpha',j'\beta',l'\mu'} 
\left(\sum_{i\alpha,j\beta,l\mu,}
g^{i\alpha,j\beta,l\mu}_{\kk,\qq} D_{i'\alpha',i\alpha}(g,\kk+\qq) D^*_{j'\beta',j\beta}(g,\kk)
D^p_{l'\mu',l\mu}(g,\qq)
\right) \cre{c}{g(\kk+\qq),i'\alpha'} \des{c}{g\kk,j'\beta'} \des{u}{g\qq,l'\mu'} \\
=& \sum_{\kk,\qq,i'\alpha',j'\beta',l'\mu'} g^{i'\alpha',j'\beta',l'\mu'}_{g\kk, g\qq} \cre{c}{g(\kk+\qq),i'\alpha'} \des{c}{g\kk,j'\beta'} \des{u}{g\qq,l'\mu'} = \hat{H}_{epc} \\
\Rightarrow
g^{i'\alpha',j'\beta',l'\mu'}_{g\kk, g\qq} =& 
\sum_{i\alpha,j\beta,l\mu,}
g^{i\alpha,j\beta,l\mu}_{\kk,\qq} D_{i'\alpha',i\alpha}(g,\kk+\qq) D^*_{j'\beta',j\beta}(g,\kk)
D^p_{l'\mu',l\mu}(g,\qq) \\
&= \sum_{i\alpha,j\beta,l\mu,}
g^{i\alpha,j\beta,l\mu}_{\kk,\qq} D_{i'\alpha',i\alpha}(g) D^*_{j'\beta',j\beta}(g)
D^p_{l'\mu',l\mu}(g)
\label{app:eq:real-space-epc-symmetry-constraint}
\end{aligned}\end{equation} 
Note that the $\kk, \qq$ dependence in the representation matrices is eliminated. 
The EPC tensor in the band basis transforms as
\begin{equation}\begin{aligned}
g\hat{H}_{epc} g^{-1} =& 
\sum_{\kk,\qq,mns} \left(
\sum_{m'n's'} \tilde{G}^{mns}_{\kk,\qq} \tilde{D}_{m'm}(g,\kk+\qq) \tilde{D}^*_{n'n}(g,\kk) \tilde{D}^p_{s's}(g,\qq)
\right) 
\cre{\gamma}{g(\kk+\qq),m'}\des{\gamma}{g\kk,n'}\des{w}{g\qq,s'} \\
=& \sum_{\kk,\qq,m'n's'} \tilde{G}^{m'n's'}_{\kk,\qq} \cre{\gamma}{g(\kk+\qq),m'}\des{\gamma}{g\kk,n'}\des{w}{g\qq,s'} = \hat{H}_{epc} \\
\Rightarrow 
\tilde{G}^{m'n's'}_{g\kk,g\qq} =& 
\sum_{mns} \tilde{G}^{mns}_{\kk, \qq} \tilde{D}_{m'm}(g,\kk+\qq) \tilde{D}^*_{n'n}(g,\kk) \tilde{D}^p_{s's}(g,\qq),
\label{app:eq:symm_trans_epc}
\end{aligned}\end{equation}
where $\tilde{D}^{(p)}(g,\kk)$ is the electron (phonon) representation matrix in the band basis, as defined in \cref{app:eq:rep_mat_el_band_basis} and \cref{app:eq:rep_mat_ph_band_basis}. 
When restricted to certain sets of bands that form IRREPs, the symmetry transformation becomes
$\tilde{G}^{s_1\gamma_1', s_2\gamma_2',s_3\gamma_3'}_{g\kk, g\qq} = 
\sum_{\gamma_1\gamma_2\gamma_3} \tilde{G}^{s_1\gamma_1, s_2\gamma_2,s_3\gamma_3}_{\kk,\qq} \tilde{D}^{s_1}_{\gamma_1'\gamma_1}(g,\kk+\qq) \tilde{D}^{s_2,*}_{\gamma_2'\gamma_2}(g,\kk) \tilde{D}^{p,s_3}_{\gamma_3'\gamma_3}(g,\qq)$, where $s_i$ denotes degenerate set of bands, $\gamma_i$ is the band index in the set, and $\tilde{D}^{(p),s_i}(g,\kk)$ is the corresponding IRREP matrix.

\subsection{Selection rule of EPC}\label{app:sec:epc_selection_rule}

The electron and phonon operators in the band basis are characterized by the little group irreducible representations (IRREPs). As a result, the EPC in the band basis \cref{app:eq:epc_ham_band_basis} is subject to the selection rule, which enforces certain EPC matrix elements to be zero. 

Consider the electrons and phonons near the high-symmetry point (generic momenta in general do not enforce selection rules). We now discuss in which situations electron operators and phonon operators are symmetry-allowed to couple with each other. Suppose we are interested in the phonon with momentum $\QQ$, with little group $\mathcal{G}_{\QQ}$. For electrons at a specific momentum $\KK$, we first collect the electron momenta generated by $\mathcal{G}_{\QQ}$ from $\KK$, \ie, 
\begin{equation}\begin{aligned} 
S_{\QQ,\KK} = \{g \KK | g\in \mathcal{G}_{\QQ} \}.
\end{aligned}\end{equation} 
We now have a list of particle-hole operators that could couple to the phonon field from momentum conservation
\begin{equation}\begin{aligned} 
O_{\KK,\QQ} =\{ \gamma_{\kk+\QQ,n}^\dag \gamma_{\kk,m} |\kk\in S_{\QQ,\KK},n,m \in N_{band}\}
\end{aligned}\end{equation} 
where $N_{band}$ denotes the set of band indices we are interested in. 
These operators will form a (closed) representation space of $\mathcal{G}_{\QQ}$, as we show in the following. We first introduce the symmetry properties of the electron operators. For a given symmetry operation $g$, we let 
\begin{equation}\begin{aligned} 
g\gamma_{\kk,n}^\dag g^{-1} = \sum_{m}\gamma_{g\kk,m}^\dag D_{mn}(g,\kk)
\end{aligned}\end{equation} 
Then 
\begin{equation}\begin{aligned} 
g\gamma_{\kk+\QQ,n}^\dag \gamma_{\kk,m}g^{-1} = \sum_{n',m'}\gamma_{g(\kk+\QQ),n'}^\dag \gamma_{g\kk,m'}D_{n'n}(g,\kk+\QQ) D_{m'm}^*(g,\kk)
\label{eq:ph_transf_little_group}
\end{aligned}\end{equation} 
We now show that $O_{\KK,\QQ}$ indeed forms a representation space of $\mathcal{G}_{\QQ}$. For a given symmetry operation $g\in \mathcal{G}_{\QQ}$, we have 
$g\QQ \equiv \QQ$,  
where $\equiv$ is the equivalence up to the reciprocal lattice vector. Then according to~\cref{eq:ph_transf_little_group}, the operation $g$ will map an operator in $O_{\KK,\QQ}$ to another operator in the linear space spanned by $O_{\KK,\QQ}$, since $\kk \in S_{\QQ,\KK}\Rightarrow g\kk \in S_{\QQ,\KK}$. 

Thus we conclude that when the representation of $O_{\KK,\QQ}$, \ie, the direct-product representation $D(g,\KK+\QQ) \otimes D^*(g,\KK)$, contains the IRREPs of the phonon $D^{p}(g,\QQ)$, the EPC is symmetry-allowed. Otherwise, the EPC is enforced to zero. 
Equivalently, starting from the symmetry transformation given in \cref{app:eq:symm_trans_epc}, the EPC tensor component is nonzero only when the trivial representation is contained in the product, \ie,
\begin{equation}
    D(g,\KK+\QQ)\otimes D^*(g,\KK) \supset D^{p}(g,\QQ) 
    \quad \Leftrightarrow \quad D(g,\KK+\QQ)\otimes D^*(g,\KK)\otimes D^{p, *}(g,\QQ) \supset \bm{1}. 
    \label{app:eq:EPC_selection_rule}
\end{equation}
where $\bm{1}$ denotes the trivial representation. The equivalence in \cref{app:eq:EPC_selection_rule} can be proven as follows. Let $D^e(g)=D(g,\KK+\QQ)\otimes D^*(g,\KK)$. The first subset relation in \cref{app:eq:EPC_selection_rule} is satisfied if
$m(D^p) = \frac{1}{|\mathcal{G}_{\QQ}|} \sum_{g\in \mathcal{G}_{\QQ}} \chi_{D^e}(g) \chi^*_{D^p}(g) >0$. The second subset relation in \cref{app:eq:EPC_selection_rule} is satisfied if 
$m(D^1) = \frac{1}{|\mathcal{G}_{\QQ}|} \sum_{g\in \mathcal{G}_{\QQ}} \chi_{D^e}(g) \chi_{D^{p,*}}(g) >0$. As $\chi^*_{D^p}(g)= \chi_{D^{p,*}}(g)$, we have the same multiplicity $m(D^p) = m(D^1)$. Thus the equivalence in \cref{app:eq:EPC_selection_rule} holds. 
 
More specifically, we consider the following cases, where we assume there are bands near $E_f$ at $\KK$ and $\KK+\QQ$. 
\begin{itemize}
\item The simplest case is $\KK=\QQ=\Gamma$, then $O_{\KK,\QQ}=\{\cre{\gamma}{\Gamma,n}\des{\gamma}{\Gamma,m}\}$, and the EPC is allowed if 
$D(g,\Gamma)\otimes D^*(g,\Gamma)\otimes D^{p,*}(g,\Gamma) \supset \bm{1}$. 

\item If $\KK=\QQ$, then $\mathcal{G}_{\KK}=\mathcal{G}_{\QQ}\subset \mathcal{G}_{\KK+\QQ}$, where the subset relation holds because $\forall g\in \mathcal{G}_{\KK}$, $g(\KK+\QQ) = g\KK + g\QQ \equiv \KK+\QQ \Rightarrow g\in \mathcal{G}_{\KK+\QQ}$. 
As a result, $D(g,\KK+\QQ)\downarrow_{\KK}$ is a representation of $\KK$. As a result, the EPC is allowed if 
$D(g,\KK+\QQ)\downarrow_{\KK} \otimes D^*(g,\KK)\otimes D^{p,*}(g,\QQ) \supset \bm{1}$. 

\item If $\KK\neq \QQ$, and $\mathcal{G}_{\KK}\supset \mathcal{G}_{\QQ}$, then $|O_{\KK,\QQ}|=1$, and $\mathcal{G}_{\KK+\QQ}\supset\mathcal{G}_{\QQ}$. In this case, both $D(g,\KK+\QQ)\downarrow_{\QQ}$ and $D(g,\KK)\downarrow_{\QQ}$ are representations at $\QQ$. As a result, the representation $\overline{D}(g)$ of $O_{\KK,\QQ}$ has the form
\begin{equation}
 \overline{D}(g) = D(g,\KK+\QQ)\downarrow_{\QQ} \otimes D^*(g,\KK)\downarrow_{\QQ}, \quad  \forall g\in \mathcal{G}_{\QQ}.
\end{equation}

\item If $\KK\neq \QQ$, and $\mathcal{G}_{\KK}\subset \mathcal{G}_{\QQ}$, then $|O_{\KK,\QQ}|=|\mathcal{G}_{\QQ}/\mathcal{G}_{\KK}|$. Assume the representation of the basis in $O_{\KK,\QQ}$ is $\overline{D}(g)$. In this case, $\mathcal{G}_{\KK+\QQ}\supset \mathcal{G}_{\KK}$ (proof is the same as in the $\KK=\QQ$ case).  Thus $D(g,\KK+\QQ)\downarrow_{\KK}$ is a representation at $\KK$. 
We have
\begin{equation}\begin{aligned}
    \overline{D}(g) = D(g,\KK+\QQ)\downarrow_{\KK} \otimes D^*(g,\KK),\quad & \forall g\in \mathcal{G}_{\KK}, \\
    \text{Tr}(\overline{D}(g)) = 0,\quad  &\forall g\in \mathcal{G}_{\QQ}, g\notin \mathcal{G}_{\KK}.
\end{aligned}\end{equation}
The second equation follows from the fact that for $g\in \mathcal{G}_{\QQ}$ but $g\notin \mathcal{G}_{\KK}$, the matrix $\overline{D}(g)$ acts as a permutation among the particle-hole operators in $O_{\KK,\QQ}$, and has no nonzero diagonal elements. Consequently, $\text{Tr}(\overline{D}(g))=0$. 
\end{itemize}
All the above cases can be summarized into a general formula, with  $|O_{\KK,\QQ}|=|\mathcal{G}_{\QQ}/(\mathcal{G}_{\KK}\cap \mathcal{G}_{\QQ})|$, and $(\mathcal{G}_{\KK}\cap \mathcal{G}_{\QQ})\subset \mathcal{G}_{\KK+\QQ}$. 
As a result, the product representation $\overline{D}(g)$ is 
\begin{equation}\begin{aligned}
    \overline{D}(g) = D(g,\KK+\QQ) \otimes D^*(g,\KK),\quad & \forall g\in \mathcal{G}_{\KK}\cap \mathcal{G}_{\QQ}, \\
    \text{Tr}(\overline{D}(g)) = 0,\quad  &\forall g\in \mathcal{G}_{\QQ}, g\notin \mathcal{G}_{\KK}\cap \mathcal{G}_{\QQ}.
\end{aligned}\end{equation}

In \cref{app:sec:MgB2_epc_selection_rule}, we will give a detailed discussion on the selection rule in \ch{MgB2}. 

In the above EPC selection-rule analysis, we neglect spin–orbit coupling (SOC), so electronic states and phonons are classified by single-group irreducible representations (IRREPs). When SOC is included, the electronic states are labeled by double-group IRREPs. Phonons still carry no spinor degree of freedom and therefore transform as single-valued representations, but can be promoted into the double group by letting the spin $2\pi$ rotation act trivially. With this replacement of single-group by double-group IRREPs, the selection-rule analysis proceeds in the same way.

\subsection{Superconducting properties from EPC}\label{app:sec:SC_properties_formalism}

In this section, we discuss the superconducting properties computed from the EPC Hamiltonian. A more detailed study is left in Ref.~\cite{EPCpaper}. 

The phonon linewidth $\gamma_{\qq\nu}$ is given by the imaginary part of the phonon self-energy $\Pi_{\qq\nu}$ in the Migdal approximation 
\begin{equation}\begin{aligned}
\gamma_{\qq\nu} = \mathfrak{Im} \Pi_{\qq\nu} &= -2\mathfrak{Im} \sum_{mn} \int_{BZ} \frac{d\kk}{\Omega_{BZ}} |G^{mn\nu}_{\kk,\qq}|^2 \frac{f_{\kk,n} - f_{\kk+\qq,m}}{\epsilon_{\kk+\qq,m} - \epsilon_{\kk,n} - \omega_{\qq,\nu} + i\eta} \\
&= 2\pi \sum_{mn} \int_{BZ} \frac{d\kk}{\Omega_{BZ}} |G^{mn\nu}_{\kk,\qq}|^2  
(f_{\kk,n} - f_{\kk+\qq,m}) \delta(\epsilon_{\kk+\qq,m} - \epsilon_{\kk,n} - \omega_{\qq\nu}),
\label{app:eq:def_ph_linewidth}
\end{aligned}\end{equation}
where $f_{\kk,n}$ is the Fermi-Dirac function, and the factor of 2 accounts for two spin components ($\epsilon_{\kk n}$ assumed to be the spinless energies), and $\mathfrak{Im}$ denotes the imaginary part. The second line in \cref{app:eq:def_ph_linewidth} results from 
$\frac{1}{x+i\eta}=\mathcal{P}(\frac{1}{x})-i\pi \delta(x)$. 
The integration can be replaced by the discrete summation, \ie, $\int_{BZ} \frac{d\kk}{\Omega_{BZ}}=\frac{1}{N_e}\sum_{\kk\in BZ}$. 
The phonon linewidth characterizes the decay rate of the phonon due to the scattering of electrons (other scattering from phonons or impurities will also contribute to the total linewidth). 
Note that it is $G^{mn\nu}_{\kk,\qq}$ instead of $\tilde{G}^{mn\nu}_{\kk,\qq}$ that appears in the phonon linewidth (see definition in \cref{app:eq:epc_ham_band_basis}, with $|\tilde{G}^{mn\nu}_{\kk,\qq}|^2=|G^{mn\nu}_{\kk,\qq}|^2 \cdot 2\omega_{\qq\nu}/\hbar$). 
A detailed derivation of the phonon self-energy is given in \cref{app:sec:phonon-el-self-energy}. 

Under the double-delta approximation (see derivation in \cref{app:sec:double_delta}, which works at zero temperature and small phonon frequency), the phonon linewidth $\gamma_{\qq\nu}$ is expressed as 
\begin{equation}\begin{aligned}
\gamma_{\qq\nu} = 2\pi \omega_{\qq,\nu} \sum_{mn} \int_{BZ} \frac{d\kk}{\Omega_{BZ}} |G^{mn\nu}_{\kk,\qq}|^2 \delta(\epsilon_{\kk,n} - E_f) \delta(\epsilon_{\kk+\qq,m} - E_f)
\label{app:eq:gamma_double_delta}
\end{aligned}\end{equation}
The electron-phonon coupling strength $\lambda_{\qq\nu}$ of a specific phonon mode $\omega_{\qq\nu}$ is defined as (see derivation in \cref{app:sec:phonon-el-self-energy}): 
\begin{equation}\begin{aligned}
\lambda_{\qq\nu} &= \frac{2}{N_f \omega_{\qq,\nu}} \sum_{mn} \int_{BZ} \frac{d\kk}{\Omega_{BZ}} |G^{mn\nu}_{\kk,\qq}|^2 \delta(\epsilon_{\kk,n} - E_f) \delta(\epsilon_{\kk+\qq,m} - E_f) \\
&= \frac{\gamma_{\qq\nu}}{\pi N_f \omega_{\qq,\nu}^2}
\label{app:eq:lambda_qv_def}
\end{aligned}\end{equation}
where $N_f$ is the (spinful) DOS at $E_f$. $\lambda_{\qq\nu}$ is proportional to the phonon linewidth $\gamma_{\qq\nu} / \omega^2_{\qq,\nu}$.  

The total EPC strength $\lambda$ is the BZ average over the phonon-resolved $\lambda_{\qq\nu}$:
\begin{equation}\begin{aligned}
\lambda = \sum_{\nu} \int_{BZ} \frac{d\qq}{\Omega_{BZ}} \lambda_{\qq \nu}
\end{aligned}\end{equation}
The Eliashberg spectral function $\alpha^2F(\omega)$ is defined as 
\begin{equation}\begin{aligned}
\alpha^2F(\omega)= \frac{1}{2} \sum_{\nu} \int_{BZ} \frac{d\qq}{\Omega_{BZ}} \omega_{\qq\nu} \lambda_{\qq\nu} \delta(\omega - \omega_{\qq\nu}).
\end{aligned}\end{equation}
An equivalent definition of total EPC $\lambda$ is
based on $\alpha^2F(\omega$:
\begin{equation}\begin{aligned}
\lambda = 2 \int_0^\infty d\omega \frac{\alpha^2F(\omega)}{\omega}
\label{app:eq:lambda_from_a2F}
\end{aligned}\end{equation}

The superconducting transition temperature $T_c$ in the Allen-Dynes modified McMillan equation is
\begin{equation}\begin{aligned}
T_c = \frac{\omega_{\log}}{1.2} \exp\left[\frac{-1.04(1+\lambda)}{\lambda (1 - 0.62\mu^*) - \mu^*}\right]
\label{app:eq: McMillian-Tc}
\end{aligned}\end{equation}
where $\omega_{\log}$ is the logarithmic averaged phonon frequency defined as 
\begin{equation}\begin{aligned}
\omega_{\log} = \exp\left[\frac{2}{\lambda}\int \frac{d\omega}{\omega} \alpha^2 F(\omega) \log\omega \right].
\end{aligned}\end{equation}
$\mu^*$ is an empirical parameter that describes the Coulomb screening, which has typical values between 0.1 and 0.16. Smaller $\mu^*$ means a stronger Coulomb screening, thus leading to a higher $T_c$. 
$\lambda$ and $\alpha^2F(\omega)$ are dimensionless. 
$\omega_{\log}$ has the same unit as $\omega$.

\paragraph{Approximated formula for $\lambda$}
We derive a useful formula that expresses $\lambda$ using FS-averaged EPC. Denote
\begin{equation}
    \langle \omega^2 \rangle \equiv \frac{
    \int_0^{\infty} d\omega\ \omega \alpha^2F(\omega)
    }{\int_0^{\infty} d\omega\ \frac{1}{\omega}\alpha^2F(\omega)},
\end{equation}
Then the total EPC constant
\begin{equation}
\begin{aligned}
    \lambda &= \frac{2}{\langle \omega^2 \rangle} \int_0^{\infty} d\omega\ \omega \alpha^2F(\omega) \\
    &= \frac{1}{\langle \omega^2 \rangle}  \sum_{\nu} \int_{BZ} \frac{d\qq}{\Omega_{BZ}} \omega_{\qq\nu} \lambda_{\qq\nu} \int_0^{\infty} d\omega\ \omega 
    \delta(\omega - \omega_{\qq\nu}) \\
    &= \frac{1}{\langle \omega^2 \rangle}  \sum_{\nu} \int_{BZ} \frac{d\qq}{\Omega_{BZ}} \omega_{\qq\nu}^2 \lambda_{\qq\nu}  \\
    &= \frac{2}{\langle \omega^2 \rangle N_f}  \sum_{\nu} \int_{BZ} \frac{d\qq}{\Omega_{BZ}} \omega_{\qq\nu} \sum_{mn} \int_{BZ} \frac{d\kk}{\Omega_{BZ}} |G^{mn\nu}_{\kk,\qq}|^2 \delta(\epsilon_{\kk,n} - E_f) \delta(\epsilon_{\kk+\qq,m} - E_f) \\
    &= \frac{\hbar}{\langle \omega^2 \rangle N_f}  \sum_{mn\nu} \int_{BZ} \frac{d\kk}{\Omega_{BZ}} \int_{BZ} \frac{d\kk'}{\Omega_{BZ}}  \frac{2\omega_{\qq \nu}}{\hbar} |G^{mn\nu}_{\kk,\qq}|^2 \delta(\epsilon_{\kk,n} - E_f) \delta(\epsilon_{\kk',m} - E_f) \\
    &= \frac{\hbar}{\langle \omega^2 \rangle N_f}  \sum_{mn\nu} \int_{BZ} \frac{d\qq}{\Omega_{BZ}} \int_{BZ} \frac{d\kk}{\Omega_{BZ}} |\tilde{G}^{mn\nu}_{\kk,\qq}|^2 \delta(\epsilon_{\kk,n} - E_f) \delta(\epsilon_{\kk+\qq,m} - E_f) \\
    &= \frac{\hbar}{\langle \omega^2 \rangle N_f}  \sum_{mn\nu} \int_{BZ} \frac{d\kk}{\Omega_{BZ}} \int_{BZ} \frac{d\kk'}{\Omega_{BZ}}  |\tilde{G}^{mn\nu}_{\kk,\kk'-\kk}|^2 \delta(\epsilon_{\kk,n} - E_f) \delta(\epsilon_{\kk',m} - E_f) \quad (\text{let} \kk'=\kk+\qq) 
\end{aligned}
\end{equation}
Denote the FS-pair averaged EPC as
\begin{equation}
    \langle |\tilde{G}|^2\rangle_{\text{FS-pair}} = \frac{1}{N_f^2} \sum_{mn\nu} \int_{BZ} \frac{d\kk}{\Omega_{BZ}} \int_{BZ} \frac{d\kk'}{\Omega_{BZ}}  |\tilde{G}^{mn\nu}_{\kk,\kk'-\kk}|^2 \delta(\epsilon_{\kk,n} - E_f) \delta(\epsilon_{\kk',m} - E_f).
\end{equation}
We have the expression~\cite{mcmillan1968transition, hopfield1969angular}: 
\begin{equation}
    \lambda = \frac{\hbar N_f}{\langle \omega^2 \rangle} \langle |\tilde{G}|^2\rangle_{\text{FS-pair}}. 
    \label{app:eq:lambda_FSpair}
\end{equation}

We then simplify \cref{app:eq:lambda_FSpair}. Assume (i) there is only a single FS $m$ and a single phonon branch $\nu_0$ with small momentum $\qq$ that is relevant to SC, (ii) EPC  $G^{mm\nu_0}_{\kk,\qq}$ is smooth for small $\qq ~(=\kk'-\kk)$. Then we have the approximation
\begin{equation}\begin{aligned}
    |\tilde{G}^{mn\nu}_{\kk,\kk'-\kk}|^2 &\approx |\tilde{G}_{\kk,\bm{0}}^{mm\nu_0}|^2 \equiv  |\tilde{G}_{\kk}|^2, \\
    \Rightarrow \quad
    \langle |\tilde{G}|^2\rangle_{\text{FS-pair}} &\approx \frac{1}{N_f} \int_{BZ} \frac{d\kk}{\Omega_{BZ}} |\tilde{G}_{\kk}|^2 \delta(\epsilon_{\kk,m} - E_f) \equiv \langle |\tilde{G}|^2\rangle_{\text{FS}}, 
\end{aligned}\end{equation}
which leads to the approximated formula for $\lambda$:
\begin{equation}
    \lambda \approx \frac{\hbar N_f}{\langle \omega^2 \rangle} \langle |\tilde{G}|^2\rangle_{\text{FS}}. 
     \label{app:eq:lambda_FS_approx}
\end{equation}
Let the frequency of phonon mode $\nu_0$ be $\omega_0$. \cref{app:eq:lambda_FS_approx} can be further massaged (using the definition of $\tilde{G}$) into
$\lambda \approx \frac{2N_f}{\omega_0} \langle |G|^2\rangle_{\text{FS}}$.

\subsubsection{Double-delta approximation of phonon linewidth}\label{app:sec:double_delta}

The double-delta approximation is a low-temperature limit for the phonon linewidth that assumes the phonon frequency $\omega_{\qq\nu}$ is small compared with the electronic energy scale over which the joint density of states and EPC matrix elements vary near the Fermi level. It is usually valid for ordinary metals, but can fail in narrow-band, small-gap, or strongly anharmonic systems. 
With no loss of generality, we assume $E_f=0$ and omit the band indices. Let 
\begin{equation}\begin{aligned}
I = \frac{f_{\kk} - f_{\kk+\qq}}{\epsilon_{\kk+\qq} - \epsilon_{\kk} - \omega_{\qq} + i\eta}.
\end{aligned}\end{equation}
Let $\Delta\epsilon=\epsilon_{\kk+\qq} - \epsilon_{\kk}$, and $\overline{\epsilon}=\frac{1}{2}(\epsilon_{\kk}+\epsilon_{\kk+\qq})$. 
The Fermi-Dirac functions can be approximated as  
\begin{equation}\begin{aligned}
f_{\kk} - f_{\kk+\qq} &= f\left(\overline{\epsilon}-\frac{\Delta\epsilon}{2}\right) - f\left(\overline{\epsilon} +\frac{\Delta\epsilon}{2}\right) \\
&= -\Delta \epsilon f'(\overline{\epsilon}) - \frac{(\Delta\epsilon)^3}{24} f^{(3)}(\overline{\epsilon}) + \mathcal{O}(\Delta\epsilon^5) \\
&\approx \Delta \epsilon \delta(\epsilon_{\bm{k}}),
\end{aligned}\end{equation}
where we assume $T\rightarrow 0$ so that $-f'(\epsilon)\approx\delta(\epsilon)$.  
Then
\begin{equation}\begin{aligned}
I &\approx \frac{\Delta \epsilon}{\Delta \epsilon - \omega_{\qq}+i\eta} \delta(\epsilon_{\kk}) 
=\left( \frac{\Delta \epsilon-\omega_{\qq}}{\Delta \epsilon - \omega_{\qq}+i\eta} + \frac{\omega_{\qq}}{\Delta \epsilon - \omega_{\qq}+i\eta}\right) \delta(\epsilon_{\kk}).
\end{aligned}\end{equation}
The imaginary part of the first term vanishes when $\eta\rightarrow 0$. Thus  
\begin{equation}\begin{aligned}
\mathfrak{Im} I &\approx -\pi \omega_{\qq} \delta(\Delta \epsilon-\omega_{\qq}) \delta(\epsilon_{\kk}) \\
&\approx -\pi \omega_{\qq} \delta(\epsilon_{\kk+\qq}) \delta(\epsilon_{\kk}).
\label{app:eq:double_delta_formula}
\end{aligned}\end{equation}
The second approximation follows when $\omega_{\qq}$ is negligible. 

We further derive a more symmetric formula by assuming $\Delta\epsilon=\omega_{\qq}$. Under this condition, we have 
\begin{equation}\begin{aligned}
f_{\kk} - f_{\kk+\qq} &= 
-\Delta\epsilon \int_{-\frac{1}{2}}^{\frac{1}{2}} f'(\overline{\epsilon } + s \omega_{\qq})ds \\
&\approx -\frac{\Delta\epsilon}{2} \left[
f'(\overline{\epsilon} + \omega_{\qq}/2) + f'(\overline{\epsilon} - \omega_{\qq}/2)
\right] \\
&\approx \frac{\Delta\epsilon}{2} \left[\delta(\overline{\epsilon} + \omega_{\qq}/2) + \delta(\overline{\epsilon} - \omega_{\qq}/2) \right].
\end{aligned}\end{equation}
Thus we have
\begin{equation}\begin{aligned}
\mathfrak{Im} I &\approx -\frac{\pi \omega_{\qq}}{2} \delta(\Delta \epsilon-\omega_{\qq}) 
\left[\delta(\overline{\epsilon} + \frac{\omega_{\qq}}{2}) + \delta(\overline{\epsilon} - \frac{\omega_{\qq}}{2}) \right].
\\
&\approx -\frac{\pi \omega_{\qq}}{2} 
\left[
\delta(\epsilon_{\kk} + \frac{\omega_{\qq}}{2}) \delta(\epsilon_{\kk+\qq}- \frac{\omega_{\qq}}{2}) + 
\delta(\epsilon_{\kk} - \frac{\omega_{\qq}}{2}) \delta(\epsilon_{\kk+\qq} + \frac{\omega_{\qq}}{2}) 
\right].
\end{aligned}\end{equation}
The same \cref{app:eq:double_delta_formula} can be obtained by setting $\omega_{\qq}\rightarrow 0$. 

By recovering the band indices, we arrive at the phonon linewidth expression \cref{app:eq:gamma_double_delta} using the double-delta approximation.

\subsubsection{Phonon and electron self-energy from EPC}\label{app:sec:phonon-el-self-energy}

\paragraph{Phonon self-energy from EPC.}
We define the Matsubara phonon Green's function as
\begin{equation}
D_\nu(\mathbf q,\tau)
= -\left\langle
T_\tau Q_{\mathbf q\nu}(\tau)Q_{-\mathbf q\nu}(0) \right\rangle ,
\end{equation}
where
\begin{equation}
Q_{\mathbf q\nu}=b_{\mathbf q\nu}+b^\dagger_{-\mathbf q\nu},
\end{equation}
the imaginary-time Heisenberg operator is
$\hat O(\tau)=e^{\tau H_0}\hat O e^{-\tau H_0}$ with $H_0$ being the free phonon Hamiltonian, and the thermal average is
$\langle \hat A\rangle
= \frac{\mathrm{Tr}\left(e^{-\beta H}\hat A\right)}{\mathrm{Tr}\left(e^{-\beta H}\right)}$. 

For the free phonon Hamiltonian defined in \cref{app:eq:free_phonon_ham}, one has
\begin{equation}
Q_{\mathbf q\nu}(\tau)
=
b_{\mathbf q\nu}e^{-\omega_{\mathbf q\nu}\tau}
+
b^\dagger_{-\mathbf q\nu}e^{\omega_{\mathbf q\nu}\tau}.
\end{equation}
Therefore, for $0<\tau<\beta$,
\begin{equation}
D_{0,\nu}(\mathbf q,\tau)
=
-\left[
(n_{\mathbf q\nu}+1)e^{-\omega_{\mathbf q\nu}\tau}
+
n_{\mathbf q\nu}e^{\omega_{\mathbf q\nu}\tau}
\right],
\end{equation}
where $n_{\mathbf q\nu}=n_B(\omega_{\mathbf q\nu})$ is the Bose distribution. Its Fourier transform is
\begin{equation}
D_{0,\nu}(\mathbf q,i\Omega_l)
=
\frac{1}{i\Omega_l-\omega_{\mathbf q\nu}}
-
\frac{1}{i\Omega_l+\omega_{\mathbf q\nu}}
=
\frac{2\omega_{\mathbf q\nu}}
{(i\Omega_l)^2-\omega_{\mathbf q\nu}^2},
\end{equation}
with bosonic Matsubara frequency $i\Omega_l=2\pi l/\beta$.

The electron-phonon coupling Hamiltonian is written as
\begin{equation}
\hat{H}_{epc}
=
\frac{1}{\sqrt{N_p}}
\sum_{\mathbf k\mathbf q mn\nu\sigma}
g^{mn\nu}_{\mathbf k,\mathbf q}
\gamma^\dagger_{\mathbf k+\mathbf q,m\sigma}
\des{\gamma}{\kk n\sigma}
Q_{\mathbf q\nu}.
\end{equation}
Note that, in this section, we use the notation $g^{mn\nu}_{\mathbf k,\mathbf q}$ for the EPC tensor in the band basis, instead of $G^{mn\nu}_{\mathbf k,\mathbf q}$ in \cref{app:eq:epc_ham_band_basis}, in order to avoid confusion with the Green’s function. 
It is useful to define
\begin{equation}
\rho_{\mathbf q\nu} = \frac{1}{\sqrt{N_p}} \sum_{\mathbf k mn\sigma} g^{mn\nu}_{\mathbf k,\mathbf q} \gamma^\dagger_{\mathbf k+\mathbf q,m\sigma} \des{\gamma}{\kk n\sigma},
\end{equation}
so that
\begin{equation}
\hat{H}_{epc}(\tau)
=
\sum_{\mathbf q\nu}
Q_{\mathbf q\nu}(\tau)\rho_{\mathbf q\nu}(\tau).
\end{equation}
Hermiticity of $\hat{H}_{epc}$ implies
$g^{nm\nu}_{\mathbf k+\mathbf q,-\mathbf q} =\left( g^{mn\nu}_{\mathbf k,\mathbf q} \right)^*$, 
and hence $\rho_{-\mathbf q\nu}=\rho^\dagger_{\mathbf q\nu}$. 

The interacting phonon Green's function is
\begin{equation}
D_\nu(\mathbf q,\tau)
=
-
\frac{
\left\langle
T_\tau
Q_{\mathbf q\nu}(\tau)
Q_{-\mathbf q\nu}(0)
e^{-S_{epc}}
\right\rangle_0
}{
\left\langle T_\tau e^{-S_{epc}}\right\rangle_0
},
\end{equation}
where
\begin{equation}
S_{epc}
=
\int_0^\beta d\tau\, \hat{H}_{epc}(\tau).
\end{equation}
Expanding $e^{-S_{epc}}$ gives
\begin{equation}
e^{-S_{epc}}
=
1-S_{epc}+\frac{1}{2}S_{epc}^2+\cdots.
\label{app:eq:expand_S_epc}
\end{equation}
The first-order term contains an odd number of phonon operators and vanishes for the usual equilibrium reference state. The connected second-order correction is therefore
\begin{equation}
D^{(2)}_\nu(\mathbf q,\tau)
=
-\frac{1}{2}
\int_0^\beta d\tau_1
\int_0^\beta d\tau_2
\left\langle
T_\tau
Q_{\mathbf q\nu}(\tau)
Q_{-\mathbf q\nu}(0)
\hat{H}_{epc}(\tau_1)
\hat{H}_{epc}(\tau_2)
\right\rangle_{0,c}.
\end{equation}

Using $\hat{H}_{epc}=\sum_{\mathbf p\lambda}Q_{\mathbf p\lambda}\rho_{\mathbf p\lambda}$, the relevant free phonon contractions are
\begin{equation}
\left\langle
T_\tau Q_{\mathbf q\nu}(\tau)Q_{\mathbf p\lambda}(\tau_1)
\right\rangle_0
=
-\delta_{\mathbf p,-\mathbf q}\delta_{\lambda\nu}
D_{0,\nu}(\mathbf q,\tau-\tau_1),
\end{equation}
and
\begin{equation}
\left\langle
T_\tau Q_{-\mathbf q\nu}(0)Q_{\mathbf p\lambda}(\tau_2)
\right\rangle_0
=
-\delta_{\mathbf p,\mathbf q}\delta_{\lambda\nu}
D_{0,\nu}(\mathbf q,\tau_2).
\end{equation}
There are two equivalent contractions between the two external phonon fields and the two internal phonon fields, which cancel the factor $1/2$. Since each phonon contraction contributes a minus sign, their product gives $D_0D_0$. Thus
\begin{equation}
D^{(2)}_\nu(\mathbf q,\tau)
=
-
\int_0^\beta d\tau_1
\int_0^\beta d\tau_2\,
D_{0,\nu}(\mathbf q,\tau-\tau_1)
D_{0,\nu}(\mathbf q,\tau_2)
\left\langle
T_\tau
\rho_{-\mathbf q\nu}(\tau_1)
\rho_{\mathbf q\nu}(\tau_2)
\right\rangle_{0,c}.
\end{equation}

The remaining connected average is the electron bubble. Using Wick's theorem,
\begin{equation}
\begin{aligned}
&
\left\langle
T_\tau
\rho_{-\mathbf q\nu}(\tau_1)
\rho_{\mathbf q\nu}(\tau_2)
\right\rangle_{0,c}
=
-\frac{2}{N_p}
\sum_{\mathbf k mn}
|g^{mn\nu}_{\mathbf k,\mathbf q}|^2
G_{0,m}(\mathbf k+\mathbf q,\tau_1-\tau_2)
G_{0,n}(\mathbf k,\tau_2-\tau_1).
\end{aligned}
\end{equation}
The minus sign is the closed-fermion-loop sign, and the factor $2$ comes from spin degeneracy for a spin-independent, nonmagnetic system.

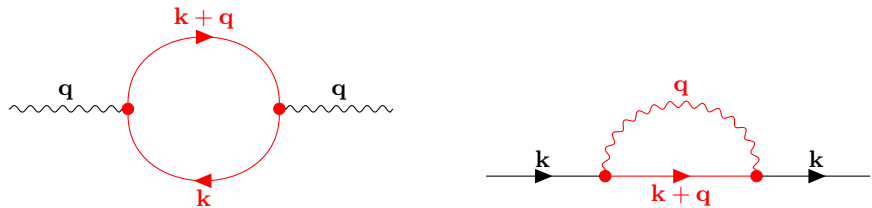
\begin{figure}[h!]
    \centering
    \begin{tikzpicture}
      \begin{feynman}
        \vertex (a);
        \vertex [right=1.5cm of a, dot, red] (b) {};
        \vertex [right=2.0cm of b, dot, red] (c) {};
        \vertex [right=1.5cm of c] (d);

        \diagram* {
          (a) -- [photon, edge label=$\mathbf{q}$] (b),
          (b) -- [fermion, half left, red, edge label=$\mathbf{k}+\mathbf{q}$] (c),
          (c) -- [fermion, half left, red, edge label=$\mathbf{k}$] (b),
          (c) -- [photon, edge label=$\mathbf{q}$] (d),
        };
      \end{feynman}
    \end{tikzpicture}
    \hspace{1cm}
    \begin{tikzpicture}
      \begin{feynman}
        \vertex (a);
        \vertex [right=1.5cm of a, dot, red] (b) {};
        \vertex [right=2.0cm of b, dot, red] (c) {};
        \vertex [right=1.5cm of c] (d);

        \diagram* {
          (a) -- [fermion, edge label=$\mathbf{k}$] (b),
          (b) -- [fermion, red, edge label'=$\mathbf{k}+\mathbf{q}$] (c),
          (c) -- [fermion, edge label=$\mathbf{k}$] (d),
          (b) -- [photon, half left, red, edge label=$\mathbf{q}$] (c),
        };
      \end{feynman}
    \end{tikzpicture}
    \caption{Feynman diagrams for phonon self-energy (left) and electron self-energy (right) from EPC.}
    \label{app:fig:self_energies_diagram}
\end{figure}

From the Dyson equation with intra-band self-energy 
\begin{equation}
D_\nu^{-1}(\mathbf q,i\Omega_l)
=
D_{0,\nu}^{-1}(\mathbf q,i\Omega_l)
-
\Pi_{\mathbf q\nu}(i\Omega_l), 
\end{equation}
the second-order contribution can be written as
\begin{equation}
D^{(2)}_\nu(\mathbf q,\tau)
=
\int_0^\beta d\tau_1
\int_0^\beta d\tau_2\,
D_{0,\nu}(\mathbf q,\tau-\tau_1)
\Pi_{\mathbf q\nu}(\tau_1-\tau_2)
D_{0,\nu}(\mathbf q,\tau_2).
\end{equation}
Therefore, we can calculate the phonon self-energy via 
\begin{equation}
\Pi_{\mathbf q\nu}(\tau_1-\tau_2)
=
-
\left\langle
T_\tau
\rho_{-\mathbf q\nu}(\tau_1)
\rho_{\mathbf q\nu}(\tau_2)
\right\rangle_{0,c}.
\end{equation}
With the Fourier convention
\begin{equation}
A(\tau)=\frac{1}{\beta}\sum_{i\Omega_l}
e^{-i\Omega_l\tau}A(i\Omega_l),
\end{equation}
the convolution becomes
\begin{equation}
D^{(2)}_\nu(\mathbf q,i\Omega_l)
=
D_{0,\nu}(\mathbf q,i\Omega_l)
\Pi_{\mathbf q\nu}(i\Omega_l)
D_{0,\nu}(\mathbf q,i\Omega_l).
\end{equation}
Consequently, we arrive at the phonon self-energy (see \cref{app:fig:self_energies_diagram})
\begin{equation}
\Pi_{\mathbf q\nu}(i\Omega_l)
=
\frac{2}{\beta N_p}
\sum_{i\omega_n}
\sum_{\mathbf k mn}
|g^{mn\nu}_{\mathbf k,\mathbf q}|^2
G_{0,m}(\mathbf k+\mathbf q,i\omega_n+i\Omega_l)
G_{0,n}(\mathbf k,i\omega_n).
\end{equation}
Here $i\omega_n=(2n+1)\pi/\beta$ is a fermionic Matsubara frequency. For bare electrons with the noninteracting electron Hamiltonian
$\hat{H}_e^0 =
\sum_{\mathbf k n\sigma}
\xi_{\kk n} \cre{\gamma}{\kk n\sigma} \des{\gamma}{\kk n\sigma}$, 
\begin{equation}
G_{0,n}(\mathbf k,i\omega_n)
=
\frac{1}{i\omega_n-\xi_{\mathbf k n}},
\qquad
\xi_{\mathbf k n}=\epsilon_{\kk n}- E_F.
\label{app:eq:bare_electron_green_func}
\end{equation}
The standard Matsubara sum gives
\begin{equation}
\frac{1}{\beta}
\sum_{i\omega_n}
\frac{1}{i\omega_n-E_1}
\frac{1}{i\omega_n+i\Omega_l-E_2}
=
\frac{f(E_1)-f(E_2)}
{i\Omega_l+E_1-E_2},
\end{equation}
where $f(E)$ is the Fermi-Dirac distribution. Hence
\begin{equation}
\Pi_{\mathbf q\nu}(i\Omega_l)
=
\frac{2}{N_p}
\sum_{\mathbf k mn}
|g^{mn\nu}_{\mathbf k,\mathbf q}|^2
\frac{
f(\xi_{\mathbf k n})
-
f(\xi_{\mathbf k+\mathbf q,m})
}{
i\Omega_l + \xi_{\mathbf k n} - \xi_{\mathbf k+\mathbf q,m}}.
\end{equation}
The retarded phonon self-energy is obtained by analytic continuation, $\Pi^R_{\mathbf q\nu}(\omega) = \Pi_{\mathbf q\nu}(i\Omega_l\rightarrow \omega+i0^+)$, and evaluating this expression on shell at $\omega=\omega_{\qq\nu}$ leading to the phonon self-energy expression in \cref{app:eq:def_ph_linewidth}.  Note that in the thermodynamic limit, one has 
$\frac{1}{N_p}\sum_{\kk} \rightarrow \int_{BZ} \frac{d^3 k}{\Omega_{BZ}}$.

\paragraph{Electron self-energy from EPC.}

We now derive the electron self-energy generated by the same EPC. The Matsubara electron Green's function is defined as
\begin{equation}
G_{n\sigma}(\mathbf k,\tau) = - \left\langle T_\tau
\des{\gamma}{\kk n\sigma}(\tau)
\cre{\gamma}{\kk n\sigma}(0)
\right\rangle .
\end{equation} 
Under the EPC Hamiltonian $\hat{H}_{epc}$, the interacting electron Green's function can be written as
\begin{equation}
G_{n\sigma}(\mathbf k,\tau)
=
-
\frac{
\left\langle
T_\tau
\des{\gamma}{\kk n\sigma}(\tau)
\cre{\gamma}{\kk n\sigma}(0)
e^{-S_{epc}}
\right\rangle_0
}{
\left\langle T_\tau e^{-S_{epc}}\right\rangle_0
},
\end{equation}
We then perform the same expansion as in \cref{app:eq:expand_S_epc}. The first-order term vanishes because it contains an odd number of phonon operators. The leading nonzero contribution is therefore second order:
\begin{equation}
G^{(2)}_{n\sigma}(\mathbf k,\tau)
=
-\frac{1}{2}
\int_0^\beta d\tau_1
\int_0^\beta d\tau_2
\left\langle
T_\tau
\des{\gamma}{\kk n\sigma}(\tau)
\cre{\gamma}{\kk n\sigma}(0)
\hat{H}_{epc}(\tau_1)
\hat{H}_{epc}(\tau_2)
\right\rangle_{0,c}.
\end{equation}
The two equivalent orderings of the interaction vertices cancel the factor $1/2$. 
From the Dyson convention, 
$G^{-1}_{n}(\mathbf k,i\omega_j)
= G^{-1}_{0,n}(\mathbf k,i\omega_j) - \Sigma_{n}(\mathbf k,i\omega_j)$ with intra-band self-energy, the second-order correction can be written as
\begin{equation}
G^{(2)}_{n}(\mathbf k,\tau)
= \int_0^\beta d\tau_1
\int_0^\beta d\tau_2\,
G_{0,n}(\mathbf k,\tau-\tau_1)
\Sigma_{n}(\mathbf k,\tau_1-\tau_2)
G_{0,n}(\mathbf k,\tau_2),
\end{equation}
where the electron self-energy is
\begin{equation}
\Sigma_{n}(\mathbf k,\tau_1-\tau_2)
= - \frac{1}{N_p}\sum_{\mathbf q\nu m}
\left|g^{mn\nu}_{\mathbf k,\mathbf q}\right|^2
D_{0,\nu}(\mathbf q,\tau_1-\tau_2)
G_{0,m}(\mathbf k+\mathbf q,\tau_1-\tau_2).
\end{equation} 
Fourier transforming gives the Matsubara electron self-energy
\begin{equation}
\Sigma_{n}(\mathbf k,i\omega_j)
= - \frac{1}{\beta N_p}
\sum_{\mathbf q\nu m}
\sum_{i\Omega_l}
\left|g^{mn\nu}_{\mathbf k,\mathbf q}\right|^2
D_{0,\nu}(\mathbf q,i\Omega_l)
G_{0,m}(\mathbf k+\mathbf q,i\omega_j-i\Omega_l).
\end{equation}
Since $D_{0,\nu}(\mathbf q,i\Omega_l)$ is even under
$i\Omega_l\rightarrow -i\Omega_l$, one may equivalently write the internal free  electron Green's function as
$G_{0,m}(\mathbf k+\mathbf q,i\omega_j+i\Omega_l)$. The overall minus sign in this expression follows from the convention
$D_0=-\langle T_\tau QQ\rangle_0$. 

We then evaluate the bosonic Matsubara sum
\begin{equation}
\frac{1}{\beta} \sum_{i\Omega_l} \left[
\frac{1}{i\Omega_l-\omega}
-
\frac{1}{i\Omega_l+\omega}
\right] \frac{1}{i\omega_j-i\Omega_l-\epsilon}.
\end{equation}
Let $A=i\omega_j-\epsilon$. For the first term,
\begin{equation}
\frac{1}{\beta} \sum_{i\Omega_l} \frac{1}{(i\Omega_l-\omega)(A-i\Omega_l)}
=
\frac{1}{\beta} \sum_{i\Omega_l} \frac{1}{A-\omega}
\left[
\frac{1}{i\Omega_l-\omega}
+
\frac{1}{A-i\Omega_l} 
\right] 
=
- \frac{n_B(\omega)+1-f(\epsilon)}{
i\omega_j-\epsilon-\omega}, 
\end{equation}
where we used
$\frac{1}{\beta} \sum_{i\Omega_l} \frac{1}{i\Omega_l-\omega}
= -n_B(\omega)$, and 
$n_B(i\omega_j-\epsilon) = f(\epsilon)-1$. 
Similarly, the second term gives
\begin{equation}
-\frac{1}{\beta}
\sum_{i\Omega_l}
\frac{1}{(i\Omega_l+\omega)(A-i\Omega_l)}
= - \frac{
n_B(\omega)+f(\epsilon)
}{
i\omega_j-\epsilon+\omega
}.
\end{equation}
Therefore
\begin{equation}
\frac{1}{\beta}
\sum_{i\Omega_l}
\left[
\frac{1}{i\Omega_l-\omega}
-
\frac{1}{i\Omega_l+\omega}
\right]
\frac{1}{i\omega_j-i\Omega_l-\epsilon}
=
-
\frac{
n_B(\omega)+1-f(\epsilon)
}{
i\omega_j-\epsilon-\omega
}
-
\frac{
n_B(\omega)+f(\epsilon)
}{
i\omega_j-\epsilon+\omega
}.
\end{equation}
The minus sign from this Matsubara sum cancels the explicit minus sign in the self-energy expression. Hence
\begin{equation}
\begin{aligned}
\Sigma_{n}(\mathbf k,i\omega_j)
=
\frac{1}{N_p} \sum_{\mathbf q\nu m}
\left|g^{mn\nu}_{\mathbf k,\mathbf q}\right|^2
\Bigg[
&
\frac{
n_{\mathbf q\nu}+1-f_{m,\mathbf k+\mathbf q}
}{
i\omega_j-\xi_{m,\mathbf k+\mathbf q}
-\omega_{\mathbf q\nu}
} + 
\frac{
n_{\mathbf q\nu}+f_{m,\mathbf k+\mathbf q}
}{
i\omega_j-\xi_{m,\mathbf k+\mathbf q}
+\omega_{\mathbf q\nu}
}
\Bigg],
\end{aligned}
\label{app:eq:el_self_energy_matsubara}
\end{equation}
where
$n_{\mathbf q\nu}=n_B(\omega_{\mathbf q\nu})$, $f_{m,\mathbf k+\mathbf q}=f(\xi_{m,\mathbf k+\mathbf q})$. 
The retarded electron self-energy $\Sigma^R_{n}(\mathbf k,\omega)$ is obtained by analytic continuation $i\omega_j\rightarrow \omega+i0^+$, and evaluate it on shell gives
$\Sigma_{n\mathbf k}
\equiv
\Sigma^R_n(\mathbf k,\xi_{n\mathbf k})$, \ie, 
\begin{equation}
\begin{aligned}
\Sigma_{n\mathbf k}
= \frac{1}{N_p} \sum_{\mathbf q\nu m}
\left|g^{mn\nu}_{\mathbf k,\mathbf q}\right|^2
\Bigg[
&
\frac{
n_{\mathbf q\nu}+1-f_{m,\mathbf k+\mathbf q}
}{
\xi_{n\mathbf k}
-\xi_{m,\mathbf k+\mathbf q}
-\omega_{\mathbf q\nu}
+i0^+} +
\frac{
n_{\mathbf q\nu}+f_{m,\mathbf k+\mathbf q}
}{
\xi_{n\mathbf k}
-\xi_{m,\mathbf k+\mathbf q}
+\omega_{\mathbf q\nu}
+i0^+
}
\Bigg].
\end{aligned}
\label{app:eq:el_self_energy}
\end{equation}
Using $\frac{1}{x+i0^+} = \mathcal P\frac{1}{x} - i\pi\delta(x)$, the imaginary part is 
\begin{equation}
\begin{aligned}
\mathfrak{Im} \Sigma_{n\mathbf k}
=
-\pi \frac{1}{N_p}
\sum_{\mathbf q\nu m} \left|g^{mn\nu}_{\mathbf k,\mathbf q}\right|^2
\Big[ \left(
n_{\mathbf q\nu}+1-f_{m,\mathbf k+\mathbf q}
\right)
\delta(
\xi_{n\mathbf k}
-\xi_{m,\mathbf k+\mathbf q}
-\omega_{\mathbf q\nu}
) + 
\left(
n_{\mathbf q\nu}+f_{m,\mathbf k+\mathbf q}
\right)
\delta(
\xi_{n\mathbf k}
-\xi_{m,\mathbf k+\mathbf q}
+\omega_{\mathbf q\nu}
) \Big].
\end{aligned}
\end{equation}
Note that in the thermodynamic limit, one has 
$\frac{1}{N_p}\sum_{\qq} \rightarrow \int_{BZ} \frac{d^3 q}{\Omega_{BZ}}$. 
The EPC-induced electron linewidth is defined as
$\Gamma_{n\mathbf k} = -2\, \mathfrak{Im}\Sigma^R_{n\mathbf k}$.

We now relate the EPC strength $\lambda$ to the electron self-energy. Let $z=i\omega_j$ on the Matsubara axis and $z=\omega+i0^+$ on the retarded real-frequency axis, we define the
Fermi-surface averaged self-energy
\begin{equation}
\Sigma(z) = \frac{1}{N_F N_p} \sum_{n\mathbf k} \delta(\xi_{n\mathbf k}) \Sigma_n(\mathbf k,z),
\end{equation}
where $N_F$ is the density of states at the Fermi level. Inserting an energy resolution for the intermediate electronic state,
$1 = \int d\xi \delta(\xi-\xi_{m,\mathbf k+\mathbf q})$ gives
\begin{equation}
\begin{aligned}
\Sigma(z)
= \frac{1}{N_F N_p^2}
\sum_{\mathbf q\nu}
\sum_{mn\mathbf k}
\left|g^{mn\nu}_{\mathbf k,\mathbf q}\right|^2
\delta(\xi_{n\mathbf k})
\int d\xi\,
\delta(\xi-\xi_{m,\mathbf k+\mathbf q})
K(z,\omega_{\mathbf q\nu},\xi),
\end{aligned}
\end{equation}
where
\begin{equation}
\begin{aligned}
K(z,\Omega,\xi)
=
\frac{
n_B(\Omega)+1-f(\xi)
}{
z-\xi-\Omega
}
+
\frac{
n_B(\Omega)+f(\xi)
}{
z-\xi+\Omega
}.
\end{aligned}
\end{equation}
Assume the EPC matrix elements and electronic states vary slowly on the phonon-energy scale. 
Therefore, the $\xi$-dependence of $\delta(\xi-\xi_{m,\mathbf k+\mathbf q})$ is evaluated at the Fermi surface, while the $\xi$ dependence in the propagator denominators and Fermi functions is retained. This gives
\begin{equation}
\begin{aligned}
\Sigma(z)
&\approx \frac{1}{N_F N_p^2}
\sum_{\mathbf q\nu}
\sum_{mn\mathbf k}
\left|g^{mn\nu}_{\mathbf k,\mathbf q}\right|^2
\delta(\xi_{n\mathbf k}) \delta(\xi_{m,\mathbf k+\mathbf q}) 
\int d\xi K(z,\omega_{\mathbf q\nu},\xi) \\
&= \frac{1}{N_p}
\sum_{\mathbf q\nu}
A_{\mathbf q\nu}
\int d\xi\,
K(z,\omega_{\mathbf q\nu},\xi),
\end{aligned}
\end{equation}
where 
$A_{\mathbf q\nu}
=
\frac{1}{N_F N_p}
\sum_{mn\mathbf k}
\left|g^{mn\nu}_{\mathbf k,\mathbf q}\right|^2
\delta(\xi_{n\mathbf k})
\delta(\xi_{m,\mathbf k+\mathbf q})$.
Equivalently, 
\begin{equation}
\Sigma(z)
\approx
\int_0^\infty d\Omega\,
\alpha^2F(\Omega)
\int d\xi\,
K(z,\Omega,\xi),
\label{app:eq:Sigma(z)}
\end{equation}
where the Eliashberg spectral function is
\begin{equation}
\alpha^2F(\Omega)
=
\frac{1}{N_F N_p^2}
\sum_{\mathbf q\nu}
\sum_{mn\mathbf k}
\left|g^{mn\nu}_{\mathbf k,\mathbf q}\right|^2
\delta(\xi_{n\mathbf k})
\delta(\xi_{m,\mathbf k+\mathbf q})
\delta(\Omega-\omega_{\mathbf q\nu}).
\end{equation}
This motivates us to define the total EPC strength $\lambda$ expressed using $\alpha^2F(\omega)$ (see \cref{app:eq:lambda_from_a2F}), \ie, 
\begin{equation}
\lambda
=
2\int_0^\infty d\Omega\,
\frac{\alpha^2F(\Omega)}{\Omega}.
\label{app:eq:lambda_from_a2F_2}
\end{equation}

In literature, $\lambda$ is also known as the mass enhancement parameter, \ie, with definition 
\begin{equation}
    \lambda=-\frac{\partial}{\partial \omega} \mathfrak{Re} \Sigma(\omega) |_{\omega=0}.
\end{equation}
We show that this definition is consistent with \cref{app:eq:lambda_from_a2F_2}. To do so, we first take $T\rightarrow 0$ limit, so that $n_B(\omega)=0,\ f(\epsilon)=\theta(-\epsilon)$, and \cref{app:eq:Sigma(z)} becomes 
\begin{equation}
\begin{aligned}
    \mathfrak{Im} \Sigma(\omega) 
    &= -\pi \int_0^\infty d\Omega \alpha^2F(\Omega) \left[
    n_B(\Omega) + 1 - f(\omega-\Omega) + n_B(\Omega) + f(\omega+\Omega) \right] \\
    &= -\pi \int_0^\infty d\Omega \alpha^2F(\Omega) [\theta(\omega-\Omega) + \theta(-\omega-\Omega)] \\
    &= -\pi \int_0^{|\omega|} d\Omega \alpha^2F(\Omega) \quad (T=0).
    \label{app:eq:ImSigmaT=0}
\end{aligned}
\end{equation}
From the Kramers–Kronig relation, the real-part of the self-energy is 
\begin{equation}\begin{aligned}
    \mathfrak{Re} \Sigma(\omega) = \frac{1}{\pi} \mathcal{P}\int_{-\infty}^{\infty} d\omega' \frac{ \mathfrak{Im} \Sigma(\omega')}{\omega'-\omega} 
    &\quad \Rightarrow\quad
    \frac{\partial}{\partial \omega} \mathfrak{Re} \Sigma(\omega) = \frac{1}{\pi} \mathcal{P}\int_{-\infty}^{\infty} d\omega' \frac{ \mathfrak{Im} \Sigma(\omega')}{(\omega'-\omega)^2}  \\
    &\quad \Rightarrow \quad 
    \left.\frac{\partial}{\partial \omega} \mathfrak{Re} \Sigma(\omega)\right|_{\omega\rightarrow 0} = \frac{1}{\pi} \int_{-\infty}^{\infty} d\omega' \frac{ \mathfrak{Im} \Sigma(\omega')}{(\omega')^2}.
\end{aligned}\end{equation}
By plugging in \cref{app:eq:ImSigmaT=0}, we arrive at
\begin{equation}\begin{aligned}
    -\left.\frac{\partial}{\partial \omega} \mathfrak{Re} \Sigma(\omega)\right|_{\omega\rightarrow 0} 
    &= \int_{-\infty}^{\infty} d\omega' \frac{\int_0^{|\omega'|} d\Omega \alpha^2F(\Omega)}{(\omega')^2} \\
    &= 2 \int_{0}^{\infty} d\omega' \frac{\int_0^{|\omega'|} d\Omega \alpha^2F(\Omega)}{(\omega')^2} \\
    &= 2 \int_{0}^{\infty} d\Omega \alpha^2F(\Omega)
    \int_{\omega'=\Omega}^{\infty} \frac{d\omega'}{(\omega')^2} \\
    &= 2 \int_{0}^{\infty} d\Omega \alpha^2F(\Omega) \frac{1}{\Omega} = \lambda. 
\end{aligned}\end{equation}
At last, to see why $\lambda$ serves as the mass enhancement parameter, we expand the electron self-energy $\Sigma(\omega) = \Sigma(0) + \omega \Sigma'(0)$, where $\Sigma'(0)=\frac{\partial}{\partial \omega}\Sigma(\omega)|_{\omega=0}$. 
Consider a single band near $E_f$ with non-interacting dispersion $\xi_{\kk}=\epsilon_{\kk}-E_f$. 
Then the interacting Green's function is
$G(\kk,\omega)\approx \frac{Z}{\omega - \overline{\xi}_{\kk} + i\eta}$, with quasiparticle weight $Z^{-1}=1 - \mathfrak{Re}\Sigma'(0)$, $\overline{\xi}_{\kk} = Z[\xi_{\kk} + \mathfrak{Re}\Sigma(0)]$, where we assume $T\rightarrow 0, \omega\rightarrow 0$ so the imaginary part of the self energy is small and absorbed into $i\eta$. 
As a result, the effective mass renormalization is 
\begin{equation}
    \frac{m^*}{m} = \frac{1}{Z} = 1 - \mathfrak{Re} \Sigma'(0).
\end{equation}
Define $\frac{m^*}{m} =1 + \lambda$, with $\lambda$ being the mass renormalization parameter, we arrive at
$\lambda=- \mathfrak{Re} \Sigma'(\omega)=-\frac{\partial}{\partial \omega} \mathfrak{Re} \Sigma(\omega)|_{\omega=0}$.

\subsection{Mean-field superconducting Hamiltonian and gap equation}

In this section, we consider the attractive electronic interacting Hamiltonian, where the attractive interaction can be generated by integrating out the EPC Hamiltonian. We derive the mean-field decoupling of the interacting Hamiltonian and the SC gap equation.  The first subsection derives the BdG Hamiltonian and solves the SC gap equation for a constant attractive interaction, while the second subsection considers a general interaction that generates spin singlet and triplet pairings.

\subsubsection{$\kk$-independent interaction}
Consider the following electron Hamiltonian with an attractive interaction (the simplest case without $\kk$-dependence)
\begin{equation}\begin{aligned}
H = \sum_{\kk,\sigma} \epsilon_{\kk} \cre{c}{\kk\sigma}\des{c}{\kk\sigma} - U \sum_{\kk,\kk'} \cre{c}{\kk,\uparrow}\cre{c}{-\kk\downarrow}\des{c}{-\kk'\downarrow}\des{c}{\kk'\uparrow}
\end{aligned}\end{equation}
where the electron operators are in the band basis with the single-particle dispersion $\epsilon_{\kk}$, and $U>0$ is the strength of the attractive interaction. We assume the (constant) attractive interaction is non-zero only for electrons within the Debye frequency, \ie, $\epsilon_{\kk}\in[-\hbar\omega_D, \hbar\omega_D]$, where $\omega_D$ represents the maximal phonon frequency of the system. 

We then perform the mean-field decoupling. Define the SC order parameter (gap) as
\begin{equation}\begin{aligned}
\Delta \equiv U \sum_{\kk} \langle \des{c}{-\kk\downarrow}\des{c}{\kk\uparrow}\rangle,\quad
\Delta^* \equiv U \sum_{\kk} \langle \cre{c}{\kk\uparrow}\cre{c}{-\kk\downarrow}\rangle.
\end{aligned}\end{equation}
The mean-field decoupling reads
\begin{equation}\begin{aligned}
\cre{c}{\kk,\uparrow}\cre{c}{-\kk\downarrow}\des{c}{-\kk'\downarrow}\des{c}{\kk'\uparrow}
&= \langle \cre{c}{\kk,\uparrow}\cre{c}{-\kk\downarrow}\rangle \des{c}{-\kk'\downarrow}\des{c}{\kk'\uparrow} +
\cre{c}{\kk,\uparrow}\cre{c}{-\kk\downarrow} \langle\des{c}{-\kk'\downarrow}\des{c}{\kk'\uparrow}\rangle -
\langle\cre{c}{\kk,\uparrow}\cre{c}{-\kk\downarrow}\rangle \langle\des{c}{-\kk'\downarrow}\des{c}{\kk'\uparrow}\rangle  \\
&+ (\cre{c}{\kk,\uparrow}\cre{c}{-\kk\downarrow} - \langle \cre{c}{\kk,\uparrow}\cre{c}{-\kk\downarrow} \rangle)(\des{c}{-\kk'\downarrow}\des{c}{\kk'\uparrow} - \langle \des{c}{-\kk'\downarrow}\des{c}{\kk'\uparrow} \rangle).
\end{aligned}\end{equation}
The last term is the fluctuation term which is ignored in the mean-field approximation. Thus the mean-field Hamiltonian reads
\begin{equation}\begin{aligned}
H_{MF} &= \sum_{\kk,\sigma} \epsilon_{\kk} \cre{c}{\kk\sigma}\des{c}{\kk\sigma} - \sum_{\kk}\left[ 
\Delta \cre{c}{\kk,\uparrow}\cre{c}{-\kk\downarrow}
+ \Delta^* \des{c}{-\kk\downarrow}\des{c}{\kk\uparrow}\right] +
\frac{|\Delta|^2}{U}
\end{aligned}\end{equation}

Define the Nambu spinor basis
\begin{equation}\begin{aligned}
\Psi_{\kk}^\dagger = \left(
\cre{c}{\kk\uparrow}, \des{c}{-\kk\downarrow}
\right), \quad
\Psi_{\kk} = \left(
\begin{array}{c}
\des{c}{\kk\uparrow}\\
\cre{c}{-\kk\downarrow}
\end{array}
\right). 
\end{aligned}\end{equation}
Then we arrive at the mean-field Hamiltonian 
\begin{equation}\begin{aligned}
H_{MF}(\kk) &= \sum_{\kk}
\Psi^\dagger_{\kk}
\begin{bmatrix}
    \epsilon_{\kk} & -\Delta \\
    -\Delta^* & -\epsilon_{\kk}
\end{bmatrix}
\Psi_{\kk}
+ \frac{|\Delta|^2}{U} - \sum_{\kk}\epsilon_{\kk}.
\end{aligned}\end{equation}
The last two terms are constant and can be absorbed into the chemical potential. 
The Bogoliubov-de Gennes (BdG) Hamiltonian is
\begin{equation}\begin{aligned}
h_{\kk}^{BdG} = \begin{bmatrix}
    \epsilon_{\kk} & -\Delta \\
    -\Delta^* & -\epsilon_{\kk}
\end{bmatrix}.
\end{aligned}\end{equation}
The eigensystem of $h_{\kk}$ is 
\begin{equation}\begin{aligned}
\lambda_{\kk,\pm} &= \pm E_{\kk},\quad E_{\kk}=\sqrt{\epsilon_{\kk}^2+\Delta^2},\\
\phi_{\kk,\pm} &= \left[ \cos(\theta_{\kk}), \pm\sin(\theta_{\kk})
\right],\quad \cos(2\theta_{\kk})=\frac{\epsilon_{\kk}}{E_{\kk}},\quad
\sin(2\theta_{\kk})=-\frac{\Delta}{E_{\kk}}.
\end{aligned}\end{equation}
Equivalently, $\cos^2(\theta_{\kk})=\frac{1}{2}(1 + \frac{\epsilon_{\kk}}{E_{\kk}})$, $\sin^2(\theta_{\kk})=\frac{1}{2}(1 - \frac{\epsilon_{\kk}}{E_{\kk}})$, $\cos(\theta_{\kk})\sin(\theta_{\kk})= -\frac{\Delta}{2E_{\kk}}$. 

The Bogoliubov quasi-particles are defined as the $\cre{a}{\kk,\pm}$:
\begin{equation}\begin{aligned}
\left(
\begin{array}{c}
\des{c}{\kk\uparrow}\\
\cre{c}{-\kk\downarrow}
\end{array}
\right) &=
S_{\kk} \left(
\begin{array}{c}
\des{a}{\kk,+} \\
\cre{a}{-\kk,-}
\end{array}\right),
\quad
\left(
\begin{array}{c}
\des{a}{\kk,+} \\
\cre{a}{-\kk,-}
\end{array}\right)
=S_{\kk}^{-1}
\left(
\begin{array}{c}
\des{c}{\kk\uparrow}\\
\cre{c}{-\kk\downarrow}
\end{array}
\right), \quad
S_{\kk} =
\begin{bmatrix}
    \cos(\theta_{\kk}) & -\sin(\theta_{\kk}) \\
    \sin(\theta_{\kk}) & \cos(\theta_{\kk})
\end{bmatrix}.
\end{aligned}\end{equation}
The BdG Hamiltonian is diagonal in $\cre{a}{\kk,\pm}$ basis: 
\begin{equation}\begin{aligned}
S^{-1} h_{\kk}^{BdG} S &= 
\begin{bmatrix}
    \epsilon_{\kk} \cos(2\theta) - \Delta \sin(2\theta) & -\Delta \cos(2\theta) - \epsilon_{\kk} \sin(2\theta) \\
    -\Delta \cos(2\theta) - \epsilon_{\kk} \sin(2\theta) & -\epsilon_{\kk} \cos(2\theta) + \Delta \sin(2\theta)
\end{bmatrix} 
= \begin{bmatrix}
    E_{\kk} & \\
     & -E_{\kk}
\end{bmatrix}.
\end{aligned}\end{equation}

The gap parameter $\Delta$ can be solved self-consistently using the gap equation
\begin{equation}\begin{aligned}
\Delta = U \sum_{\kk} \langle \des{c}{-\kk\downarrow}\des{c}{\kk\uparrow}\rangle.
\end{aligned}\end{equation}
As
\begin{equation}\begin{aligned}
\des{c}{-\kk\downarrow}\des{c}{\kk\uparrow} &= (\cos(\theta_{\kk}) \des{a}{-\kk,-} + \sin(\theta_{\kk}) \cre{a}{\kk,+}) (\cos(\theta_{\kk}) \des{a}{\kk,+} - \sin(\theta_{\kk}) \cre{a}{-\kk,-}) \\
&= \cos^2(\theta_{\kk}) \des{a}{-\kk,-}\des{a}{\kk,+} + \sin^2(\theta_{\kk}) \cre{a}{\kk,+}\cre{a}{-\kk,-} + \cos(\theta_{\kk})\sin(\theta_{\kk})(\cre{a}{\kk,+}\des{a}{\kk,+} - \des{a}{-\kk,-}\cre{a}{-\kk,-}),
\end{aligned}\end{equation}
where only the $\langle \cre{a}{\kk,-}\des{a}{\kk,-}\rangle$ term is nonzero at zero temperature, \ie,
\begin{equation}\begin{aligned}
\langle \des{c}{-\kk\downarrow}\des{c}{\kk\uparrow}\rangle &= -\cos(\theta_{\kk})\sin(\theta_{\kk})=\frac{\Delta}{2E_{\kk}}. 
\label{app:eq:gap_eq_T=0}
\end{aligned}\end{equation}
Then the gap equation becomes 
\begin{equation}\begin{aligned}
\Delta= U\sum_{\kk} \frac{\Delta}{2E_{\kk}}.
\end{aligned}\end{equation}
Thus at $T=0$, if the gap $\Delta$ is nonzero, it can be solved using \cref{app:eq:gap_eq_T=0}.

At finite temperature, we have
\begin{equation}\begin{aligned}
\langle \cre{a}{\kk,+}\des{a}{\kk,+}\rangle &= f(E_{\kk}),\quad 
\langle \cre{a}{\kk, -}\des{a}{\kk, -}\rangle = 1 - f(E_{\kk}),\quad 
f(E_{\kk}) = \frac{1}{e^{\frac{E_{\kk}}{k_BT}}+1}.
\end{aligned}\end{equation}
Thus
\begin{equation}\begin{aligned}
\langle \des{c}{-\kk\downarrow}\des{c}{\kk\uparrow}\rangle &= \cos(\theta_{\kk})\sin(\theta_{\kk}) (f(E_{\kk}) - (1 - f(E_{\kk}))) = -\cos(\theta_{\kk})\sin(\theta_{\kk}) \tanh\left(\frac{E_{\kk}}{2k_BT}\right), 
\end{aligned}\end{equation}
where we have used: 
$1-2f(E_{\kk})=\frac{e^{E_{\kk}/k_BT} - 1}{e^{E_{\kk}/k_BT}+1} = \tanh\left(\frac{E_{\kk}}{2k_BT}\right)$. 

Thus the gap equation at finite temperature becomes
\begin{equation}\begin{aligned}
\Delta &= U\sum_{\kk} \frac{\Delta}{2E_{\kk}} \tanh\left(\frac{E_{\kk}}{2k_BT}\right), \\
\Rightarrow 
1 &= U\sum_{\kk} \frac{1}{2E_{\kk}} \tanh\left(\frac{E_{\kk}}{2k_BT}\right), \quad \text{when } \Delta\neq 0.
\end{aligned}\end{equation}
We then solve the gap equation. When $T=T_c$, $\Delta \rightarrow0$, and $E_{\kk}=\epsilon_{\kk}$. Then the linearized gap equation is
\begin{equation}\begin{aligned}
1 &= U\sum_{\kk} \frac{1}{2\epsilon_{\kk}} \tanh\left(\frac{\epsilon_{\kk}}{2k_B T_c}\right).
\end{aligned}\end{equation}
Assume the DOS at $E_f$ is a constant  $N_{E_f}$, and we only consider the electron within Debye frequency, \ie, $\epsilon_{\kk}\in [-\omega_D,\omega_D]$. 
Let $f(\epsilon_{\kk})=\frac{1}{2\epsilon_{\kk}} \tanh\left(\frac{\epsilon_{\kk}}{2k_B T_c}\right)$. As the DOS is $N(E)=\sum_{\kk} \delta(E-\epsilon_{\kk})$, we have
\begin{equation}\begin{aligned}
\sum_{\kk} f(\epsilon_{\kk}) &= \sum_{\kk} \int_{-\infty}^{\infty} dE f(E) \delta(E-\epsilon_{\kk}) = \int_{-\infty}^{\infty} dE f(E) N(E) \approx N_{E_f} \int_{-\omega_D}^{\omega_D} dE f(E).
\end{aligned}\end{equation}
Thus we have
\begin{equation}\begin{aligned}
1&= U N_{E_f} \int_{-\omega_D}^{\omega_D} \frac{d\epsilon}{2\epsilon} \tanh(\frac{\epsilon}{2k_B T_c}) \\
&= U N_{E_f} \int_{0}^{\omega_D} \frac{d\epsilon}{\epsilon} \tanh(\frac{\epsilon}{2k_B T_c}) \\
&= U N_{E_f} \int_{0}^{\omega_D/2k_BT_c} \frac{dx}{x} \tanh(x). \\
\end{aligned}\end{equation} 
The integral can be evaluated as 
\begin{equation}\begin{aligned}
\int_{0}^{\omega_D/2k_BT_c} \frac{dx}{x} \tanh(x) &= 
\left.\ln x \tanh x \right|_{0}^{\omega_D/2k_BT_c} - 
\int_{0}^{\omega_D/2k_BT_c} dx \ln x \text{ sech}^2 x \\
&\approx \ln(\frac{\omega_D}{2k_BT_c})-\int_0^{\infty} dx \ln x \text{ sech}^2 x,
\end{aligned}\end{equation}
where in the last approximated equation, we have assumed $\frac{\omega_D}{2k_BT_c}\gg 1$ (i.e., $\omega_D\gg k_BT_c$, so $\tanh(\frac{\omega_D}{2k_BT_c})=1$). Note that for high-$T_c$ superconductors (\eg, $T_c=100$ K, for which $k_B T_c\approx 10$ meV has the same order as $\omega_D$), this assumption is not valid. 
The second term is 
$\int_0^{\infty} dx \ln x \text{ sech}^2 x=-\int_0^{\infty} \frac{dx}{x} \tanh(x)= -\ln(\frac{4e^\gamma}{\pi})$, with $\gamma\approx 0.577$ being the Euler–Mascheroni constant.
As a result, we have
\begin{equation}\begin{aligned}
1 &\approx  U N_{E_f}\left(
\ln(\frac{\omega_D}{2k_B T_c}) + \ln(\frac{4e^\gamma}{\pi})
\right) = U N_{E_f} \ln \left(\frac{2e^\gamma \omega_D}{\pi k_B T_c}\right). 
\end{aligned}\end{equation}
Then the mean-field $T_c$ formula is given by
\begin{equation}\begin{aligned}
k_B T_c &= \frac{2e^\gamma}{\pi} \omega_D \exp\left(-\frac{1}{U N_{E_f}}\right) \\
&\approx 1.14 \hbar\omega_D \exp\left(-\frac{1}{UN_{E_f}}\right). 
\label{app:eq:Tc_meanfield_eq}
\end{aligned}\end{equation}
We remark that the mean-field expression for $T_c$ is not applicable when a flat band lies at $E_f$, where the density of states $N_{E_f}$ diverges.

\subsubsection{General interaction and spin singlet and triplet pairing}

Consider a general momentum- and spin-dependent attractive interacting Hamiltonian for a single band
\begin{equation}\begin{aligned}
H = \sum_{\kk,\sigma} \epsilon_{\kk} \cre{c}{\kk\sigma}\des{c}{\kk\sigma} - \sum_{\kk\sigma,\kk'\sigma'} V_{\kk,\kk',s_1s_2s_3s_4}
\cre{c}{\kk,s_1}\cre{c}{-\kk,s_2}\des{c}{-\kk',s_3}\des{c}{\kk',s_4}.
\end{aligned}\end{equation}
As we only consider a single band with spin degeneracy, there is an effective spin $SU(2)$ symmetry and no spin-orbital coupling (SOC) in the band basis. 
The interaction satisfies
$
V_{\kk,\kk',s_1s_2s_3s_4} =
-V_{-\kk,\kk',s_2s_1 s_3s_4} = 
-V_{\kk,-\kk',s_1s_2s_4s_3} = 
V_{-\kk, -\kk',s_2s_1s_4s_3}$, from the fermionic anticommutation relation.  
Define the mean-field $b_{\kk,ss'}=\langle \des{c}{-\kk,s}\des{c}{\kk,s'} \rangle$, and the gap function
\begin{equation}\begin{aligned}
\Delta_{\kk,ss'} &= -\sum_{\kk',s_3s_4} V_{\kk,\kk',ss's_3s_4} b_{\kk',s_3s_4}, \\
\Delta^*_{\kk,ss'} &= -\sum_{\kk',s_1s_2} V_{\kk,\kk',s_1s_2 s's}^* b_{\kk',s_1s_2},
\end{aligned}\end{equation}
where the sign of $-1$ is immaterial, as the gap function can always be made positive by a global $U(1)$ gauge transformation. 
The mean-field Hamiltonian has the form
\begin{equation}\begin{aligned}
H_{MF}= 
\sum_{\kk,\sigma} \epsilon_{\kk} \cre{c}{\kk\sigma}\des{c}{\kk\sigma} + \sum_{\kk,s_1s_2} \left( 
\Delta_{\kk,s_1s_2} \cre{c}{\kk,s_1}\cre{c}{-\kk,s_2} +
\Delta^*_{\kk,s_1s_2} \des{c}{-\kk,s_2}\des{c}{\kk,s_1}\right) + K,
\end{aligned}\end{equation}
where $K=\sum_{\kk,\kk',s_1s_2s_3s_4}V_{\kk,\kk',s_1s_2s_3s_4} \langle \cre{c}{\kk,s_1}\cre{c}{-\kk,s_2}\rangle \langle \des{c}{-\kk',s_3}\des{c}{\kk',s_4} \rangle$.

The gap function can be separated into spin singlet and triplet channels:
\begin{equation}\begin{aligned}
\text{Singlet: } 
\Delta_{\kk} &=
\left(
\begin{matrix}
    0 & d_0(\kk) \\
    -d_0(\kk) & 0
\end{matrix}
\right)=  d_0(\kk) \cdot i\sigma_y, \\
\text{Triplet: } 
\Delta_{\kk} &= 
\left(
\begin{matrix}
    -d_x(\kk)+id_y(\kk) & d_z(\kk) \\
    d_z(\kk) & d_x(\kk)+id_y(\kk)
\end{matrix}
\right) = (\mathbf{d}(\kk)\cdot \vec{\sigma}) i\sigma_y,
\end{aligned}\end{equation}
where $\dd(\kk)=(d_x(\kk), d_y(\kk), d_z(\kk)), \vec{\sigma}=(\sigma_x,\sigma_y,\sigma_z)$. 
For singlet channel, we have 
$\Delta_{\kk}\Delta_{\kk}^\dagger = |d_0(\kk)|^2 \sigma_0$, while for triplet channel, 
$\Delta_{\kk}\Delta_{\kk}^\dagger = |\mathbf{d}(\kk)|^2\sigma_0+i(\mathbf{d}(\kk) \times \mathbf{d}^*(\kk))\cdot\vec{\sigma}$. 
We observe the following symmetry properties from fermion anticommutation:
\begin{equation}\begin{aligned}
d_0(\kk) &= d_0(-\kk),\quad \mathbf{d}(\kk) = -\mathbf{d}(-\kk), \\
\hat{T} d_0(\kk) \hat{T}^{-1} &= d_0^*(-\kk),\quad
\hat{T}\mathbf{d}(\kk)\hat{T}^{-1} = -\mathbf{d}^*(-\kk),
\end{aligned}\end{equation}
where $\hat{T}$ is the time-reversal symmetry (TRS). Thus the gap function (for both singlet and triplet channels) satisfies
\begin{equation}\begin{aligned}
\Delta_{\kk} = -\Delta_{-\kk}^T,
\end{aligned}\end{equation}
\ie, singlet pairing is even in momentum, while triplet pairing is odd.

\paragraph{\textbf{Nambu basis. }}
Define the Nambu spinor basis
\begin{equation}\begin{aligned}
\hat{\Psi}_{\kk} &=
\left(
\begin{array}{c}
\des{c}{\kk\uparrow}\\
\des{c}{\kk\downarrow}\\
\cre{c}{-\kk\uparrow}\\
\cre{c}{-\kk\downarrow}
\end{array}
\right),\quad
\hat{A}_{\kk} = \left(
\begin{array}{c}
\des{a}{\kk,+} \\
\des{a}{\kk,-} \\
\cre{a}{-\kk,+} \\
\cre{a}{-\kk,-}
\end{array}\right),\quad
\hat{\Psi}_{\kk} = U_{\kk} \hat{A}_{\kk},\quad
U_{\kk} =
\left(
\begin{array}{cc}
u_{\kk} & v_{\kk} \\
-v_{\kk}^\dagger & u_{\kk}^\dagger
\end{array}
\right),
\end{aligned}\end{equation}
where $U_{\kk}$ is unitary and satisfies $u_{\kk} u_{\kk}^\dagger+v_{\kk} v_{\kk}^\dagger=1$. 
Assume the pairing is unitary, i.e., 
$\Delta_{\kk}\Delta_{\kk}^\dagger\propto \sigma_0$, or equivalently, $\dd(\kk)\times \dd^*(\kk)=0$ (\ie, $\dd(\kk)\parallel \dd^*(\kk)$). 
Let $E_{\kk}=\sqrt{\epsilon_{\kk}^2+|\Delta_{\kk}|^2}$, with $|\Delta_{\kk}|^2=\frac{1}{2}\text{Tr}(\Delta_{\kk}\Delta_{\kk}^\dagger)$. Then we can take 
\begin{equation}\begin{aligned}
u_{\kk} &= \frac{(E_{\kk}+\epsilon_{\kk})\sigma_0}{\sqrt{2E_{\kk}(E_{\kk}+\epsilon_{\kk})}}, \quad
v_{\kk} = \frac{-\Delta_{\kk}}{\sqrt{2E_{\kk}(E_{\kk}+\epsilon_{\kk})}},
\label{app:eq:uv_parameter}
\end{aligned}\end{equation}
which satisfies $u_{\kk} v_{\kk}=-\frac{\Delta_{\kk}}{2E_{\kk}}$.

Under \cref{app:eq:uv_parameter}, let
\begin{equation}\begin{aligned}
\mathcal{E}_{\kk} &=
\left(
\begin{matrix}
    \epsilon_{\kk} \sigma_0 & \Delta_{\kk} \\
    \Delta_{\kk}^\dagger & -\epsilon_{\kk}\sigma_0
\end{matrix}\right),\\
E_{\kk} &= U_{\kk}^\dagger\mathcal{E}_{\kk}U_{\kk} = \text{Diag}\left(
E_{\kk}, E_{\kk}, -E_{\kk}, -E_{\kk}
\right).
\label{app:eq:HMF_eigval}
\end{aligned}\end{equation}
The mean-field Hamiltonian has the form 
\begin{equation}\begin{aligned}
H_{MF} &= \sum_{\kk} \hat{\Psi}_{\kk}^\dagger \mathcal{E}_{\kk} \hat{\Psi}_{\kk} + K -\sum_{\kk}\epsilon_{\kk} \\
&= \sum_{\kk} \hat{A}_{\kk}^\dagger E_{\kk} \hat{A}_{\kk}+K- \sum_{\kk}\epsilon_{\kk},
\end{aligned}\end{equation}
In \cref{app:eq:HMF_eigval}, we have assumed the system to have time-reversal symmetry (TRS), so that $\epsilon_{\kk}=\epsilon_{-\kk}$.

\paragraph{\textbf{Gap equation. }}
We observe that the $SU(2)$-symmetric interaction can be parameterized as
\begin{equation}\begin{aligned}
V_{\kk,\kk',s_1s_2s_3s_4} &=
J_{\kk,\kk'}^0 \sigma_{0,s_1s_4} \sigma_{0,s_2s_3} + J_{\kk,\kk'} \vec{\sigma}_{s_1s_4}\cdot\vec{\sigma}_{s_2s_3},
\end{aligned}\end{equation}
where the first term is the charge-charge channel (spin-independent), while the second term is the spin-spin channel (spin-dependent). 
Such a decomposition is the most general form that respects the spin-rotation SU(2) invariance. This is because the interaction is defined in the two-particle spin space $\mathbb{C}^2\otimes\mathbb{C}^2$, and $\sigma_0\otimes\sigma_0$ and $\vec{\sigma}\cdot\vec{\sigma}$ are the only two spin-rotation-invariant terms. Thus $J^0_{\kk,\kk'}$ and $J_{\kk,\kk'}$ correspond to the spin-independent and spin-dependent isotropic interactions. 
Note that 
$\vec{\sigma}\cdot\vec{\sigma}=\sum_{i=x,y,z}\sigma_i\otimes\sigma_i$, and the completeness relation (Fierz identity) $\vec{\sigma}_{s_1s_4}\cdot\vec{\sigma}_{s_2s_3}=2\delta_{s_1s_3}\delta_{s_2s_4}-\delta_{s_1s_4}\delta_{s_2s_3}$. 

We then derive the gap equation. First, 
\begin{equation}\begin{aligned}
\langle \des{c}{-\kk,s_3}\des{c}{\kk,s_4}\rangle &= 
\langle
\sum_{s'} (u_{\-kk,s_3s'}\des{a}{-\kk,s'}+v_{-\kk,s_3s'}\cre{a}{\kk,s'})
\sum_{s''} (u_{-kk,s_4s''}\des{a}{\kk,s''}+v_{\kk,s_4s''}\cre{a}{-\kk,s''})
\rangle \\
&= \sum_{s'} u_{\kk,s_3s'}v_{\kk,s_4s'} \langle \des{a}{-\kk,s'} \cre{a}{-\kk,s'}\rangle
+ v_{-\kk,s_3s'} u_{\kk,s_4s'} \langle \cre{a}{\kk,s'} \des{a}{\kk,s'} \rangle \\
&= \frac{-\Delta_{\kk,s_4s_3}}{2E_{\kk}} (1-f(E_{-\kk})) + \frac{-\Delta_{-\kk,s_3s_4}}{2E_{\kk}} f(E_{\kk}) \\
&= -\frac{\Delta_{\kk,s_4s_3}}{2E_{\kk}} (1-2f(E_{\kk})) \\
&= -\frac{\Delta_{\kk,s_4s_3}}{2E_{\kk}} \tanh\left(\frac{E_{\kk}}{2k_BT}\right),
\end{aligned}\end{equation}
Then the gap equation has the form
\begin{equation}\begin{aligned}
\Delta_{\kk,s_1s_2} &= -\sum_{\kk',s_3s_4} V_{\kk,\kk',s_1s_2s_3s_4} \langle \des{c}{-\kk',s_3}\des{c}{\kk',s_4}\rangle \\
&=  \sum_{\kk',s_3s_4} V_{\kk,\kk',s_1s_2s_3s_4} \frac{\Delta_{\kk',s_4s_3}}{2E_{\kk'}} \tanh\left(\frac{E_{\kk'}}{2k_B T}\right).
\label{app:eq:gap_eq_1}
\end{aligned}\end{equation}
Note that if we pair the indices $\kk,s_1s_2$ and $\kk,s_3s_4$, respectively, the right-hand side of \cref{app:eq:gap_eq_1} becomes a matrix product. 
For singlet and triplet channels, the gap equation is given by
\begin{equation}\begin{aligned}
\text{Singlet: } d_0(\kk) &= \sum_{\kk'} (J^0_{\kk,\kk'} - 3 J_{\kk,\kk'}) \frac{d_0(\kk')}{2E_{\kk'}}\tanh\left(\frac{E_{\kk'}}{2k_B T}\right), \\
\text{Triplet: } \dd(\kk) &= \sum_{\kk'} (J^0_{\kk,\kk'} + J_{\kk,\kk'}) \frac{\dd(\kk')}{2E_{\kk'}}\tanh\left(\frac{E_{\kk'}}{2k_B T}\right).
\end{aligned}\end{equation}

\paragraph{\textbf{Spin-independent interaction. }} 
When the interaction is momentum-dependent but spin-independent, i.e., $J_{\kk,\kk'}=0$, with 
\begin{equation}\begin{aligned}
V_{\kk,\kk',s_1s_2s_3s_4} = U_{\kk,\kk'} \delta_{s_1s_4}\delta_{s_2s_3},
\end{aligned}\end{equation}
then
\begin{equation}\begin{aligned}
H = \sum_{\kk,\sigma} \epsilon_{\kk} \cre{c}{\kk\sigma}\des{c}{\kk\sigma} - \sum_{\kk\sigma,\kk'\sigma'} U_{\kk,\kk'}
\cre{c}{\kk,\sigma}\cre{c}{-\kk,\sigma'}\des{c}{-\kk',\sigma'}\des{c}{\kk',\sigma}.
\end{aligned}\end{equation}
Hermitian requires that $U_{\kk,\kk'}=U_{\kk',\kk}^*$.  
In this case, $J^0_{\kk,\kk'}=U_{\kk,\kk'}, J_{\kk,\kk'}=0$. Thus the gap equation is simplified into
\begin{equation}\begin{aligned}
\text{Singlet: } d_0(\kk) &= \sum_{\kk'} U_{\kk,\kk'} \frac{d_0(\kk')}{2E_{\kk'}}\tanh\left(\frac{E_{\kk'}}{2k_B T}\right), \\
\text{Triplet: } \dd(\kk) &= \sum_{\kk'} U_{\kk,\kk'} \frac{\dd(\kk')}{2E_{\kk'}}\tanh\left(\frac{E_{\kk'}}{2k_B T}\right).
\label{app:eq:gap_eq_kdependent_ham_singlet}
\end{aligned}\end{equation}
We note that the singlet and triplet channels have the same gap equation. However, the most favored channel is determined by the momentum dependence of $U_{\kk,\kk'}$. If the attractive interaction is even in momentum ($U_{\kk,\kk'}=U_{-\kk,\kk'}$, \eg, $s$-wave), then the singlet channel is favored, while if the interaction is odd in momentum ($U_{\kk,\kk'}=-U_{-\kk,\kk'}$, \eg, $p$-wave),  the triplet channel is favored and the singlet channel only has zero gap.

\subsubsection{Solve the gap equation numerically}\label{app:sec:solve_gap_eq_numerical}
The (singlet) gap equation in \cref{app:eq:gap_eq_kdependent_ham_singlet} can be solved numerically. Denote
\begin{equation}\begin{aligned}
\tilde{d}_0(\kk, T) = d_{0}(\kk) \sqrt{\frac{\tanh(E_{\kk}/2k_B T)}{2E_{\kk}}}. 
\end{aligned}\end{equation}
We can rewrite the gap equation into
\begin{equation}\begin{aligned}
\tilde{d}_{0}(\kk,T) &= \sum_{\kk'} M_{\kk,\kk'}(T) \tilde{d}_{0}(\kk',T),\quad
M_{\kk,\kk'}(T) = U_{\kk,\kk'}
\sqrt{\frac{\tanh(E_{\kk}/2k_B T)}{2E_{\kk}}} \sqrt{\frac{\tanh(E_{\kk'}/2k_B T)}{2E_{\kk'}}}
\end{aligned}\end{equation}
$M_{\kk,\kk'}(T)$ is Hermitian and has real eigenvalues as long as the interaction $U_{\kk,\kk'}$ is Hermitian. 
At the superconducting critical temperature $T_c$, the gap amplitude becomes infinitesimal with $E_{\kk}\to |\epsilon_{\kk}|$, so the $M_{\kk,\kk'}(T)$ becomes independent of $d_0(\kk)$. The gap equation then reduces to a linearized eigenvalue problem. A nontrivial solution exists only when $M(T)$ has an eigenvalue equal to 1. Thus, at $T=T_c$, the function $\tilde d_0(\kk,T)$ is the eigenvector of $M_{\kk,\kk'}(T_c)$ with eigenvalue 1. 
In practice, $T_c$ can be obtained by lowering the temperature from the normal state and monitoring the leading eigenvalue $\lambda_{\max}(T)$ of $M(T)$. For an attractive pairing kernel, 
$\frac{\tanh(|\epsilon_{\kk}|/2k_B T)}{2|\epsilon_{\kk}|}$ increases as $T$ decreases, so the leading attractive eigenvalue grows. The critical temperature is therefore the highest temperature at which $\lambda_{\max}(T_c)=1$. The corresponding eigenvector determines the momentum dependence of the superconducting order parameter, while the overall gap magnitude is fixed only below $T_c$ by the nonlinear gap equation.
A more detailed study on the property of the superconducting gap from $M_{\kk,\kk'}$ is left to Ref.~\cite{EPCpaper}.

\subsection{EPC from Gaussian approximation}\label{app:sec:EPC_from_GA}

We review the formalism of Gaussian approximation (GA)~\cite{yu2024non, hu2023kagome}, which allows us to obtain the analytic EPC from the electron Hamiltonian. 
GA assumes the hopping is two-center (see discussion in \cref{app:sec:two-center-approx}) and can be approximated by a Gaussian function. In this section, we discuss the property of EPC derived from the GA form hopping, focusing on the $s$-orbitals for simplicity.  

In the Gaussian approximation, the \textit{direct} hopping between two ($s$-like isotropic) orbitals is assumed to take the Gaussian form 
\begin{equation}\begin{aligned}
t_{i\alpha,j\beta}(\Delta\RR) &= t_{i\alpha,j\beta}^0 e^{-\gamma_{i\alpha,j\beta}|\Delta\RR+\rr_j-\rr_i|^2/2},
\end{aligned}\end{equation}
where $i,j$ are the atom index, $\alpha,\beta$ are the orbital index, and $\gamma_{i\alpha, j\beta}<0$ is the Gaussian decaying factor. 
The derivative (w.r.t. the displacement vector $\dd=\Delta\RR+\rr_j-\rr_i$) of the hopping can be computed analytically 
\begin{equation}\begin{aligned}
\partial_{\dd^\mu} t_{i\alpha,j\beta}(\Delta\RR) &= 
-t_{i\alpha,j\beta}^0 \gamma_{i\alpha,j\beta} (\Delta\RR+\rr_j-\rr_i)^\mu e^{-\gamma_{i\alpha,j\beta}|\Delta\RR+\rr_j-\rr_i|^2/2} \\
&= -\gamma_{i\alpha,j\beta} (\Delta\RR+\rr_j-\rr_i)^\mu t_{i\alpha,j\beta}(\Delta\RR)
\end{aligned}\end{equation}
Consider the electronic Hamiltonian with atomic displacement $\uu_{i}(\RR)$ for atom $i$:
\begin{equation}\begin{aligned}
\hat{H}' &=
\sum_{\RR,\Delta\RR,i\alpha j\beta} t_{ij}(\Delta\RR+\uu_{j}(\RR+\Delta\RR) - \uu_i(\RR)) \cre{c}{\RR,i\alpha}\des{c}{\RR+\Delta\RR,j\beta}
\end{aligned}\end{equation}
We then expand $\hat{H}'$ to the first order of the displacement:
\begin{equation}\begin{aligned}
\hat{H}' \approx&
\sum_{\RR,\Delta\RR,i\alpha j\beta} t_{ij}(\Delta\RR) \cre{c}{\RR,i\alpha}\des{c}{\RR+\Delta\RR,j\beta} \\
&+\sum_{\RR,\Delta\RR,i\alpha j\beta} \partial_{\dd^\mu} t_{ij}(\Delta\RR) \left(\uu_{j}(\RR+\Delta\RR) - \uu_i(\RR)\right)^{\mu} \cre{c}{\RR,i\alpha}\des{c}{\RR+\Delta\RR,j\beta},
\label{app:eq:expand_H_linear_order}
\end{aligned}\end{equation} 
where the zeroth order term is the original TB Hamiltonian, while the linear term is the electron-phonon coupling $\hat{H}_g$, which can be rewritten as
\begin{equation}\begin{aligned}
\hat{H}_g &=
\sum_{\RR,\Delta\RR,i\alpha j\beta} \sum_{\RR_p, l\mu} g^{i\alpha j\beta,l\mu}_{\RR_e,\RR_p} \cre{c}{\RR,i\alpha}\des{c}{\RR+\RR_e,j\beta} \des{u}{\RR+\RR_p, l\mu}, \\
g^{i\alpha j\beta,l\mu}_{\RR_e,\RR_p} 
&= -\gamma_{i\alpha,j\beta} (\RR_e + \rr_j-\rr_i)^\mu t_{i\alpha,j\beta}(\RR_e) (\delta_{j,l}\delta_{\RR_e,\RR_p} - \delta_{i,l}\delta_{\RR_p,\bm{0}}).
\end{aligned}\end{equation}
We observe that:
\begin{itemize}
\item Within the Gaussian approximation, the onsite terms in the electronic Hamiltonian cannot generate EPC because the distance $\Delta\RR+\rr_j-\rr_i=\bm{0}$. However, the onsite EPC terms $\cre{c}{\RR,i\alpha}\des{c}{\RR,j\beta} \des{u}{\RR+\RR_p, l\mu}$ are in general non-zero in the first-principle calculation, which is beyond the Gaussian approximation. In \ch{MgB2}, the largest onsite EPC terms have the same order of magnitude as the NN bond EPC terms. 
\item The EPC terms that involve three atoms are beyond the Gaussian approximation, \ie, only the EPC terms that involve the phonon from the atom of either of the two electrons $\cre{c}{\RR,i\alpha}$ and $\des{c}{\RR+\RR_e,j\beta}$ are nonzero. 
In \ch{MgB2}, such 3-center EPC terms (\ie, $\RR_e\neq \RR_p, \RR_e\neq \bm{0}$) have typical values smaller than \SI{0.15}{eV/\angstrom}. 
The onsite EPC terms (\ie, $\RR_e\neq \bm{0}$) are also beyond the Gaussian approximation, which can be large in realistic materials, including \ch{MgB2} (see more details in \cref{app:sec:EPC_3s_basis}). 

\item Once the EPC is known, one can use \cref{app:eq:expand_H_linear_order} to consider the EPC contribution to the electron Hamiltonian for a given (small) phonon displacement field. This is known as the ``frozen phonon'' approach~\cite{an2001superconductivity} in the tight-binding context, \ie, use EPC to update the tight-binding Hamiltonian from a given phonon mode. 

\item It is straightforward to see that the EPC from the Gaussian approximation automatically satisfies the acoustic sum rule of the EPC tensor, as discussed in \cref{app:eq:EPC_sum_rule}: $\forall i,j, \RR_e, \mu$, we have
$\sum_{\RR_{p},l} g_{\RR_e,\RR_{p}}^{ij,l\mu}=
-\gamma_{i\alpha,j\beta} (\RR_e + \rr_j-\rr_i)^\mu t_{i\alpha,j\beta}(\RR_e) \sum_{\RR_{p},l} (\delta_{j,l}\delta_{\RR_e,\RR_p} - \delta_{i,l}\delta_{\RR_p,\bm{0}})=0$.
\end{itemize}

Similarly, we derive the EPC in momentum space from the Gaussian approximation. From
\begin{equation}\begin{aligned}
\sum_{\RR_e} -\gamma_{i\alpha,j\beta} (\RR_e + \rr_j-\rr_i)^\mu t_{i\alpha,j\beta}(\RR_e) e^{i\kk\cdot(\RR_e+\rr_j-\rr_i)}
&= i\gamma_{i\alpha,j\beta} \partial_{k^\mu} t_{i\alpha,j\beta}(\kk), 
\end{aligned}\end{equation}
we have
\begin{equation}\begin{aligned}
g_{\kk,\qq}^{i\alpha j\beta,l\mu} &= \sum_{\RR_{e},\RR_{p}} g_{\RR_e,\RR_{p}}^{i\alpha j\beta,l\mu} e^{i\kk\cdot(\RR_e+\rr_j-\rr_i)+i\qq\cdot(\RR_{p}+\rr_l-\rr_i)} \\
&= -\gamma_{i\alpha,j\beta} 
\sum_{\RR_{e},\RR_{p}} 
(\RR_e + \rr_j-\rr_i)^\mu t_{i\alpha,j\beta}(\RR_e) (\delta_{j,l}\delta_{\RR_e,\RR_p} - \delta_{i,l}\delta_{\RR_p,\bm{0}})
e^{i\kk\cdot(\RR_e+\rr_j-\rr_i)+i\qq\cdot(\RR_{p}+\rr_l-\rr_i)} \\
&= i\gamma_{i\alpha,j\beta} \left(
\partial_{k^\mu} t_{i\alpha,j\beta}(\kk+\qq)\delta_{jl} - \partial_{k^\mu} t_{i\alpha, j\beta}(\kk)\delta_{il} 
\right)
\end{aligned}\end{equation}

Following Ref.~\cite{yu2024non}, the Gaussian form of EPC is defined as
\begin{equation}\begin{aligned}
f^{i\alpha j\beta}_{\mu}(\kk) &= i\gamma_{i\alpha,j\beta}\partial_{k^\mu} t_{i\alpha,j\beta}(\kk).
\end{aligned}\end{equation}
By assuming all hoppings have the same decaying factor $\gamma_{i\alpha,j\beta}:= \gamma$, we have
\begin{equation}\begin{aligned}
f_{\mu}(\kk) &= i\gamma \partial_{k^\mu} h(\kk)\\
&:= f_{\mu}^{\text{E}}(\kk) + f_{\mu}^{\text{geo}}(\kk),
\end{aligned}\end{equation}
where we rewrite the Hamiltonian as $h(\kk)=\sum_{n}\epsilon_{n\kk}P^{n}(\kk)$, and separate the EPC into the energetic part and geometric parts:
\begin{equation}\begin{aligned}
f_{n\mu}^{\text{E}}(\kk) &= i\gamma \left(\partial_{k^\mu}\epsilon_{n\kk}\right) P^n(\kk),\\
f_{n\mu}^{\text{geo}}(\kk) &= i\gamma \epsilon_{n\kk} \partial_{k^\mu} P^n(\kk).
\label{app:eq:f_E_geo}
\end{aligned}\end{equation}
The momentum-space EPC tensor can be expressed as (omitting the orbital indices $\alpha\beta$ for simplicity)
\begin{equation}\begin{aligned}
g_{\kk,\qq}^{ij,l\mu} &= f^{ij}_{\mu}(\kk+\qq)\delta_{jl} - f^{ij}_{\mu}(\kk)\delta_{il} \\
&= f^{il}_{\mu}(\kk+\qq) - f^{lj}_{\mu}(\kk).
\label{app:eq:EPC_GA_s_orbital}
\end{aligned}\end{equation}
The EPC tensor projected onto the band basis is
\begin{equation}\begin{aligned}
G^{mn,\nu}_{\kk,\qq} &= \sum_{ijl\mu} \sqrt{\frac{\hbar}{2M_l \omega_{\qq\nu}}} g_{\kk,\qq}^{ij,l\mu} U^*_{i,m}(\kk+\qq)  U_{j,n}(\kk) U^p_{l\mu,\nu}(\qq).\\
|G^{mn,\nu}_{\kk,\qq}|^2 
&= \frac{\hbar}{2 \omega_{\qq\nu}} \sum_{ijl\mu}\sum_{i'j'l'\mu'} \frac{1}{\sqrt{M_l M_{l'}}}
g_{\kk,\qq}^{ij,l\mu} g_{\kk,\qq}^{i'j',l'\mu',*} U^*_{i,m}(\kk+\qq) U_{i',m}(\kk+\qq)  U_{j,n}(\kk)  U_{j',n}^*(\kk) U^p_{l\mu,\nu}(\qq) U^{p,*}_{l'\mu',\nu}(\qq) \\
&= \frac{\hbar}{2 \omega_{\qq\nu}} \sum_{ijl\mu}\sum_{i'j'l'\mu'} \frac{1}{\sqrt{M_l M_{l'}}}
g_{\kk,\qq}^{ij,l\mu} g_{\kk,\qq}^{i'j',l'\mu',*} 
P^{m}_{i'i}(\kk+\qq) P^{n}_{jj'}(\kk) P^{p,\nu}_{l\mu,l'\mu'}(\qq)\\
&= \frac{\hbar}{2 \omega_{\qq\nu}} 
\sum_{l\mu,l\mu'} \frac{P^{p,\nu}_{l\mu,l'\mu'}(\qq)}{\sqrt{M_l M_{l'}}} \text{Tr}\left[
g^{l\mu}_{\kk,\qq} P^{n}(\kk) \left(g^{l'\mu'}_{\kk,\qq} \right)^\dagger P^{m}(\kk+\qq) \right],
\label{app:eq:EPC_band_basis_expression}
\end{aligned}\end{equation}
where we use the projector $P^n_{ij}(\kk)=U_{i,n}(\kk)U^{*}_{j,n}(\kk)$, and $\text{Tr}$ is the trace over the electron indices. Note that \cref{app:eq:EPC_band_basis_expression} does not depend on the Gaussian approximation.

We next specialize to $\qq=0$ with $m=n$, relevant when only small-momentum phonons matter, e.g., for a small Fermi surface, or when evaluating intraband EPC. In this limit, under the Gaussian approximation, we obtain
\begin{equation}\begin{aligned}
G^{nn,\nu}_{\kk,\bm{0}} &= \sqrt{\frac{\hbar}{2\omega_{\qq\nu}}}
\sum_{ij,l\mu} \frac{U^p_{l\mu,\nu}(\bm{0})}{\sqrt{M_l}}  g_{\kk,\bm{0}}^{ij,l\mu} U^*_{i,n}(\kk)  U_{j,n}(\kk) \\
&= \sqrt{\frac{\hbar}{2\omega_{\qq\nu}}}
\sum_{ij,l\mu} \frac{U^p_{l\mu,\nu}(\bm{0})}{\sqrt{M_l}} \left(f^{ij}_{\mu}(\kk)\delta_{jl} - f^{ij}_{\mu}(\kk)\delta_{il} \right) P^n_{ji}(\kk) \\
&= \sqrt{\frac{\hbar}{2\omega_{\qq\nu}}}
\sum_{ij,l\mu} \frac{U^p_{l\mu,\nu}(\bm{0})}{\sqrt{M_l}} \left(\chi_{l} f_{\mu}(\kk) - f_{\mu}(\kk)\chi_{l} \right)_{ij} P^n_{ji}(\kk) \\
&= \sqrt{\frac{\hbar}{2\omega_{\qq\nu}}}
\sum_{ij,l\mu} \frac{U^p_{l\mu,\nu}(\bm{0})}{\sqrt{M_l}} \text{Tr}\left[\left(\chi_{l} f_{\mu}(\kk) - f_{\mu}(\kk)\chi_{l} \right) P^n(\kk)\right],
\end{aligned}\end{equation}
where 
\begin{equation}\begin{aligned}
(\chi_l)_{ij}=\delta_{li}\delta_{lj}.
\end{aligned}\end{equation}
We further separate the EPC into energetic and geometric parts using \cref{app:eq:f_E_geo}:
\begin{equation}\begin{aligned}
(G^{nn,\nu}_{\kk,\bm{0}})^{\text{E}} &=
\sqrt{\frac{\hbar}{2\omega_{\qq\nu}}}
\sum_{ij,l\mu} \frac{U^p_{l\mu,\nu}(\bm{0})}{\sqrt{M_l}} \text{Tr}\left[\sum_{m}\left(\chi_{l} f_{m\mu}^{\text{E}}(\kk) - f_{m\mu}^{\text{E}}(\kk)\chi_{l} \right) P^n(\kk)\right] \\
&= \sqrt{\frac{\hbar}{2\omega_{\qq\nu}}}
\sum_{ij,l\mu} \frac{U^p_{l\mu,\nu}(\bm{0})}{\sqrt{M_l}} \sum_{m} i\gamma \partial_{k^\mu} \epsilon_m \text{Tr}\left[\left(\chi_{l} P^m(\kk) - P^m(\kk) \chi_{l} \right) P^n(\kk)\right],\\ 
&= \sqrt{\frac{\hbar}{2\omega_{\qq\nu}}}
\sum_{ij,l\mu} \frac{U^p_{l\mu,\nu}(\bm{0})}{\sqrt{M_l}} \sum_{m} i\gamma \partial_{k^\mu} \epsilon_m \delta_{mn} \left[\text{Tr}(\chi_{l} P^n(\kk)) - \text{Tr}(\chi_{l} P^n(\kk))\right] =0 ,\\ 
(G^{nn,\nu}_{\kk,\bm{0}})^{\text{geo}} 
&=
\sqrt{\frac{\hbar}{2\omega_{\qq\nu}}}
\sum_{ij,l\mu} \frac{U^p_{l\mu,\nu}(\bm{0})}{\sqrt{M_l}} \text{Tr}\left[\sum_{m}\left(\chi_{l} f_{m\mu}^{\text{geo}}(\kk) - f_{m\mu}^{\text{geo}}(\kk)\chi_{l} \right) P^n(\kk)\right] \\
&= \sqrt{\frac{\hbar}{2\omega_{\qq\nu}}}
\sum_{ij,l\mu} \frac{U^p_{l\mu,\nu}(\bm{0})}{\sqrt{M_l}} \sum_{m} i\gamma \epsilon_m \text{Tr}\left[\left(\chi_{l} \partial_{k^\mu}P^m(\kk) - \partial_{k^\mu}P^m(\kk) \chi_{l} \right) P^n(\kk)\right]\\
&= \sqrt{\frac{\hbar}{2\omega_{\qq\nu}}}
\sum_{ij,l\mu} \frac{U^p_{l\mu,\nu}(\bm{0})}{\sqrt{M_l}} \sum_{m} i\gamma \epsilon_m \text{Tr} \big[ \chi_{l} \left[\partial_{k^\mu}P^m(\kk), P^n(\kk)\right] \big].
\end{aligned}\end{equation}
For general $\qq\neq \bm{0}$, $(G^{nn,\nu}_{\kk,\qq})^{\text{E}}$ and $(G^{nn,\nu}_{\kk,\qq})^{\text{geo}}$ can be defined similarly. 
The mode-specific EPC strength
\begin{equation}\begin{aligned}
    \lambda_{\qq\nu} &= \frac{2}{N_f \omega_{\qq\nu}^2} \sum_{mn} \int_{BZ} \frac{d\kk}{\Omega_{BZ}} |G^{mn\nu}_{\kk,\qq}|^2  
(f_{\kk,n} - f_{\kk+\qq,m}) \delta(\epsilon_{\kk+\qq,m} - \epsilon_{\kk,n} - \omega_{\qq\nu}),\quad 
G^{mn\nu}_{\kk,\qq}=(G^{mn\nu}_{\kk,\qq})^{\text{E}} + (G^{mn\nu}_{\kk,\qq})^{\text{geo}} 
\end{aligned}\end{equation}
can be separated into: 
\begin{equation}\begin{aligned}
\lambda_{\qq\nu}^{\text{E}} &= \frac{2}{N_f \omega_{\qq\nu}^2} \sum_{mn} \int_{BZ} \frac{d\kk}{\Omega_{BZ}} |(G^{mn\nu}_{\kk,\qq})^{\text{E}}|^2  
(f_{\kk,n} - f_{\kk+\qq,m}) \delta(\epsilon_{\kk+\qq,m} - \epsilon_{\kk,n} - \omega_{\qq\nu}), \\
\lambda_{\qq\nu}^{\text{geo}} &= \frac{2}{N_f \omega_{\qq\nu}^2} \sum_{mn} \int_{BZ} \frac{d\kk}{\Omega_{BZ}} |(G^{mn\nu}_{\kk,\qq})^{\text{geo}}|^2  
(f_{\kk,n} - f_{\kk+\qq,m}) \delta(\epsilon_{\kk+\qq,m} - \epsilon_{\kk,n} - \omega_{\qq\nu}), \\
\lambda_{\qq\nu}^{\text{geo-E}} &= \frac{4}{N_f \omega_{\qq\nu}^2} \sum_{mn} \int_{BZ} \frac{d\kk}{\Omega_{BZ}} |(G^{mn\nu}_{\kk,\qq})^{\text{geo}}(G^{mn\nu}_{\kk,\qq})^{\text{E}}|  
(f_{\kk,n} - f_{\kk+\qq,m}) \delta(\epsilon_{\kk+\qq,m} - \epsilon_{\kk,n} - \omega_{\qq\nu}). 
\end{aligned}\end{equation}
As a result, the total EPC strength $\lambda=\sum_{\qq\nu}\lambda_{\qq\nu}$ is decomposed into three parts
\begin{equation}\begin{aligned}
\lambda &= \lambda^{\text{geo}} + \lambda^{\text{E}} + \lambda^{\text{geo-E}},
\label{app:eq:lambda_separation}
\end{aligned}\end{equation}
where 
\begin{equation}
    \lambda^{\text{geo}}=\sum_{\qq\nu} \lambda^{\text{geo}}_{\qq\nu},\quad
    \lambda^{\text{E}}=\sum_{\qq\nu} \lambda^{\text{E}}_{\qq\nu},\quad
    \lambda^{\text{geo-E}}=\sum_{\qq\nu} \lambda^{\text{geo-E}}_{\qq\nu}. 
\end{equation}

\subsubsection{Two-center approximation of EPC}\label{app:sec:two-center-approx}

We briefly discuss the two-center approximation applied to both the electron hopping and the EPC tensor~\cite{mitra1969electron, yu2024non}, which is an inherent property of the EPC derived from GA. 

Starting from a general single-particle Kohn-Sham Hamiltonian 
\begin{equation}
    \hat{H}=\hat{T}+ V_{ext}(\rr) + V_H[n](\rr) + V_{xc}[n](\rr),
\end{equation}
where $\hat{T}=-\frac{\hbar^2}{2m}\nabla^2$ is the the kinetic term, $V_{ext}(\rr)= \sum_{i\RR'} v^{ion}_i(\rr-\RR'-\bm{\tau}_i)$, $v^{ion}_i(\rr-\RR'-\bm{\tau}_i)=\frac{-Z_i}{|\rr-\RR'-\bm{\tau}_i|}$ 
is the Coulomb interaction from the nuclei of the $i$-th atom with effective charge $Z_i$ at position $\RR' + \bm{\tau}_i$, with $\bm{\tau}_i$ being the position of the $i$-th atom in the unit cell. 
$V_H[n](\rr)=\int d\rr' \frac{n(\rr')}{|\rr-\rr'|}$ is the Hartree term, and $V_{xc}[n](\rr)=\frac{\delta E_{xc}[n]}{\delta n(\rr)}$ is the exchange-correlation term, with $n(\rr)$ being the ground state charge density which is solved self-consistently in DFT. Note that the Hartree and exchange-correlation terms are cell-periodic as $n(\rr)$ is cell-periodic, \ie, 
$V_{H/xc}[n](\rr)=V_{H/xc}[n](\rr+\RR)$, so does $V_{ext}(\rr)$.
Then the hopping matrix element between Wannier orbitals $\alpha,\beta$ is
\begin{equation}
    t_{i\alpha,j\beta}(\RR,\{\bm{\tau}\}) = \int_{\rr} W^*_{i\alpha}(\rr-\bm{\tau}_i) \hat{H} W_{j\beta}(\rr-\bm{\tau}_j-\RR),
\end{equation} 
where the subscript $i\alpha,j\beta$ denotes the hopping between $i\alpha,j\beta$ orbitals, and $t_{i\alpha,j\beta}$ is a function of $(\RR,\{\bm{\tau}\})$, with $\{\bm{\tau}\}$ being the set of atomic positions of all atoms in the unit cell. 
In the usual TB formalism, $\{\bm{\tau}\}$ are omitted as atoms are assumed to be fixed. Here, we include them explicitly as we need to consider phonon displacements.

The two-center hopping involves the kinetic term and the Coulomb potential of the nuclei of the $i,j$ atoms, irrelevant to other electrons:
\begin{equation}\begin{aligned} 
    t_{i\alpha,j\beta}^{\text{2-center}}(\RR, \{\bm{\tau}\}) &= \int_{\rr} W^*_{i\alpha}(\rr-\bm{\tau}_i) \left(\hat{T} + v^{ion}_i(\rr-\bm{\tau}_i) + v^{ion}_j(\rr-\bm{\tau}_j-\RR)\right) W_{j\beta}(\rr-\bm{\tau}_j-\RR) \\
    &= t_{i\alpha,j\beta}^{\text{2-center}}(\RR+\bm{\tau}_j-\bm{\tau}_i),
\end{aligned}\end{equation}  
In this case, the hopping is a function of the relative displacement $\RR+\bm{\tau}_j-\bm{\tau}_i$ of $ij$ atoms, but does not depend on the specific value of $\bm{\tau}_{i}$ and $\bm{\tau}_{j}$. Note that two-center hopping becomes the one-center onsite term when $(i,\bm{0})= (j,\RR)$. In this case, we need to modify the integrand to
$\int_{\rr} W^*_{i\alpha}(\rr-\bm{\tau}_i) \left(\hat{T} + v^{ion}_i(\rr-\bm{\tau}_i)\right) W_{j\beta}(\rr-\bm{\tau}_j)$ to avoid double counting.  

The three-center terms, however, involve the contribution from the nuclei' Coulomb potential of other atoms 
\begin{equation}
    t_{i\alpha,j\beta}^{\text{3-center}}(\RR,\{\bm{\tau}\}) = \int_{\rr} W^*_{i\alpha}(\rr-\bm{\tau}_i) \left(\sum_{(k,\RR')\neq (i,\bm{0}),(j,\RR)} v^{ion}_k(\rr-\bm{\tau}_k-\RR') \right) W_{j\beta}(\rr-\bm{\tau}_j-\RR). 
\label{app:eq:3-center-hopping}
\end{equation} 
Note that there are no four-center or higher terms from $V_{ext}(\rr)$ in the hopping.
Compared with two-center terms, three-center terms cannot be written as a function of the relative displacement of the $ij$ atoms, but instead depend on the position of $\bm{\tau}_{i}$ and $\bm{\tau}_{j}$ in the lattice, due to the sum over the Coulomb potential of other atoms. The three-center terms are still translationally invariant in $\RR$. 

The Hartree and exchange-correlation terms contribute to the remaining part of the hopping (which is not two-center in general, as they are determined by the full self-consistent density $n(\rr)$, which depends on all electrons in the system):
\begin{equation}
    t_{i\alpha,j\beta}^{H+xc}(\RR,\{\bm{\tau}\}) = \int_{\rr} W^*_{i\alpha}(\rr-\bm{\tau}_i) \left(V_H[n](\rr) + V_{xc}[n](\rr) \right) W_{j\beta}(\rr-\bm{\tau}_j-\RR). 
\end{equation}
The onsite energy term is the special case of $i=j, \alpha=\beta$ and $\RR=\bm{0}$, \ie,
\begin{equation}
\begin{aligned}
\epsilon_{i\alpha} = t_{i\alpha,i\alpha}(\bm{0},\{\bm{\tau}\}),
\label{app:eq:onsite-energy-integral}
\end{aligned}
\end{equation}
which comes from two parts: one is the atomic part from $\hat{T}+v^{ion}_i(\rr-\bm{\tau}_i)$, and the crystal-field part from $\sum_{(k,\RR') \neq (i,\bm{0})} v_i^{ion}(\rr-\bm{\tau}_k-\RR') + V_H[n](\rr) + V_{xc}[n](\rr)$.

As discussed in \cref{app:eq:expand_H_linear_order}, once a small displacement $\uu_i$ is added to atom $i$, the first-order derivative of the two-center hopping gives the (two-center approximated) EPC, \ie,
\begin{equation}\begin{aligned}
\partial_{\mu} t_{i\alpha,j\beta}^{\text{2-center}}(\RR+\bm{\tau}_j-\bm{\tau}_i) (\uu_i - \uu_j)
\end{aligned}\end{equation}
The EPCs from the $ij$ atoms have opposite signs, and there is no EPC to other atoms. 
Thus, the two-center approximation of EPC states that for any two orbitals from two distinct atoms in the system, their EPC is solely determined by the relative motions of these two atoms~\cite{yu2024non}. The property is equivalent to assuming the EPC tensor satisfies
\begin{equation}\begin{aligned}
g^{i\alpha j\beta,l\mu}_{\RR_e,\RR_p} &= 0, \quad \text{for } \RR_p \neq \bm{0} \text{ and } \RR_p\neq \RR_e,\\
g^{i\alpha j\beta,l\mu}_{\RR_e,\RR_e} &= - g^{i\alpha j\beta,l\mu}_{\RR_e, \bm{0}}. 
\end{aligned}\end{equation}

We remark that the two-center approximation may not hold well in the \textit{ab initio} EPC tensor, due to the hoppings beyond two-center. As seen from the three-center hopping in \cref{app:eq:3-center-hopping}, $t_{i\alpha,j\beta}^{\text{3-center}}(\RR, \{\bm{\tau}\})$ is also a function of the atomic positions of $k~(\neq i,j)$ atoms. Thus, the corresponding EPC is also a function of $\uu_k$ ($k\neq i,j$), which is beyond the two-center approximation of EPC. These terms are typically small (\eg, $<1$ eV/\SI{}{\angstrom} in \ch{MgB2}). 
Moreover, we ignore the Pulay correction terms~\cite{baroni2001phonons, li2024deep} (\ie, derivative w.r.t the Wannier basis) when computing the EPC from the electron hoppings for simplicity. 
These terms can also be expressed in terms of the electron Hamiltonian and orbital overlap matrices.

\section{Ab initio calculation details}
The \textit{ab-initio} electronic band structures in this work are computed using the Vienna ab-initio Simulation Package (VASP)~\cite{kresse1996efficiency, kresse1993ab1, kresse1993ab2, kresse1994ab, kresse1996efficient} with generalized gradient approximation of Perdew-Burke-Ernzerhof (PBE) exchange-correlation functional~\cite{perdew1996generalized}. An energy cutoff of 400 eV is used for self-consistency computations. Spin-orbital coupling (SOC) is not considered. The maximally localized Wannier functions (MLWFs) are obtained using WANNIER90~\cite{marzari1997maximally, souza2001maximally, marzari2012maximally, pizzi2020wannier90}.  
Quantum ESPRESSO~\cite{giannozzi2009quantum, giannozzi2017advanced, giannozzi2020quantum} and EPW~\cite{giustino2007electron,noffsinger2010epw,margine2013anisotropic,ponce2016epw,lee2023electron} are used to compute the phonon, electron-phonon, and superconducting properties, with the Perdew-Burke-Ernzerhof for solids~\cite{perdew2008restoring} (PBEsol) pseudopotentials from the PSEUDODOJO project (stringent, scalar-relativistic, norm-conserving set)~\cite{van2018pseudodojo}. A strict energy cutoff of 96 Ry is used for \ch{MgB2} in QE. A uniform $24\times 24\times 24$ electron k-mesh and $6\times 6\times 6$ phonon q-mesh are used in QE, and a finer $120\times 120\times 48$ k-mesh and $30\times 30 \times 24$ q-mesh are used in the Wannier interpolation in EPW. Further increasing the fine k-mesh to $144 \times 144 \times 60$ and q-mesh to $36\times 36\times 30$ only changes $T_c$ by less than 0.3 K.

\section{Electronic property of \ch{MgB2}}

In this section, we first present the first-principles electronic properties of \ch{MgB2}. We then develop analytic models, emphasizing the symmetry perspective, which allow us to match many key features of \ch{MgB2} even with minimal inputs from \textit{ab initio} calculations.

\subsection{Crystal structure}

\begin{figure}[htbp]
    \centering
    \includegraphics[width=0.7\textwidth]{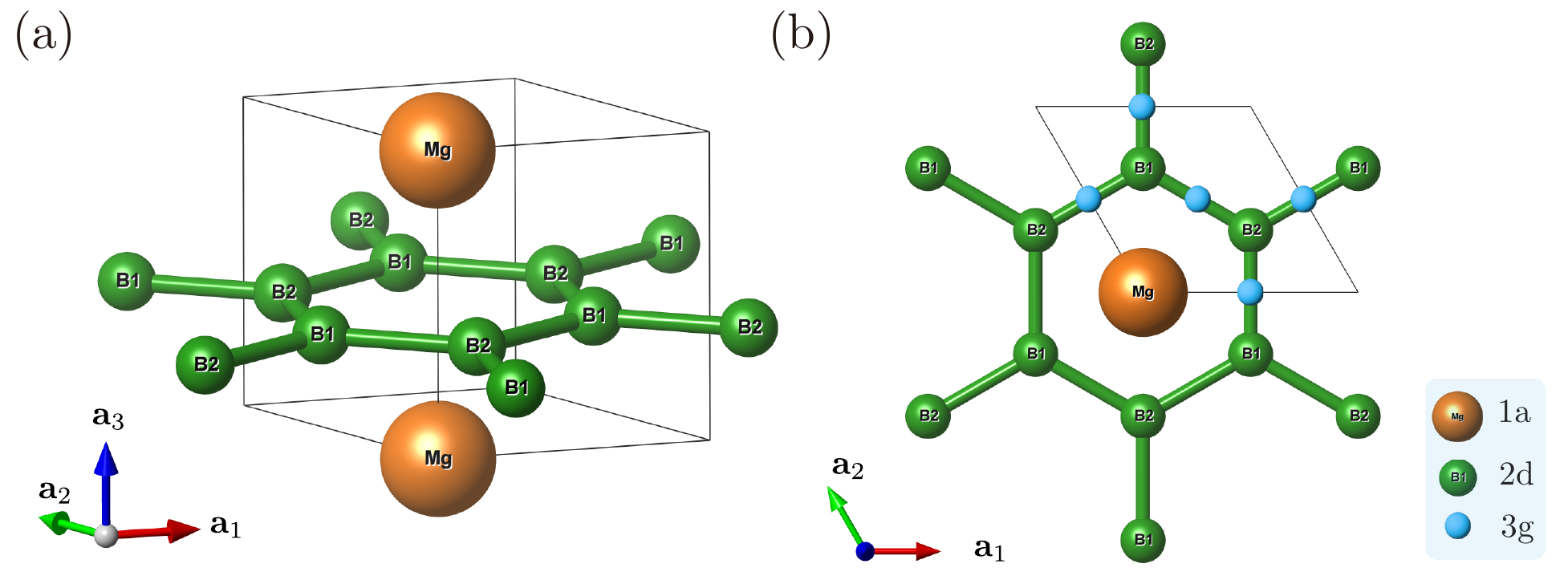}
    \caption{\label{app:fig:crystal_structure} The crystal structure of \ch{MgB2} from (a) side view and (b) top view. The Mg atom forms a triangular lattice while the B atoms form a honeycomb lattice. In (b), three blue stars mark the $sp^2$ hybrid states formed by the $s,p_x,p_y$ orbitals of B in the unit cell. They are located at the bond center of two neighboring B atoms and form an effective kagome lattice.  
    }
\end{figure}

The crystal structure of \ch{MgB2} is shown in \cref{app:fig:crystal_structure}. It has space group (SG) 191 $P6/mmm$ symmetry. 
The unit cell basis is defined as
\begin{equation}\begin{aligned}
\bm{a}_1=a(1,0,0),\ \bm{a}_2=a(-\frac{1}{2}, \frac{\sqrt{3}}{2}, 0),\ \bm{a}_3=c(0,0,1),
\label{app:eq:MgB2_unit_cell_basis}
\end{aligned}\end{equation}
with the lattice constants $a=$\SI{3.083}{\angstrom} and $c=$\SI{3.521}{\angstrom}~\cite{agrestini2004sc}. The atoms are located at
\begin{equation}\begin{aligned}
\tt_{Mg}= (0,0,0),\ 
\tt_{B_1}= (\frac{1}{3},\frac{2}{3},\frac{1}{2}),\
\tt_{B_2}= (\frac{2}{3}, \frac{1}{3}, \frac{1}{2}) ,
\label{app:eq:atomic_coord}
\end{aligned}\end{equation}
where the direct coordinates are written under the three unit cell bases \cref{app:eq:MgB2_unit_cell_basis}. 
The Mg atom forms a triangular lattice on the $z=0$ plane,  while the two B atoms form a honeycomb lattice on the $z=\frac{c}{2}$ plane.

\begin{table}[htbp]
\centering
\begin{tabular}{c|c|c|c|c|c|c}
\hline\hline
EBR                          & $\Gamma$                 & M                          & K            & A            & L                          & H             \\ \hline
$A_1^{\prime}(s)@2d$         & $\Gamma_1^+, \Gamma_4^-$ & $M_1^+, M_4^-$               & $K_5$          & $A_2^-, A_3^+$ & $L_2^-, L_3^+$               & $H_6$           \\ \hline
$E^\prime (p_x,p_y)@2d$      & $\Gamma_5^+, \Gamma_6^-$ & $M_1^+, M_2^+, M_3^-, M_4^-$ & $K_1,K_4, K_5$ & $A_5^-, A_6^+$ & $L_1^-, L_2^-, L_3^+, L_4^+$ & $H_2, H_3, H_6$ \\ \hline
$A_2^{\prime\prime}(p_z)@2d$ & $\Gamma_2^-, \Gamma_3^+$ & $M_2^-, M_3^+$               & $K_6$          & $A_1^+, A_4^-$ & $L_1^+, L_4^-$               & $H_5$           \\\hline 
$A_g (s) @ 3g$               & $\Gamma_1^+, \Gamma_5^+$ & $M_1^+,M_3^-,M_4^-$          & $K_1,K_5$      & $A_2^-,A_5^-$  & $L_2^-, L_3^+, L_4^+$        & $H_3,H_6$       \\\hline
$B_{2u} (p_y) @ 3g$               & $\Gamma_4^-, \Gamma_6^-$ & $M_1^+,M_2^+,M_4^-$          & $K_4,K_5$      & $A_3^+,A_6^+$  & $L_1^-, L_2^-, L_3^+$        & $H_2,H_6$       \\\hline
$A_1 (s) @6m$ & $\Gamma_1^+, \Gamma_4^-, \Gamma_5^+, \Gamma_6^-$ & $2M_1^+, M_2^+, M_3^-, 2M_4^-$  & $K_1,K_4, 2K_5$ & $A_2^-, A_3^+, A_5^-, A_6^+$ & $ L_1^-, 2L_2^-, 2L_3^+, L_4^+$ & $H_2, H_3, 2H_6$ \\
\hline\hline
\end{tabular}
\caption{\label{table: EBR_SG191}
The elementary band representations (EBRs)~\cite{bradlyn2017topological, cano2018building, elcoro2021magnetic} in SG 191 for different orbitals on kagome Wyckoff position $3g=(\frac{1}{2}, 0, \frac{1}{2}), (\frac{1}{2}, \frac{1}{2}, \frac{1}{2}), (0, \frac{1}{2}, \frac{1}{2})$, the honeycomb site $2d=(\frac{1}{3}, \frac{2}{3}, \frac{1}{2}), (\frac{2}{3}, \frac{1}{3}, \frac{1}{2})$, and the non-maximal site $6m=(x,2x,\frac{1}{2}),(-2x,-x,\frac{1}{2}),(x,-x,\frac{1}{2}),(-x,-2x,\frac{1}{2}),(2x,x,\frac{1}{2}),(-x,x,\frac{1}{2})$. 
The coordinates are given in unit cell basis \cref{app:eq:MgB2_unit_cell_basis}. Note that at K, the $K_5$ and $K_6$ IRREPs are 2D, while others are 1D. At $\Gamma$, $\Gamma_5^{\pm}$ and $\Gamma_6^{\pm}$ are 2D. At M, all IRREPs are 1D. }
\end{table}

\subsection{First-principle results for electronic structure}\label{app:sec:DFT_electron}

In this section, we discuss the electronic properties of \ch{MgB2}. We start by presenting the \textit{ab initio} results and then show that most of the properties of these results can be obtained with \emph{minimal} help from the \textit{ab initio}. Comparison with graphene is also included. 

\begin{figure}[htbp]
    \centering
    \includegraphics[width=\textwidth]{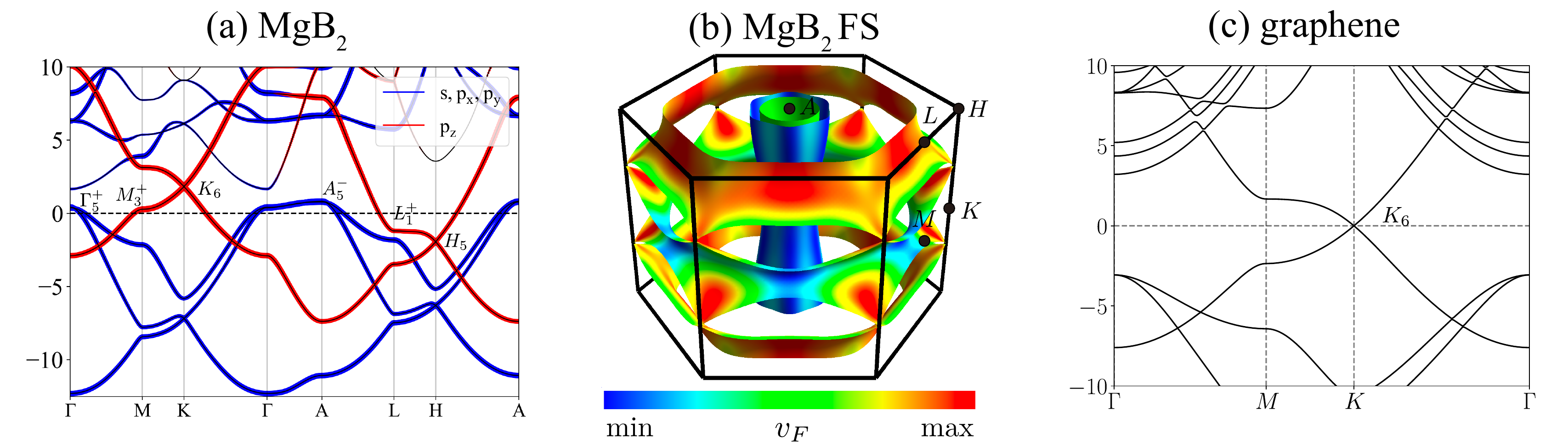}
    \caption{\label{Fig: MgB2-bands} (a) The band structure of \ch{MgB2}. Blue bands are from the $s, p_x, p_y$ orbitals of B, and red bands are from the $p_z$ orbitals of B. (b) The Fermi surface (FS) of \ch{MgB2}, where the color denotes the Fermi velocity (red denotes large velocity). (c) The band structure of graphene. 
    }
\end{figure}

\begin{figure}[htbp]
    \centering
    \includegraphics[width=0.8\textwidth]{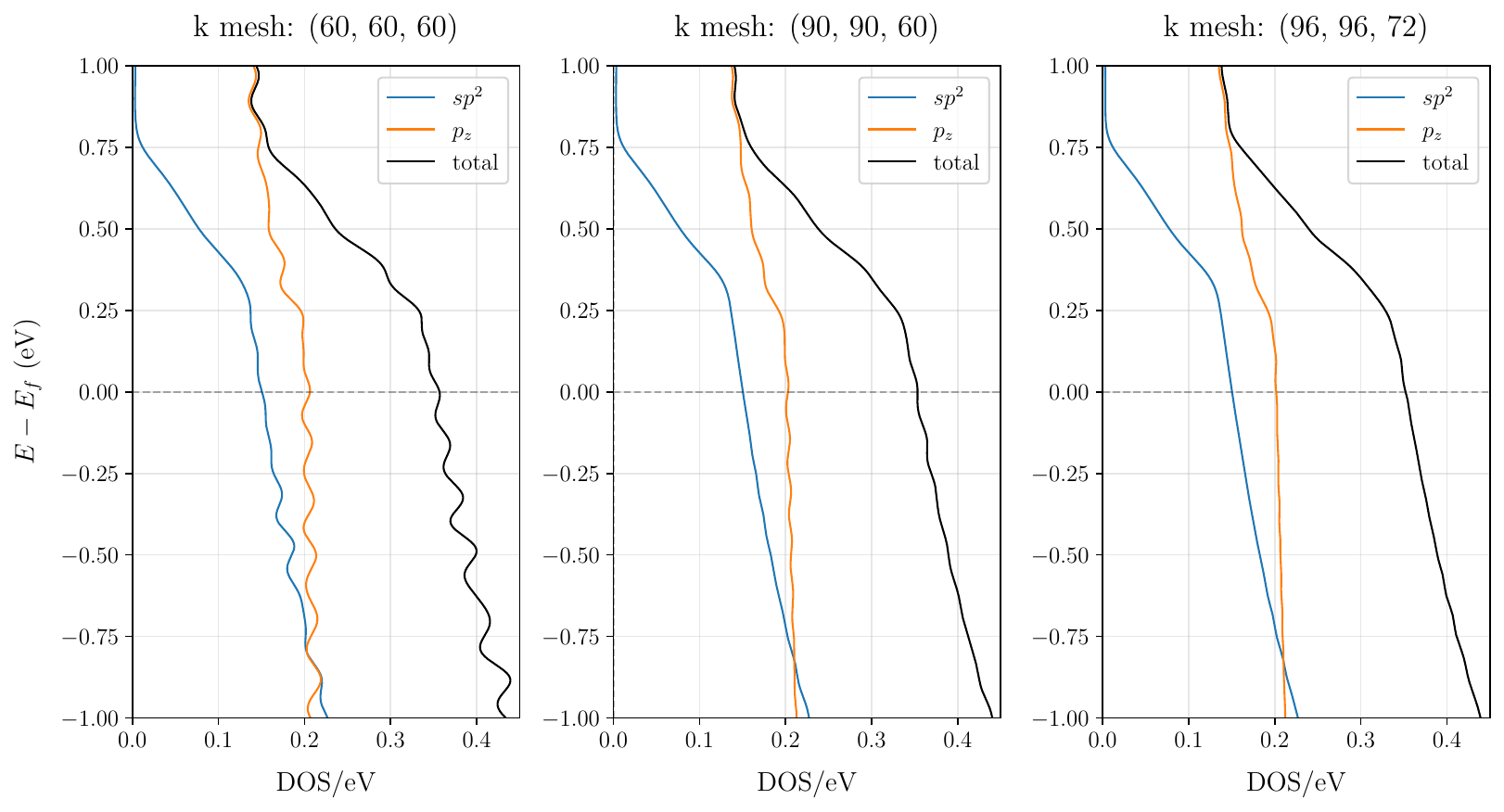}
    \caption{\label{app:fig:DOS_test_convergence} DOS of \ch{MgB2}. We test the convergence of DOS w.r.t the $\kk$ mesh sampling $(N_{k_1}, N_{k_2}, N_{k_3})$ in the BZ, as given in the caption of each subplot. As the $\sigma$ FS from the $sp^2$ bands is small, a dense $\kk$ mesh is required to converge the DOS. We find a $(96,96,72)$ mesh gives smooth enough DOS in \ch{MgB2}. A Gaussian smearing of 50 meV is used when calculating the DOS. We remark that a larger smearing parameter will require a smaller $\kk$ mesh. 
    }
\end{figure}

In \cref{Fig: MgB2-bands} (a), the band structure of \ch{MgB2} is shown with orbital weights and irreducible representations (IRREPs). In \cref{Fig: MgB2-bands} (b), the Fermi surface (FS) of \ch{MgB2} is shown. We observe that:

\begin{itemize}
\item The $s,p_x,p_y$ orbitals of Boron form the $sp^2$ bonding and anti-bonding states. The three bonding states are equivalent to the $s$ orbitals on a kagome lattice formed at the B--B bond centers, as shown in \cref{app:fig:crystal_structure}(b). The bands of the bonding states have a large bandwidth ranging from about -12 to 1 eV, as shown by the three lower blue bands in \cref{Fig: MgB2-bands}(a). They contribute to two cylinder FSs close to the $\Gamma$-$A$ line, which are denoted as the $\sigma$ FSs. These $sp^2$ bonding states have weak $z$-directional hoppings and the $\sigma$ FSs from them are quasi-2D. The three anti-bonding states are located high above $E_f$. We will show that the hoppings of these $sp^2$ electrons can be obtained without \textit{ab initio}, but just from hopping integrals of hydrogen-like orbitals. 

\item The $p_z$ orbital of Boron forms a Dirac cone at K and H, as marked by the two red bands in \cref{Fig: MgB2-bands}(a). The Dirac point is above $E_f$ on the $k_z=0$ plane while it is below $E_f$ on the $k_z=\pi$ plane. The FS given by the $p_z$ bands is denoted as the $\pi$ FS, which has a strong $k_z$-dispersion. Being a 3D bulk system, the $p_z$ bands in \ch{MgB2} are different from those in graphene, where the system is 2D and the Dirac point is located exactly at $E_f$, as shown in \cref{Fig: MgB2-bands}(c). In \ch{MgB2}, the Dirac nodes in-plane form line nodes in the 3D BZ. The dispersion of $p_z$ bands can also be obtained without \textit{ab initio}. 

\item The band from the $s$ orbital of Mg lies above the $E_f$, which is less relevant to the FS property. 

\item The density of states (DOS) of \ch{MgB2}, resolved into the $sp^{2}$ and $p_{z}$ manifolds, is shown in \cref{app:fig:DOS_test_convergence}. With a dense $\kk$-mesh $(N_{k_1},N_{k_2},N_{k_3})=(96,96,72)$, the $sp^{2}$ DOS falls almost linearly near $E_{f}$ when the energy is increased. The slope steepens beyond $\sim0.35$ eV, and the DOS drops to zero at $\sim 0.8$ eV, marking the upper edge of the $sp^{2}$ bands on the $k_z=\pi$ plane. In contrast, the $p_{z}$ DOS is flatter and larger compared with the $sp^2$ DOS near $E_f$. The behaviour of the $sp^{2}$ DOS will be discussed analytically in \cref{app:sec:DOS_analytic} using the $\kk\cdot\pp$ model.
\end{itemize}

\subsubsection{Wannier tight-binding models}\label{Sec:wannier_TB}

To analyze the electronic properties of \ch{MgB2}, we construct three tight-binding models, each in a different orbital basis derived from maximally localized Wannier functions. These models are used for different purposes, as we stated in the following.

The first model contains the $s$ and $p$ orbitals of both boron and magnesium, i.e., 12 orbitals in total. This model can faithfully reproduce all bands within $\pm5$ eV of the Fermi level $E_f$, and give the couplings between boron and magnesium. It is also useful to study the splitting between the $sp^2$ bonding and anti-bonding states of boron, which is vital for the large electron-phonon coupling (to be studied in \cref{app:sec:spxpy2sp2_basis}). 

The second model contains the $s$ and $p$ orbitals of boron only, \ie, 8 orbitals in total. This model can faithfully reproduce the bands near $E_f$, and the $(s,p_x,p_y)$ and $p_z$ orbitals are decoupled as they have opposite $M_z$ eigenvalues. Thus this model is block-diagonal for the $(s,p_x,p_y)$ and $p_z$ orbitals. 

The third model contains only the $sp^2$ bonding states and $p_z$ orbitals of boron, i.e., 5 orbitals in total. This simple model can faithfully reproduce the bands relevant to the Fermi surface. The $sp^2$ bonding states and $p_z$ orbitals are decoupled in this model as they have opposite $M_z$ eigenvalues. 
In this model, the onsite energy and the nearest-neighbor (NN) hopping of the boron-$p_z$ orbitals are $\epsilon_{p_z}=0.00, t_{p_z}^{NN}=-2.04$, while the onsite energies, NN, and the next-NN (NNN) hoppings for the $sp^2$ bonding states are $\epsilon_{b}=-5.38, t_{b}^{NN}=-1.56, t_b^{NNN}=-0.43$, respectively, with numbers given in eV. 
An approximated value for these hoppings can be obtained without \textit{ab initio} from atomic orbital overlap, which we show in \cref{app:sec:hopping_integral}.

\begin{figure}[htbp]
    \centering
    \includegraphics[width=1\textwidth]{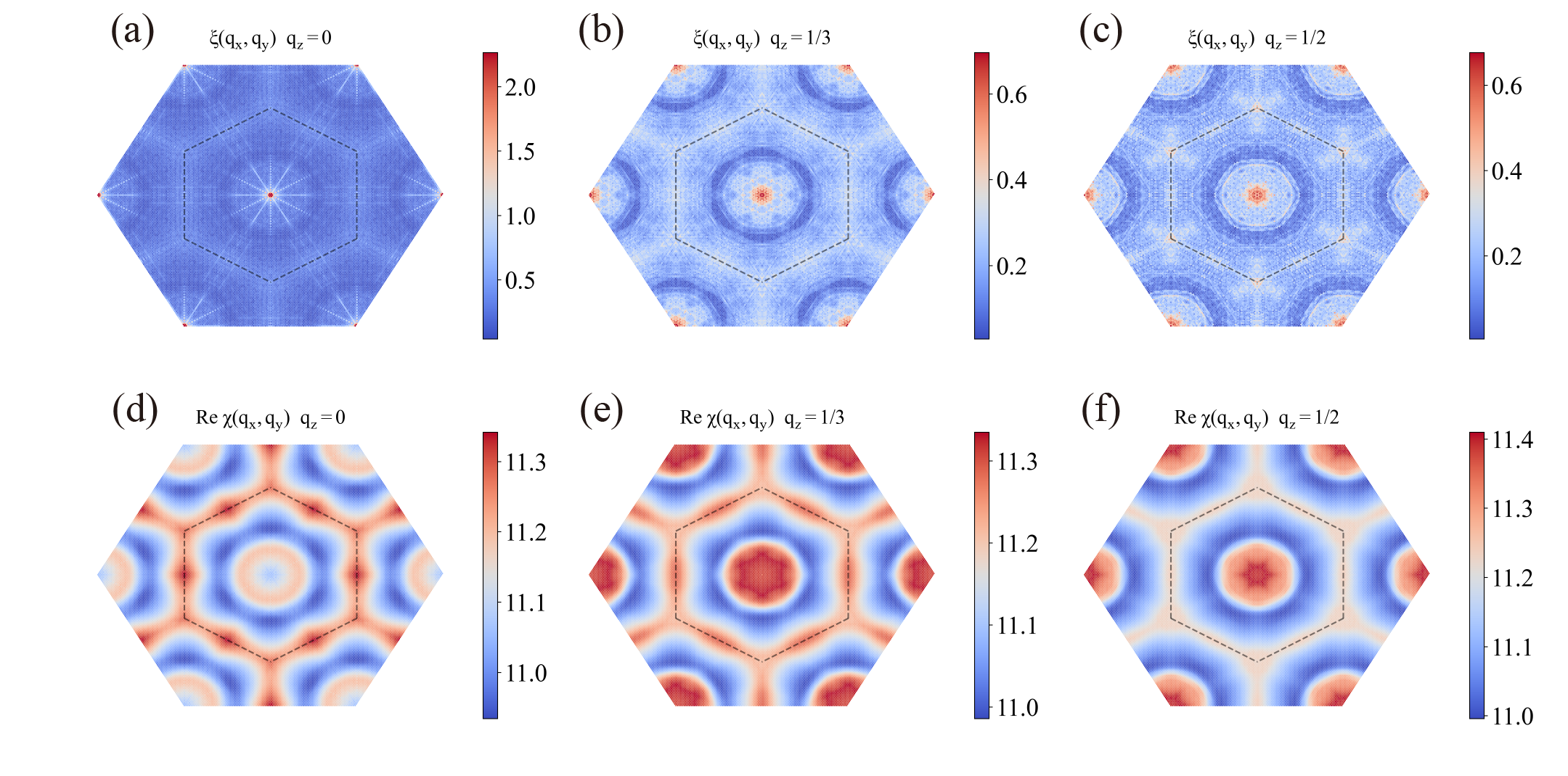}
    \caption{\label{Fig: MgB2-nesting} (a)-(c) The Fermi surface (FS) nesting function and (d)-(f) the real part of the total susceptibility for fixed $q_z/2\pi=0,\frac{1}{3},\frac{1}{2}$ planes in \ch{MgB2}. The strongest FS nesting is observed near $\Gamma$, while weak nesting peaks are shown close to K and M. 
    }
\end{figure}

\subsubsection{Fermi surface nesting function}

In \cref{Fig: MgB2-nesting}, we show the Fermi surface nesting function and the real part of the total susceptibility for fixed $q_z/2\pi=0,\frac{1}{3},\frac{1}{2}$ (in direct coordinates) planes in \ch{MgB2}. The strongest FS nesting is observed near $\Gamma$, while weak nesting peaks are shown close to K and M. The results are obtained using the faithful 12-orbital Wannier TB model. The FS nesting function is useful to discuss the EPC strength, as will be shown in \cref{app:sec:EPC_strength}. 
In an upcoming paper~\cite{EPCpaper}, we will show that the susceptibility can also be obtained analytically.

\subsection{Analytic understanding of the electronic property of \ch{MgB2}}\label{app:sec:electron_analytic}

In this section, we construct minimal analytic models for the electronic Hamiltonian of \ch{MgB2}. These models capture the essential physics of \ch{MgB2} while requiring only very minimal input from density-functional theory (DFT).  
\cref{app:sec:hopping_integral} demonstrates that the \textit{ab initio} hopping parameters in \ch{MgB2} can be well approximated using simple hopping integrals of hydrogen-like orbitals.  
\cref{app:sec:spxpy2sp2_basis} presents a convenient basis transformation from the boron $s,p_x,p_y$ orbitals to three effective $s$-like orbitals located at non-maximal Wyckoff positions near boron, and then to the $sp^2$ bonding and anti-bonding basis located at the $sp^2$ bond centers. 
\cref{app:sec:simple_Hel_model} develops minimal tight-binding (TB) models for the boron $s,p_x,p_y$ orbitals, while \cref{app:sec:emergence_sigma_FS} analyzes the reason for the emergence of the $\sigma$ Fermi surface in \ch{MgB2} based on analytic TB couplings between Mg and B atoms.

\subsubsection{Hopping integral of electrons}\label{app:sec:hopping_integral}

In this section, we show that the hoppings obtained from DFT in \ch{MgB2} can be faithfully reproduced using the hopping integral of hydrogen-like orbitals. 

We start with a two-nuclei system. The Hamiltonian can be written as 
\begin{equation}\begin{aligned} 
\hat{H}= \hat{K} + \hat{V}_1 + \hat{V}_2
\end{aligned}\end{equation} 
where $\hat{K}$ is the kinetic term of an electron with $m_e$ mass, $\hat{V}_{i}$ is the Coulomb potential provided by the $i$'th nucleus. We now aim to solve this two-particle problem in order to find the hopping between the electrons on the two atoms. 
Note that we assume no self-consistent Hartree or exchange potential. $V_i$ will contain the effective charge $Z_i$, a parameter used to model hydrogen-like orbitals. 

First, we assume 
\begin{equation}\begin{aligned} 
&(\hat{K}+\hat{V}_i)|\phi_{i,\alpha}\rangle = \epsilon_{i,\alpha}|\phi_{i,\alpha}\rangle  
\end{aligned}\end{equation} 
where $|\phi_{i,\alpha}\rangle$ is the $\alpha$-th eigen wavefunction, obtained by solving the single-atom problems at atom $i$. Note that $|\phi_{i,\alpha} \rangle $ are not necessarily orthogonal when $i\ne j$.  We define $S_{i\alpha, j\beta}=\langle \phi_{i,\alpha} |\phi_{j,\beta}\rangle$ as the orbital overlap matrix, with $S_{i\alpha, i\beta}=\delta_{\alpha\beta}$. 
In the $|\phi_{i,\alpha}\rangle$ basis, the Hamiltonian is defined as
\begin{equation}\begin{aligned} 
H_{i\alpha, j\beta} &= \langle \phi_{i,\alpha}| \hat{H} |\phi_{j,\beta}\rangle \\ 
H_{1\alpha, 1\beta} &= \epsilon_{1, \alpha} \delta_{\alpha\beta} + \langle \phi_{1,\alpha}|\hat{V}_2|\phi_{1,\beta}\rangle \\ 
H_{2\alpha, 2\beta} &= \epsilon_{2, \alpha} \delta_{\alpha\beta} + \langle \phi_{2,\alpha} |\hat{V}_1|\phi_{2, \beta}\rangle \\ 
H_{1\alpha, 2\beta} &= \langle \phi_{1,\alpha} |\hat{V}_1 |\phi_{2,\beta}\rangle +\epsilon_{2,\beta} \langle \phi_{1,\alpha} |\phi_{2,\beta} \rangle
\\
H_{2\alpha, 1\beta} &= \langle \phi_{2,\alpha} |\hat{V}_2|\phi_{1,\beta} \rangle +\epsilon_{1,\beta} \langle \phi_{2,\alpha} |\phi_{1, \beta}\rangle \\
&= \langle \phi_{2,\alpha} |\hat{V}_1 |\phi_{1,\beta} \rangle +\epsilon_{2,\alpha} \langle \phi_{2,\alpha} |\phi_{1, \beta}\rangle. 
\label{app:eq:Hij_orbital_integral}
\end{aligned}\end{equation} 
$H_{1\alpha, 2\beta}$ and $H_{2\beta, 1\alpha}$ are the hopping integral between orbitals from two atoms. The generalized eigen equation in the $|\phi_{i,\alpha} \rangle$ basis reads 
\begin{equation}\begin{aligned}
\sum_{j\beta} H_{i\alpha, j\beta} \psi_{n, j\beta}= \epsilon_n S \psi_{n, i\alpha},
\label{app:eq:hopping_integral_H_eq}
\end{aligned}\end{equation}
where $\psi_n$ and $\epsilon_n$ are the $n$-th eigenvector and eigenvalue. 
We can define a new orthogonal basis
\begin{equation}\begin{aligned}
\tilde{\psi}_n &= S^{\frac{1}{2}} \psi_n, \quad
\tilde{H} &= S^{-\frac{1}{2}} H S^{-\frac{1}{2}}\quad
\Rightarrow \quad 
\sum_{j\beta} \tilde{H}_{i\alpha, j\beta} \tilde{\psi}_{n, j\beta} &= \epsilon_n \tilde{\psi}_{n, i\alpha}. 
\label{app:eq:hopping_orthonal_basis}
\end{aligned}\end{equation}
Note that when the basis $|\phi_{i,\alpha}\rangle$ is incomplete, $H$ in \cref{app:eq:hopping_integral_H_eq} is obtained by projecting into the subspace we consider, \ie, $\hat{P}\hat{H}\hat{P}$, where
$\hat{P}=\sum_{i,\alpha,j,\beta} |\phi_{i,\alpha}\rangle (S^{-1})_{i\alpha,j\beta}\langle \phi_{j,\beta}|$ is a projector instead of the identity operator.

Consider a special case where the two atoms are the same and the atomic orbitals have the same eigenvalues, so that $\epsilon_{i,\alpha}=E_0$ is the same for all orbitals. This holds for \ch{MgB2} when we consider the hydrogen-like $2s,2p$ orbitals with the same principal quantum number $n=2$, since in the hydrogenic Schrödinger equation the energy depends only on $n$ (the general solution will be discussed in an upcoming paper~\cite{EPCpaper}). 
In this case, we can define 
\begin{equation}
    h_{1\alpha, 1\beta}= \langle \phi_{1,\alpha} |\hat{V}_2| \phi_{1,\beta} \rangle, \quad 
    h_{2\alpha, 2\beta}= \langle \phi_{2,\alpha} |\hat{V}_1| \phi_{2,\beta} \rangle, \quad 
    h_{1\alpha, 2\beta}= \langle \phi_{1,\alpha} |\hat{V}_1| \phi_{2,\beta} \rangle, \quad
    h_{2\beta, 1\alpha}= h_{1\alpha, 2\beta}^\dagger,
    \label{app:eq:Hbar_orbital_overlap}
\end{equation}
and the generalized eigen equation becomes
\begin{equation}\begin{aligned}
    \sum_{j\beta} h_{i\alpha, j\beta}  \psi_{n, j\beta} &= (\epsilon_n - E_0) S\psi_{n, i\alpha}, \\
    \Rightarrow 
    \sum_{j\beta} \tilde{h}_{i\alpha, j\beta}  \tilde{\psi}_{n, j\beta} &= (\epsilon_n - E_0) \tilde{\psi}_{n, i\alpha},
    \label{app:eq:Hbar_eigenequation}
\end{aligned}\end{equation}
where $\tilde{\psi}_n = S^{\frac{1}{2}} \psi_n, \tilde{h}= S^{-\frac{1}{2}} h S^{-\frac{1}{2}}$. 
It can be seen that $E_0$ acts as a global shift of the eigenvalues. Consequently, we can use \cref{app:eq:Hbar_orbital_overlap} to evaluate the hopping integrals and then apply \cref{app:eq:Hbar_eigenequation} to obtain the hoppings in the orthonormal basis $\tilde{\psi}_n$, which can be directly compared with those obtained from DFT. Remark that if different orbitals have different onsite energies, then hoppings in the orthonormal basis $\tilde{\psi}_n$ will have contributions from both onsite energies and hoppings in the original non-orthogonal basis $\psi_n$. 

More generally, write $S=I+\Sigma$, with
\begin{equation}
    \Sigma = \begin{bmatrix}
        0 & S_{12} \\
        S_{21} & 0 
    \end{bmatrix}.
\end{equation}
Assume the off-diagonal terms in $S$ are small, then
\begin{equation}
    S^{-\frac{1}{2}}=(I+\Sigma)^{-\frac{1}{2}}\approx
    I - \frac{1}{2}\Sigma.
\end{equation}
Thus
\begin{equation}
    \tilde{H} \approx H - \frac{1}{2}(\Sigma H + H \Sigma),
\end{equation}
where we ignore the second-order term $\frac{1}{4}\Sigma H \Sigma$. By using $H_{ij}$ in \cref{app:eq:Hij_orbital_integral}, we obtain
\begin{equation}
    \tilde{H}_{1\alpha, 2\beta} = h_{1\alpha, 2\beta} 
    - \frac{1}{2}\left[ (S_{12}h_{22})_{\alpha\beta} + (h_{11} S_{12})_{\alpha\beta} \right]
    + \frac{1}{2}(\epsilon_{2,\beta} - \epsilon_{1,\alpha}) S_{1\alpha, 2\beta}.
\end{equation}

\paragraph{Orbital integral from hydrogen-like orbitals.}
The Coulomb potential from the nucleus at position $\RR$ is
\begin{equation}\begin{aligned} 
V_i(\rr-\RR) = -\frac{Z_e e^2}{4\pi \epsilon_0|\rr-\RR|}
\end{aligned}\end{equation} 
where $\epsilon_0=5.5263\times 10^{-3}$ \SI{}{e^2 eV^2 A^{-1}}, and $Z_e$ is the nuclear charge.

We consider the hydrogen-like $s$ and $p$ orbitals~\cite{griffiths2018introduction, PhysRev.36.57} with \textit{cubic harmonic form}, defined in the following. The hydrogen-like orbitals in spherical harmonics form are $\psi_{nlm}(\rr)=R_{nl}(r) Y^m_l(\theta,\phi)$, where $R_{nl}(r)$ is the radial function and $Y^m_l(\theta,\phi)$ is the spherical harmonics. The cubic harmonics are more convenient for solid states with point group symmetry instead of spherical symmetry. Cubic harmonics $X_{lc}(\rr)$ are defined as linear combinations of spherical harmonics, e.g.,
$s=X_{00}(\rr)=Y_0^0=\frac{1}{\sqrt{4\pi}}$, $p_z=\sqrt{\frac{3}{4\pi}} \frac{z}{r}=Y^0_1, p_x=\sqrt{\frac{3}{4\pi}} \frac{x}{r}=\frac{1}{\sqrt{2}}(Y^{-1}_1 - Y^{1}_{1}),  p_y=\sqrt{\frac{3}{4\pi}} \frac{y}{r}=\frac{1}{\sqrt{2}}(Y^{-1}_1 + Y^{1}_{1})$.

Thus the hydrogen-like wavefunctions of $2s, 2p$ orbitals (e.g., boron valence shell) read 
\begin{equation}\begin{aligned} 
&\psi_{2s}(\rr) = \frac{(2-Z_e r/a_0)}{4\sqrt{2\pi} (a_0/Z_e)^{3/2}}e^{-\frac{Z_e r}{2a_0}}\nonumber\\ 
&\psi_{2 p_{i}}(\rr) = \frac{r_i}{ 8\sqrt{\pi} (a_0/Z_e)^{5/2}}e^{-\frac{Z_e r}{2 a_0}}
\end{aligned}\end{equation}
where $Z_e$ is the effective nuclear charge, and $a_0$ is a decaying parameter. 
The hopping between two orbitals is given by the integration (\ie, using \cref{app:eq:Hbar_orbital_overlap}, in non-orthogonal basis): 
\begin{equation}\begin{aligned} 
t_{ij}(\RR-\RR') &= \int_{\rr}d\rr \frac{-Z_e e^2}{4\pi \epsilon_0 |\rr-\RR| } \psi^*_{i}(\rr-\RR) \psi_{j}(\rr-\RR') \\
&= \int_{\rr}d\rr \frac{-Z_e e^2}{4\pi \epsilon_0 |\rr| } \psi^*_{i}(\rr) \psi_{j}(\rr+\RR-\RR').
\end{aligned}\end{equation} 

We then evaluate the hopping integral using hydrogen-like orbitals from boron. Boron has an effective nuclear charge $Z_e=3$ if the two $1s$ electrons are treated as core electrons, and we use $a_0=$\SI{0.529}{\angstrom}, \ie, the Bohr radius. We place the $s,p$ orbital on two boron atoms located at two honeycomb sites, as defined in \cref{app:eq:atomic_coord}, with the lattice constants $a=$\SI{3.08}{\angstrom} for \ch{MgB2} (which is the only constant needed in the calculation from experiments or from \textit{ab initio}). Notice that the orbitals from two borons have non-zero overlaps, thus the overlap matrix $S$ of the basis needs to be considered (e.g., the overlap of the $s$ orbitals from two borons is 0.34, and the bare hopping integral without considering the overlap matrix can reach \SI{10}{eV}, much larger than the value with overlap matrix, which is about \SI{3}{eV}). Moreover, in order to be consistent with the gauge used in the Wannier TB model from DFT, we add a minus sign to the $p_x,p_y$ orbitals. Such gauge signs have no bearing on the physics. 

The computed hopping integrals (in orthogonal basis, \ie, \cref{app:eq:hopping_orthonal_basis}) are tabulated in \cref{app:table:hopping-integral-values}, with a comparison with the hoppings in Wannier TB. A very good agreement is observed, which means that the electronic model of the $s,p$ orbitals can be simply obtained with high accuracy without DFT information. 
We remark that the value of the hopping integrals depends on the value of the spread $a_0$. Larger spreads lead to larger hoppings. We use the Bohr radius for simplicity, which assumes the orbitals are the same as atomic orbitals. The corresponding spreads of the $s$ and $p_{x,y}$ orbitals are 1.30 and 0.93 \SI{}{\angstrom}$^2$, respectively, which are close to the spread of the corresponding Wannier functions in \ch{MgB2}, \ie, 1.20 and 1.04 \SI{}{\angstrom}$^2$ for $s$ and $p_{x,y}$ orbitals, respectively, in the 8-orbital Wannier model. Thus we take the Bohr radius for simplicity.

\begin{table}[htbp]
\centering
\begin{tabular}{c|c|c|c|c c|c|c|c|c|c}
\hline\hline
Wannier TB & $B_{s}^2$ & $B_{p_x}^2$ & $B_{p_y}^2$ & $B_{p_z}^2$ &  & Hopping integral & $B_{s}^2$ & $B_{p_x}^2$ & $B_{p_y}^2$ & $B_{p_z}^2$ \\ \hline
$B_{s}^1$ & -2.7 & 3.4 & -2.0 & 0.0 &  & $B_{s}^1$ & -3.0 & 3.3 & -1.9 & 0.0 \\ \hline
$B_{p_x}^1$ & -3.4 & 2.9 & -3.0 & 0.0 &  & $B_{p_x}^1$ & -3.3 & 3.6 & -3.3 & 0.0\\ \hline
$B_{p_y}^1$ & 2.0 & -3.0 & -0.5 & 0.0 &  & $B_{p_y}^1$ & 1.9 & -3.3 & -0.2 & 0.0 \\ \hline
$B_{p_z}^1$ & 0.0 & 0.0 & 0.0 & -2.14 &  & $B_{p_z}^1$ & 0.0 & 0.0 & 0.0 & -2.13 \\ \hline\hline
\end{tabular}
\caption{\label{app:table:hopping-integral-values}
Comparison of hopping values in the Wannier TB model (left half) and the hydrogen-like orbital integration (right half). All numbers are given in \SI{}{eV}. 
}
\end{table}

We also consider the hopping integral between boron and Magnesium. 
For hydrogen-like orbitals with effective charge $Z_e$ and effective Bohr radius $a_0^e$, the eigenenergies are 
\begin{equation}
    E_n = -\frac{Z_e^2}{n^2} \left(\frac{a_0}{a_0^{e}}\right)^2 \text{Ry},
\end{equation}
where Ry$\approx -13.606$ eV is the Rydberg constant, and $n$ is the principal quantum number. 
Mg has $3s$ and $3p$ orbitals, with the form
\begin{equation}\begin{aligned} 
&\psi_{3s}(\rr) = \frac{1}{\sqrt{27\pi} (a_0^{\text{Mg}}/Z_e^{\text{Mg}})^{3/2}} \left(1- \frac{2Z_e^{\text{Mg}} r}{3a_0^{\text{Mg}}} + \frac{1}{9} (\frac{2Z_e^{\text{Mg}} r}{3a_0^{\text{Mg}}})^2\right) e^{-\frac{Z_e^{\text{Mg}}  r}{3a_0^{\text{Mg}}}} \\ 
&\psi_{3 p_{i}}(\rr) = \frac{4 r_i}{ 27\sqrt{2\pi} (a_0^{\text{Mg}}/Z_e^{\text{Mg}})^{5/2}} \left(1-\frac{Z_e^{\text{Mg}} r}{6a_0^{\text{Mg}}}\right) e^{-\frac{Z_e^{\text{Mg}} r}{3a_0^{\text{Mg}}}},
\end{aligned}\end{equation}
where we take $Z_e^{\text{Mg}}=2$, and $a_0^{\text{Mg}}=$\SI{0.23}{\angstrom}. The spreads of the $3s$ and $3p$ orbitals are 2.74 and \SI{2.38}{\angstrom}$^2$, respectively, close to the spreads in the \textit{ab initio} Wannier model of around \SI{2.40}{\angstrom}$^2$. By using \cref{app:eq:Hij_orbital_integral}, the hoppings between $B^1_{p_z}$ at $(\frac{1}{3}, \frac{2}{3}, \frac{1}{2})$ and Mg orbitals at $(0, 0, 0)$ are obtained and tabulated in \cref{app:table:hopping-integral-values-Mg-Bpz}, which agrees reasonably with DFT values (note that we add a $-1$ sign to $\psi_{3s},\psi_{3p_z}$ when compute \cref{app:table:hopping-integral-values-Mg-Bpz}, in order to match the phase in DFT Wannier model).

\begin{table}[htbp]
\centering
\begin{tabular}{c|c|c|c|c |c|c|c|c|c}
\hline\hline
Wannier TB & $\text{Mg}_{s}$ & $\text{Mg}_{p_x}$ & $\text{Mg}_{p_y}$ & $ \text{Mg}_{p_z}$ & Hopping integral & $\text{Mg}_{s}$ & $\text{Mg}_{p_x}$ & $\text{Mg}_{p_y}$ & $ \text{Mg}_{p_z}$ \\ \hline
$B_{p_z}^1$ & 1.08 & 0.0 & 1.55 & 1.00  & $B_{p_z}^1$ & 0.62 & 0.0 & 1.20 & 0.73 \\ \hline\hline
\end{tabular}
\caption{\label{app:table:hopping-integral-values-Mg-Bpz}
Comparison of hopping values between $B^1_{p_z}$ and Mg $s,p$ orbitals in the Wannier TB model (left half) and the hydrogen-like orbital integration (right half). All numbers are given in \SI{}{eV}. 
}
\end{table}

\subsubsection{From boron $(s,p_x,p_y)$ basis to $sp^2$ basis }\label{app:sec:spxpy2sp2_basis}

In this section, we discuss the basis transformation from boron $(s,p_x,p_y)$ at honeycomb sites to the $3s$ basis at non-maximal Wyckoff position $6m=(x,2x,\frac{1}{2}),(-2x,-x,\frac{1}{2}),(x,-x,\frac{1}{2}),(-x,-2x,\frac{1}{2}),(2x,x,\frac{1}{2}),(-x,x,\frac{1}{2})$, and $sp^2$-bonding/anti-bonding basis at kagome sites $3g=(\frac{1}{2},0,\frac{1}{2}), (\frac{1}{2},\frac{1}{2},\frac{1}{2}), (0,\frac{1}{2},\frac{1}{2})$, where the coordinates are fractional coordinates given under the hexagonal lattice basis \cref{app:eq:MgB2_unit_cell_basis}. 
In the language of topological quantum chemistry (TQC) and elementary band representations (EBRs)~\cite{bradlyn2017topological, cano2018building, zak1980symmetry, zak1981band}, the basis transformation corresponds to the equivalence of EBRs
\begin{equation}
\begin{aligned}
    \{s, p_x,p_y\} &@ 2d \text{ (honeycomb)}\\ \Leftrightarrow \ \{s\} &@ 6m \text{ (non-maximal)}\\
    \Leftrightarrow \ \{s,p_y\} &@ 3g \text{ (kagome)},
\end{aligned}
\end{equation}
where the bonding (antibonding) orbitals are effective local $s$ ($p_y$) orbitals.

\begin{figure}[htbp]
    \centering
    \includegraphics[width=0.4\textwidth]{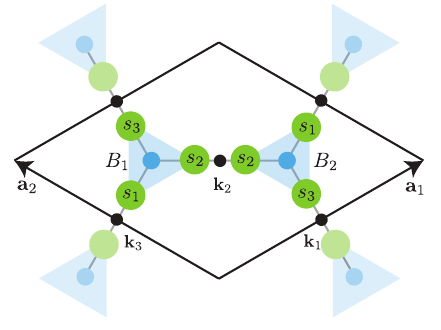}
    \caption{\label{Fig: sp2-basis} The $3s@6m$ basis (green circles) transformed from the $s,p_x,p_y$ orbitals of two boron atoms $B_1, B_2$ (blue circles, located at two honeycomb sites $\mathbf{t}_{h_1}=(\frac{1}{3},\frac{2}{3}), \mathbf{t}_{h_2}=(\frac{2}{3},\frac{1}{3})$, respectively). The black circles denote the kagome sites at the B--B bond center ($\mathbf{t}_{k_1}=(\frac{1}{2},0), \mathbf{t}_{k_2}=(\frac{1}{2}, \frac{1}{2}), \mathbf{t}_{k_3}=(0, \frac{1}{2})$), where the $sp^2$ bonding and anti-bonding states form.
    }
\end{figure}

Starting from the 12-orbital Wannier TB model introduced in \cref{Sec:wannier_TB}, we have the following representation matrix for the $(s,p_x,p_y,p_z)$ orbitals of each boron at the $2b$ wyckoff position:
\begin{equation}\begin{aligned}
D_{0}(C_6) &= 
\begin{bmatrix}
1 & 0 & 0 & 0 \\
0 & \frac{1}{2} & -\frac{\sqrt{3}}{2} & 0 \\
0 & \frac{\sqrt{3}}{2} & \frac{1}{2} & 0 \\
0 & 0 & 0 & 1
\end{bmatrix},\quad
D_{0}(C_3) = 
\begin{bmatrix}
1 & 0 & 0 & 0 \\
0 & -\frac{1}{2} & -\frac{\sqrt{3}}{2} & 0 \\
0 & \frac{\sqrt{3}}{2} & -\frac{1}{2} & 0 \\
0 & 0 & 0 & 1
\end{bmatrix},\\
D_0(M_x) &= 
\text{Diag}[1,1,-1,1],\quad
D_0(P) = \text{Diag}[1,-1,-1,-1], \quad
D_0(M_z) &= 
\text{Diag}[1,1,1,-1]
\end{aligned}\end{equation} 
Note that $C_6$ exchanges two boron atoms while $C_3$ does not. 
Moreover, since $s,p_x,p_y$, and $p_z$ have opposite $M_z$ eigenvalues and are decoupled, the representation matrices are block diagonal.
We then perform a basis transformation to the $3s$ basis at non-maximal position $6m$ (with $S_{B_i}$ left multiply on the $(s,p_x,p_y,p_z)$ basis):
\begin{equation}\begin{aligned}
S_{B_1}=
\begin{bmatrix}
-\frac{1}{\sqrt{3}} & \frac{1}{\sqrt{2}} & \frac{1}{\sqrt{6}} & 0 \\
-\frac{1}{\sqrt{3}} & -\frac{1}{\sqrt{2}} & \frac{1}{\sqrt{6}} & 0 \\
-\frac{1}{\sqrt{3}} & 0 & -\frac{\sqrt{2}}{\sqrt{3}} & 0 \\
0 & 0 & 0 & 1
\end{bmatrix}
\quad
S_{B_2}=
\begin{bmatrix}
-\frac{1}{\sqrt{3}} & -\frac{1}{\sqrt{2}} & -\frac{1}{\sqrt{6}} & 0 \\
-\frac{1}{\sqrt{3}} & \frac{1}{\sqrt{2}} & -\frac{1}{\sqrt{6}} & 0 \\
-\frac{1}{\sqrt{3}} & 0 & \frac{\sqrt{2}}{\sqrt{3}} & 0 \\
0 & 0 & 0 & 1
\end{bmatrix}.
\label{app:eq:S_spxpy_to_sp2}
\end{aligned}\end{equation}
After the transformation, i.e., 
$D_{B_i}(g)=S_{B_i} D_0(g) S_{B_i}^{-1}$, 
we have
\begin{equation}\begin{aligned}
D_{B_i}(C_6) &= 
\begin{bmatrix}
\frac{2}{3} & -\frac{1}{3} & \frac{2}{3} & 0 \\
\frac{2}{3} & \frac{2}{3} & -\frac{1}{3} & 0 \\
-\frac{1}{3} & \frac{2}{3} & \frac{2}{3} & 0 \\
0 & 0 & 0 & 1
\end{bmatrix},\quad
D_{B_i}(C_3) = 
\begin{bmatrix}
0 & 0 & 1 & 0 \\
1 & 0 & 0 & 0 \\
0 & 1 & 0 & 0 \\
0 & 0 & 0 & 1
\end{bmatrix},\\
D_{B_i}(M_x) &= 
\begin{bmatrix}
0 & 1 & 0 & 0 \\
1 & 0 & 0 & 0 \\
0 & 0 & 1 & 0 \\
0 & 0 & 0 & 1
\end{bmatrix},\quad
D_{B_i}(P) = 
\begin{bmatrix}
-\frac{1}{3} & \frac{2}{3} & \frac{2}{3} & 0 \\
\frac{2}{3} & -\frac{1}{3} & \frac{2}{3} & 0 \\
\frac{2}{3} & \frac{2}{3} & -\frac{1}{3} & 0 \\
0 & 0 & 0 & -1
\end{bmatrix}
\label{Eq: Dg_sp2}
\end{aligned}\end{equation}
Note that although $S_{B_1}\neq S_{B_2}$, they give the same $D_{B_i}(g)$. The specific form of $S_{B_i}$ is taken to transform the original $(s,p_x,p_y)$ orbitals into three effective $s$ orbitals, such that the three $s$ orbitals are located along the bond directions (\ie, at $6m$ non-maximal WP) to form the $sp^2$ hybrid states. In \cref{Fig: sp2-basis}, we plot the schematic positions of these effective orbitals, which respect the representation matrices in \cref{Eq: Dg_sp2}.

After transforming into the $3s@6m$ basis, we analyze the hopping magnitudes in this basis. 
For the three $s$ orbitals from the same boron, the hopping between them is -1.482 eV. This hopping mainly determines the bandwidth of the $sp^2$ bonding states and is given by the onsite difference of the original $s$ and $(p_x,p_y)$ orbitals of each boron.  
The NN coupling between the orbitals of two boron atoms also contributes to the bandwidth. More detailed derivation will be shown in the TB model in \cref{app:sec:simple_Hel_model}. 
Between two borons, the dominant hopping comes from the two $s$ orbitals along the bond direction, i.e., $s_2@B_1$ to $s_2@B_2$ (see \cref{Fig: sp2-basis} for the label of $s$ orbitals), which has a large value of $t_{NN}= -7.754$ eV. The other three inequivalent hoppings from $B_1$ to $B_2$ are $t_{s_1@B_1, s_1@B_2}=1.300$ eV,  $t_{s_1@B_1, s_2@B_2}=-0.298$ eV, and  $t_{s_1@B_1, s_1@B_3}=-0.930$ eV, as tabulated in \cref{app:table:hopping-sp-3s}. 
We remark that these hopping values can be faithfully reproduced using the hydrogen-like orbital integrals of boron $s,p_x,p_y$ orbitals and transform into the $3s@6m$ basis, as discussed in \cref{app:sec:hopping_integral}.

\begin{table}[htbp]
\centering
\begin{tabular}{c|c|c|c c|c|c|c|c}
\hline\hline
$(s,p)$ basis & $B_{s}^2$ & $B_{p_x}^2$ & $B_{p_y}^2$ &  & $3s$ basis  & $B_{s_1}^2$ & $B_{s_2}^2$ & $B_{s_3}^2$ \\ \hline
$B_{s}^1$ & -2.7 & 3.4 & -2.0 &  & $B_{s_1}^1$ & 1.3 & -0.3 & -0.9 \\ \hline
$B_{p_x}^1$ & -3.4 & 2.9 & -3.0  &  & $B_{s_2}^1$ & -0.3 & -7.8 & -0.3 \\ \hline
$B_{p_y}^1$ & 2.0 & -3.0 & -0.5 &  & $B_{s_3}^1$ & -0.9 & -0.3 & 1.3 \\ \hline\hline
\end{tabular}
\caption{\label{app:table:hopping-sp-3s}
Comparison of hopping values in the Wannier TB model in the $(s,p)$ basis (left half) and the $3s$ basis (right half). All numbers are given in \SI{}{eV}. 
}
\end{table}

From the hopping magnitude, it can be seen that the most dominant hopping is between two $s$ orbitals along the $B_1$-$B_2$ bond (denoted as $t_{NN}$). This large hopping has two implications. First, it splits the $s$ orbitals from $B_1$ and $B_2$ into the bonding and anti-bonding states by a large gap at the order of $2t_{NN}$. If we only consider two $s$ orbitals on the same bond, the local Hamiltonian can be transformed into the bonding ($\frac{1}{\sqrt{2}}(s_{B_1} + s_{B_2})$ and anti-bonding basis ($\frac{1}{\sqrt{2}}(s_{B_1} - s_{B_2})$), i.e., 
\begin{equation}\begin{aligned}
H_{\text{bond}} = 
\begin{bmatrix}
\epsilon_s & t_{NN} \\
t_{NN} & \epsilon_s
\end{bmatrix},\quad
S = \frac{1}{\sqrt{2}}
\begin{bmatrix}
1 & 1 \\
1 & -1
\end{bmatrix},
\Rightarrow
H_{bond}^{\prime} = S H_{\text{bond}} S^{-1} =
\begin{bmatrix}
\epsilon_s + t_{NN} & 0 \\
0 & \epsilon_s - t_{NN}
\end{bmatrix}
\label{app:eq:local_3s_to_sp2}
\end{aligned}\end{equation}
Thus the bonding state is formed by two $s$ orbitals on a bond with the form $\frac{1}{\sqrt{2}}(s_{B_1}+s_{B_2})$ and energy $\epsilon_{\text{bonding}}=\epsilon_s+t_{NN}=-3.7$ eV, where $\epsilon_s=3.9$ eV (from Wannier TB model). The bonding states form an obstructed atomic insulator (OAI) with EBR $A_g@3g$, i.e., $s$ orbitals placed at the bond centers (the $3g$ Wyckoff position).

To summarize, the onsite energy difference between the $s$ and $(p_x,p_y)$ orbitals determines the (kagome-like) bandwidth of the bonding states. However, this difference cannot be too large; otherwise, the low-lying $s$ orbital remains largely inert, and the three equivalent in-plane $sp^2$-like hybrids are no
longer energetically favored. 
For instance, in DFT pseudopotentials of B, C, N, O, and F, the energy separations between the $2s$ and $2p$ orbitals are approximately 6, 9, 11, 15, and 19 eV, respectively. While B, C, and N readily form $sp^2$ hybrid orbitals, O
and F generally do not form extended three-connected $sp^2$-bonded frameworks, as their deeper $2s$ levels and larger valence-electron counts favor $p$-dominated bonding. 
On the other hand, the hopping $t_{NN}$ between the two effective $s$ orbitals of two borons determines the gap between bonding and anti-bonding states, which can be used to characterize the strength of the $sp^2$ bond. As a result, the onsite energies and hopping show an interesting ``duality'' when transforming the basis from the $3s@6m$ to the $sp^2$ bonding/anti-bonding basis at kagome sites. In the next subsection \cref{app:sec:simple_Hel_model}, we discuss this duality quantitatively using a minimal TB model.

Second, the large nearest-neighbor hopping $t_{NN}$ implies a large EPC between the bonding states and the phonons of boron that stretch the bond. Because of the strong covalent bonding, even a small change in bond length significantly alters the total energy, leading to a large bond-related EPC—namely, the coupling between the two effective $s$ orbitals along an $sp^2$ bond and the in-plane phonon modes of the two boron atoms forming that bond. A detailed analytic derivation of the EPC Hamiltonian in \ch{MgB2} is presented in \cref{app:sec:EPC_hamiltonian}. 
From the symmetry analysis given in \cref{app:sec:EPC_symm_constraints}, we know that the bonding states at $\Gamma$ only couple to the $\Gamma_5^+$ phonon from the in-plane movements of boron, but cannot couple to the $\Gamma_6^-$ phonon. This also agrees with the bonding analysis, because the $\Gamma_6^-$ phonon is formed by the unidirectional movements of borons that do not change the bond length, while the $\Gamma_5^+$ phonon is the opposite movements of two borons, \ie, the two bond-stretching modes along $x$ and $y$ directions. Thus the $\Gamma_5^+$ phonon couples strongly to the $\sigma$ FS formed by the bonding states.

\subsubsection{Simple electron model of \ch{MgB2}}\label{app:sec:simple_Hel_model}
We first build a 6-band electron model based on the two borons' $s,p_x,p_y$ orbitals at the honeycomb site (Wyckoff position $2d$). They form $sp^2$ bonding and anti-bonding states. The bonding states give a small quasi-2D $\sigma$ FS that contributes most to SC. 

\begin{figure}[htbp]
    \centering
    \includegraphics[width=0.4\textwidth]{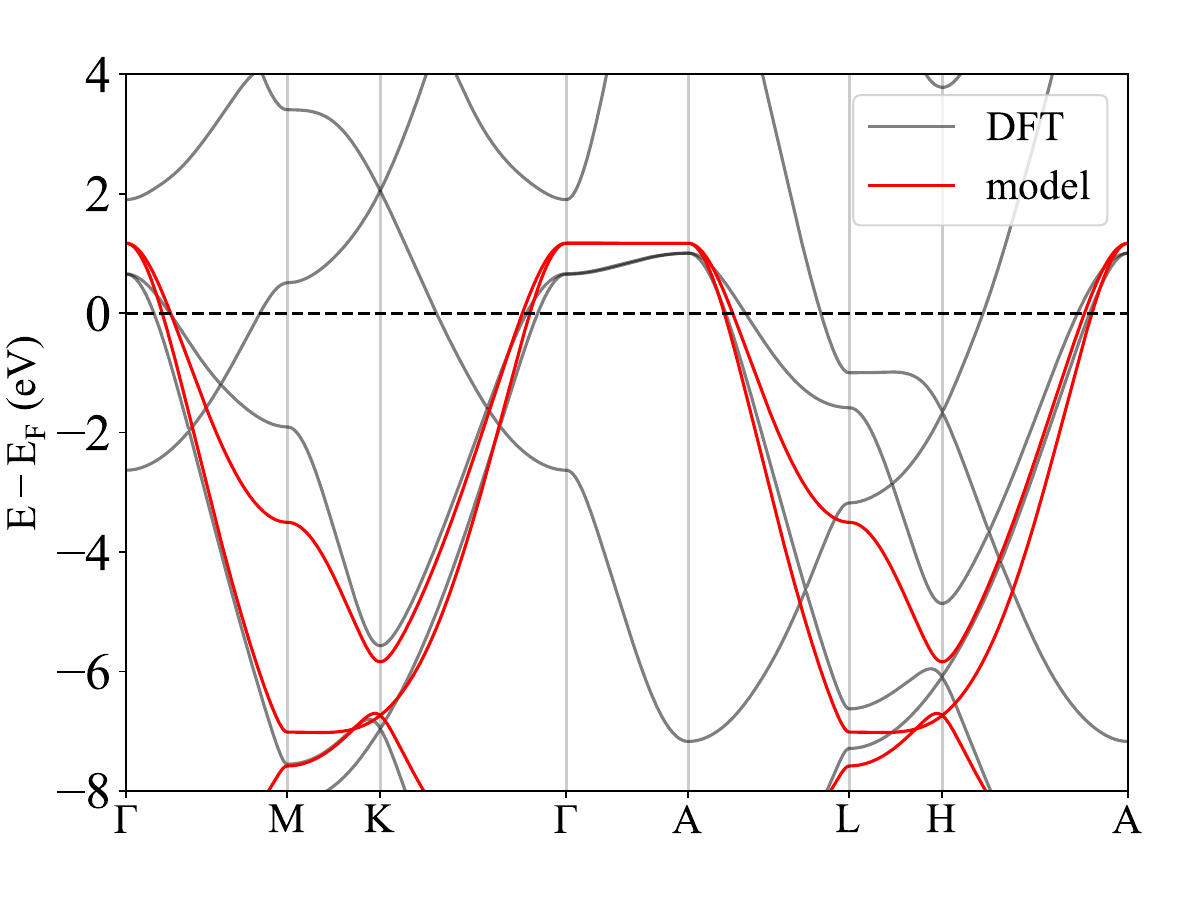}
    \caption{\label{Fig: minimal-model} Minimal electron models for \ch{MgB2} from the $(s, p_x, p_y)$ orbitals of two borons. The three bands from the anti-bonding states have very high energies and are not shown in the plot. 
    }
\end{figure}

Define the real-space electron creation operator $\cre{c}{\RR,i}$, where $i$ is a composite index for both boron sites and orbitals. The corresponding momentum space operator is 
$c^\dag_{\kk,i,\sigma} = \frac{1}{\sqrt{N}} \sum_{\RR} c^\dag_{\RR,i,\sigma} e^{i \kk \cdot (\RR+\mathbf{t}_i)}$,
where $\mathbf{t}_i$ is the sublattice shift for each orbital, as introduced in \cref{app:eq:atomic_coord}. The spin degree of freedom will be ignored as the SOC is negligible in \ch{MgB2}. 
The single-particle Hamiltonian has the form
\begin{equation}\begin{aligned} 
& \hat{\mathcal{H}}^{sp_xp_y} = \sum_{\kk,\alpha\beta} \hat{\psi}^{sp_xp_y, \dag}_{\kk,\alpha} h^{sp_xp_y}_{\kk,\alpha\beta} \hat{\psi}^{s,p}_{\kk,\beta} \nonumber\\ 
&\hat{\psi}^{sp_xp_y}_{\kk} = 
\begin{bmatrix}
    \des{c}{\kk, B^1_{s}} & \des{c}{\kk, B^1_{p_x}} & \des{c}{\kk, B^1_{p_y}} &
    \des{c}{\kk, B^2_{s}} & \des{c}{\kk, B^2_{p_x}} & \des{c}{\kk, B^2_{p_y}}
\end{bmatrix}
\end{aligned}\end{equation}
where $h^{sp_xp_y}_{\kk}$ is defined as
\begin{equation}\begin{aligned}
&h^{sp_xp_y}_{\kk} = 
\begin{bmatrix}
    \text{diag}(\epsilon_s, \epsilon_p,\epsilon_p)& T_{\kk}^\dag \\
    T_{\kk} &    \text{diag}(\epsilon_s, \epsilon_p,\epsilon_p)
\end{bmatrix} \\ 
&T_{\kk} = \\
&
\left[
\begin{array}{ccc}
t_1 e^{-i\frac{{k_x+2k_y}}{3}}+t_1 e^{i\frac{{k_y-k_x}}{3}} + t_1 e^{i \frac{2k_x+k_y}{3} } 
& \frac{1}{2} \sqrt{3} {t_3} e^{i\frac{2 {k_x} + k_y}{3} }-\frac{1}{2} \sqrt{3} {t_3} e^{i
\frac{{k_y} - k_x}{3}}  
& \frac{t_3}{2} \left(-2e^{-i\frac{{k_x}+2k_y}{3}} + e^{i\frac{{k_y}-k_x}{3}} + e^{i
\frac{2 {k_x}+k_y}{3}}\right) 
\\
\frac{1}{2} \sqrt{3} {t_3} e^{i\frac{{k_y-k_x}}{3}}-\frac{1}{2} \sqrt{3} {t_3} e^{i\frac{2{k_x} + k_y}{3} } & 
\frac{{t_2}+3{t_4}}{4} \left(e^{i\frac{{k_y}-k_x}{3}} + e^{i\frac{2 {k_x} +k_y}{3}} \right) +{t_2} e^{i-\frac{{k_x}+2 {k_y}}{3}} &
\frac{{t_2}-{t_4}}{4} \sqrt{3} \left( e^{i\frac{{k_y}-k_x}{3}}- e^{i \frac{2 {k_x}+k_y}{3}}\right)
\\
\frac{t_3}{2} \left(2e^{-i\frac{{k_x}+2 {k_y}}{3}} - e^{i\frac{{k_y}-k_x}{3}}- e^{i\frac{2 {k_x}+k_y}{3}}\right) &
\frac{t_2-{t_4}}{4} \sqrt{3} \left(e^{i\frac{{k_y}-k_x}{3}} - e^{i\frac{2{k_x}+k_y}{3}}\right) &
\frac{3{t_2}+{t_4}}{4} \left(e^{i\frac{{k_y}-k_x}{3}} + e^{i\frac{2 {k_x}+k_y}{3}}\right) + {t_4} e^{-i\frac{{k_x}+2k_y}{3}}
\\
\end{array}\right]
\label{app:eq:H_spxpy}
\end{aligned}\end{equation}
In the model, $t_1$ denotes hopping between $B_s^1$ and $B_s^2$ in the unit cell, $t_3$ denotes the hopping between $B_s^1$ and $B_{p_x}^2$, 
$\frac{1}{4}(t_2+3t_4)$ is the hopping between $B_{p_x}^1$ and $B_{p_x}^2$, and $\frac{1}{4}(3t_2+t_4)$ is the hopping between $B_{p_y}^1$ and $B_{p_y}^2$. 
For now, we ignore the $z$-directional hoppings and treat the system as 2D. The values of these parameters are obtained by fitting to DFT dispersion, as tabulated in \cref{app:table:fitted_TB_parameters_spxpy}. The dispersion is given in \cref{Fig: minimal-model}(a).
We note that the hopping values from hydrogen-like orbital integrals given in \cref{app:table:hopping-integral-values} correspond to $t_1=-3.0, t_2=-2.1, t_3=3.3, t_4=5.5$ eV, which are close to the fitted hoppings given in \cref{app:table:fitted_TB_parameters_spxpy}. 

\begin{table}[htb]
\centering
\begin{tabular}{c|c|c|c|c|c|c}
\hline\hline
Parameter & $\epsilon_s$ & $\epsilon_p$ & $t_1$ & $t_2$ & $t_3$ & $t_4$ \\ \hline
Value (\SI{}{eV}) & 0.321 & 4.619 & -3.265 & -2.361 & 4.241 & 4.735 \\ \hline
Parameter & $\epsilon_{3s}$ & $t_{s}$ & $c_1$ & $c_2$ & $c_3$ & $c_4$ \\ \hline
Value (\SI{}{eV}) & 3.186 & -1.433 & -8.243 & 1.302 & -1.059 & -0.510 \\ 
\hline\hline
\end{tabular}
\caption{The fitted values of TB parameters in Hamiltonian \cref{app:eq:H_spxpy} and \cref{app:eq:el_model_3s_basis}.}
\label{app:table:fitted_TB_parameters_spxpy}
\end{table}

We then use \cref{app:eq:S_spxpy_to_sp2} to transform the $(s,p_x,p_y)$ basis to three effective $s$ orbital basis at non-maximal WP $6m$, i.e., 
\begin{equation}
\hat{\psi}^{3s}_{\kk} = 
\begin{bmatrix}
    \des{c}{\kk, B^1_{s_1}} & \des{c}{\kk, B^1_{s_2}} & \des{c}{\kk, B^1_{s_3}} &
    \des{c}{\kk, B^2_{s_1}} & \des{c}{\kk, B^2_{s_2}} & \des{c}{\kk, B^2_{s_3}}
\end{bmatrix}
\end{equation}
Let $S_{3s}=S_{B_1}\oplus S_{B_2}$, 
\begin{equation}\begin{aligned}
\left[\hat{\psi}^{3s}_{\kk}\right]^T &=  S_{3s} \left[\hat{\psi}^{sp_xp_y}_{\kk} \right]^T, \\
h^{3s}_{\kk} &= S_{3s} h^{sp_xp_y}_{\kk}  S_{3s}^{-1}
\end{aligned}\end{equation}
where
\begin{equation}\begin{aligned}
h^{3s}_{\kk} &=
\begin{bmatrix}
    T_0^{3s} & T_{\kk}^{3s,\dag} \\
    T_{\kk}^{3s} &  T_0^{3s}
\end{bmatrix}\\ 
T_0^{3s} &= S_{B_1} \text{diag}(\epsilon_s,\epsilon_p,\epsilon_p) S_{B_1}^{-1}=
\begin{bmatrix}
    \epsilon_{3s} & t_s & t_s \\
    t_s & \epsilon_{3s} & t_s \\
    t_s & t_s & \epsilon_{3s} \\
\end{bmatrix},\\
T_{\kk}^{3s} &= S_{B_2} T_{\kk} S_{B_1}^{-1} \\
&=e^{-\frac{i}{3}(k_1+2k_2)}
\begin{bmatrix}
    c_2+e^{ik_y}(c_2+c_1 e^{ik_x}) & 
    c_3+c_4 e^{ik_y}(1+ e^{ik_x}) &
    c_4+e^{ik_y}(c_3+c_4 e^{ik_x}) \\
    c_3+c_4 e^{ik_y}(1+ e^{ik_x}) &
    c_2+e^{ik_y}(c_1+c_2 e^{ik_x}) & 
    c_4e^{ik_y}(c_4+c_3 e^{ik_x}) \\
    c_4+ e^{ik_y}(c_3 + c_4 e^{ik_x}) &
    c_4+ e^{ik_y}(c_4+c_3 e^{ik_x}) &
    c_1+ c_2 e^{ik_y}(1 + e^{ik_x}) 
\end{bmatrix}
\label{app:eq:Hk_3s_basis}
\end{aligned}\end{equation}
where $\epsilon_{3s}=\frac{1}{3}(\epsilon_s+2\epsilon_p), t_s=\frac{1}{3}(\epsilon_s-\epsilon_p), c_1=\frac{1}{3}(t_1 - 2 \sqrt{2} t_3 - 2 t_4),
c_2 = \frac{1}{6} (2 t_1 - 3 t_2 + 2 \sqrt{2} t_3 - t_4), c_3 =\frac{1}{6} (2 t_1 + 3 t_2 + 2 \sqrt{2} t_3 - t_4), c_4 =\frac{1}{6}(2 t_1 - \sqrt{2} t_3 + 2 t_4)$. 
We observe that $c_1$ corresponds to the NN coupling between two effective $s$ orbitals at $6m$ along the same B--B bond, which has a dominant hopping value. The transformed parameters are given in \cref{app:table:fitted_TB_parameters_spxpy}. Notice that $T_0^{3s}$ is just the on-site B $s$ vs $p_{xy}$ energy difference, and equals the hopping between the new orbitals.

We then further transform the effective $3s$ orbital basis at $6m$ to the $sp^2$ bonding and anti-bonding states basis at the kagome site $3g$. 
The bonding and anti-bonding states are effective $s$ and $p_y$ orbitals, respectively, at kagome $3g$. We first shift the three effective $s$ orbitals from honeycomb sites to kagome sites, and then combine the two $s$ orbitals on the same kagome site to form bonding and anti-bonding basis. These two steps are performed using the following basis transformation: 
\begin{equation}\begin{aligned}
S_{sp^2}(\kk) &= S_{3s\rightarrow sp^2} S_{\text{shift}}(\kk),\\
S_{\text{shift}}(\kk) &=\text{Diag}\left[
e^{i(\kk\cdot (\mathbf{t}_{h_1} -\mathbf{t}_{k_3})},
e^{i(\kk\cdot (\mathbf{t}_{h_1} -\mathbf{t}_{k_2})},
e^{i(\kk\cdot (\mathbf{t}_{h_1} -\mathbf{t}_{k_1}-\mathbf{a}_2)},
e^{i(\kk\cdot (\mathbf{t}_{h_2} -\mathbf{t}_{k_3}-\mathbf{a}_1)},
e^{i(\kk\cdot (\mathbf{t}_{h_2} -\mathbf{t}_{k_2})},
e^{i(\kk\cdot (\mathbf{t}_{h_2} -\mathbf{t}_{k_1})}
\right]
,\\
S_{3s\rightarrow sp^2}
&=
\frac{1}{\sqrt{2}}
\begin{bmatrix}
0 & 0 & 1 & 0 & 0 & 1 \\
0 & 1 & 0 & 0 & 1 & 0 \\
1 & 0 & 0 & 1 & 0 & 0 \\
0 & 0 & 1 & 0 & 0 & -1 \\
0 & 1 & 0 & 0 & -1 & 0 \\
1 & 0 & 0 & -1 & 0 & 0 
\end{bmatrix},
\label{app:eq:S_3s_to_sp2}
\end{aligned}\end{equation}
where $\mathbf{t}_{h_1}=(\frac{1}{3},\frac{2}{3}), \mathbf{t}_{h_2}=(\frac{2}{3},\frac{1}{3})$ are the sublattice shifts of two honeycomb sites, and $\mathbf{t}_{k_1}=(\frac{1}{2},0), \mathbf{t}_{k_2}=(\frac{1}{2}, \frac{1}{2}),
\mathbf{t}_{k_3}=(0, \frac{1}{2})$ are the sublattice shifts of three kagome sites, as shown in \cref{Fig: sp2-basis}.  
$S_{3s\rightarrow sp^2}$ transformation is $\frac{1}{\sqrt{2}}(\cre{c}{s_i,B_1}\pm \cre{c}{s_i,B_2})$, \ie, the bonding (anti-bonding) basis is formed by the symmetric (anti-symmetric) combination of the two effective $s$ orbitals at $6m$ along a $sp^2$ bond. 

Denote the electron operators in the $sp^2$ bonding/anti-bonding basis as
\begin{equation}\begin{aligned}
\hat{\psi}^{sp^2}_{\kk} &= 
\begin{bmatrix}
    \des{c}{\kk, b^+_{1}} & \des{c}{\kk, b^+_{2}} & \des{c}{\kk, b^+_{3}} &
    \des{c}{\kk, b^-_{1}} & \des{c}{\kk, b^-_{2}} & \des{c}{\kk, b^-_{3}}  
\end{bmatrix}, \\
\left[\hat{\psi}^{sp^2}_{\kk} \right]^T &=  S_{sp^2}(\kk) \left[\hat{\psi}^{3s}_{\kk} \right]^T,
\end{aligned}\end{equation}
where $\des{c}{\kk, b^{\pm}_{i}}$ denotes the bonding (anti-bonding) states at kagome site $\tt_{k_i}$. 
With $S_{sp^2}(\kk)$, we transform 
The Hamiltonian in the $sp^2$ basis reads:
\begin{equation}\begin{aligned}
&h^{sp^2}(\kk) = S_{sp^2}(\kk) h^{3s}_{\kk}  S_{sp^2}^{-1}(\kk) 
=\begin{bmatrix}
    T_{+,\kk}^{sp^2} & T_{\kk}^{sp^2 \dag} \\
    T_{\kk}^{sp^2} &  T_{-,\kk}^{sp^2}
\end{bmatrix}\\ 
&T_{\pm,\kk}^{sp^2} = \\
&
\begin{bmatrix}
\epsilon_{3s} \pm \left(c_1 + c_2 (\cos(k_2)+\cos(k_1+k_2)) \right) 
& (t_s \pm 2c_4) \cos(\frac{k_2}{2}) \pm c_3 \cos(k_1+\frac{k_2}{2}) 
& (t_s \pm 2c_4) \cos(\frac{k_1+k_2}{2}) \pm c_3 \cos(\frac{k_1-k_2}{2}) \\
& \epsilon_{3s} \pm \left(c_1 + c_2 (\cos(k_1)+\cos(k_2)) \right) 
& (t_s \pm 2c_4) \cos(\frac{k_1}{2}) \pm c_3 \cos(\frac{k_1}{2}+k_2) \\
h.c. & & \epsilon_{3s} \pm \left(c_1 + c_2 (\cos(k_1)+\cos(k_1+k_2)) \right) 
\end{bmatrix},\\
& T_{\kk}^{sp^2} = 
\begin{bmatrix}
 i c_2 (\sin (k_1+k_2)+\sin (k_2)) & i (c_3 \sin (k_1+\frac{k_2}{2})-t_{s} \sin (\frac{k_2}{2})) & -i (c_3 \sin (\frac{k_1-k_2}{2})+t_{s} \sin (\frac{k_1+k_2}{2})) \\
 i (c_3 \sin (k_1+\frac{k_2}{2})+t_{s} \sin (\frac{k_2}{2})) & i c_2 (\sin (k_1)-\sin (k_2)) & -i (c_3 \sin (\frac{k_1}{2}+k_2)+t_{s} \sin (\frac{k_1}{2})) \\
 -i (c_3 \sin (\frac{k_1-k_2}{2})-t_{s} \sin (\frac{k_1+k_2}{2})) & i (t_{s} \sin (\frac{k_1}{2})-c_3 \sin (\frac{k_1}{2}+k_2)) & -i c_2 (\sin (k_1+k_2)+\sin (k_1)) \\
\end{bmatrix}
\label{app:eq:el_model_3s_basis}
\end{aligned}\end{equation}
where $+$ ($-$) denotes the Hamiltonian for the bonding (anti-bonding) states. This Hamiltonian has 6 parameters, i.e., $\epsilon_{3s}, t_s, c_{i=1,2,3,4}$ (defined in $h_{\kk}^{3s}$ in \cref{app:eq:Hk_3s_basis}). 
We observe that:
\begin{itemize}
\item The onsite energy of bonding/anti-bonding states is given by $\epsilon_{3s}\pm c_1$, where $c_1$ is the dominant hopping between two effective $s$ orbitals at $6m$ along the B--B bond. The large value of $c_1$ leads to the large splitting of bonding and anti-bonding states, at the order of 16 eV. 

\item If we only keep $\epsilon_{3s}\pm c_1$ and set all other parameters to zero, then $h^{sp^2}(\kk)= (\epsilon_{3s}+c_1)\bm{1}_3 \oplus (\epsilon_{3s}-c_1)\bm{1}_3$ is $\kk$-independent. This is an OAI limit where both bonding and anti-bonding states are perfect flat bands, with a gap $2c_1$ determined by the hopping between two effective $s$ orbitals at $6m$. To achieve this limit, one needs $t_s=\frac{1}{3}(\epsilon_s-\epsilon_p)=0$, \ie, the boron atomic $s$ and $p_x,p_y$ orbitals have the same onsite, and $c_{i=2,3,4}=0$. 

\item The bonding and anti-bonding states both form a kagome lattice, with NN hopping given by $t_s\pm 2c_4$. Recall that $t_s=\frac{1}{3}(\epsilon_s-\epsilon_p)$ is determined by the onsite energy difference between $s$ and $p_x,p_y$ orbitals of boron. $c_4=\frac{1}{6}(2t_1-\sqrt{2}t_3+2t_4)=-0.51$ eV is small, resulting from the cancellation of the original hoppings in the $s,p_x,p_y$ basis.

\item $c_3$ and $c_2$ correspond to the NNN and 4th NN kagome hoppings, and also the coupling between bonding and anti-bonding states. 
\end{itemize}

The bonding and anti-bonding states form EBR $A_g@3g$ (\ie, an $s$ orbital at $3g$ position) and $B_{2u}@3g$ (\ie, a $p_y$ orbital at $3g$, respectively, with IRREPs shown in \cref{app:table:EBR_sp2_bonding}. At $\Gamma$, the eigenvalues and IRREPs (following the convention in \textit{Bilbao Crystallographic server}~\cite{aroyo2006bilbao1, aroyo2006bilbao2, aroyo2011crystallography}) of the model \cref{app:eq:el_model_3s_basis} are 
\begin{equation}\begin{aligned}
\text{bonding: } & \Gamma_5^+ (2D): ~
(\epsilon_{3s}+c_1) - (t_s+2c_4) + (2c_2-c_3), \quad
\Gamma_1^+ (1D): ~
(\epsilon_{3s}+c_1) + 2(t_s+2c_4) + 2c_2+2c_3,\\
\text{anti-bonding: } & 
\Gamma_6^- (2D): ~
(\epsilon_{3s}-c_1) - (t_s - 2c_4) + (-2c_2+c_3), \quad
\Gamma_4^- (1D): ~
(\epsilon_{3s}-c_1) + 2(t_s - 2c_4) - (2c_2+2c_3),\\
\end{aligned}\end{equation}
It can be seen that the bandwidth of $sp^2$ bonding states is
$3(t_s+2c_4)+3c_3$. Using the values of the fitted parameters in \cref{app:table:fitted_TB_parameters_spxpy}, we find that $t_s$, \ie, the onsite difference of $s$ and $p_x,p_y$, gives about half of the kagome bandwidth (measured at $\Gamma$), with the rest given by long-range hoppings.

\begin{table}[htb]
\centering
\begin{tabular}{c|c|c|c}
\hline\hline
EBR & $\Gamma$ & $\mathrm{M}$ & $\mathrm{K}$ \\ \hline
$A_g@3g$ & $\Gamma_1^+,\Gamma_5^+$ & $M_1^+,M_3^-,M_4^-$ & $K_1, K_5$ \\
$B_{2u}@3g$ & $\Gamma_4^-,\Gamma_6^-$ & $M_1^+,M_2^+,M_4^-$ & $K_4,K_5$ \\
\hline\hline
\end{tabular}
\caption{The irreducible representations at high-symmetry points $\Gamma, \mathrm{M}, \mathrm{K}$ from EBR $A_g@3g$ and $B_{2u}@3g$.}
\label{app:table:EBR_sp2_bonding}
\end{table}

\subsubsection{Emergence of $\sigma$ FS in \ch{MgB2}}\label{app:sec:emergence_sigma_FS}

The key distinction between \ch{MgB2} and graphene lies in the emergence of the $\sigma$ Fermi surface in \ch{MgB2}, whereas in graphene the $\sigma$ bands lie far below $E_f$, despite both materials having the same number of valence electrons. In this section, we employ analytic tight-binding models and symmetry arguments to show that the intercalated Mg layers are responsible for the appearance of the $\sigma$ Fermi surface. Mg effectively pushes the boron $p_z$ Dirac bands downward\cite{an2001superconductivity, miao2015finding}, and the $sp^2$ bonding states move upward to maintain the total number of filled electrons. The approximated values of hoppings between Mg and boron orbitals can also be obtained accurately from the hydrogen-like orbital overlap integral (see \cref{app:sec:hopping_integral}).

We begin by comparing the valence counts in \ch{MgB2} and graphene. Carbon has $2s^2 2p^2$ valence electrons. Ignoring spin, a graphene unit cell (two C atoms) yields four valence bands: three $sp^2$–bonding bands that are fully filled and two $p_z$–derived bands forming a Dirac cone, which together contribute one half-filled (valence) band. Boron has one fewer (p) electron than carbon ($2s^2 2p^1$), but Mg donates two $s$ electrons to the two B atoms, so the total valence electron count in \ch{MgB2} matches that of graphene. One might therefore expect a similar band structure. This holds qualitatively, except that (i) the $p_z$ Dirac points are not pinned exactly at $E_f$ but are compensated between the $k_z=0$ and $k_z=\pi$ planes, and (ii) the $sp^2$ bonding manifold is not fully filled, producing a small $\sigma$ Fermi surface near $\Gamma$. We now explain both features analytically. 

Because the boron $p_z$ and $sp^2$ orbitals belong to opposite $M_z$ sectors, they cannot hybridize. Their relative fillings can therefore only be influenced by the intercalated Mg layers. Since the boron $p_z$ orbitals are out of plane, they are expected to hybridize more strongly with Mg states.

From \textit{ab initio} calculations, the filling of the boron $p_z$ Dirac cones is $1.05$, while the $sp^2$ manifold has filling $2.95$ (spin ignored for simplicity). The latter can be estimated as follows: the two $\sigma$ Fermi surfaces have radii $\approx 0.13\pi$ and $0.21\pi$. Assuming a 2D Brillouin-zone area of $1$, their total area is
$A_\sigma=\pi\left(0.13^2+0.21^2\right)\approx 0.05$, so the $sp^2$ filling is $3-A_\sigma\simeq 2.95$.

\begin{figure}
    \centering
    \includegraphics[width=1\linewidth]{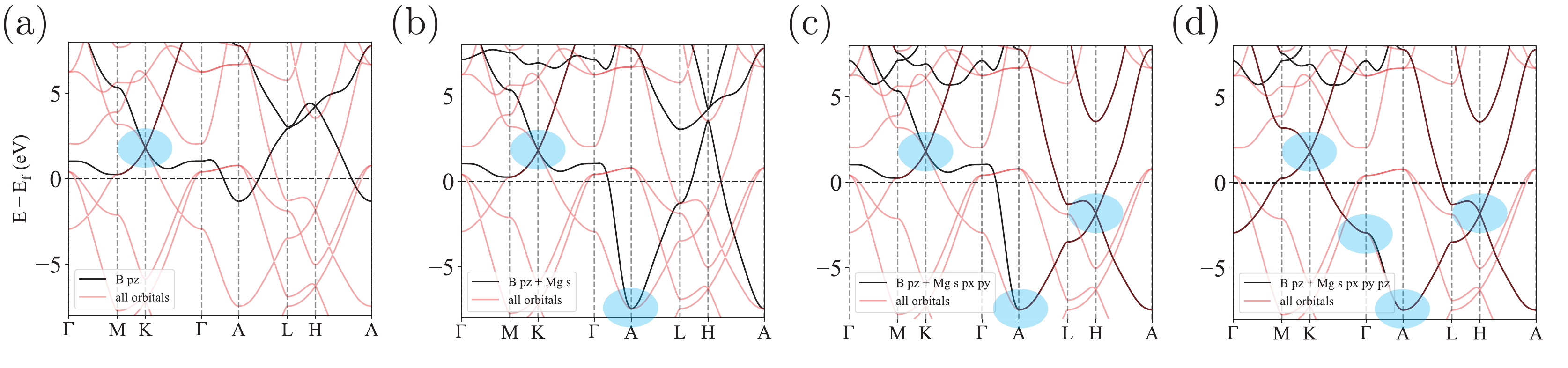}
    \caption{Boron $p_z$ bands under Mg hybridization. We extract \textit{ab initio} Wannier Hamiltonian blocks for B $p_z$ together with selected Mg orbitals, and compare the resulting bands (black) with the full \textit{ab initio} bands (red). Blue shading marks the matched portions at high-symmetry points. (a) shows the bands from B $p_z$ only, which form a distorted, largely unoccupied Dirac cone. Only the Dirac crossing at K coincides with the \textit{ab initio} bands (blue circle). (b) shows the bands from boron $p_z$ and Mg $s$ orbitals. (c) shows the bands from boron $p_z$ and Mg $s,p_x,p_y$ orbitals. (d) shows the bands from boron $p_z$ and Mg $s,p_x,p_y,p_z$ orbitals. In this case, the boron $p_z$ bands match perfectly with \textit{ab initio} bands. The filling and matching in each situation are summarized in \cref{app:table:Bpz-filling-perturbation-effect}. 
    }
    \label{app:fig:Bpz-perturbation-effect}
\end{figure}

To quantify the Mg effect on the boron $p_z$ bands, we use an \textit{ab initio} Wannier model. As shown in \cref{app:fig:Bpz-perturbation-effect} and \cref{app:table:Bpz-filling-perturbation-effect}, progressively including all Mg $s$ and $p$ orbitals (boron $s$ orbitals are perturbed out) is essential to reproduce both the total filling of the boron $p_z$ Dirac bands and their dispersions at high-symmetry points. In the following, we construct simple analytic tight-binding models to show explicitly how the Mg orbitals modify the boron $p_z$ bands.

\begin{table}[htbp]
\centering
\begin{tabular}{c|c|c|c|c|c}
\hline\hline
Wannier orbitals & Filling & $\Gamma$ & K & A & H \\ \hline
B $p_z$ & 0.05 & $\times$ & \checkmark & $\times$ & $\times$ \\ \hline
B $p_z$ + Mg $s$ & 0.40 & $\times$ & \checkmark & \checkmark & $\times$ \\ \hline
B $p_z$ + Mg $s,p_x,p_y$ & 0.97 & $\times$ & \checkmark & \checkmark & \checkmark \\ \hline
B $p_z$ + Mg $s,p_x,p_y,p_z$ & 1.05 & \checkmark & \checkmark & \checkmark & \checkmark \\ \hline\hline
\end{tabular}
\caption{\label{app:table:Bpz-filling-perturbation-effect} 
Boron $p_z$ bands under the influence of Mg orbitals, corresponding to the four cases in \cref{app:fig:Bpz-perturbation-effect}. The table reports the filling of the boron $p_z$ bands and their agreement with the \textit{ab initio} bands at high-symmetry points, where \checkmark indicates agreement and $\times$ indicates mismatch.
}
\end{table}

We first consider the Mg $s$ orbital. 
The nearest-neighbor (NN) coupling Hamiltonian between the two $p_z$ orbitals of B and the $s$ of Mg has the form 
\begin{equation}\begin{aligned}
S_{p_z^B,s^{Mg}}^{NN}(\kk)= 
2 t_{p_z^B,s^{Mg}} i\sin\left(\frac{k_3}{2}\right)
\begin{bmatrix}
-e^{-\frac{i}{3}(k_1+2k_2)} (1+e^{ik_2} + e^{i(k_1+k_2)}) \\
-e^{-\frac{i}{3}(2k_1+k_2)} (1+e^{ik_2} + e^{i(k_1+k_2)}) \\
\end{bmatrix}
\label{app:eq:S_pz_s}
\end{aligned}\end{equation}
where $|t_{p_z^B,s^{Mg}}|\approx1$ eV. 
Through a second-order perturbation (e.g., see Supplemental Material IX of Ref.~\cite{jiang2025fege}), these coupling terms give a $z$-directional coupling term between the $p_z$ orbitals on different layers, i.e.,
\begin{equation}\begin{aligned}
&H^{(2)}_{p_z^B, s^{Mg}}(\kk) = \frac{S_{p_z^B,s^{Mg}}^{NN}(\kk) S_{p_z^B,s^{Mb}}^{NN}(\kk)^\dagger}{\mu_{p_z}^{B}-\mu_s^{Mg}}\\
&= \frac{4t_{p_z^B,s^{Mb}}^2}{\mu_{p_z}^{B}-\mu_s^{Mg}} \sin^2\left(\frac{k_3}{2}\right)
\left(
\begin{array}{cc}
 2 \cos (k_1+k_2)+2 \cos (k_1)+2 \cos (k_2)+3 & e^{-\frac{2i}{3}(k_1+2 k_2)} \left(e^{i (k_1+k_2)}+e^{i k_2}+1\right)^2 \\
 e^{-\frac{2i}{3}(2 k_1+k_2)} \left(e^{i (k_1+k_2)}+e^{i k_1}+1\right)^2 & 2 \cos (k_1+k_2)+2 \cos (k_1)+2 \cos (k_2)+3 \\
\end{array}
\right).
\label{app:eq:Mgs-Bpz-2ndpertH}
\end{aligned}\end{equation} 
In \ch{MgB2}, the Mg atoms donate their $s$ electrons to the boron layers, thus the Mg $s$ band is pushed well above the Fermi level $E_f$ and becomes an empty conduction band, leading to 
$\mu_{p_z}^B-\mu_s^{Mg} <0$. 
From \textit{ab initio} Wannier model, $\mu_{p_z}^B-\mu_s^{Mg} =-1.5$ eV. 
Thus $H^{(2)}_{pz}(\kk)$ is zero when $k_3=0$, while negative when $k_3=\pi$ and shift the Dirac band lower. The filling of $p_z$ bands will increase due to this coupling term. More explicitly, we have
\begin{equation}\begin{aligned}
k_3=0:&\quad H^{(2)}_{p_z^B, s^{Mg}}(\kk)=\bm{0}, \\
(k_1,k_2) = 2\pi(\frac{1}{3},\frac{1}{3}):&\quad H^{(2)}_{p_z^B, s^{Mg}}(\kk)=\bm{0}, \\
(k_1,k_2,k_3) = (0,0,\pi):&\quad H^{(2)}_{p_z^B, s^{Mg}}(\kk)= \frac{4t_{p_z^B,s^{Mb}}^2}{\mu_{p_z}^{B}-\mu_s^{Mg}} 
\begin{bmatrix}
    9 & 9 \\
    9 & 9 \\
\end{bmatrix}.
\end{aligned}\end{equation}
As a result, among the four high-symmetry points $\Gamma$, K, A, and H, Mg $s$ orbitals will only affect dispersion at $A$, but cannot change the position of the Dirac crossing along the K-H line, in agreement with \cref{app:fig:Bpz-perturbation-effect}.

We next consider Mg $p_{x,y}$ orbitals (\cref{app:fig:Bpz-perturbation-effect} (b)), which have the following NN coupling Hamiltonian with B $p_z$ orbital:
\begin{equation}
\begin{aligned}
S^{NN}_{p_z^B, p_{x,y}^{Mg}}(\kk) &=
2t_{p_z^B, p_{x,y}^{Mg}} i\sin\left(\frac{k_3}{2}\right)
\left(
\begin{array}{cc}
 2i \sqrt{3} e^{\frac{1}{6} i (k_1+2 k_2)} \sin \left(\frac{k_1}{2}\right) & e^{-\frac{1}{3} i (k_1+2 k_2)} \left(e^{i (k_1+k_2)}+e^{i k_2}-2\right) \\
 \sqrt{3} \left(-1+e^{i k_1}\right) e^{-\frac{1}{3} i (2 k_1+k_2)} &  e^{-\frac{1}{3} i (2 k_1+k_2)} \left(-1+e^{i k_1} \left(-1+2 e^{i k_2}\right)\right) \\
\end{array}
\right)
\end{aligned}
\end{equation}
After the second-order perturbation, we obtain the effective Hamiltonian on B $p_z$ from Mg $p_{x,y}$:
\begin{equation}
\begin{aligned}
&H^{(2)}_{p_z^B, p_{x,y}^{Mg}}(\kk) = \frac{S_{p_z^B,p_{x,y}^{Mg}}^{NN}(\kk) S_{p_z^B,p_{x,y}^{Mg}}^{NN}(\kk)^\dagger}{\mu_{p_z}^{B}-\mu_{p_{x,y}}^{Mg}}\\
&= \frac{16 t_{p_{z}^B,p_{x,y}^{Mg}}^2}{\mu_{p_z}^{B}-\mu_{p_{x,y}}^{Mg}} \sin^2\left(\frac{k_3}{2}\right)
\left(
\begin{array}{c}
 3-\cos (k_1+k_2)-\cos (k_1)-\cos (k_2)  \\
 e^{\frac{-2i}{3} (2 k_1+k_2)} \left(e^{i (k_1+k_2)}-e^{2 i (k_1+k_2)}+e^{i (2 k_1+k_2)}+e^{i k_1}-e^{2 i k_1}-1\right)  \\
\end{array}
\right.\\
&\quad\quad\quad \left.
\begin{array}{c}
e^{-\frac{2i}{3} (k_1+2 k_2)} \left(e^{i k_2} \left(e^{i (k_1+k_2)}+e^{i k_1}+1\right)-2 e^{i (k_1+2 k_2)} \cos (k_1)-1\right) \\
3-\cos (k_1+k_2)-\cos (k_1)-\cos (k_2)
\end{array}
\right).
\label{app:eq:H2_pz_Mgxy}
\end{aligned}\end{equation} 
From DFT, we have $\mu_{p_z}^{B}-\mu_{p_{x,y}}^{Mg}= -6.2$ eV, and $t_{p_{z}^B,p_{x,y}^{Mb}}= 0.77$ eV. We observe that 
\begin{equation}\begin{aligned}
k_3=0:&\quad H^{(2)}_{p_z^B, p_{x,y}^{Mg}}(\kk)=\bm{0}, \\
(k_1,k_2) = (0, 0):&\quad H^{(2)}_{p_z^B, p_{x,y}^{Mg}}(\kk)=\bm{0}, \\
(k_1,k_2) = 2\pi(\frac{1}{3},\frac{1}{3}):&\quad 
H^{(2)}_{p_z^B, p_{x,y}^{Mg}}(\kk)= \frac{72 t_{p_{z}^B,p_{x,y}^{Mg}}^2}{\mu_{p_z}^{B}-\mu_{p_{x,y}}^{Mg}} \sin^2\left(\frac{k_3}{2}\right)\cdot \bm{1}.
\label{app:eq:Mgpxpy-pert-pz}
\end{aligned}\end{equation}
Thus Mg $p_{x,y}$ orbitals do not change the energy of B $p_z$ bands along $\Gamma$--A line and at K, but only affect the high-symmetry point $H$, due to the $\sin\left(\frac{k_3}{2}\right)$ factor in \cref{app:eq:H2_pz_Mgxy}. By inserting the \textit{ab initio} hoppings in \cref{app:eq:Mgpxpy-pert-pz}, we obtain that the Dirac crossing at $H$ is 6 eV lower than that at K, in agreement with \cref{app:fig:Bpz-perturbation-effect}.

Lastly, we consider the NN coupling Hamiltonian between B $p_z$ and Mg $p_z$:
\begin{equation}
\begin{aligned}
    S^{NN}_{p_z^B, p_{z}^{Mg}}(\kk)  &=
    2 t_{p_z^B, p_{z}^{Mg}} \cos\left(\frac{k_3}{2}\right) \left(
\begin{array}{c}
 e^{-\frac{1}{3} i (k_x+2 k_y)} \left(e^{i (k_x+k_y)}+e^{i k_y}+1\right) \\
 e^{-\frac{1}{3} i (2 k_x+k_y)} \left(e^{i (k_x+k_y)}+e^{i k_x}+1\right) \\
\end{array}
\right)
\end{aligned}
\end{equation}
After the second-order perturbation, we obtain the effective Hamiltonian on B $p_z$ from Mg $p_{x,y}$:
\begin{equation}
\begin{aligned}
&H^{(2)}_{p_z^B, p_{z}^{Mg}}(\kk) = \frac{S_{p_z^B,p_{z}^{Mg}}^{NN}(\kk) S_{p_z^B,p_{z}^{Mg}}^{NN}(\kk)^\dagger}{\mu_{p_z}^{B}-\mu_{p_{z}}^{Mg}}\\
&= 
4 t^2_{p_z^B, p_{z}^{Mg}} \cos^2\left(\frac{k_3}{2}\right) 
\left(
\begin{array}{cc}
 2 \cos (k_x+k_y)+2 \cos (k_x)+2 \cos (k_y)+3 & e^{-\frac{2i}{3} (k_x+2 k_y)} \left(e^{i (k_x+k_y)}+e^{i k_y}+1\right)^2 \\
 e^{-\frac{2i}{3} (2 k_x+k_y)} \left(e^{i (k_x+k_y)}+e^{i k_x}+1\right)^2 & 2 \cos (k_x+k_y)+2 \cos (k_x)+2 \cos (k_y)+3 \\
\end{array}
\right).
\end{aligned}
\end{equation}
This Hamiltonian has the same form as that of the Mg $s$ orbital in \cref{app:eq:Mgs-Bpz-2ndpertH}, except that the $k_3$-dependence changes from $\sin$ to $\cos$, reflecting the opposite $M_z$ eigenvalues of the Mg $s$ and $p_z$ orbitals. 
From DFT, we have $\mu_{p_z}^{B}-\mu_{p_{z}}^{Mg}=-8.0$ eV, and $t_{p_{z}^B,p_{x,y}^{Mb}}\approx 1.0 $ eV. We observe that 
\begin{equation}\begin{aligned}
k_3=\pi:&\quad H^{(2)}_{p_z^B, p_z^{Mg}}(\kk)=\bm{0}, \\
(k_1,k_2) = 2\pi(\frac{1}{3},\frac{1}{3}):&\quad H^{(2)}_{p_z^B, p_z^{Mg}}(\kk)=\bm{0}, \\
(k_1,k_2,k_3) = (0,0,0):&\quad H^{(2)}_{p_z^B, p_z^{Mg}}(\kk)= \frac{4t_{p_z^B,p_z^{Mb}}^2}{\mu_{p_z}^{B}-\mu_{p_z^{Mg}}} 
\begin{bmatrix}
    9 & 9 \\
    9 & 9 \\
\end{bmatrix}.
\end{aligned}\end{equation}
Thus, among the four high-symmetry points $\Gamma,K,A,H$, Mg $p_z$ orbitals will only affect dispersion $\Gamma$ (while Mg $x,p_x,p_y$ cannot due to the $\sin(\frac{k_3}{2})$) factor), but cannot change the position of the Dirac crossing along K-H line, in agreement with \cref{app:fig:Bpz-perturbation-effect}. 

In summary, we find that different Mg orbitals affect distinct features of the boron $p_z$ Dirac cone: Mg $s$ fixes the Dirac dispersion at $A$, Mg $p_x,p_y$ fix the dispersion at $H$, and Mg $p_z$ fixes the dispersion at $\Gamma$, as summarized in \cref{app:table:Bpz-filling-perturbation-effect}. The dispersion on $k_z=\pi$ is fixed by Mg $(s,p_x,p_y)$. The Dirac crossing at K remains unchanged under all Mg orbitals. 
Taken together, all Mg $s$ and $p$ orbitals increase the total filling of the boron $p_z$ states by 0.05, thereby driving the emergence of the boron $sp^2$ Fermi surface.

The $sp^2$ bonding states form a kagome lattice with a negative nearest-neighbor (NN) hopping of approximately $-1.5\,\mathrm{eV}$. This effective hopping originates mainly from the onsite-energy difference between the original boron $s$ orbital and the $(p_x,p_y)$ orbitals. Since the boron $s$ orbital lies lower in energy than the $(p_x,p_y)$ orbitals, the resulting NN hopping in the $sp^2$ bonding basis is negative (see the tight-binding model in \cref{app:sec:spxpy2sp2_basis} for details). 
The irreducible representations (IRREPs) of both the $sp^2$ bonding states and the $p_z$ orbitals can be obtained directly from their elementary band representations (EBRs)~\cite{bradlyn2017topological,cano2018building,elcoro2021magnetic}. In particular, the $sp^2$ bonding states realize the EBR $A_g@3g$, whose IRREPs are summarized in \cref{table: EBR_SG191}. Because of the negative NN hopping on the kagome lattice, the two-dimensional IRREP $\Gamma_5^+$ appears at the top of the kagome bands formed by the $sp^2$ bonding states.

\section{Phonon property of \ch{MgB2}}

\subsection{First-principle results for phonon}

\begin{figure}[htbp]
    \centering
    \includegraphics[width=0.7\textwidth]{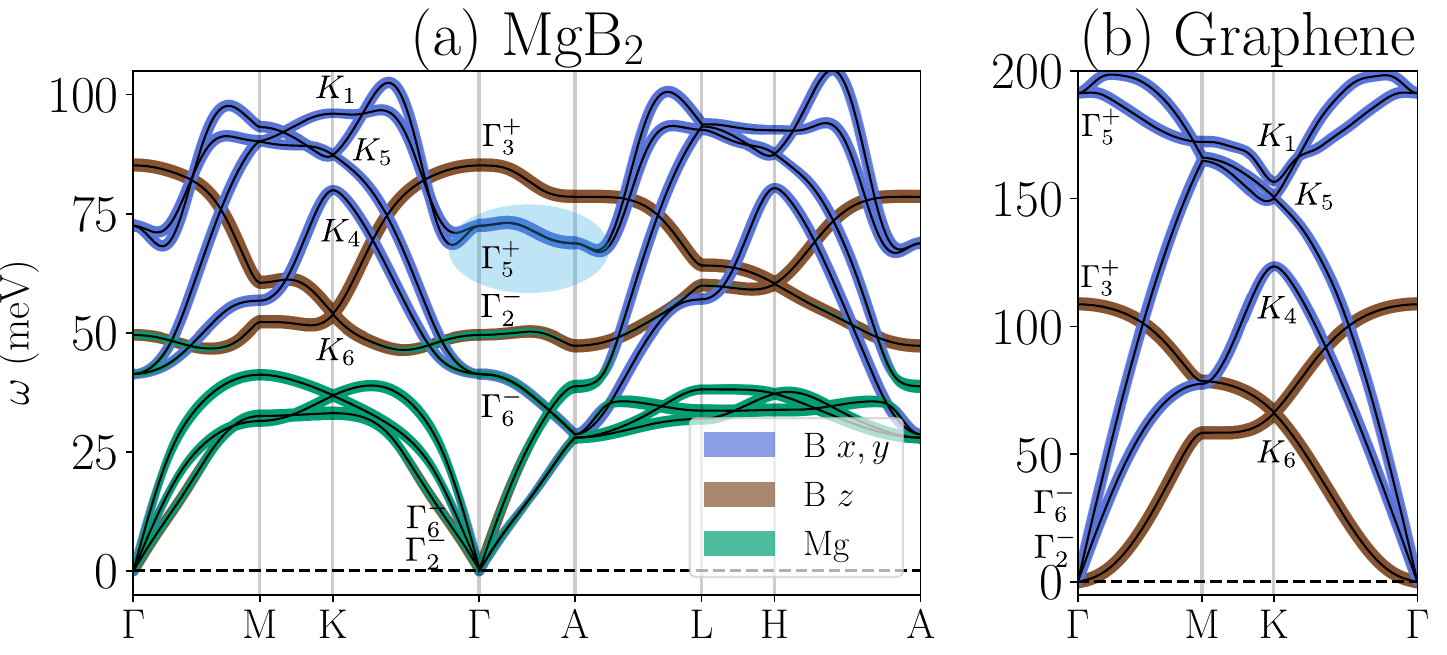}
    \caption{\label{app:fig: MgB2-DFT-phonon} (a) The phonon spectrum of \ch{MgB2} with B and Mg phonon weights. The blue circle marks the phonon bands with the largest electron-phonon coupling to the Fermi surface (see \cref{Fig: MgB2-lambda_a2f}). (b) The phonon spectrum of graphene. IRREPs are marked in the plot.
    }
\end{figure}

In this section, we discuss the phonon spectrum of \ch{MgB2}. In \cref{app:fig: MgB2-DFT-phonon}(a), we show the phonon and their orbital weights in \ch{MgB2}. In \cref{app:fig: MgB2-DFT-phonon}(b), the phonons of graphene with orbital weights are shown for comparison. We observe that:
\begin{itemize}
\item The three acoustic modes are mainly given by magnesium, as the mass of magnesium (i.e., 24.305 u) is about 2.5 times the mass of boron (i.e., 10.811 u). This atom-dependent separation of acoustic vs optical modes can be simply analytically understood as a mass effect using simple analytic spring-ball models. 

\item The six optical modes are mainly given by boron, where the $z$-directional and $xy$-directional modes are decoupled on $k_z=0,\pi$ planes because they have the opposite $M_z$ eigenvalues. These $z$- and $xy$-phonons share, approximately, the same frequency range. 

\item Contrary to the $p_z$ electron bands which have a strong $k_z$-directional dispersion, the $z$-directional boron phonon has a weak $k_z$-dispersion (\ie, with almost the same mean frequency on $k_z=0$ and $\pi$ planes). This difference between the electronic and phonon spectra is understood because the electron $p_z$ orbitals are extended along the $z$-direction and have strong hybridization with the $s$ orbital of Mg. However, the $z$-directional phonons are local movements of boron and have weak coupling to Mg. 

\item For graphene, the in-plane $xy$- and out-of-plane $z$-directional phonons are decoupled owing to the $M_z$ symmetry. Their dispersions are similar to those of boron, except that graphene exhibits three acoustic modes enforced by the acoustic sum rule. If Mg were removed from \ch{MgB2}, the three lowest optical modes at $\Gamma$ originating from boron would likewise become acoustic.
The optical phonon modes of \ch{MgB2}, which mainly arise from boron vibrations, share the same symmetry eigenvalues and exhibit dispersions closely resembling those of graphene. Indeed, for layered systems where the interlayer atom is much heavier than the intralayer atoms, the optical branches correspond to the full phonon spectrum of the lighter-layer material, shifted upward by the energy associated with the acoustic modes of the heavy atom. A more detailed study of this mass separation property in phonon is left in Ref.~\cite{PaperMassSeparation}.

\item The $xy$-phonons of boron form the EBR $E^{\prime} @2d$, while the $z$-phonon of boron form $A_2^{\prime\prime}@2d$ (same as the $p_z$ orbital), with IRREPs of the EBRs shown in \cref{table: EBR_SG191}. 
As shown in Ref.~\cite{xu2024catalog}, the symmetry eigenvalues of the phonon bands are identical to those of the EBRs induced from the Wyckoff positions of the atoms for the in-plane vector representation ($p_x,p_y$) and the out-of-plane vector representation ($p_z$). These data are tabulated for all space groups on the \textit{Bilbao Crystallographic server}~\cite{aroyo2006bilbao1, aroyo2006bilbao2, aroyo2011crystallography}. Hence, the IRREPs at each high-symmetry momentum point, including $\Gamma$, can be directly inferred from the corresponding EBR. 
For the symmetry group of \ch{MgB2}, the mirror-even IRREPs at $\Gamma$ are $\Gamma_6^-$ and $\Gamma_5^+$ (both two-dimensional), while the mirror-odd IRREPs are $\Gamma_2^-$ and $\Gamma_3^+$ (both one-dimensional).

In general, symmetry alone does not determine which of these representations lies higher or lower in frequency. However, in \ch{MgB2}, the large mass hierarchy between Mg and B allows such an ordering to be inferred. If Mg were removed, the boron phonon spectrum would resemble that of graphene and contain acoustic modes. In \ch{MgB2}, these modes become optical due to the upward energy shift of the boron phonon branches discussed previously. Since acoustic modes near $\qq=\bm{0}$ must transform as vector representations, they are odd under inversion. Therefore, the $\Gamma_6^-$ IRREP of the in-plane phonons and the $\Gamma_2^-$ IRREP of the out-of-plane phonon must correspond to the lower-frequency modes, while the even-parity phonons cannot. Hence we can reproduce the phonon spectrum analytically.

\item Alternatively, the higher frequency of the $\Gamma_5^+$ phonon compared with the $\Gamma_6^-$ can also be understood as a result of the $sp^2$ bonding. $\Gamma_5^+$ corresponds to opposite-directional movements of two borons that stretch the B--B bond, while the $\Gamma_6^-$ corresponds to the unidirectional movements of borons. Since the two neighboring borons form the strong $sp^2$ bond, the bond-stretching mode $\Gamma_5^+$ is expected to have a higher frequency (i.e., it costs a large energy to change the bond length). $\Gamma_6^-$ mode, in the absence of Mg, would be the acoustic mode of boron. 

\end{itemize}

\begin{figure}
    \centering
    \includegraphics[width=0.4\linewidth]{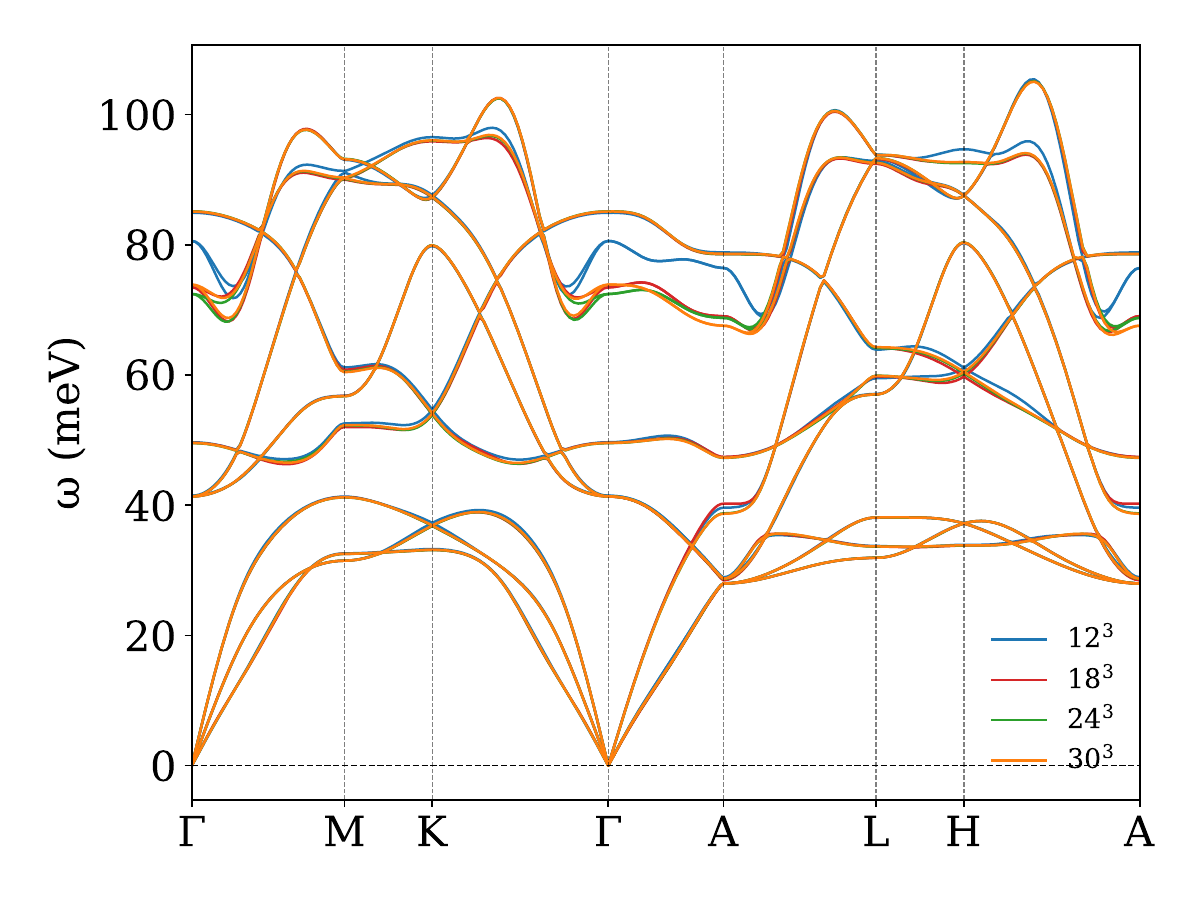}
    \caption{Convergence of the \ch{MgB2} phonon spectrum as a function of the electron self-consistency k-mesh. The phonon q-mesh is fixed at $6\times 6\times 6$. }
    \label{app:fig:phonon_kmesh_convergence_test}
\end{figure}

In \cref{app:fig:phonon_kmesh_convergence_test}, we examine the convergence of the \ch{MgB2} phonon spectrum with respect to the electron self-consistency k-mesh. The results indicate that once the k-mesh exceeds $18\times18\times18$, the spectrum stabilizes to within 2 meV. Accordingly, we employ a $24\times 24\times 24$ mesh throughout this work.

\subsection{Analytic understanding of phonon property}

\subsubsection{Simplified model of boron $xy$ phonon}
\begin{figure}[htbp]
    \centering
    \includegraphics[width=0.4\textwidth]{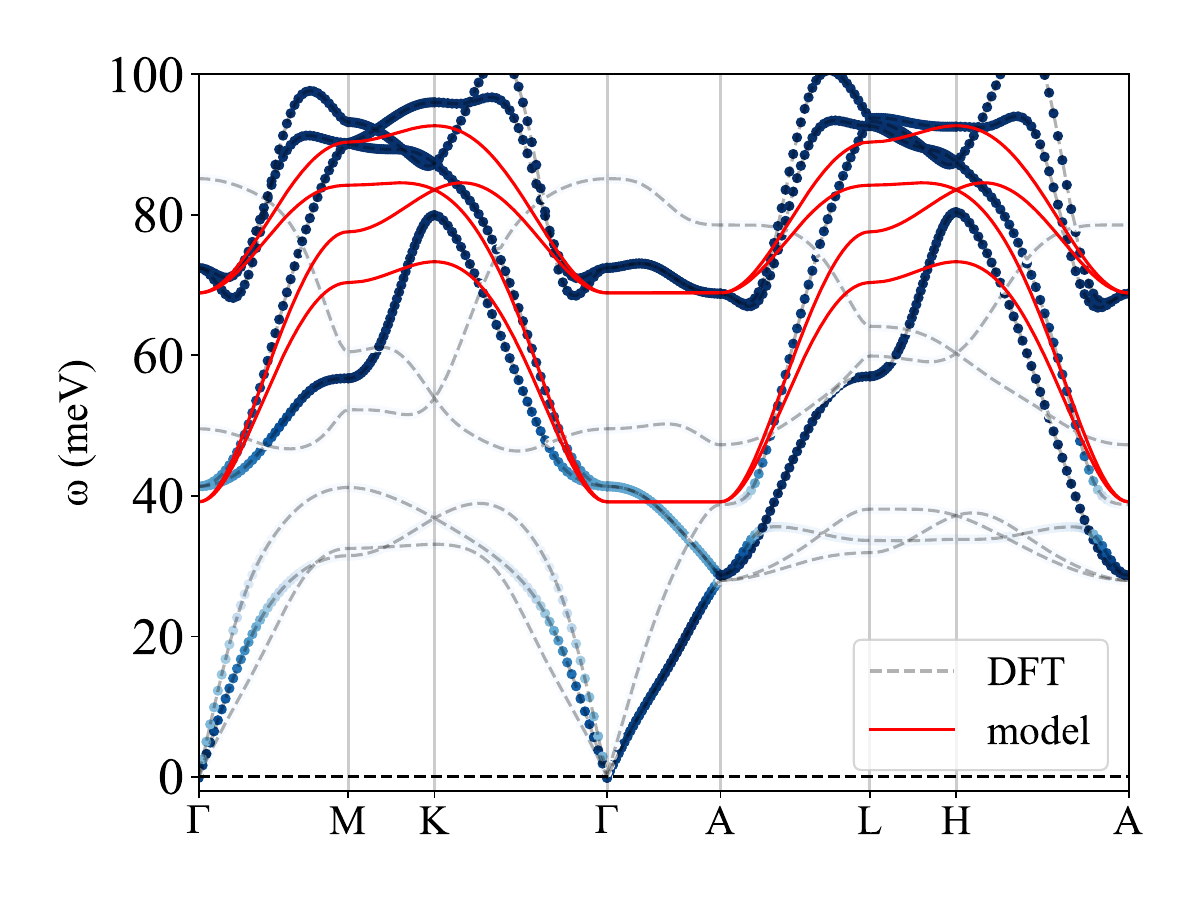}
    \caption{\label{Fig: minimal-ph-model} Minimal phonon models \cref{app:eq:H-phonon-Bxy} for \ch{MgB2} from the $xy$ phonon of boron, where the blue dots mark the weight of the in-plane phonons of boron in DFT. 
    }
\end{figure}

In this section, we build a simple model for the dynamical matrix of the $xy$ phonon of two borons in \ch{MgB2}. The orbital basis is $(B_{1x}, B_{1y}, B_{2x}, B_{2y})$, where $x,y$ denote phonons in cartesian coordinates. Consider the onsite terms, NN, and NNN coupling, the dynamical matrix has the form 
\begin{equation}\begin{aligned} 
D^{B, xy}_{\kk} &= 
\begin{bmatrix}
    \epsilon^{B, xy}_{\kk} \mathbf{1}_2 & T_{\kk}^{B, xy\dag} \\
    h.c. &   \epsilon^{B, xy}_{\kk} \mathbf{1}_2 
\end{bmatrix} \\ 
\epsilon^{B, xy}_{\kk} &= \epsilon^{B, xy} + 2 t_{NNN}^{B,xy} \left(
\cos(k_1) + \cos(k_2) + \cos(k_1+k_2)
\right)
\\
T_{\kk}^{B, xy} &= \frac{1}{4} t_{NN}^{B,xy}
\begin{bmatrix}
    3 e^{-\frac{i}{3}(k_1-k_2)} (1 + e^{i k_1}) &
    \sqrt{3} e^{-\frac{i}{3}(k_1-k_2)} (-1 + e^{i k_1}) \\
    \sqrt{3} e^{-\frac{i}{3}(k_1-k_2)} (-1 + e^{i k_1}) &
    e^{-\frac{i}{3}(k_1-k_2)} (1 + e^{i k_1} + 4e^{-i k_2}) \\
\end{bmatrix}
\label{app:eq:H-phonon-Bxy}
\end{aligned}\end{equation} 
By fitting to the \textit{ab initio} phonon spectrum, we find $\epsilon^{B, xy}=5700.51$ \SI{}{meV}$^2$, $t_{NN}^{B,xy} =-1070.78$ \SI{}{meV}$^2$, $t_{NNN}^{B,xy} = -812.57$ \SI{}{meV}$^2$. The fitted dispersion is shown in \cref{Fig: minimal-ph-model}(b).

\subsubsection{Full phonon model}

In this section, we consider the full phonon model to analyze the relationship between the phonon structures of \ch{MgB2} and graphene. Specifically, we observe that the inclusion of Mg atoms in \ch{MgB2} shifts the energy of the B phonons upwards, while the dispersion of the boron phonons closely resembles that in graphene. 

We build a full symmetry-allowed phonon dynamical matrix of \ch{MgB2} to illustrate this rule, which contains up to NNN in the boron $xy$ block and NN in all other blocks. The dynamical matrix has the form
\begin{equation}\begin{aligned} 
D^{full}_{\kk} &= 
\begin{bmatrix}
   D^{B, xy}_{\kk} & \mathbf{0} & T_{\kk}^{B,xy; Mg, xy} & T_{\kk}^{B,xy; Mg, z} \\
   \mathbf{0} & D^{B, z}_{\kk} & T_{\kk}^{B,z; Mg, xy} & T_{\kk}^{B,z; Mg, z}\\
    & & \epsilon^{Mg, xy} \mathbf{1}_2 & \mathbf{0} \\
   h.c. & & & \epsilon^{Mg, z} \mathbf{1}_1
\end{bmatrix} \\ 
D^{B,z}_{\kk} &= 
\epsilon_{z}^B \mathbf{1}_2 + t_{NN}^{B,z}
\begin{bmatrix}
    0 & e^{-\frac{i}{3}(2k_1+k_2)} (1 + e^{i k_1} + e^{i (k_1 + k_2)}) \\
    c.c. & 0 \\
\end{bmatrix} 
\label{app:eq:H-phonon-full}
\end{aligned}\end{equation} 
$D^{B,xy}_{\kk}$ is the same as in \cref{app:eq:H-phonon-Bxy}, and other blocks are 
\begin{equation}
\footnotesize
\begin{aligned}
T_{\kk}^{B,xy; Mg, xy} &= t_{NN}^{B,xy;Mg,xy}
\left[
\begin{array}{cc}
 -\frac{1}{2 \sqrt{3}}\left(1+e^{i k_3}\right) \left(-1+2 \left(1+e^{i k_1}\right) e^{i k_2}\right) e^{-\frac{1}{6} i (2 k_1+4 k_2+3 k_3)} & -2 i e^{\frac{1}{6} i
   (k_1+2 k_2)} \sin \left(\frac{k_1}{2}\right) \cos \left(\frac{k_3}{2}\right) \\
 -2 i e^{\frac{1}{6} i (k_1+2 k_2)} \sin \left(\frac{k_1}{2}\right) \cos \left(\frac{k_3}{2}\right) & -\frac{1}{2} \sqrt{3} \left(1+e^{i k_3}\right) e^{-\frac{1}{6} i (2
   k_1+4 k_2+3 k_3)} \\
\frac{1}{2 \sqrt{3}} \left(1+e^{i k_3}\right) \left(-2+e^{i k_1} \left(-2+e^{i k_2}\right)\right) e^{-\frac{1}{6} i (4 k_1+2 k_2+3 k_3)} & \frac{1}{2} \left(-1+e^{i
k_1}\right) \left(1+e^{i k_3}\right) e^{-\frac{1}{6} i (4 k_1+2 k_2+3 k_3)} \\
 \frac{1}{2} \left(-1+e^{i k_1}\right) \left(1+e^{i k_3}\right) e^{-\frac{1}{6} i (4 k_1+2 k_2+3 k_3)} & -\frac{1}{2} \sqrt{3} \left(1+e^{i k_3}\right) e^{\frac{1}{6} i
   (2 k_1+4 k_2-3 k_3)} \\
\end{array}
\right] \\
T_{\kk}^{B, xy; Mg, z} &= t_{NN}^{B,xy;Mg,z}
\left[
\begin{array}{c}
 -2 \sqrt{3} e^{\frac{1}{6} i (k_1+2 k_2)} \sin \left(\frac{k_1}{2}\right) \sin \left(\frac{k_3}{2}\right) \\
 \frac{1}{2} \left(-1+e^{i k_3}\right) \left(e^{i (k_1+k_2)}+e^{i k_2}-2\right) e^{-\frac{1}{6} i (2 k_1+4 k_2+3 k_3)} \\
 \frac{1}{2} \sqrt{3} \left(-1+e^{i k_1}\right) \left(-1+e^{i k_3}\right) e^{-\frac{1}{6} i (4 k_1+2 k_2+3 k_3)} \\
 \frac{1}{2} \left(-1+e^{i k_3}\right) \left(-1+e^{i k_1} \left(-1+2 e^{i k_2}\right)\right) e^{-\frac{1}{6} i (4 k_1+2 k_2+3 k_3)} \\
\end{array}
\right] \\
T_{\kk}^{B,z; Mg, xy} &= t_{NN}^{B,z;Mg,xy}
\left[
\begin{array}{cc}
 -2 \sqrt{3} e^{\frac{1}{6} i (k_1+2 k_2)} \sin \left(\frac{k_1}{2}\right) \sin \left(\frac{k_3}{2}\right) & \frac{1}{2} \left(-1+e^{i k_3}\right) \left(e^{i
   (k_1+k_2)}+e^{i k_2}-2\right) e^{-\frac{1}{6} i (2 k_1+4 k_2+3 k_3)} \\
 \frac{1}{2} \sqrt{3} \left(-1+e^{i k_1}\right) \left(-1+e^{i k_3}\right) e^{-\frac{1}{6} i (4 k_1+2 k_2+3 k_3)} & \frac{1}{2} \left(-1+e^{i k_3}\right) \left(-1+e^{i
   k_1} \left(-1+2 e^{i k_2}\right)\right) e^{-\frac{1}{6} i (4 k_1+2 k_2+3 k_3)} \\
\end{array}
\right]
\\
T_{\kk}^{B,z; Mg, z} &= t_{NN}^{B,z;Mg,z}
\left[
\begin{array}{c}
 \frac{1}{2} \left(1+e^{i k_3}\right) \left(e^{i (k_1+k_2)}+e^{i k_2}+1\right) e^{-\frac{1}{6} i (2 k_1+4 k_2+3 k_3)} \\
 \frac{1}{2} \left(1+e^{i k_3}\right) \left(e^{i (k_1+k_2)}+e^{i k_1}+1\right) e^{-\frac{1}{6} i (4 k_1+2 k_2+3 k_3)} \\
\end{array}
\right]
\end{aligned}\end{equation}

\begin{figure}[htbp]
\centering
\includegraphics[width=\textwidth]{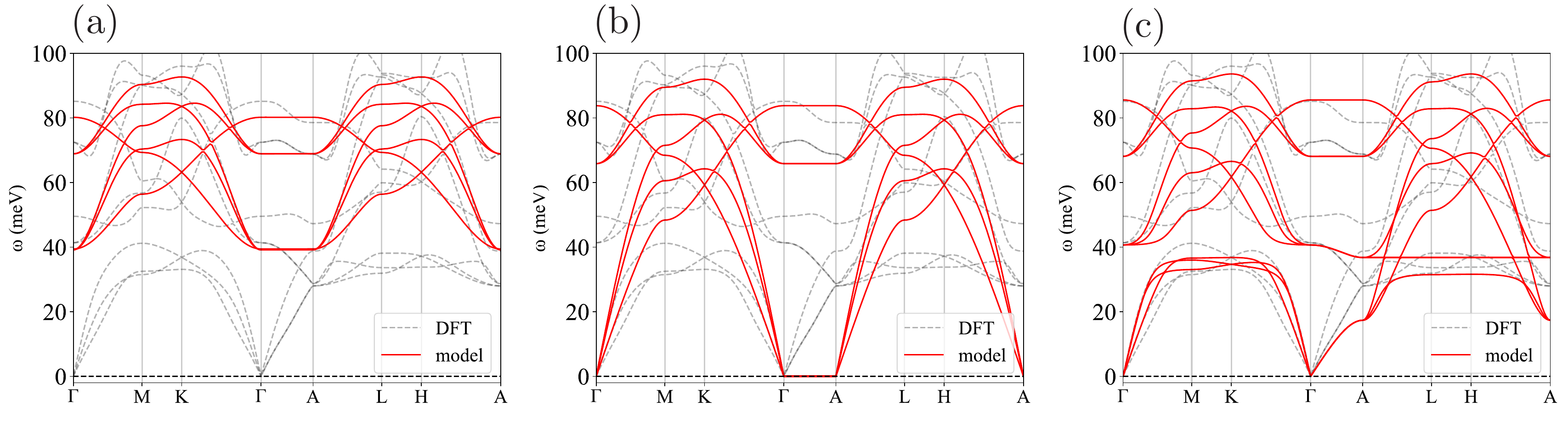}
\caption{\label{Fig: fit-phonon-asr} Phonon models from \cref{app:eq:H-phonon-full} for \ch{MgB2} are considered under different setups:
(a) The boron $xyz$ phonon block is extracted from \cref{app:eq:H-phonon-full} and fitted directly to DFT phonons, with the parameters given in \cref{app:eq:param_boron_only}.
(b) The acoustic sum rule is applied to case (a), followed by further fitting of the force constants, which yields three acoustic modes. The corresponding parameters are provided in \cref{app:eq:param_boron_only_asr} and \cref{app:eq:param_boron_only_asr2}.
(c) The full model from \cref{app:eq:H-phonon-full} is used, which adds the NN force constants between Mg and B atoms and enforces the acoustic sum rule. The parameters are given in \cref{app:eq:param_full_phonon} and \cref{app:eq:param_full_phonon2}.}
\end{figure}

In the full phonon mode, we first take out the boron phonon block and fit the parameters directly to the DFT phonon spectrum. Since magnesium is more than two times heavier than boron, the acoustic modes have small weights from boron, as shown in the orbital projection in \cref{app:fig: MgB2-DFT-phonon}. 
Thus we first fit boron phonons without enforcing the acoustic sum rules (a hypothetical case with sum rules enforced will be discussed shortly). 
The fitted parameters in $h_{\kk}^{B,xy}$ (\cref{app:eq:H-phonon-Bxy}) and $h_{\kk}^{B,z}$ (\cref{app:eq:H-phonon-full}) are 
\begin{equation}\begin{aligned}
\epsilon^{B, xy}=5700.51,\
\epsilon^{B, z}=3992.09,\quad 
t_{NN}^{B,xy}=-1070.78, \ 
t_{NNN}^{B,xy}=-427.01,\
t_{NN}^{B,z}=-812.57 \ 
\SI{}{meV}^2
\label{app:eq:param_boron_only}
\end{aligned}\end{equation}
The fitted dispersion is shown in \cref{Fig: fit-phonon-asr} (a).

We then consider a hypothetical case where only boron atoms exist, similar to graphene. As an approximation, we assume the same force constants in \cref{app:eq:param_boron_only}, but with the following additional constraints from the acoustic sum rule added on the boron part, which modify the onsite energies $\epsilon^{B,xy/z}$ and create three acoustic modes: 
\begin{equation}\begin{aligned}
\sum_j h^{B,xyz}_{\kk=\Gamma, i\mu,j\nu}/\sqrt{M_j} = 0,\quad \Rightarrow \quad
\epsilon^{B,xy} = 4917.735, \quad
\epsilon^{B,z} = 3509.376~ \SI{}{meV}^2. 
\label{app:eq:param_boron_only_asr}
\end{aligned}\end{equation}
After obtaining the onsite energies $\epsilon^{B,xy/z}$, the force constants are further refitted as
\begin{equation}
    t_{NN}^{B,xy}=-1444.824,\quad t_{NNN}^{B,xy}=-1169.795,\quad 
    t_{NN}^{B,z}=-458.415 \quad \SI{}{meV}^2.
\label{app:eq:param_boron_only_asr2}
\end{equation}
This refitting (to DFT phonon) is necessary; otherwise, merely enforcing the acoustic sum rule would lower the overall phonon dispersion and fail to reproduce the DFT results accurately. 
The corresponding dispersion is shown in \cref{Fig: fit-phonon-asr} (b). Three modes are enforced to pass the $\Gamma$ point due to the acoustic sum rule. This phonon spectrum is very close to the phonon of graphene shown in \cref{app:fig: MgB2-DFT-phonon}(c), except for the energy scale due to different magnitudes of force constants.

We then include the magnesium atoms. The force constants between magnesium and boron phonon modes are obtained by fitting to the DFT phonon spectrum:
\begin{equation}\begin{aligned}
t_{NN}^{B,xy;Mg,xy} = 130.238,\quad
t_{NN}^{B,xy;Mg,z} = -91.809,\quad
t_{NN}^{B,z;Mg,z} = -75.184~ \SI{}{meV}^2. 
\label{app:eq:param_full_phonon}
\end{aligned}\end{equation}
$t_{NN}^{B,z;Mg,xy}=0$ is set for simplicity.
The phonon onsite energies are determined from the acoustic sum rule enforced on $h^{full}_{\kk}$:
\begin{equation}\begin{aligned}
\epsilon^{Mg,xy} = 1352.851,\quad 
\epsilon^{Mg,z} = 1352.680, \quad
\epsilon^{B,xy} = 5218.638,\quad
\epsilon^{B,z} = 3810.259 ~\SI{}{meV}^2.
\label{app:eq:param_full_phonon2}
\end{aligned}\end{equation}
With these parameters, the phonon spectrum is shown in \cref{Fig: fit-phonon-asr} (c). Compared to \cref{Fig: fit-phonon-asr} (b), the inclusion of magnesium atoms primarily raises the boron phonon frequencies, while the three acoustic modes are dominated by magnesium vibrations due to the relatively large atomic mass of magnesium.

This analysis highlights a deep connection between the phonon spectra of graphene and \ch{MgB2}. Starting from a 2D graphene-like (boron) layer of light atoms, we compute the phonon spectrum of this layer. One then inserts the heavy Mg atoms between each pair of layers and incorporates the force constants between phonon modes. By enforcing the acoustic sum rule for the 3D system, the resulting spectrum closely resembles that of \ch{MgB2}, \ie, the light atom (boron) phonons almost rigidly shift up in energy and become the optical modes, due to the mass separation between magnesium and boron.

\section{First-principle results on electron-phonon coupling and superconducting property of \ch{MgB2}}\label{app:sec:DFT_doping_results}

\subsection{EPC and $T_c$}\label{app:sec:DFT_EPC_Tc}

In this section, we discuss the electron-phonon coupling (EPC) properties in \ch{MgB2} from DFT. The analytical EPC Hamiltonian will be derived in \cref{app:sec:EPC_hamiltonian} and will be shown to match the DFT results. 

\begin{figure}[htbp]
    \centering
    \includegraphics[width=0.6\textwidth]{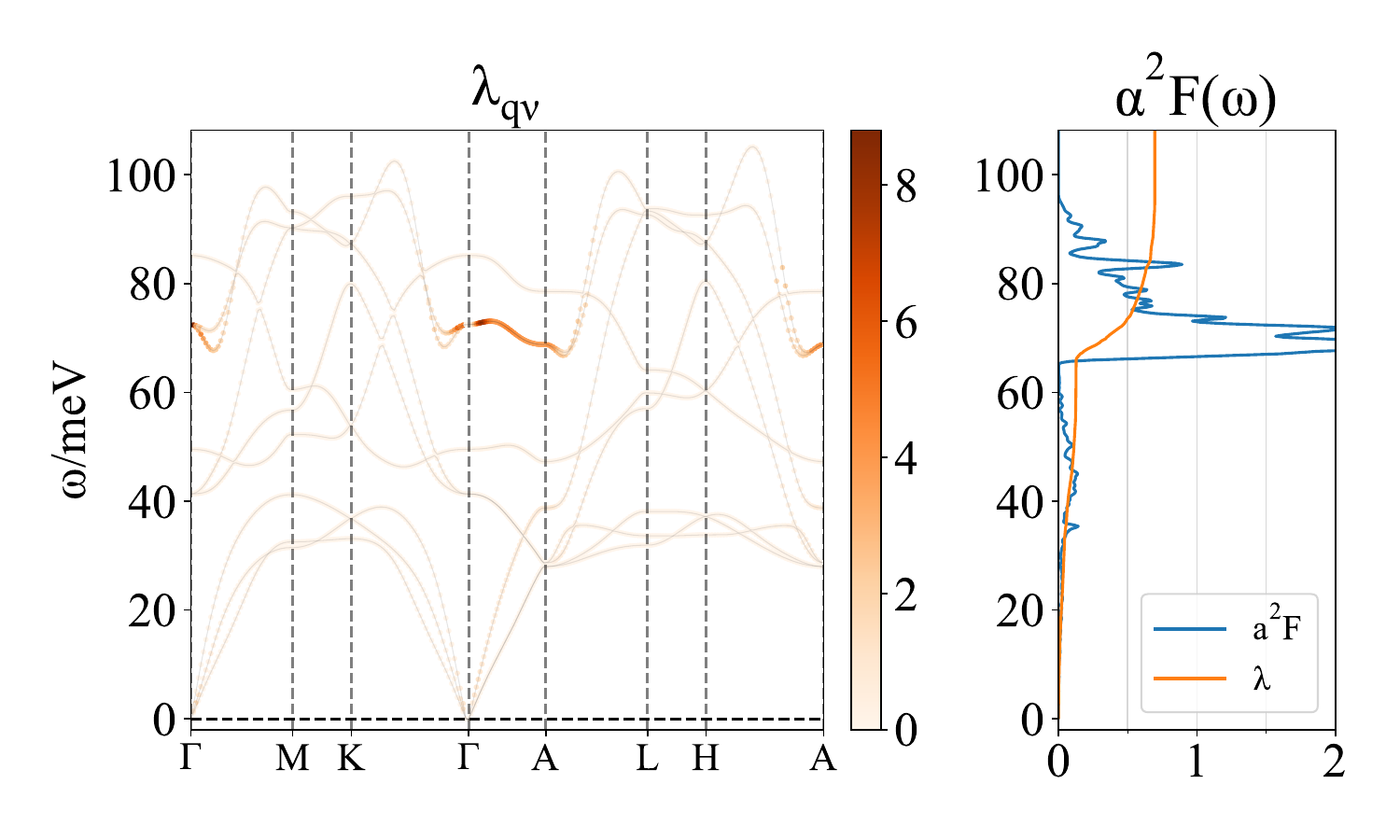}
    \caption{\label{Fig: MgB2-lambda_a2f} The electron-phonon coupling (EPC) strength $\lambda_{\qq\nu}$ on the phonon spectrum of \ch{MgB2}, together with the isotropic Eliashberg spectral function $\alpha^2 F(\omega)$ and the integrated $\lambda(\omega)$. The dominant EPC is given by an optical phonon along $\Gamma$-$A$ from the bond-stretching models of boron (see \cref{app:fig: MgB2-DFT-phonon}). 
    }
\end{figure}

We restrict to the $sp^2$ sector and perform the isotropoic $T_c$ calculation. 
As shown in \cref{Fig: MgB2-lambda_a2f}, the dominant EPC $\lambda_{\qq,\nu}$ is given by an optical phonon along $\Gamma$--A close to 70 meV. This optical phonon is mainly given by the in-plane phonon of boron with $\Gamma_5^+$ IRREP, as seen from the phonon orbital weights in \cref{app:fig: MgB2-DFT-phonon}. The boron $z$-phonon with a Dirac crossing at K at about 60 meV has approximately 10 times smaller EPC compared with the bond-stretching mode along $\Gamma$--A. 
To quantitatively estimate the contribution from the bond-stretching mode, we compute $\lambda$ from the 7th and 8th phonon branches (corresponding to the $\Gamma_5^+$ mode) with in-plane momentum $|k_{\parallel}| < \frac{1}{2} |\Gamma\text{-M}|$ (which occupies about $20\%$ of the Brillouin zone), yielding a $72.5\%$ contribution to the total $\lambda$.

In \cref{app:sec:MgB2_epc_selection_rule}, we will understand the EPC first from a selection rule of the EPC by finding which coupling between the electrons and phonons is symmetry-allowed. Then we use the obstructed atomic insulator (OAI) formed by the bonding states of boron to explain the origin of the dominant EPC in \cref{app:sec:EPC_strength}.

\subsection{Doping effect on $T_c$}\label{app:sec:doping_effect}

We investigate how to further increase $T_c$ in \ch{MgB2}. We first adopt the so-called rigid-band approximation in \textit{ab initio} in \cref{app:sec:rigid-band-doping} to investigate the behavior of $T_c$, followed by a full \textit{ab initio} treatment of doping that goes beyond rigid-band approximation, where the system is relaxed under doping, with electron, phonon, and EPC recalculated for the relaxed structure. 
In both approaches, we find a competition between the DOS and the EPC strength upon electron doping: $T_c$ initially increases at small electron doping even though the DOS actually \textit{decreases}. We show that this behavior arises from a subtle \textit{quantum-geometry} effect of the EPC projected onto the $\Gamma$-centered Fermi surface, where the EPC in the band basis is maximized at the center of each $k_z$ plane.

\subsubsection{Rigid-band doping}\label{app:sec:rigid-band-doping}

We first consider the rigid-band approximation in the $sp^2$ bonding basis, \ie, the electron hopping, the phonon force constants, and the real-space EPC tensor are all obtained from the undoped \textit{ab initio} calculation. We only shift the Fermi level $E_f$ and evaluate the SC-related properties, including $\lambda$ and $T_c$, based on the shifted $E_f$. The $\pi$-FS from Boron $p_z$ orbitals is ignored for simplicity, as it has a small contribution (20\% from Ref.~\cite{yu2024non}) to the SC properties.

\begin{figure}[htbp]
    \centering
    \includegraphics[width=0.8\textwidth]{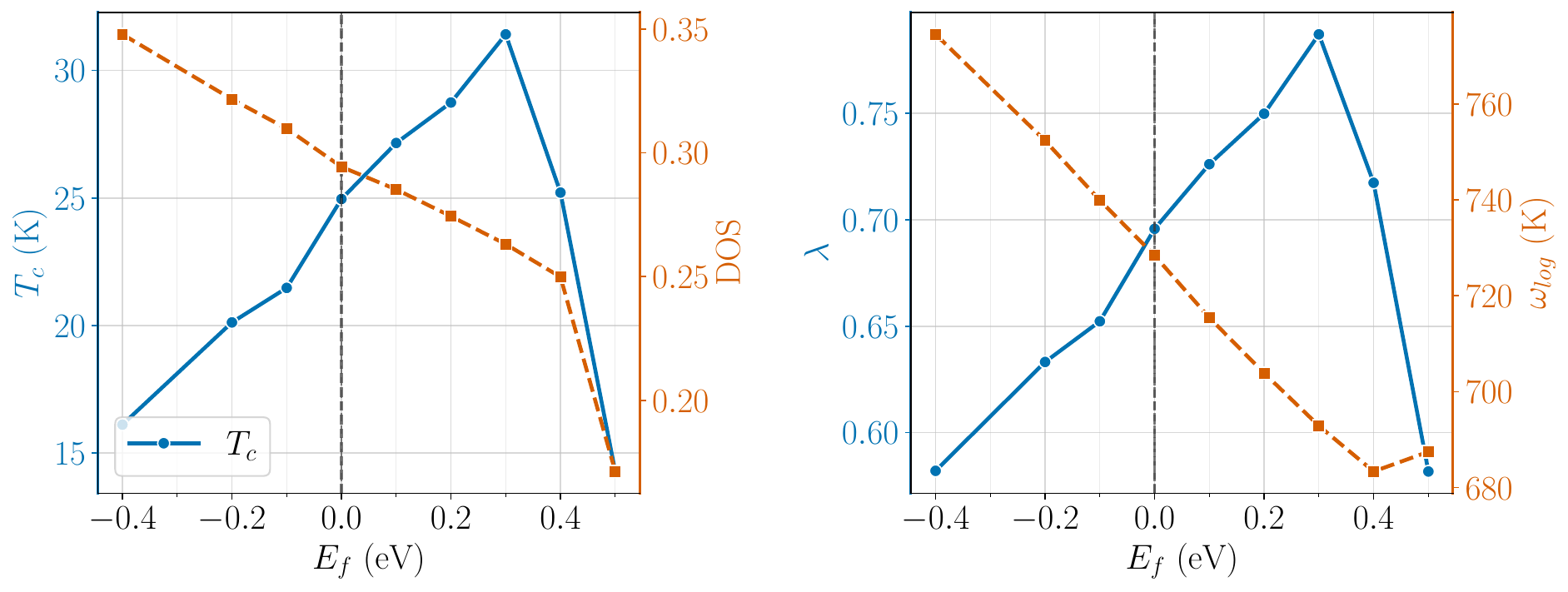}
    \caption{\label{app:fig:sp2-Tc-rigid-doping-effect} Doping effect on SC properties in \ch{MgB2}, where we show $T_c$, DOS, $\lambda$, and $\omega_{\text{log}}$ as a function of chemical potential. The results are computed using the rigid-band approximation in the $sp^2$ bonding basis, where the electron hopping, phonon force constants, and real-space EPC tensor are all obtained from the undoped \textit{ab initio} calculation, and only the Fermi level is shifted rigidly to simulate the doping effect. 
    }
\end{figure}

The resultant $T_c$, DOS, $\lambda$, and $\omega_{log}$ as a function of $E_f$ are shown in \cref{app:fig:sp2-Tc-rigid-doping-effect}. We observe that:
\begin{itemize}
    \item Near charge neutrality, the DOS decreases linearly when doping electrons (increasing $E_c$), and then drops rapidly after around $0.4$ eV, as the $sp^2$ bands on $k_z=0$ plane have the maximum around this energy and drop out of the Fermi surface upon further doping. Upon further electron doping, the entire $\Gamma- A$ Fermi surface disappears. 
    \item Opposite to the decreasing DOS, $T_c$ increases approximately linearly near the charge neutrality when doping electrons, and then drops rapidly after around $E_f=0.3$ eV, concomitant with the disappearance of the Fermi surface at $\Gamma$. 
    \item $\lambda$ shows a similar trend as $T_c$. 
    \item $\omega_{log}$ decreases linearly as a function of $E_f$.  
\end{itemize}

\begin{figure}[htbp]
    \centering
    \includegraphics[width=0.8\textwidth]{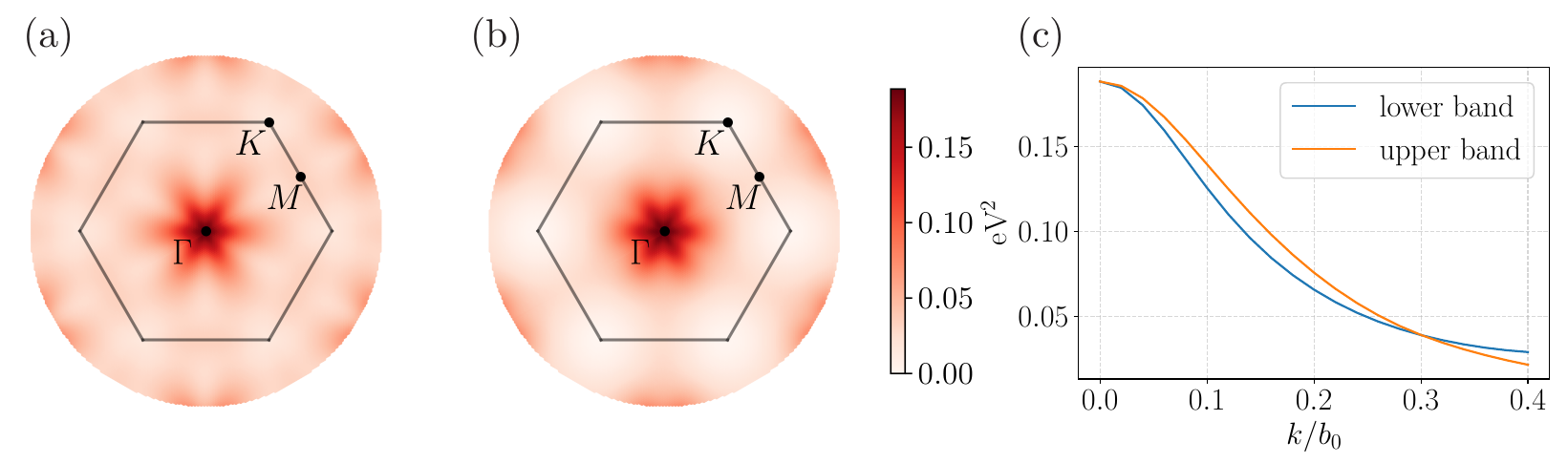}
    \caption{\label{app:fig:DFT_EPC_band_basis} Behavior of EPC in the band basis. 
    (a,b) Band-resolved EPC $G^m_{\mathbf{k}}$ ($m=1,2$ for the two $sp^2$ bands contributing to the $\sigma$ FS, defined in \cref{app:eq:Gk_band_basis_sp2_bands}, with panel (a) being the lower band) on the $k_z=0$ plane for phonon momentum $\mathbf{q}=0$, summed over the two bond-stretching modes, for the two $sp^2$ bands that contribute to the $\sigma$ FS. $G_{\mathbf{k}}$ peaks at $\Gamma$ and stays high along the $\Gamma$-K line. (c) Radial average of $G_{\kk}$ versus distance $k$ from $\Gamma$, decaying approximately as $k^{2}$.
    }
\end{figure}

The rise of $\lambda$ and $T_c$ at the small electron doping regime implies that the increase in EPC strength outweighs the decrease in the DOS. To verify this, we compute the EPC tensor projected to the band basis. In \cref{app:fig:DFT_EPC_band_basis} (a), (b), we show the band-basis EPC on the $k_z=0$ plane with phonon momentum $\qq=\bm{0}$, for the two $sp^2$ bands contributing to the $\sigma$ FS, and sum over the two bond-stretching phonon modes of $\Gamma_5^+$ (\ie, $u_1=\frac{1}{\sqrt{2}}(u_{B_{1x}} - u_{B_{2x}}), u_2=\frac{1}{\sqrt{2}}(u_{B_{1y}} - u_{B_{2y}})$), which is 
\begin{equation}\begin{aligned}
G^m_{\kk} &= |G^{m,m,u_1}_{\kk,\bm{0}}|^2 + |G^{m,m,u_2}_{\kk,\bm{0}}|^2,
\label{app:eq:Gk_band_basis_sp2_bands}
\end{aligned}\end{equation}
where $m=1,2$ label the two $sp^2$ bands and $u_{1,2}$ the two bond-stretching phonons. We observe that $G^m_{\kk}$ peaks at $\Gamma$ and decays away from it. $G^m_{\kk}$ is also large along the $\Gamma$-K line when compared with other directions, which results from the wavefunction of the $sp^2$ bonding states (see the analytic EPC model discussed in \cref{app:sec:EPC_sp2_basis_analytic_model}). 
\cref{app:fig:DFT_EPC_band_basis}(c) shows the angular average of $G^m_{\kk}$, \ie, 
$\frac{1}{2\pi}\int_{0}^{2\pi} G^m_{\kk} d\theta$, for circles of radius $k$ around $\Gamma$. The angle-averaged EPC decays approximately quadratically away from $\Gamma$. In \cref{app:sec:EPC_sp2_basis_analytic_model}, we will use an analytic model to understand the behavior of $G^m_{\kk}$. In \cref{app:sec:analytic_estimation_lambda}, we will adopt analytic modes of EPC and DOS to show that enhanced EPC when electron-doping towards $\Gamma$ wins over the linear-decreasing DOS, which leads to the rise of $\lambda$ and $T_c$.

\subsubsection{\textit{Ab initio} doping}

To validate the rigid-band doping results, we further consider the full \textit{ab initio} treatment of doping. To do so, we change the number of valence electrons in the system and perform the lattice relaxation. The relaxed structure is used to compute the electron, phonon, and EPC properties.

\begin{figure}[htbp]
    \centering
    \includegraphics[width=0.8\textwidth]{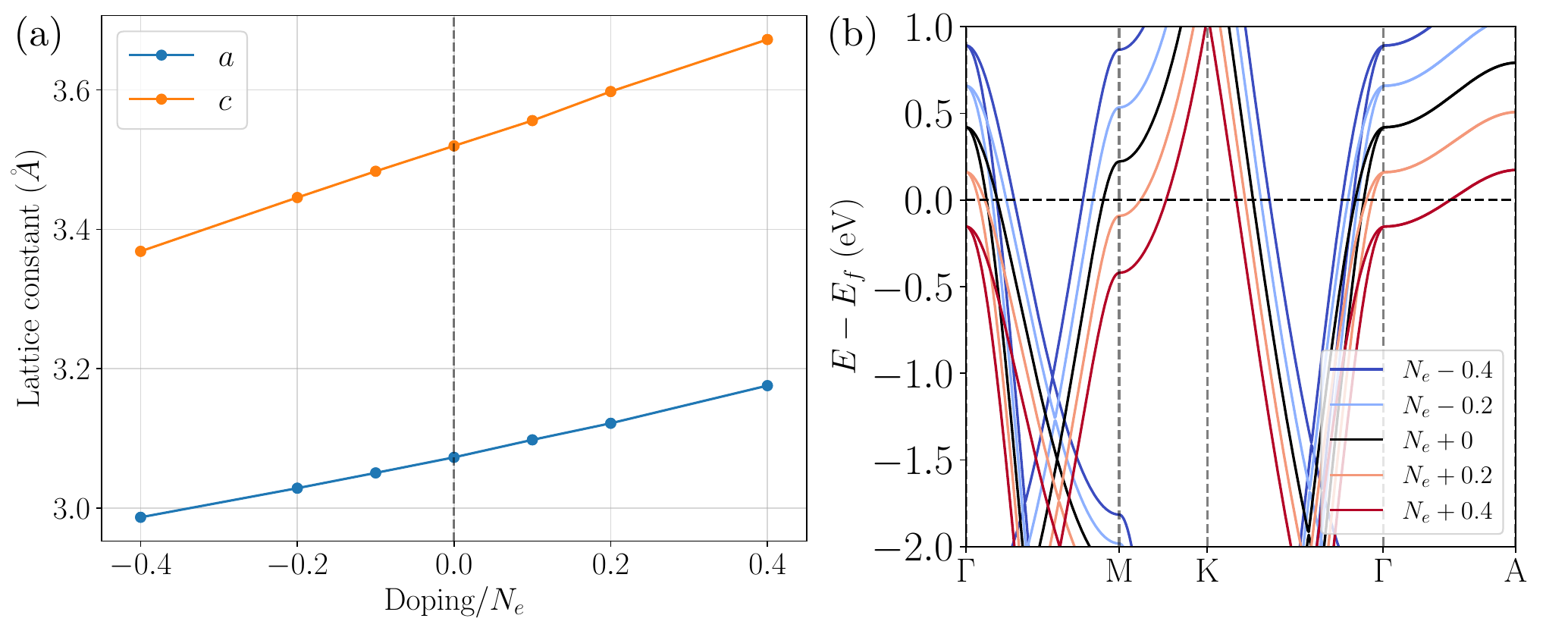}
    \caption{\label{app:fig:DFT-doping-lattice-bands} Doping effects on (a) lattice constants and (b) electronic band structures. The lattice constants increase (decrease) with electron (hole) doping. The $\sigma$ FS expands (shrinks) with hole (electron) doping, and disappears after around $0.4$ electron doping on the $k_z=0$ plane. The number of doped electrons is per unit cell and takes into account spin. 
    }
\end{figure}

\begin{figure}[htbp]
    \centering
    \includegraphics[width=\textwidth]{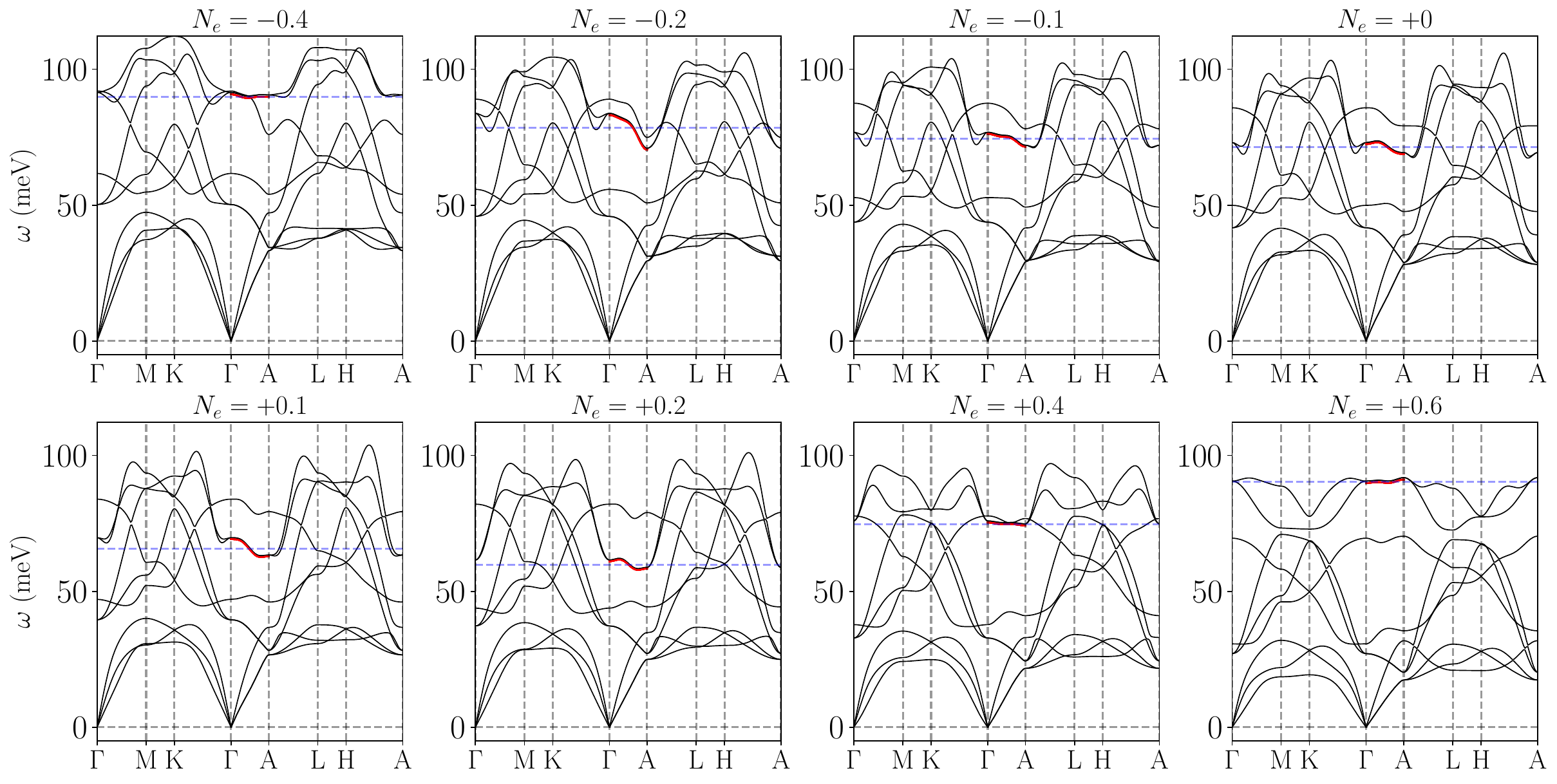}
    \caption{\label{app:fig:DFT-doped-phonon} Doping effects on the phonon spectrum in \ch{MgB2}, where the B--B bond-stretching mode along $\Gamma$--A is marked in red. A blue dashed line denotes the averaged frequency of this mode along $\Gamma$--A. 
    }
\end{figure}

\begin{figure}[htbp]
    \centering
    \includegraphics[width=0.3\textwidth]{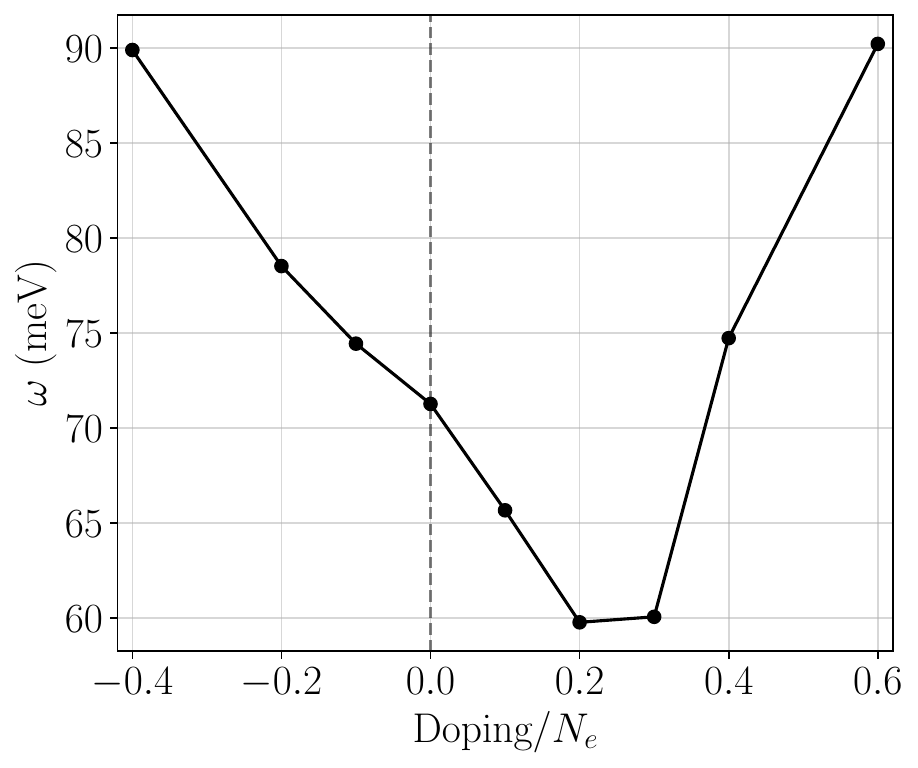}
    \caption{\label{app:fig:DFT-doped-phonon-GM5p} Doping effects on the frequency of the B--B in-plane bond-stretching mode, which is averaged along the $\Gamma$--A line (see the red bands in \cref{app:fig:DFT-doped-phonon}).
    }
\end{figure}

\cref{app:fig:DFT-doping-lattice-bands} (a) shows the doping effect on the relaxed lattice constants of \ch{MgB2}. The lattice constants increase (decrease) with electron (hole) doping. 
This is a new effect, not taken into consideration by the rigid doping approximation in the previous section \cref{app:sec:rigid-band-doping}. 
Using the relaxed crystal structure, the corresponding electronic bands are computed and shown in \cref{app:fig:DFT-doping-lattice-bands} (b). We observe that the $\sigma$ FS expands (shrinks) with hole (electron) doping, and disappears at around $0.3$ electron doping on the $k_z=0$ plane, close to the undoped band where the $\Gamma_5^+$ node at $\Gamma$ has an energy of 0.42 eV.

The doping effect on the phonon spectrum is shown in \cref{app:fig:DFT-doped-phonon}, with the averaged frequency of the B--B bond-stretching mode along $\Gamma$--A extracted in \cref{app:fig:DFT-doped-phonon-GM5p}. The averaged frequency increases approximately linearly towards \SI{90}{meV} at the hole-doped side. At the electron-doping side, the frequency first drops to around \SI{60}{meV} at 0.2 electron doping, and then stiffens to \SI{90}{meV} for 0.6 electron doping, where the $sp^2$ FS is completely below $E_f$, and the $\Gamma_5^+$ phonon mode has no softening, similar to that in graphene. 
We observe that in the $+0.6$ electron-doped phonon system, the four modes dominated by the boron $xy$ phonon split into two disconnected sets (i.e., the four phonon bands among the top six phonon bands, after excluding the two bands mainly from the boron $z$ phonon that form a Dirac crossing, similar to the orbital projection observed in the undoped phonon case shown in \cref{app:fig: MgB2-DFT-phonon}(b)). The upper set exhibits the IRREPs $\Gamma_5^+, M_3^-,M_4^-, K_1,K_4$, and its elementary band representation (EBR) decomposition is $E'@2d - B_{2u}@3g + B_{1u}@1b$. In contrast, the lower set features the IRREPs $\Gamma_6^-, M_1^+,M_2^+, K_5$, and has an EBR decomposition $B_{2u}@3g - B_{1u}@1b$. Consequently, both sets host a fragile topology.

\begin{figure}[htbp]
    \centering
    \includegraphics[width=0.8\textwidth]{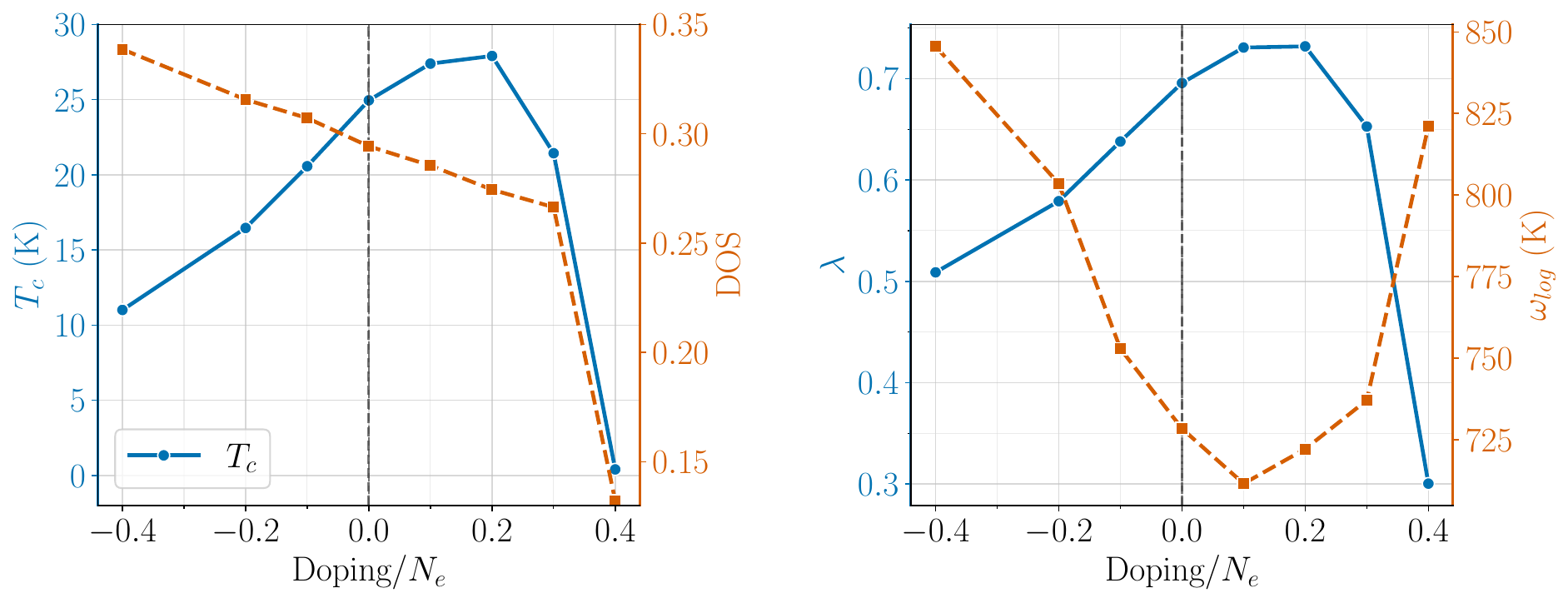}
    \caption{\label{app:fig:sp2-Tc-DFT-doping-effect} Doping effect on SC properties in \ch{MgB2} with full \textit{ab initio} treatment. $T_c$, DOS, $\lambda$, and $\omega_{\text{log}}$ are shown as a function of chemical potential. We only consider three $sp^2$ bonding states of boron when computing $T_c$. 
    }
\end{figure}

The softening of the bond-stretching phonon mode in the light electron doping region can be understood from the one-loop correction to the phonon force constants from the EPC~\cite{hu2023kagome}:
\begin{equation}
    \Phi_{i\mu,j\nu}^{eff}(\qq) =  \Phi_{i\mu,j\nu}^0(\qq) - \Phi_{i\mu,j\nu}^{corr}(\qq),
\end{equation}
where $\Phi_{i\mu,j\nu}^{0}(\qq)$ is the bare phonon FCs, and $\Phi_{i\mu,j\nu}^{corr}(\qq)$ is the one-loop correction term
\begin{equation}
     \Phi_{i\mu,j\nu}^{corr}(\qq) = 
     \sum_{mn,\sigma,\kk} \frac{\tilde{G}_{\kk,\qq}^{nm,j\nu} \tilde{G}_{\kk+\qq,-\qq}^{mn,i\mu}}{N} \frac{-f_{\kk+\qq,n} + f_{\kk,m}}{\epsilon_{\kk+\qq,n} - \epsilon_{\kk, m}}. 
\end{equation}
where $\tilde{G}_{\kk,\qq}^{nm,j\nu}$ is the EPC tensor defined in electron band basis and phonon in atomic basis (see \cref{app:eq:EPC_band_basis}). 
Qualitatively, for $\qq\sim 0$ and small FS near $\Gamma$, we have $\Phi_{i\mu,j\nu}^{corr}(\qq)\propto \langle |\tilde{G}_{\kk,\qq}|^2 \rangle_{FS} N(E_f)$, where $\langle |\tilde{G}_{\kk,\qq}|^2 \rangle_{FS}$ is the FS-averaged EPC strength, and $N(E_f)$ is the DOS at $E_f$. 
Thus, doping induces opposite trends in the DOS and EPC matrix elements: upon electron (hole) doping, the DOS decreases (increases) while the EPC strength (projected to the band basis) increases (decreases), leading to a direct competition between DOS and coupling strength. Consequently, the averaged frequency of the B--B bond-stretching mode along $\Gamma$--A \cref{app:fig:DFT-doped-phonon} follows the same doping dependence as the EPC constant $\lambda$ but with the opposite sign, as we will show in the following \cref{app:fig:sp2-Tc-DFT-doping-effect}(b).

The superconducting properties under \textit{ab initio} doping are presented in \cref{app:fig:sp2-Tc-DFT-doping-effect}, where we plot the trend of $T_c$, DOS, $\lambda$, and $\omega_{log}$ as a function of doping.
We only consider the $sp^2$ bonding states sector in the calculation, with $T_c$ computed from the Allen-Dynes modified McMillian equation in \cref{app:eq: McMillian-Tc}. 
The evolution of $T_c$ under doping is similar to the rigid-band doping results in \cref{app:fig:sp2-Tc-rigid-doping-effect}, where the $T_c$ is peaked at small electron doping around 0.1 electron per unit cell. $\lambda$ has a similar trend as $T_c$. $\omega_{log}$, however, first decreases and then increases for electron doping, resulting from the hardening of the $\Gamma_5^+$ mode at large electron doping. 
In summary, the \textit{ab initio} doping effect on $T_c$ qualitatively agrees with the rigid-band doping result.

\subsubsection{Tensile-strained structures}\label{app:sec:strained_Tc}

\begin{figure}[htbp]
    \centering
    \includegraphics[width=0.8\textwidth]{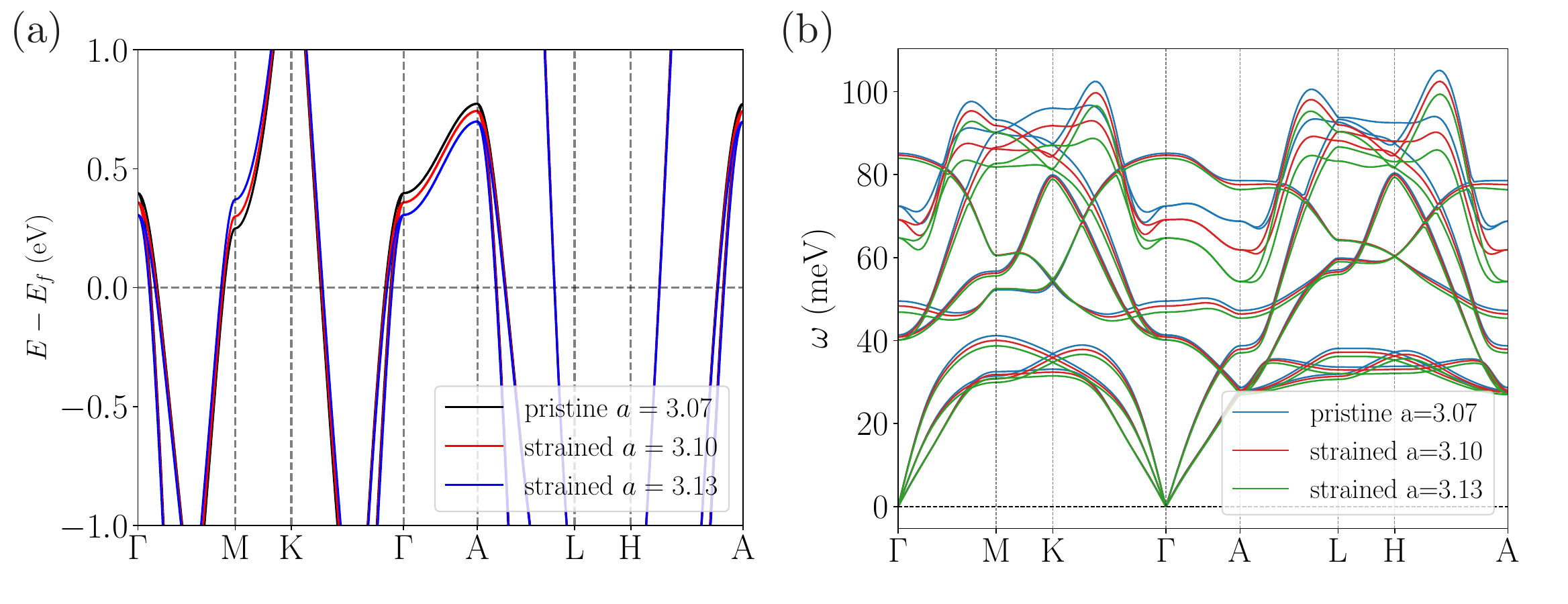}
    \caption{\label{app:fig:compare-strained-cell}
    (a) Electronic band structures of pristine ($a=3.073$ \SI{}{\angstrom}, $c=3.520$ \SI{}{\angstrom}) and tensile-strained \ch{MgB2} (red bands have $a=3.100$ \SI{}{\angstrom}, $c=3.515$ \SI{}{\angstrom}, blue bands have $a=3.130$ \SI{}{\angstrom}, $c=3.515$ \SI{}{\angstrom}). Adding tensile strain introduces a monotonic reduction in the size of the $\sigma$ FS cylinders, which corresponds to an effective light electron doping of the $sp^2$ bands. (b) Phonon spectrum of pristine and tensile-strained \ch{MgB2}, where a stronger softening of the bond stretching mode is observed under tensile strain. 
    }
\end{figure}

At last, we note that our doping results at light electron-doping are qualitatively consistent with the tensile-strain experiments of Refs.~\cite{pogrebnyakov2004enhancement, xi2007mgb2}, which reported an enhanced $T_c$ up to $41.8$~K in \ch{MgB2} thin films. Using the strained lattice constants from Ref.~\cite{pogrebnyakov2004enhancement}, $a=\SI{3.100}{\angstrom}$ and $c=\SI{3.515}{\angstrom}$ (corresponding to $\sim1$\% in-plane lattice constant stretch), we compute the corresponding electronic structure, shown in the red bands in \cref{app:fig:compare-strained-cell} (a). Relative to the pristine cell, the $\sigma$ FS shrinks slightly, indicating that the strained structure is effectively electron-doped within the $sp^2$ sector. \cref{app:fig:compare-strained-cell} (b) compares the phonon spectrum in the pristine and strained cells, where a clear softening of about 4 meV from the bond-stretching mode is observed. We further compute the SC property (considering only the $sp^2$ bonding sector for simplicity) for this tensile strain structure, as shown in \cref{app:table:Tc_comparsion_strained_cell}, which is about 3 K higher than the pristine cell. This qualitatively agrees with the experimentally observed softening of the bond-stretching mode under tensile strain and the observed $T_c$ increase.

\begin{table}[htbp]
\centering
\begin{tabular}{c|c|c|c}
\hline
Lattice parameter/\SI{}{\angstrom} & $\lambda$ & $\omega_{\text{log}}$/K & $T_c/K$ \\\hline\hline
$a=3.07$ & 0.696 & 727.89 & 24.96  \\\hline
$a=3.10$ & 0.733 & 704.21 & 27.31 \\\hline
$a=3.13$ & 0.799 & 672.63 & 31.44 \\\hline
\end{tabular}
\caption{\label{app:table:Tc_comparsion_strained_cell} Comparison of $T_c$ in the pristine ($a=3.07$ \SI{}{\angstrom}) and tensile-strained structure of \ch{MgB2}.}
\end{table}

We further explore a larger in-plane tensile strain of approximately 2\%, corresponding to a lattice constant of $a=$ \SI{3.130}{\angstrom} while keeping $c=$ \SI{3.515}{\angstrom}, as shown by the electron and phonon bands in \cref{app:fig:compare-strained-cell} (a) and (b). Under this lattice parameter, a more pronounced reduction in the $sp^2$ Fermi surface (FS) and softening in the bond stretching mode are observed, suggesting that further enhancement of $T_c$ in \ch{MgB2} (see \cref{app:table:Tc_comparsion_strained_cell}), albeit such strain levels remain experimentally challenging to realize.

\subsection{Superfluid weight}\label{app:sec:superfluid_weight}

Here we estimate the superfluid weight, or superfluid stiffness, tensor $D_s=D_\text{geom} + D_\text{conv}$ in \ch{MgB2}. 
The superfluid weight quantifies the phase stiffness of the superconducting condensate, governing the Meissner response, penetration depth, and the BKT transition temperature in 2D. In multiband systems, it acquires an additional geometric contribution from interband current matrix elements—equivalently, from the quantum geometry of Bloch states. 

We compute the superfluid weight from DFT data using the mean-field and uniform-pairing limit expressions of superfluid weight~\cite{hiorth2026ab, Liang2017, Huhtinen2022} at zero temperature:
\begin{align}
    D_{conv}^{ij} &= \frac{1}{V} \sum_{ \mathbf{k}m }
    \frac{ \Delta^2 }{ \sqrt{\epsilon_{\kk m}^{2} + \Delta^2}^{3}}
    \partial_i \epsilon_{\kk m} \partial_j \epsilon_{\kk m} 
    \label{eq:dconv-orig} \\
    D_{geom}^{ij} &= \frac{1}{V} \sum_{\kk m} \sum_{n\neq m} 
\frac{ \epsilon_{\kk n} - \epsilon_{\kk m} }{ \epsilon_{\kk m} + \epsilon_{\kk n}} 
    \left[ \frac{ \Delta^2}{ \sqrt{ \epsilon_{\kk m}^2 + \Delta^2 }} - \frac{ \Delta^2}{ \sqrt{\epsilon_{\kk n}^2 + \Delta^2 } } \right] \times (\langle\partial_i \kk m|\kk n \rangle \langle \kk n|\partial_j \kk m \rangle + \text{H.c.}).
    \label{eq:dgeom-orig}
\end{align}
Here $i, j \in {x,y,z}$ are Cartesian components, $\Delta$ is the superconducting gap, and $\epsilon_m(\mathbf{k})$ is the band dispersion (measured relative to the Fermi level) for band index $m$. The dispersions $\epsilon_m$ and Bloch states $\ket{m}$ are obtained from DFT, while the (FS-dependent but $\kk$-independent) gap values are taken from experiment, with $\Delta_{\sigma}=6.8$ meV and $\Delta_{\pi}=1.8$ meV.

\begin{figure}
    \centering
    \includegraphics[width=0.95\linewidth]{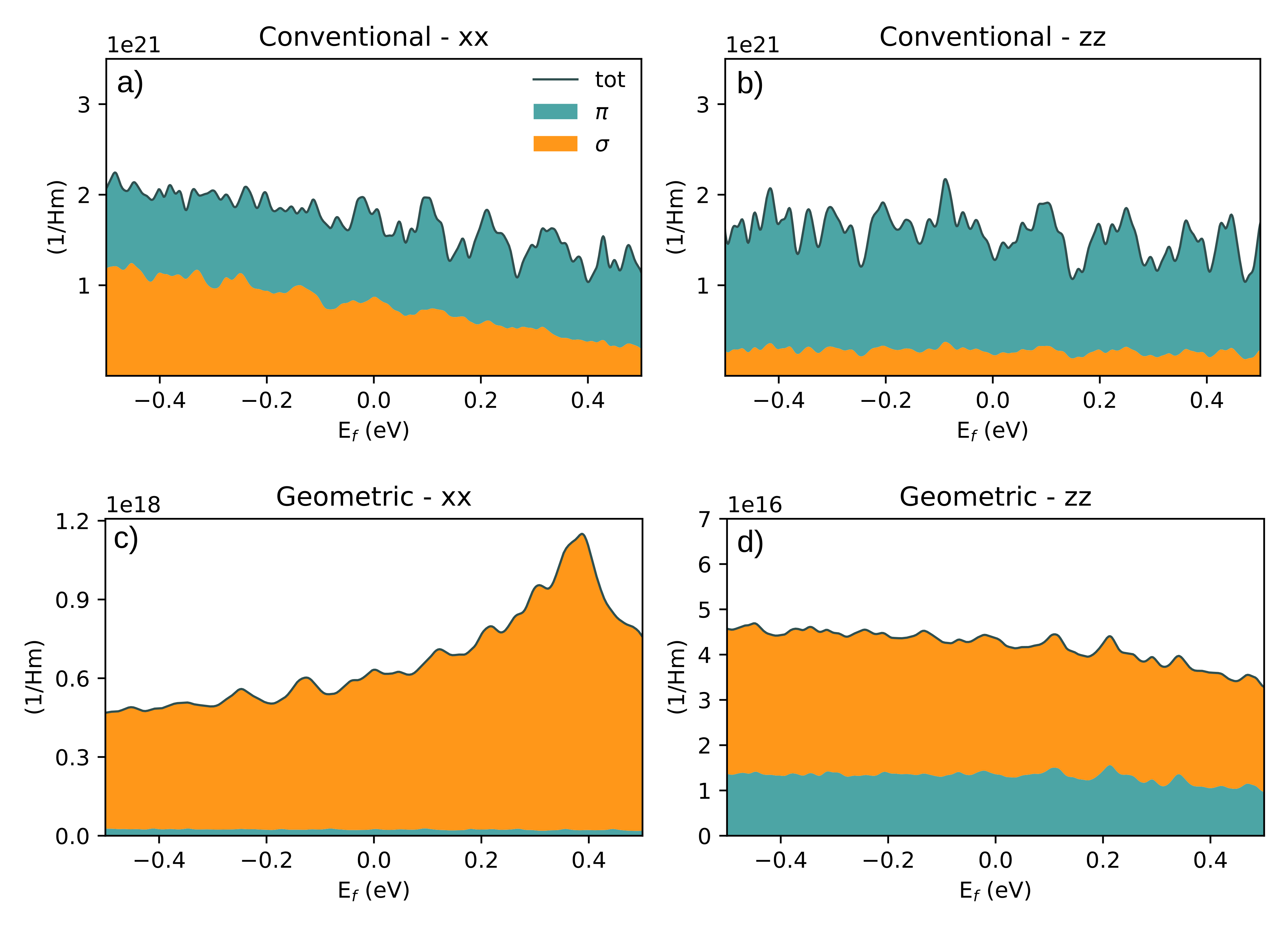}
    \caption{The superfluid weight as a function of the Fermi level, split into the conventional (a, b) and the geometric (c, d) contributions. The left column shows the $xx$ component (equivalently $yy$), and the right column shows the $zz$ component. The conventional term is dominated by the $\pi$ bands, particularly along $z$, while the geometric one is dominated by the $\sigma$ bands.}
    \label{app:fig:sfw-sigma-pi-contr}
\end{figure}

We evaluate the superfluid weight for both gaps of \ch{MgB2}. 
\cref{app:fig:sfw-sigma-pi-contr} shows the superfluid weight as a function of the Fermi-level shift, separated into the conventional contribution, panels (a) and (b), and the geometric contribution, panels (c) and (d). The left column displays the $xx$ component, which is equivalent to the $yy$ component by symmetry, and the right column shows the $zz$ component

The conventional term is three to five orders of magnitude larger than the geometric term, as the bands crossing the Fermi level are very dispersive. In the conventional term, the $\pi$ bands dominate, especially in the $zz$ component. In contrast, the geometric term is primarily governed by the $\sigma$ bands. In the geometric $xx/yy$ component, we observe a pronounced peak at \SI{0.39}{eV}, corresponding to the twofold-degenerate 2D states at $\Gamma$. 
We note that, although the quantum metric diverges near a symmetry-protected degeneracy, as discussed in \cref{app:sec:quantum_metric_deg_point}, this divergence does not directly translate into a divergent geometric contribution to the superfluid weight. In \cref{eq:dgeom-orig}, the geometric term is weighted by an energy- and SC gap-dependent prefactor, which suppresses the singular contribution from the degenerate point.

\begin{figure}
    \centering
    \includegraphics[width=0.95\linewidth]{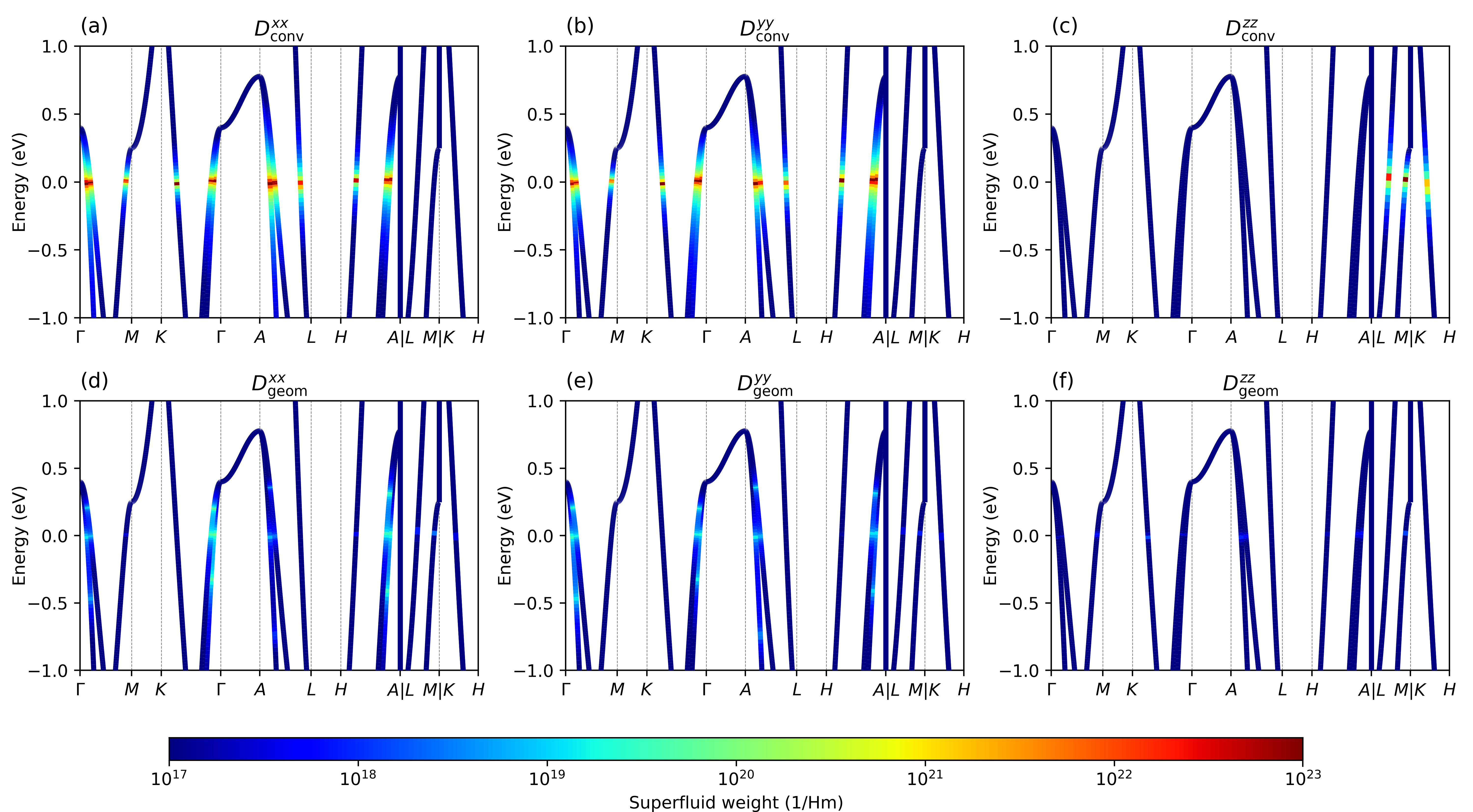}
    \caption{The superfluid weight integrand $D^{ij}(\mathbf{k},m)$ of \ch{MgB2} along the high-symmetry path for all three Cartesian directions, split into the conventional (a)-(c) and the geometric (d)-(f) terms. The $\sigma$ and $\pi$ orbitals have been marked in (a).
    }
    \label{app:fig:sfw-path-mgb2}
\end{figure}

The superfluid weight integrand $D^{ij}_{conv/geom}(\kk, m)$ (\ie, the summed quantities under $\sum_{\kk m}$ in Eqs.~\ref{eq:dconv-orig} and \ref{eq:dgeom-orig}) is defined as
\begin{align}
    D_{conv}^{ij}(\kk,m) &= 
    \frac{ \Delta^2 }{ \sqrt{\epsilon_{\kk m}^{2} + \Delta^2}^{3}}
    \partial_i \epsilon_{\kk m} \partial_j \epsilon_{\kk m} 
    \label{eq:dconv-orig-km} \\
    D_{geom}^{ij}(\kk,m) &= \sum_{n\neq m} 
\frac{ \epsilon_{\kk n} - \epsilon_{\kk m} }{ \epsilon_{\kk m} + \epsilon_{\kk n}} 
    \left[ \frac{ \Delta^2}{ \sqrt{ \epsilon_{\kk m}^2 + \Delta^2 }} - \frac{ \Delta^2}{ \sqrt{\epsilon_{\kk n}^2 + \Delta^2 } } \right] \times (\langle\partial_i \kk m|\kk n \rangle \langle \kk n|\partial_j \kk m \rangle + \text{H.c.}).
    \label{eq:dgeom-orig-km}
\end{align}
\cref{app:fig:sfw-path-mgb2} shows $D^{ij}_{conv/geom}(\kk, m)$ of \ch{MgB2} along the high-symmetry path, with the $\sigma$ and $\pi$ orbitals indicated in panel (a). The conventional contribution is primarily concentrated at the Fermi surface, whereas the geometric contribution exhibits a broader distribution in energy.
Consistent with our previous observations, the geometric superfluid weight is largely associated with the $\sigma$ bands, and its $zz$ component remains small. The conventional one is fairly evenly divided between the $\pi$ and $\sigma$ bands in the $xx/yy$ components, while the $\pi$ bands clearly dominate in the $zz$ component. 

Our calculations for \ch{MgB2} also reveal a slight anisotropy: within the Fermi-level window $(-0.6, 0.3),\mathrm{eV}$, the $xx/yy$ component of the superfluid weight is generally larger than the $zz$ component. Since the superfluid weight is related to the London penetration depth via $\lambda_L = (\mu_0 D_s)^{-1/2}$, we can qualitatively compare our results with penetration-depth measurements. Anisotropy in \ch{MgB2} has indeed been reported in several experiments \cite{loudon_measurement_2015, eltsev_anisotropic_2002}, although the material becomes nearly isotropic in the low-field, clean limit, in good agreement with our findings \cite{klein_magnetic_2006}. Moreover, \textcite{tan_enhancement_2015} reported an enhanced $\pi$-band contribution to the penetration depth in thicker samples; this is consistent with the strong $\pi$-band contribution to the $zz$ component found in our calculations.

\section{Analytic Electron-phonon coupling Hamiltonian in \ch{MgB2}}\label{app:sec:EPC_hamiltonian}

\subsection{Analysis of EPC tensor in \ch{MgB2}}\label{app:sec:MgB2_epc_selection_rule}

In this section, we analyze the EPC in \ch{MgB2} and find that (i) symmetry constraints enforce certain EPC terms to be zero, and (ii) strong $sp^2$ bonding leads to strong EPC to the boron in-plane phonon. 

\subsubsection{Symmetry-constraints of EPC}\label{app:sec:EPC_symm_constraints}

In \ch{MgB2}, there are two Fermi surfaces (FSs): the $\sigma$ FS near GM with IRREP $\Gamma_5^+$, and the $\pi$ FS near M with IRREP $M_3^+$ (\ie, the von-Hove singularity at M from the Dirac band of $p_z$). The $\sigma$ FS is given by the $sp^2$ bonding states of B, while the $\pi$ FS is given by the $p_z$ orbitals of B. The convention of the IRREPs follows the \textit{Bilbao Crystallographic server}~\cite{aroyo2006bilbao1, aroyo2006bilbao2, aroyo2011crystallography}. 

We now consider which EPC is allowed by symmetry for these two FSs. We make the approximation by assuming these two FSs are located at $\Gamma$ and M, respectively. This approximation is great for the $\sigma$ FS, which is very small: the phonons that couple that Fermi surface to cause superconductivity have to scatter between $k$ and $-k$. Since $k$ is small, the phonon wavevector is also small, and we approximate it to zero. 
\begin{itemize}
\item For $\QQ=\Gamma, \KK=\Gamma$, we only need to consider the particle-hole operation $O_{\Gamma,\Gamma}=\{\gamma_{\Gamma,n}^\dagger \gamma_{\Gamma,m} \}$. This operation has the direct product IRREP: 
$\Gamma_5^+\otimes\Gamma_5^+=\Gamma_1^+ \oplus \Gamma_2^+ \oplus \Gamma_5^+$. 
Thus it can only couple to phonons with $\Gamma_1^+, \Gamma_2^+,\Gamma_5^+$ IRREPs.There is \emph{only one} boron phonon mode at $\Gamma$ corresponding to these representations, namely the $\Gamma_5^+$ phonon.

\item For $\QQ=\Gamma, \KK=M$, we need to consider three equivalent M points $M^{(1)}=\frac{1}{2}\bm{b}_1, M^{(2)}=\frac{1}{2}(-\bm{b}_1+\bm{b}_2), M^{(3)}=-\frac{1}{2}\bm{b}_2$, \ie, $O_{M,\Gamma}=\{ (\gamma_{M^{(1)},n}^\dagger \gamma_{M^{(1)},n}, \gamma_{M^{(2)},n}^\dagger \gamma_{M^{(2)},n}, \gamma_{M^{(3)},n}^\dagger \gamma_{M^{(3)},n}) \}$, where $n$ denote the (only one) $\pi$ band that cross $E_f$ close to $M^{(i)}$, with IRREP $M_3^+$. For each $M^{(i)}$, the direct product IRREP is the trivial IRREP, i.e., $M_3^+\otimes M_3^+ = M_1^+$. The three particle-hole operators with $M_1^+$ IRREP in $O_{M,\Gamma}$ induce a reducible representation $\overline{D}_{O_{M,\Gamma}}(g)$ at $\Gamma$, with 
\begin{equation}
    \overline{D}_{O_{M,\Gamma}}(C_3) = \begin{bmatrix}
        0 & 1 & 0 \\
        0 & 0 & 1 \\
        1 & 0 & 0
    \end{bmatrix}, 
\end{equation}
\ie, the $C_3$ representation being a permutation matrix that permutes $M^{(i)}$. 
$\overline{D}_{O_{M,\Gamma}}(g)$ can be decomposed into $\Gamma_1^+ \oplus \Gamma_5^+$. Thus the $\pi$ FS at M can couple to $\Gamma$ point phonon with IRREP $\Gamma_1^+$ and $\Gamma_5^+$. In \ch{MgB2}, only $\Gamma_5^+$ exists.

\item For $\QQ=M, \KK=\Gamma$, we have $O_{\Gamma, M}=\{\gamma_{M,n}^\dagger \gamma_{\Gamma, m} \}$. As phonon momentum $\QQ=M$, we need to express the particle-hole operator using IRREPs at M. To do so, we first reduce the $\Gamma$ IRREPs to M, \ie, $\Gamma_5^+\downarrow_{M} = M_1^+ \oplus M_2^+$ (where $M_1^+$ and $M_1^+$ have the same inversion but opposite in-plane eigenvalues). Then we obtain the product representation $(\Gamma_5^+\downarrow_{M}) \otimes M_3^+= M_3^+ \oplus M_4^+$. As a result, the phonon modes with $M_3^+$ and $M_4^+$ IRREPs are symmetry-allowed to couple.

\item For $\QQ=M, \KK=M$, we first consider a specific case when $\QQ=M^{(1)}, \KK=M^{(2)}$. The set $S_{\QQ,\KK}=\{g\KK|g\in \mathcal{G}_{\QQ}\}=\{M^{(2)}, M^{(3)}\}$, generated by $\mathcal{G}_{\QQ}$ (see definition in \cref{app:sec:epc_selection_rule}). 
The particle-hole operators set is $O_{\KK=M^{(2)}, \QQ=M^{(1)}}=\{\gamma_{M^{(2)},n}^\dagger \gamma_{M^{(3)}, n}, \gamma_{M^{(3)},n}^\dagger \gamma_{M^{(2)}, n}\}$, where $n$ denote the $\pi$ bnad that cross $E_f$ near M. 
Assume the particle-hole operators form the representation $D$ under $g\in \mathcal{G}_{\QQ}$, and we need to find the representation matrix. To do so, we separate the operations $g\in \mathcal{G}_{\QQ}$ into the following two sets. 
For an operation $g$ in the intersection set $\cap_{i=1,2,3} \mathcal{G}_{M^{(i)}}=\{E,C_{2z},P,M_z\}$, we have
\begin{equation}\begin{aligned}
\overline{D}(g)=M_3^+(g)\otimes M_3^+(g)=M_1^+(g)=I_2
\end{aligned}\end{equation}
where $I_2$ denotes the identity matrix. 
For $g\in\mathcal{G}_{\QQ=M^{(1)}}$ but $g \notin \cap_{i=2,3} \mathcal{G}_{M^{(i)}}$, i.e., $g\in\{C_{010},C_{210}, M_{010}, M_{210}\}$, $g$ flips the two particle hole operators, and we have
\begin{equation}\begin{aligned}
\overline{D}(g)=
\begin{bmatrix}
    0 & e^{i \phi_g} \\
    e^{-i\phi_g} & 0
\end{bmatrix}
\end{aligned}\end{equation}
where $\phi_g$ is a phase depending on the basis. Then it is straightforward (using the character table) to obtain the decomposed representation $\overline{D} \sim M_1^+\oplus M_2^+$, where $\sim$ denotes equivalent representations. As a result, phonons with $M_1^+, M_2^+$ IRREPs are allowed.

We then consider $\KK=\QQ=M^{(1)}$, then $S_{\QQ,\KK}=\{M^{(1)}\}$, and $O_{\KK,\QQ}=\{\cre{\gamma}{\Gamma,n}\des{\gamma}{M^{(1)},m}\}$. To obtain which phonon IRREPs are allowed to couple to the FSs, we only need to consider operations in $\mathcal{G}_{\QQ=M^{(1)}}$. The result is the same as in the case of $\QQ=M, \KK=\Gamma$, \ie, only $M_3^+$ and $M_4^+$ are allowed (which have opposite $C_{2z}$ eigenvalues compared with $M_1^+$ and $M_2^+$). 

\end{itemize}

The symmetry analysis above shows that only symmetry-allowed EPCs can exist. In practice, the strength of the symmetry-allowed EPCs is determined by the specific electron, phonon, and their coupling properties of the system, as we discuss in the following two sections. We also note that, in general, Fermi surfaces are located at generic momenta where symmetry imposes little constraint on the EPC—particularly in the case of large Fermi surfaces, where one must examine the specific form of the band-basis EPC tensor.

\subsubsection{EPC strength}\label{app:sec:EPC_strength}

In \cref{app:sec:EPC_symm_constraints}, we use symmetry analysis to show that the $\sigma$ FS is only allowed to couple to the $xy$-phonon of boron with $\Gamma_5^+$ IRREP at $\Gamma$. However, the strength of the EPC cannot be known from symmetry. In \cref{app:sec:spxpy2sp2_basis}, by transforming the boron $(s,p_x,p_y)$ basis to the $sp^2$-bonding/anti-bonding basis, we showed that this EPC is strong due to the strong $sp^2$ bonding. 

Except for the dominant EPC $\lambda_{\qq\nu}$ along the $\Gamma$-$A$ line given by the bond-stretching mode of boron, there is also a weak EPC given by the $z$-phonons of boron that form a Dirac crossing at K and von Hove singularity at M. In the following, we discuss why the EPC of the $z$-phonon of boron is weak.

From the definition of $\lambda_{\qq\nu}$ in \cref{app:eq:lambda_qv_def}, we can see that it is given by the product of the EPC strength and the FS nesting function
\begin{equation}
    \xi(\qq) = \frac{2}{N_{\kk}}\sum_{\kk, m,n} \delta(\epsilon_{\kk,n}-E_f) \delta(\epsilon_{\kk+\qq,m}-E_f).
\end{equation}
In \cref{Fig: MgB2-nesting}, the FS nesting function shows dominant peaks near $\Gamma$, and weak peaks near both K and M on different $q_z$ planes. The nesting peaks near K and M have two contributions: one is the nesting inside the $\pi$ FS, and the other is the nesting between the $\sigma$ and $\pi$ FS. We then consider the EPC from these two types of FSs to the $z$-phonon of boron:
\begin{itemize}
\item Within the $\pi$ FS, we observe that the direct hoppings between the $p_z$ orbitals of boron from different layers are very weak ($<0.1$ eV) due to the large interlayer distance. The $z$-directional coupling between the $p_z$ orbitals of boron is mainly given by the assistant hopping from the $s$ orbital of Mg. However, since the orbital weight from the $s$ orbital of Mg is almost zero close to $E_f$, they cannot give a strong EPC to the $z$-phonon of boron. From DFT, the EPC between $p_z@B_1, s@Mg$ and $B_1$ $z$ phonon is $-1.56$ eV/\AA, which is relatively weak compared with the EPC from the bond-stretching modes, which is about 7 eV/\AA. 

\item For the nesting between the $\sigma$ and $\pi$ FSs, we observe that the direct hopping between the boron $p_z$ orbital and the effective $s$ orbitals (i.e., the $sp^2$ bonding states) is forbidden by the $M_z$ symmetry because they are on the same plane but have opposite $M_z$ eigenvalues. However, the EPC between the $p_z$ and $s$ orbitals to the $z$-phonon of boron is not forbidden by $M_z$ (as the $z$ phonon breaks $M_z$). Nonetheless, this EPC is very small, i.e., $0.57$ eV/\AA, as the out-of-plane $p_z$ orbitals have small overlaps with the in-plane effective $s$ orbitals. 
\end{itemize}
Thus, we conclude that the EPC on the $z$-phonon of boron is weak.

\subsection{EPC Hamiltonian in \ch{MgB2}}

\subsubsection{EPC in $3s@6m$ orbital basis}\label{app:sec:EPC_3s_basis}

To obtain the \textit{ab initio} hopping parameters and the real-space EPC tensor, we perform DFT on \ch{MgB2} and build a faithful 8-orbital Wannier TB model from the $s,p$ orbitals of two boron atoms. The hopping parameters and real-space EPC strength are extracted in this basis. The dominant EPC matrix elements in the $s,p$ basis are tabulated in \cref{app:table:epc_DFT_spxpy} and \cref{app:table:epc_DFT_spxpy_onsite}. We label in red the EPC terms that violate the two-center form (see definition in \cref{app:sec:two-center-approx}), \ie, which require $g^{i\alpha j\beta,l\mu}_{\RR_e,\RR_p}= 0$, for $\RR_p \neq \bm{0} \text{ and } \RR_p\neq \RR_e$. Physically, these two-center-breaking terms arise from assisted hoppings that go beyond the two-center approximation for electron hoppings.

\begin{table}[htbp]
\centering
\begin{tabular}{c|c|c|c|c|c|c|c|c|c}
\hline\hline
EPC (eV/\AA) 
 & $(B_s^1, B_s^2)$ & 
 {\color{red}$(B_s^1, B_{p_x}^2)$} & 
 {\color{red}$(B_s^1, B_{p_y}^2)$} & 
 {\color{red}$(B_{p_x}^1, B_{s}^2)$} & 
 $(B_{p_x}^1, B_{p_x}^2)$ & 
 {\color{red}$(B_{p_x}^1, B_{p_y}^2)$} & 
 {\color{red}$(B_{p_y}^1, B_{s}^2)$} & 
 {\color{red}$(B_{p_y}^1, B_{p_x}^2)$} & $(B_{p_y}^1, B_{p_y}^2)$ \\ \hline
 $B_{x}^1$ & 
-1.77 &  1.18 & -1.58 & -3.85 &  3.37 & -4.15 &  2.16 & -3.21 &  0.43 \\ \hline 
$B_{y}^1$ & 
1.02 & -1.58 & -0.64 &  2.16 & -3.08 &  0.66 & -1.35 &  2.28 &  0.89 \\ \hline 
$B_{x}^2$ & 1.77 & -3.85 &  2.16 &  1.18 & -3.37 &  3.21 & -1.58 &  4.15 & -0.43 \\ \hline
$B_{y}^2$ & -1.02 &  2.16 & -1.35 & -1.58 &  3.08 & -2.28 & -0.64 & -0.66 & -0.89 \\
 \hline\hline
\end{tabular}
\caption{\label{app:table:epc_DFT_spxpy} The \textit{ab initio} real-space EPC, obtained from an 8-orbital Wannier model of Boron $s,p$ orbitals. We only tabulate the intra-home-unit-cell NN EPC terms (\ie, $\RR_e=\RR_p=\bm{0}$), where $B^1$ and $B^2$ denotes the two boron atoms at two honeycomb sites $(\frac{1}{3},\frac{2}{3})$ and $(\frac{2}{3},\frac{1}{3})$, respectively, as shown in \cref{Fig: sp2-basis}. The block with the electron operator $(B^1_j, B^2_j)$ in the column and phonon operator $B^l_\mu$ denotes the EPC element $g^{ij,l\mu}_{\bm{0},\bm{0}}$. 
We mark in red for EPC columns that break the two-center form (see \cref{app:sec:two-center-approx}), where the EPC terms to $B_{\mu}^1$ phonon are not the opposite of the EPC to $B_{\mu}^2$ phonon.}
\end{table}

\begin{table}[htbp]
\centering
\begin{tabular}{c|c|c|c|c|c|c|c|c|c}
\hline\hline
EPC (eV/\AA) & 
 {\color{red}$(B_s^1, B_s^1)$} & 
 {\color{red}$(B_s^1, B_{p_x}^1)$} & 
 {\color{red}$(B_s^1, B_{p_y}^1)$} & 
 {\color{red}$(B_{p_x}^1, B_{s}^1)$} & 
 {\color{red}$(B_{p_x}^1, B_{p_x}^1)$} & 
 {\color{red}$(B_{p_x}^1, B_{p_y}^1)$} & 
 {\color{red}$(B_{p_y}^1, B_{s}^1)$} & 
 {\color{red}$(B_{p_y}^1, B_{p_x}^1)$} & {\color{red}$(B_{p_y}^1, B_{p_y}^1)$} 
 \\ \hline 
 $B_{x}^1$ & -0.00 &  4.85 &  0.00 &  4.85 &  0.00 & -5.00 &  0.00 & -5.00 &  0.00 \\ \hline
$B_{y}^1$ & 
  0.00 &  0.00 &  4.85 &  0.00 & -5.00 &  0.00 &  4.85 &  0.00 &  5.00 \\ 
\hline
&{\color{red}$(B_s^2, B_s^2)$} & 
 {\color{red}$(B_s^2, B_{p_x}^2)$} & 
 {\color{red}$(B_s^2, B_{p_y}^2)$} & 
 {\color{red}$(B_{p_x}^2, B_{s}^2)$} & 
 {\color{red}$(B_{p_x}^2, B_{p_x}^2)$} & 
 {\color{red}$(B_{p_x}^2, B_{p_y}^2)$} & 
 {\color{red}$(B_{p_y}^2, B_{s}^2)$} & 
 {\color{red}$(B_{p_y}^2, B_{p_x}^2)$} & {\color{red}$(B_{p_y}^2, B_{p_y}^2)$} 
 \\ \hline 
$B_{x}^2$ & 
0.00 &  4.85 &  0.00 &  4.85 & -0.00 &  5.00 &  0.00 &  5.00 & -0.00 \\ \hline 
$B_{y}^2$ &
-0.00 &  0.00 &  4.85 &  0.00 &  5.00 & -0.00 &  4.85 & -0.00 & -5.00 \\ 
 \hline\hline
\end{tabular}
\caption{\label{app:table:epc_DFT_spxpy_onsite} 
The \textit{ab initio} electron onsite-type EPC in \ch{MgB2} in the $(s, p_x, p_y)$ orbital basis. 
The notation is the same as in \cref{app:table:epc_DFT_spxpy}. 
The electron onsite-type EPC term has the two electron operators coming from the same atom, which are beyond the two-center approximation. Thus, we mark all columns in red. We remark that the B $p_z$ orbital has almost zero onsite EPCs (similarly in graphene). 
}
\end{table}

We transform the EPC from the boron atomic $s,p$ basis into the $3s@6m$ orbital basis (see definition in \cref{app:eq:S_spxpy_to_sp2} in \cref{app:sec:spxpy2sp2_basis}). 
The corresponding NN EPC terms are tabulated in \cref{app:table:epc_DFT_3s}. 
We observe that the dominant EPC is the coupling along the $sp^2$ bond, with electron operator $(B_{s_2}^1, B_{s_2}^2)$ (\ie, the 6th column in \cref{app:table:epc_DFT_3s}). Although there exist EPC terms that break the two-center form, the dominant bond EPC terms satisfy the two-center form, enforced by the inversion symmetry. 

Moreover, there exist large electron onsite-type EPC terms in \textit{ab initio}, as tabulated in \cref{app:table:epc_DFT_3s_onsite}. These onsite-type EPC terms have the two-electron operators coming from the same atom. 
Explicitly, we give the expression for the onsite-type and NN bond EPC in the $3s@6m$ basis, using the EPC in the atomic $s,p$ basis, based on the transformation matrix defined in \cref{app:eq:S_spxpy_to_sp2}:
\begin{equation}\begin{aligned}
\text{Onsite EPC: } g^{B^1_{s_2},B^1_{s_2}} = \frac{1}{6}(
& 2 g^{B^1_{s}, B^1_{s}} + 
\sqrt{6} g^{B^1_{s}, B^1_{p_x}} 
- \sqrt{2} g^{B^1_{s}, B^1_{p_y}} \\
& +\sqrt{6} g^{B^1_{p_x}, B^1_{s}}
+ 3 g^{B^1_{p_x}, B^1_{p_x}} - \sqrt{3} g^{B^1_{p_x}, B^1_{p_y}} \\
& -\sqrt{2} g^{B^1_{p_y}, B^1_{s}} 
-\sqrt{3} g^{B^1_{p_y}, B^1_{p_x}} + g^{B^1_{p_y}, B^1_{p_y}} ),
\\
\text{Bond EPC: } g^{B^1_{s_2},B^2_{s_2}} = \frac{1}{6}(
& 2 g^{B^1_{s}, B^2_{s}} - 
\sqrt{6} g^{B^1_{s}, B^2_{p_x}} 
+ \sqrt{2} g^{B^1_{s}, B^2_{p_y}} \\
& +\sqrt{6} g^{B^1_{p_x}, B^2_{s}}
- 3 g^{B^1_{p_x}, B^2_{p_x}} + \sqrt{3} g^{B^1_{p_x}, B^2_{p_y}} \\
& -\sqrt{2} g^{B^1_{p_y}, B^2_{s}} 
+\sqrt{3} g^{B^1_{p_y}, B^2_{p_x}} - g^{B^1_{p_y}, B^2_{p_y}} ), 
\label{app:eq:g_sp2_basis_transformation}
\end{aligned}\end{equation}
where we omitted the phonon $\mu$ and $\RR_e=\bm{0},\RR_p=\bm{0}$ indices as they take the same value on both the left- and right-hand sites. The first term in \cref{app:eq:g_sp2_basis_transformation} is the electron onsite-type EPC (linear combination of onsite EPC terms in the atomic $s,p$ basis), while the second term is the NN bond EPC (linear combination of NN EPC terms between $B_1$ and $B_2$ in the atomic $s,p$ basis). 
We observe that the onsite EPC terms in the $3s@6m$ basis are beyond the two-center approximation, as they come from the linear combination of onsite EPC terms in the $s,p$ basis (which are all beyond the two-center approximation). 
We also observe that the dominant onsite-type EPC terms are from electron pair $(B_{s_j}^i, B_{s_j}^i)$ coupled to $B_{\mu}^i$ (\ie, the 2st, 6th, and the last columns in \cref{app:table:epc_DFT_3s_onsite}), which have similar magnitudes as the NN bond EPC tabulated in \cref{app:table:epc_DFT_3s}.

\begin{table}[htbp]
\centering
\begin{tabular}{c|c|c|c|c|c|c|c|c|c}
\hline\hline
EPC (eV/\AA) 
 & $(B_{s_1}^1, B_{s_1}^2)$ & 
 {\color{red}$(B_{s_1}^1, B_{s_2}^2)$} & 
 {\color{red}$(B_{s_1}^1, B_{s_3}^2)$} & 
 {\color{red}$(B_{s_2}^1, B_{s_1}^2)$} & 
 $(B_{s_2}^1, B_{s_2}^2)$ & 
 {\color{red}$(B_{s_2}^1, B_{s_3}^2)$} & 
 {\color{red}$(B_{s_3}^1, B_{s_1}^2)$} & 
 {\color{red}$(B_{s_3}^1, B_{s_2}^2)$} & 
 $(B_{s_3}^1, B_{s_3}^2)$ \\ \hline
 $B_{x}^1$ & 
0.95 &  1.50 & -1.03 & -1.22 & -7.41 &  0.61 & -1.17 &  1.58 &  0.88 \\ \hline
$B_{y}^1$ & 
-0.47 & -0.95 &  0.75 & -1.41 &  4.28 &  1.76 &  0.52 & -0.82 & -0.59 \\ \hline
$B_{x}^2$ & 
-0.95 &  1.22 &  1.17 & -1.50 &  7.41 & -1.58 &  1.03 & -0.61 & -0.88 \\ \hline 
$B_{y}^2$ &
0.47 &  1.41 & -0.52 &  0.95 & -4.28 &  0.82 & -0.75 & -1.76 &  0.59 \\ 
 \hline\hline
\end{tabular}
\caption{\label{app:table:epc_DFT_3s} 
The \textit{ab initio} real-space EPC, obtained from an 8-orbital Wannier model of Boron $s,p$ orbitals, transformed into the $3s@6m$ orbital basis (note that the two $p_z$ orbitals are ignored for simplicity),
as defined in \cref{app:sec:spxpy2sp2_basis}. We only tabulate the intra-home-unit-cell NN EPC terms (\ie, $\RR_e=\RR_p=\bm{0}$), where $B^1$ and $B^2$ denotes the two boron atoms at two honeycomb sites $(\frac{1}{3},\frac{2}{3})$ and $(\frac{2}{3},\frac{1}{3})$, respectively, as shown in \cref{Fig: sp2-basis}. The block with the electron operator $(B^1_j, B^2_j)$ in the column and phonon operator $B^l_\mu$ denotes the EPC element $g^{ij,l\mu}_{\bm{0},\bm{0}}$. 
We mark in red for EPC columns that break the two-center approximation (see \cref{app:sec:two-center-approx}). 
The dominant EPC is the EPC along the $sp^2$ bond, with electron operator $(B_{s_2}^1, B_{s_2}^2)$, which satisfies two-center form. 
}
\end{table}

\begin{table}[htbp]
\centering
\begin{tabular}{c|c|c|c|c|c|c|c|c|c}
\hline\hline
EPC (eV/\AA) & 
 {\color{red}$(B_{s_1}^1, B_{s_1}^1)$} & 
 {\color{red}$(B_{s_1}^1, B_{s_2}^1)$} & 
 {\color{red}$(B_{s_1}^1, B_{s_3}^1)$} & 
 {\color{red}$(B_{s_2}^1, B_{s_1}^1)$} & 
 {\color{red}$(B_{s_2}^1, B_{s_2}^1)$} & 
 {\color{red}$(B_{s_2}^1, B_{s_3}^1)$} & 
 {\color{red}$(B_{s_3}^1, B_{s_1}^1)$} & 
 {\color{red}$(B_{s_3}^1, B_{s_2}^1)$} & 
 {\color{red}$(B_{s_3}^1, B_{s_3}^1)$} \\ \hline
 $B_{x}^1$ & -6.84 & -0.00 &  0.91 & -0.00 &  6.84 & -0.91 &  0.91 & -0.91 &  0.00 \\ \hline
$B_{y}^1$ & 
 -3.95 &  1.05 & -0.52 &  1.05 & -3.95 & -0.52 & -0.52 & -0.52 &  7.90 \\ 
\hline
&{\color{red}$(B_{s_1}^2, B_{s_1}^2)$} & 
 {\color{red}$(B_{s_1}^2, B_{s_2}^2)$} & 
 {\color{red}$(B_{s_1}^2, B_{s_3}^2)$} & 
 {\color{red}$(B_{s_2}^2, B_{s_1}^2)$} & 
 {\color{red}$(B_{s_2}^2, B_{s_2}^2)$} & 
 {\color{red}$(B_{s_2}^2, B_{s_3}^2)$} & 
 {\color{red}$(B_{s_3}^2, B_{s_1}^2)$} & 
 {\color{red}$(B_{s_3}^2, B_{s_2}^2)$} & 
 {\color{red}$(B_{s_3}^2, B_{s_3}^2)$} \\ \hline
$B_{x}^2$ & 
6.84 &  0.00 & -0.91 &  0.00 & -6.84 &  0.91 & -0.91 &  0.91 & -0.00 \\ \hline 
$B_{y}^2$ &
3.95 & -1.05 &  0.52 & -1.05 &  3.95 &  0.52 &  0.52 &  0.52 & -7.90 \\ 
 \hline\hline
\end{tabular}
\caption{\label{app:table:epc_DFT_3s_onsite} 
The \textit{ab initio} electron onsite-type EPC in \ch{MgB2} in the $3s$ orbital basis. The notation is the same as in \cref{app:table:epc_DFT_3s}. We mark all columns in red as the onsite-type EPC is beyond the two-center approximation. 
The dominant onsite-type EPC terms are from electron pair $(B_{s_j}^i, B_{s_j}^i)$ coupled to $B_{\mu}^i$, which have similar magnitudes as the NN bond EPC tabulated in \cref{app:table:epc_DFT_3s}. 
}
\end{table}

We extract the dominant EPC along the $sp^2$ bond from DFT and transform it into the band basis. We first consider the EPC in the $3s$ orbital basis. 

\begin{figure}[htbp]
    \centering
    \includegraphics[width=0.4\linewidth]{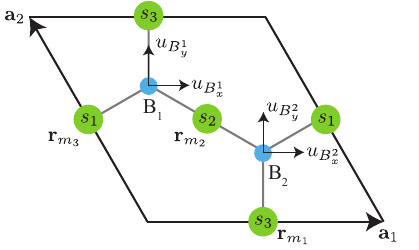}
    \caption{The unit cell with electron and phonon. In the figure, two boron atoms $B_{1,2}$ are placed at two honeycomb sites $\rr_{h_{1,2}}$, with phonon modes $u_{B^{1/2}_{x/y}}$ in global cartesian coordinates. The effective $s$ orbitals from two boron atoms (three $s$ per boron) are placed at three kagome sites $\rr_{m_{1,2,3}}$. Note that, unlike \cref{Fig: sp2-basis} where the effective $s$ orbitals are at non-maximal $6m$ WP, here we manually shift them to kagome sites, so that $B^1_{s_i}$ share the same kagome site as $B^2_{s_i}$, up to a lattice translation. Under the site symmetry group of the kagome site, these two overlapping $s$ orbitals transform as effective $s$ and $p$ orbitals, formed from the bonding and anti-bonding combinations of the two original $s$ orbitals at the non-maximal $6m$ position.}
    \label{app:fig:unitcell_honeycomb_kagome_phonon}
\end{figure}

Following the definition of $3s$-orbital basis in \cref{Fig: sp2-basis}, where $B_1$ and $B_2$ are at two in-plane honeycomb sites $(\frac{1}{3},\frac{2}{3)}$ and $(\frac{2}{3}, \frac{1}{3})$, respectively, we consider the following electron operators:
\begin{equation}\begin{aligned}
\left( c_{B^1_{s_1}, \RR_e}^\dagger, c_{B^1_{s_2}, \RR_e}^\dagger, c_{B^1_{s_3}, \RR_e}^\dagger, c_{B^2_{s_1}, \RR_e}^\dagger, c_{B^2_{s_2}, \RR_e}^\dagger, c_{B^2_{s_3}, \RR_e}^\dagger
\right),
\end{aligned}\end{equation}
and phonon operators:
\begin{equation}\begin{aligned}
\left(u_{B^1_{x}, \RR_p}, u_{B^1_{y}, \RR_p}, u_{B^2_{x}, \RR_p}, u_{B^2_{y}, \RR_p}
\right)
\end{aligned}\end{equation}
We assume the phonon operators are located at honeycomb sites $\mathbf{r}_{h_1}=(\frac{1}{3},\frac{2}{3}), \mathbf{r}_{h_2}=(\frac{2}{3},\frac{1}{3})$, while the electron operators in $3s$ orbital basis are located at kagome sites 
$\mathbf{r}_{m_1}=(\frac{1}{2},0), \mathbf{r}_{m_2}=(\frac{1}{2}, \frac{1}{2}),
\mathbf{r}_{m_3}=(0, \frac{1}{2})$, as shown in \cref{app:fig:unitcell_honeycomb_kagome_phonon}. 

Let $g_0=$ \SI{8.56}{eV/\angstrom}, $g_1=$\SI{7.90}{eV/\angstrom} be the dominant bond and onsite EPC in the $3s@6m$ basis, with $\frac{\sqrt{3}}{2} g_0=g^{B^1_{s_1}, B^2_{s_2}, B^2_x}_{\bm{0}, \bm{0}}$, $g_1=g^{B^1_{s_3}, B^1_{s_3}, B^1_y}_{\bm{0}, \bm{0}}$, where $B^{i}_{x/y}$ denotes the $x/y$ phonon mode of the $i$-th boron atom.  
We transform the real-space EPC matrix elements $g_{\RR_e,\RR_p}^{ijl\mu}$ of operator $c_{\RR,i}^\dagger c_{\RR+\RR_e,j} u_{\RR_p,l\mu}$ to momentum space using \cref{app:eq:epc_FT_formula}, rewritten here for clarity: 
\begin{equation}\begin{aligned} 
g_{\kk,\qq}^{ij,l\mu}= \sum_{\RR_{e},\RR_{p}} g_{\RR_e,\RR_{p}}^{ij,l\mu} e^{i\kk\cdot(\RR_e+\rr_j-\rr_i)+i\qq\cdot(\RR_{p}+\rr_l-\rr_i)}.
\label{app:eq:EPC_FT_eq}
\end{aligned}\end{equation} 
We note that the $\kk$-dependence in EPC comes from the FT factor $e^{i\kk\cdot(\RR_e+\rr_j-\rr_i)}$. When the two electron operators $c_{\RR,i}^\dagger$ and $c_{\RR+\RR_e,j}$ are at the same site, \ie, $\rr_i=\rr_j+\RR_e$, the $\kk$-dependence disappears. 
To achieve this, we intentionally place the effective $s$ orbitals on kagome sites instead of the non-maximal $6m$ position, so that the $B^1_{s_i}$ share the same kagome site as $B^2_{s_i}$ up to a lattice translation. As a result, the bond and onsite EPC terms all come from the electron operators at the same site, and the momentum-space EPC tensor $g_{\kk,\qq}$ is $\kk$-independent. On the contrary, if we use the original $6m$ position, $g_{\kk,\qq}$ will be $\kk$-dependent.

Explicitly, the momentum-space EPC tensor with bond and onsite EPC terms is
\begin{equation}
\small
\begin{aligned}
g_{\kk,\qq}^{ij,B_x^1} &=g_0
\left[
\begin{array}{ccc|ccc}
& & & \frac{\sqrt{3}}{2} e^{i\qq\cdot(\rr_{h_1}-\rr_{m_3})} & & \\
& & & & -\frac{\sqrt{3}}{2}e^{i\qq\cdot(\rr_{h_1}-\rr_{m_2})} & \\
& & & & & 0 \\\hline
* & & & & &  \\
& * & & & &  \\
& & * &  & &
\end{array}\right]_{ij}
+ g_1 e^{i\qq\cdot \rr_{h_1}} \text{Diag}\left[-\frac{\sqrt{3}}{2}e^{-i\qq\cdot\rr_{m_3}}, \frac{\sqrt{3}}{2}e^{-i\qq\cdot\rr_{m_2}}, 0,0,0,0
\right]_{ij}
,\\
g_{\kk,\qq}^{ij,B_x^2} &= g_0
\left[
\begin{array}{ccc|ccc}
& & & -\frac{\sqrt{3}}{2} e^{i\qq\cdot (\rr_{h_2}-\rr_{m_3}-\mathbf{a}_1)} & & \\
& & & & \frac{\sqrt{3}}{2} e^{i\qq\cdot(\rr_{h_2}-\rr_{m_2})} & \\
& & & & & 0 \\\hline
* & & & & &  \\
& * & & & &  \\
& & * &  & &
\end{array}\right]_{ij}
+ g_1 e^{i\qq\cdot\rr_{h_2}} \text{Diag}\left[0,0,0, \frac{\sqrt{3}}{2}e^{-i\qq\cdot(\rr_{m_3}+\mathbf{a}_1)}, -\frac{\sqrt{3}}{2}e^{-i\qq\cdot\rr_{m_2}}, 0
\right]_{ij}
,\\
g_{\kk,\qq}^{ij,B_y^1} &= g_0
\left[
\begin{array}{ccc|ccc}
& & & \frac{1}{2} e^{i\qq\cdot(\rr_{h_1}-\rr_{m_3})}  & & \\
& & & &  \frac{1}{2} e^{i\qq\cdot(\rr_{h_1}-\rr_{m_2})}  & \\
& & & & & -e^{i\qq\cdot(\rr_{h_2}-\rr_{m_1}-\mathbf{a}_2)} \\\hline
* & & & & &  \\
& * & & & &  \\
& & * &  & &
\end{array}\right]_{ij}
+ g_1 e^{i\qq\cdot\rr_1} \text{Diag}\left[-\frac{1}{2}e^{-i\qq\cdot\rr_{m_3}}, -\frac{1}{2}e^{-i\qq\cdot\rr_{m_2}}, e^{-i\qq\cdot(\rr_{m_1}+\mathbf{a}_2)},0,0,0
\right]_{ij}
,\\
g_{\kk,\qq}^{ij,B_y^2} &= g_0
\left[
\begin{array}{ccc|ccc}
& & & -\frac{1}{2} e^{i\qq\cdot(\rr_{h_2}-\rr_{m_3}-\mathbf{a}_1)} 
 & & \\
& & & &  -\frac{1}{2} e^{i\qq\cdot(\rr_{h_2}-\rr_{m_2})} & \\
& & & & &  e^{i\qq\cdot(\rr_{h_2}-\rr_{m_1})} \\\hline
* & & & & &  \\
& * & & & &  \\
& & * &  & &
\end{array}\right]_{ij}
+ g_1 e^{i\qq\cdot\rr_2} \text{Diag}\left[0,0,0, \frac{1}{2}e^{-i\qq\cdot(\rr_{m_3}+\mathbf{a}_1)}, \frac{1}{2}e^{-i\qq\cdot\rr_{m_2}}, -e^{-i\qq\cdot\rr_{m_1}} \right]_{ij}
,
\label{app:eq:EPC_3s_bond_onsite_term_mom_space}
\end{aligned}
\end{equation}
where $*$ denotes transposed complex conjugate matrix elements. Note that there is no $\kk$-dependence in $g_{\kk,\qq}$. 
We take the EPC element $g_{\kk,\qq}^{B^1_{s_1}, B^2_{s_1},B_x^1}$ as an example to show how it is obtained. For this bond EPC term, only one real-space EPC $g_{\RR_e=-\bm{a}_1,\RR_p=\bm{0}}^{B^1_{s_1}, B^2_{s_1},B_x^1}=\frac{\sqrt{3}}{2}g_0$ contributes, with $\rr_i=\rr_{m_3}, \rr_j=\rr_{m_3}+\bm{a}_1, \rr_l=\rr_{h_1}$. As a result, by using the FT in \cref{app:eq:EPC_FT_eq}, we arrive at $g_{\kk,\qq}^{B^1_{s_1}, B^2_{s_1},B_x^1} = \frac{\sqrt{3}}{2}g_0 e^{i\qq\cdot(\rr_{h_1}-\rr_{m_3})}$.

With the momentum space EPC in \cref{app:eq:EPC_3s_bond_onsite_term_mom_space}, we then transform it into the band basis. 
As an approximation, we use the phonon model of boron $xy$ modes defined in \cref{app:eq:H-phonon-Bxy} and consider only the $\Gamma_5^+$ mode at $\qq=\Gamma$, with the wavefunction:
\begin{equation}\begin{aligned}
u_{1,\Gamma}=\frac{1}{\sqrt{2}}(u_{B_x^1,\Gamma} - u_{B_x^2,\Gamma}),\quad
u_{2,\Gamma}=\frac{1}{\sqrt{2}}(u_{B_y^1,\Gamma} - u_{B_y^2,\Gamma}),
\label{app:eq:phonon_GM5+_basis}
\end{aligned}\end{equation}
which are the opposite-directional movements of two boron atoms. $\Gamma_5^+$ phonon mode is a 2D IRREP as $C_3$ and $C_{2x}$ do not commute. 
Note that the phonon $u_{B_\mu^i,\Gamma}$ has odd parity and $u_{B_\mu^1}$ is mapped to $u_{B_\mu^2}$ under inversion, thus wavefunctions in \cref{app:eq:phonon_GM5+_basis} both have even parity. We then transform the EPC into the $\Gamma_5^+$ basis:
\begin{equation}\begin{aligned}
g_{\kk,\Gamma}^{ij,u_1} &= \frac{1}{\sqrt{2}}(g_{\kk,\Gamma}^{ij,B_x^1} - g_{\kk,\Gamma}^{ij,B_x^2})
=\frac{\sqrt{3}}{\sqrt{2}} g_0
\left[
\begin{array}{ccc|ccc}
& & & 1 & & \\
& & & & -1 & \\
& & & & & 0 \\\hline
* & & & & &  \\
& * & & & &  \\
& & * &  & &
\end{array}\right]_{ij}
+ g_1 \text{Diag}\left[-\frac{\sqrt{3}}{2}, \frac{\sqrt{3}}{2}, 0, -\frac{\sqrt{3}}{2}, \frac{\sqrt{3}}{2}, 0
\right]_{ij}
,\\
g_{\kk,\Gamma}^{ij,u_2} &= \frac{1}{\sqrt{2}}(g_{\kk,\Gamma}^{ij,B_y^1} - g_{\kk,\Gamma}^{ij,B_y^2})
=\frac{g_0}{\sqrt{2}}
\left[
\begin{array}{ccc|ccc}
& & & 1 & & \\
& & & & 1 & \\
& & & & & -2 \\\hline
* & & & & &  \\
& * & & & &  \\
& & * &  & &
\end{array}\right]_{ij}
+ g_1 \text{Diag}\left[-\frac{1}{2}, -\frac{1}{2}, 1, -\frac{1}{2}, -\frac{1}{2}, 1 \right]_{ij}.
\end{aligned}\end{equation}

We then transform the electronic operators of $3s@6m$ into the $sp^2$ bonding/anti-bonding states basis at kagome sites to obtain $c^\dagger_{b_i,\kk}$ and $c^\dagger_{ab_i,\kk}$, where $i=1,2,3$ denotes three bonding states, with
\begin{equation}\begin{aligned}
\cre{c}{b_1,\kk} &= \frac{1}{\sqrt{2}}(\cre{c}{B_{s_3}^1,\kk} + \cre{c}{B_{s_3}^2,\kk}),\quad
\cre{c}{b_2,\kk} = \frac{1}{\sqrt{2}}(\cre{c}{B_{s_2}^1,\kk} + \cre{c}{B_{s_2}^2,\kk}),\quad
\cre{c}{b_3,\kk} = \frac{1}{\sqrt{2}}(\cre{c}{B_{s_1}^1,\kk} + \cre{c}{B_{s_1}^2,\kk}),\\
\cre{c}{ab_1,\kk} &= \frac{1}{\sqrt{2}}(\cre{c}{B_{s_3}^1,\kk} - \cre{c}{B_{s_3}^2,\kk}),\quad
\cre{c}{ab_2,\kk} = \frac{1}{\sqrt{2}}(\cre{c}{B_{s_2}^1,\kk} - \cre{c}{B_{s_2}^2,\kk}),\quad
\cre{c}{ab_3,\kk} = \frac{1}{\sqrt{2}}(\cre{c}{B_{s_1}^1,\kk} - \cre{c}{B_{s_1}^2,\kk}).\\
\end{aligned}\end{equation}
The EPC tensor is transformed to the same bonding/anti-bonding state basis, \ie,
\begin{equation}\begin{aligned}
g_{\kk,\Gamma}^{b_i,b_j,u_1} &= \frac{\sqrt{3}}{2}(\sqrt{2}g_0-g_1) 
\text{Diag}\left[0, -1, 1 \right]_{ij}
,\quad
g_{\kk,\Gamma}^{b_i,b_j,u_2}= (\sqrt{2}g_0-g_1)
\text{Diag}\left[-1, \frac{1}{2}, \frac{1}{2}
\right]_{ij},\\
g_{\kk,\Gamma}^{ab_i,ab_j,u_1} &= \frac{\sqrt{3}}{2}(\sqrt{2}g_0+g_1) 
\text{Diag}\left[0, 1, -1 \right]_{ij}
,\quad
g_{\kk,\Gamma}^{ab_i,ab_j,u_2}= (\sqrt{2}g_0 + g_1)
\text{Diag}\left[1, -\frac{1}{2}, -\frac{1}{2}
\right]_{ij}.\\
\label{app:eq:epc_3s_bonding_basis}
\end{aligned}\end{equation}
There are no onsite EPC terms between bonding and anti-bonding states due to the symmetry.

\subsubsection{EPC in $sp^2$ bonding basis}\label{app:sec:EPC_sp2_basis_analytic_model}

In the previous section \cref{app:sec:EPC_3s_basis}, we start from the EPC in the original $3s$-orbital basis (at the kagome sites) and transform it into the $sp^2$ bonding and anti-bonding basis. The final EPC tensor in the bonding state basis \cref{app:eq:epc_3s_bonding_basis} is proportional to $\sqrt{2}g_0 - g_1$, where $g_0\ (g_1)$ is the dominant bond (onsite) EPC term in the $3s$ basis as defined in \cref{app:eq:EPC_3s_bond_onsite_term_mom_space}. In the following, we start directly from the $sp^2$ bonding basis and relate the resultant EPC to \cref{app:eq:epc_3s_bonding_basis} from the $3s@6m$ basis. We also use the wavefunction from the NN kagome model to obtain the EPC in the band basis. 

\begin{figure}[htbp]
    \centering
    \includegraphics[width=0.5\linewidth]{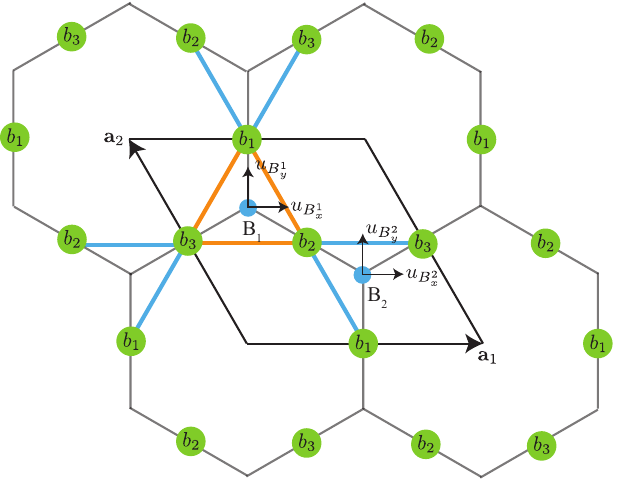}
    \caption{The unit cell with two boron atoms $B_{1,2}$ placed at two honeycomb sites $\rr_{h_{1,2}}$ with phonon modes $u_{B^{1/2}_{x/y}}$, and three effective $sp^2$ bonding orbitals $b_{i=1,2,3}$ placed at three kagome sites $\rr_{m_{1,2,3}}$. The three orange bonds denote the NN EPC terms that couple to the closest boron phonon modes, while the six blue bonds denote the NN EPC terms that couple to the NNN boron atom, as discussed in \cref{app:eq:EPC_ham_3H}. }
    \label{app:fig:unit-cell-sp2-bonding-epc}
\end{figure}

The electron operators in the $sp^2$ bonding basis are
\begin{equation}\begin{aligned}
\left(\cre{c}{b_1,\RR_e},\cre{c}{b_2,\RR_e},\cre{c}{b_3,\RR_e} \right),
\end{aligned}\end{equation}
located at three kagome sites $\mathbf{t}_{k_1}=(\frac{1}{2},0), \mathbf{t}_{k_2}=(\frac{1}{2}, \frac{1}{2}),
\mathbf{t}_{k_3}=(0, \frac{1}{2})$ (see \cref{app:fig:unit-cell-sp2-bonding-epc}). 
In this basis, we only consider the onsite EPC between two electron operators, which is the dominant term (the NN EPC terms in this basis arise from couplings involving more than two boron atoms and have smaller magnitudes). From DFT, $g_0^b=4.43 eV/$\AA. We find the momentum space EPC matrix elements read
\begin{equation}\begin{aligned}
g_{\kk,\qq}^{b_i,b_i,B_x^1} &= g_0^b
\text{Diag}\left[0, 
-\frac{\sqrt{3}}{2}e^{i\qq\cdot(\rr_{h_1}-\rr_{m_2})}, 
\frac{\sqrt{3}}{2}e^{i\qq\cdot(\rr_{h_1}-\rr_{m_3})} \right]_{ii},\\
g_{\kk,\qq}^{b_i,b_i,B_y^1} &= g_0^b
\text{Diag} \left[
-e^{i\qq\cdot(\rr_{h_1}-\rr_{m_1}-\mathbf{a}_2)}, \frac{1}{2}e^{i\qq\cdot(\rr_{h_1}-\rr_{m_2})}, 
\frac{1}{2}e^{i\qq\cdot(\rr_{h_1}-\rr_{m_3})} \right]_{ii},\\
g_{\kk,\qq}^{b_i,b_i,B_x^2} &= -g_0^b
\text{Diag}\left[0, 
-\frac{\sqrt{3}}{2}e^{i\qq\cdot(\rr_{h_2}-\rr_{m_2})}, 
\frac{\sqrt{3}}{2} e^{i\qq\cdot(\rr_{h_2}-\rr_{m_3}-\mathbf{a}_1))} \right]_{ii},\\
g_{\kk,\qq}^{b_i,b_i,B_y^2} &= -g_0^b
\text{Diag} \left[
-e^{i\qq\cdot(\rr_{h_2}-\rr_{m_1}-\mathbf{a}_2)}, \frac{1}{2}e^{i\qq\cdot(\rr_{h_2}-\rr_{m_2})}, 
\frac{1}{2}e^{i\qq\cdot(\rr_{h_2}-\rr_{m_3}-\mathbf{a}_1)} \right]_{ii},
\label{app:eq:epc_sp2_bonding_ham}
\end{aligned}\end{equation}
We then transform to the phonon $\Gamma_5^+$ basis defined in \cref{app:eq:phonon_GM5+_basis}, \ie,
\begin{equation}\begin{aligned}
g_{\kk,\Gamma}^{b_i,b_i,u_1} &=  \sqrt{2}g_0^b \frac{\sqrt{3}}{2} \text{Diag}\left[0, -1, 1 \right]_{ii}
,\quad
g_{\kk,\Gamma}^{b_i,b_i,u_2}= \sqrt{2} g_0^b
\text{Diag}\left[-1, \frac{1}{2}, \frac{1}{2}
\right]_{ii}.
\label{app:eq:epc_momentum_sp2_bonding}
\end{aligned}\end{equation}
Comparing with \cref{app:eq:epc_3s_bonding_basis}, we find that the EPC for the bonding states is the same if $\sqrt{2}g_0^b=\sqrt{2}g_0-g_1$. Numerically, $\sqrt{2}g_0^b\approx 6.26$ eV/\AA, $\sqrt{2}g_0-g_1\approx 4.21$ eV/\AA. This mismatch arises from the presence of the anti-bonding states in the $3s$ basis. 
In the $3s$ description, there are 6 orbitals in total: three bonding and three anti-bonding combinations. In contrast, the $sp^2$ basis contains only the three bonding states. When we project from the full $3s$ space onto the bonding subspace (i.e., integrate out the anti-bonding manifold), an enhanced effective deformation potential for the bonding bands is obtained, as seen from the \textit{ab initio} values. 
As a result, the EPC parameter extracted directly in the $sp^2$ bonding basis, $g_0^b$, is larger than the naive combination $\sqrt{2} g_0 - g_1$ obtained from the $3s$ basis without accounting for these anti-bonding contributions.

More generally, we consider both onsite EPC and NN bond EPC in the $sp^2$ bonding basis at kagome sites, \ie, $\mathbf{r}_{m_1}=(\frac{1}{2},0), \mathbf{r}_{m_2}=(\frac{1}{2}, \frac{1}{2}),
\mathbf{r}_{m_3}=(0, \frac{1}{2})$, with boron atoms at $\mathbf{r}_{h_1}=(\frac{1}{3},\frac{2}{3}), \mathbf{r}_{h_2}=(\frac{2}{3},\frac{1}{3})$, as shown in \cref{app:fig:unitcell_honeycomb_kagome_phonon}, where the coordinates are written under the hexagonal basis \cref{app:eq:MgB2_unit_cell_basis}. Let $\cre{c}{\RR,a}=\cre{c}{\RR+\rr_{m_a}}, u_{\RR,s}^\mu=u_{\RR+\rr_{h_s}}^\mu$. 

We consider the SG 191 symmetries to constrain the symmetry-allowed EPC terms, with the following symmetry representations for electron and phonon operators in real space, \ie, 
$\left(\cre{c}{\RR+\rr_{m_1}}, \cre{c}{\RR+\rr_{m_2}}, \cre{c}{\RR+\rr_{m_3}}\right)$, and 
$\left(u_{\RR+\rr_{h_1}}^x, u_{\RR+\rr_{h_1}}^y, 
u_{\RR+\rr_{h_2}}^x,
u_{\RR+\rr_{h_2}}^y\right)$, following the convention defined in \cref{app:sec:symmetry_constraints}: 
\begin{equation}\begin{aligned}
    D^{el}(C_6)&=\begin{bmatrix}
        0 & 1 & 0 \\
        0 & 0 & 1 \\
        1 & 0 & 0 \\
    \end{bmatrix},\quad
    D^{el}(P) = \bm{1}_3,\quad
    D^{el}(C_{2y}) = \begin{bmatrix}
        1 & 0 & 0 \\
        0 & 0 & 1 \\
        0 & 1 & 0 \\
    \end{bmatrix},\quad
    D^{el}(T)=\mathcal{K}\bm{1}_3,
    \\
    D^{ph}(C_6)&=\begin{bmatrix}
        0 & 0 & \frac{1}{2} & -\frac{\sqrt{3}}{2} \\
        0 & 0 & \frac{\sqrt{3}}{2} & \frac{1}{2} \\
        \frac{1}{2} & -\frac{\sqrt{3}}{2} & 0 & 0 \\
        \frac{\sqrt{3}}{2} & \frac{1}{2} & 0 & 0
    \end{bmatrix},\quad
    D^{ph}(P) = \begin{bmatrix}
        0 & 0 & -1 & 0 \\
        0 & 0 & 0 & -1 \\
        -1 & 0 & 0 & 0 \\
        0 & -1 & 0 & 0 \\
    \end{bmatrix},\quad
    D^{ph}(C_{2y}) = \text{Diag}[-1, 1, -1, 1],\quad
    D^{ph}(T)=\mathcal{K}\bm{1}_4\\
\end{aligned}\end{equation}
Let $\omega=e^{i\frac{2\pi}{3}}$, denote $C_{3, h_s}$ as the local $C_3$ rotation centered at $\rr_{h_s}$ site, \ie, $C_{3, h_1}=\{C_3|\bm{a}_1+\bm{a}_2\},\ 
C_{3, h_2}=\{C_3|\bm{a}_1\}$, and define the following phonon $C_3$-eigen operators:
\begin{equation}\begin{aligned}
    u_{s,\pm}(\RR) &= \frac{1}{\sqrt{2}} (u_{\RR,s}^x \pm i u_{\RR,s}^y), \quad
    D(C_{3, h_s})=\omega^{\mp 1},
\end{aligned}\end{equation}
Similarly, for other symmetries, we have
$P: u_{s,\pm}(\RR) \rightarrow -u_{\overline{s},\pm}(-\RR-\rr_s-\rr_{\overline{s}})$, 
$T: u_{s,\pm}(\RR) \rightarrow -u_{{s},\mp}(\RR)$,  
$C_{2y}: u_{s,\pm}(\RR) \rightarrow -u_{{s},\mp}(C_{2y}(\RR+\rr_{s})-\rr_{\overline{s}})$, where $\overline{s}$ denotes the opposite honeycomb site. 

For the electron density operators on kagome sites $\rr_{m_i}$ (assume no SOC and omit the spin index), define the following local $C_3$ eigen basis
\begin{equation}
\begin{aligned}
    N_{s=1,\pm}(\RR) &= n_{\RR+\bm{a}_2,1} + \omega^{\pm 1} n_{\RR,2} + \omega^{\mp 1} n_{\RR, 3}, \quad
    D(C_{3, h_1})=\omega^{\pm 1} \\
    N_{s=2,\pm}(\RR) &= n_{\RR,1} + \omega^{\pm 1} n_{\RR,2} + \omega^{\mp 1} n_{\RR+\bm{a}_1, 3},\quad
    D(C_{3, h_2})=\omega^{\pm 1},
    \label{app:eq:epc_model_N_def}
\end{aligned}
\end{equation}
where $\bm{a}_i$ are added so that the three density operators are centered at the same boron site. The local $C_3$ eigenvalues of $N_{s,\pm}(\RR)$ result from the fact that local $C_3$ permutes the three kagome sites $\rr_{m_1}\rightarrow \rr_{m_3}\rightarrow \rr_{m_2}\rightarrow \rr_{m_1}$, as can be seen from \cref{app:fig:unit-cell-sp2-bonding-epc}. For example, $C_{3, h_1} N_{s=1,\pm}(\bm{0}) C_{3, h_1}^{-1}= n_{\bm{0}, 3} + \omega^{\pm 1} n_{\bm{a}_2, 1} + \omega^{\mp 1} n_{\bm{0}, 2}=\omega^{\pm 1} N_{s=1,\pm}(\bm{0})$. For other symmetries, we have
$P: N_{s,\pm}(\RR) \rightarrow N_{\overline{s},\pm}(-\RR-\rr_s-\rr_{\overline{s}})$, 
$T: N_{s,\pm}(\RR) \rightarrow N_{{s},\mp}(\RR)$, 
$C_{2y}: N_{s,\pm}(\RR) \rightarrow N_{{s},\mp}(C_{2y}(\RR+\rr_{s})-\rr_{\overline{s}})$. Note that in \cref{app:eq:epc_model_N_def}, we set $n_{\RR,1}$ terms with no $\omega$ phases, so that $C_{2y}$ representation has no additional phases, as $C_{2y}$ leaves the site $r_{m_1}$ unchanged while switches $r_{m_2}, r_{m_3}$.

We then consider the NN bond EPC terms,\ eg, the electrons forming the orange bonds that couple to $B_1$, as shown in \cref{app:fig:unit-cell-sp2-bonding-epc}. Denote the three NN kagome bond displacement vectors as 
$\bm{\delta}_{12}=\rr_{m_2}-\rr_{m_1}=(0,\frac{1}{2}),\ \bm{\delta}_{23}=\rr_{m_3}-\rr_{m_2}=(-\frac{1}{2}, 0),\ \bm{\delta}_{31}=\rr_{m_1}+\bm{a}_2-\rr_{m_3}=(\frac{1}{2},\frac{1}{2})$. Define the NN bond operators at two honeycomb sites as
\begin{equation}\begin{aligned}
    B_{s=1,23}(\RR) &= \cre{c}{\RR,2} \des{c}{\RR,3} + h.c.,\quad
    B_{s=1,31}(\RR) = \cre{c}{\RR,3} \des{c}{\RR+\bm{a}_2, 1} + h.c.,\quad 
    B_{s=1,12}(\RR) = \cre{c}{\RR+\bm{a}_2,1} \des{c}{\RR,2} + h.c.,\\
    B_{s=2,23}(\RR) &= \cre{c}{\RR,2} \des{c}{\RR+\bm{a}_1,3} + h.c.,\quad
    B_{s=2,31}(\RR) = \cre{c}{\RR+\bm{a}_1,3} \des{c}{\RR, 1} + h.c.,\quad
    B_{s=2,12}(\RR) = \cre{c}{\RR,1} \des{c}{\RR,2} + h.c..
\end{aligned}\end{equation}
The local $C_3$ eigen bases formed by the NN bond operators are
\begin{equation}\begin{aligned}
    B_{s=1,\pm}(\RR) &= B_{1,23}(\RR) + \omega^{\pm 1} B_{1,31}(\RR) + \omega^{\mp 1} B_{1, 12}(\RR), \quad
    D(C_{3, h_1})=\omega^{\pm 1} \\
    B_{s=2,\pm}(\RR) &= B_{2,23}(\RR) + \omega^{\pm 1} B_{2,31}(\RR) + \omega^{\mp 1} B_{2, 12}(\RR), \quad
    D(C_{3, h_2})=\omega^{\pm 1}. 
    \label{app:eq:epc_model_B_def}
\end{aligned}\end{equation}
Note that the local $C_3$ permutes the NN bond operators by $B_{s,12}\rightarrow B_{s,31} \rightarrow B_{s,23} \rightarrow B_{s,12}$. For example, $C_{3, h_1} B_{s,\pm}(\bm{0}) C_{3, h_1}^{-1}= B_{1,31}(\bm{0}) + \omega^{\pm 1} B_{1,12}(\bm{0}) + \omega^{\mp 1} B_{1,23}(\bm{0})=\omega^{\pm 1} B_{s,\pm}(\bm{0})$. 
For other symmetries, we have
$P: B_{s,\pm}(\RR) \rightarrow B_{\overline{s},\pm}(-\RR-\rr_s-\rr_{\overline{s}})$, 
$T: B_{s,\pm}(\RR) \rightarrow B_{{s},\mp}(\RR)$, 
$C_{2y}: B_{s,\pm}(\RR) \rightarrow B_{{s},\mp}(C_{2y}(\RR+\rr_{s})-\rr_{\overline{s}})$. 
Note that in \cref{app:eq:epc_model_B_def}, we set $B_{s,23}(\RR)$ terms with no $\omega$ phases, so that $C_{2y}$ representation has no additional phases, as $C_{2y}$ leaves the bond $B_{s,23}(\RR)$ unchanged while switches the two bonds $B_{s,31}(\RR),B_{12}(\RR)$.

We further consider the NN bond EPC terms that couple to the NNN atoms, \eg, the electrons forming the blue bonds that couple to $B_1$, as shown in \cref{app:fig:unit-cell-sp2-bonding-epc}. Denote the corresponding NN bond operators at two honeycomb sites as
\begin{equation}\begin{aligned}
    B^{(2)}_{s=1,\pm}(\RR) &= 
    \cre{c}{\RR-\bm{a}_1,2}\des{c}{\RR,3} + \omega^{\pm} \cre{c}{\RR,1}\des{c}{\RR,2} + \omega^{\mp} \cre{c}{\RR+\bm{a}_1+\bm{a}_2, 3}\des{c}{\RR+\bm{a}_2,1} \\
    &+ \cre{c}{\RR+\bm{a}_1,3}\des{c}{\RR,2} + \omega^{\pm} \cre{c}{\RR+\bm{a}_2, 2}\des{c}{\RR+\bm{a}_2, 1} + \omega^{\mp} \cre{c}{\RR-\bm{a}_1,1}\des{c}{\RR,3} + h.c., 
    \quad
    D(C_{3, h_1})=\omega^{\pm 1} \\
    B^{(2)}_{s=2,\pm}(\RR) &= 
    \cre{c}{\RR,3}\des{c}{\RR,2} + \omega^{\pm} \cre{c}{\RR-\bm{a}_2,2} \des{c}{\RR,1} + \omega^{\mp} \cre{c}{\RR+\bm{a}_1+\bm{a}_2, 1}\des{c}{\RR+\bm{a}_1, 3} \\
    &+ \cre{c}{\RR+\bm{a}_1, 2}\des{c}{\RR+\bm{a}_1, 3} + \omega^{\pm} \cre{c}{\RR+\bm{a}_2, 1}\des{c}{\RR, 2} + \omega^{\mp} \cre{c}{\RR-\bm{a}_2, 3}\des{c}{\RR,1} + h.c., 
    \quad
    D(C_{3, h_1})=\omega^{\pm 1}
\label{app:eq:epc_model_B2_def}
\end{aligned}\end{equation}
For other symmetries, we have
$P: B^{(2)}_{s,\pm}\rightarrow B^{(2)}_{\overline{s},\pm}$, 
$T: B^{(2)}_{s,\pm}\rightarrow B^{(2)}_{{s},\mp}$, 
$C_{2y}: B^{(2)}_{s,\pm}\rightarrow B^{(2)}_{{s},\mp}$.

By using the symmetry constraint \cref{app:eq:real-space-epc-symmetry-constraint} for the EPC tensor, we obtain the following onsite and NN bond EPC terms
\begin{equation}\begin{aligned}
    H_{\text{onsite}} &= \frac{ig^b_0}{\sqrt{2}} \sum_{\RR} \left[
    u_{1,+}(\RR) N_{1,+}(\RR) - u_{1,-} N_{1,-}(\RR) - u_{2,+}(\RR) N_{2,+}(\RR) + u_{2,-} N_{2,-}(\RR)
    \right],  \\
    H_{\text{NN}} &= \frac{ig^b_1}{\sqrt{2}} \sum_{\RR} \left[
    u_{1,+}(\RR) B_{1,+}(\RR) - u_{1,-} B_{1,-}(\RR) - u_{2,+}(\RR) B_{2,+}(\RR) + u_{2,-} B_{2,-}(\RR)
    \right], \\
    H_{\text{NN2}} &= \frac{(g^b_{2r}+ig^b_{2i})}{\sqrt{2}} \sum_{\RR} \left[
    u_{1,+}(\RR) B^{(2)}_{1,+}(\RR) - u_{1,-} B^{(2)}_{1,-}(\RR) - u_{2,+}(\RR) B^{(2)}_{2,+}(\RR) + u_{2,-} B^{(2)}_{2,-}(\RR) 
    \right]. 
    \label{app:eq:EPC_ham_3H}
\end{aligned}\end{equation}
We note that all these onsite and bond EPC terms are expressed in terms of the effective $sp^2$ bonding states at the kagome sites, away from the phonon modes at the honeycomb sites. Consequently, they go beyond the two-center approximation described in \cref{app:sec:two-center-approx}. 
The phonon and electron particle-hole operators are matched so that their $C_3$ eigenvalues cancel, leading to $ C_3$-invariant terms. 
Inversion $P$ enforces the coupling constant to be opposite on two honeycomb sites, while $T$ and $C_{2y}$ enforce the coupling constant in $H_{\text{onsite}}$ and $H_{\text{NN}}$ to be purely imaginary and $u_{s,\pm}$ terms with opposite signs. In $H_{\text{NN2}}$, the coupling constant is complex. 
From \textit{ab initio} in \ch{MgB2}, we have 
\begin{equation}
    g^b_0=4.43, \ g^b_1=0.06,\ g^b_{2r}=0.59, g^b_{2i}=0.57\ \text{eV/\SI{}{\angstrom}}.
    \label{app:eq:MgB2_epc_values_sp2}
\end{equation}
We find that $g_1^b$ is roughly two orders of magnitude smaller than the onsite term $g_0^b$ and can therefore be neglected. Remark that $g_1^b$ is not symmetry-forbidden. Its smallness instead arises from an accidental cancellation among EPC matrix elements in the $3s@6m$ basis, where the underlying NN bond EPC terms are of order $\sim 0.5$ eV/\SI{}{\angstrom}.

We then transform the EPC Hamiltonian to momentum space (see FT convention in \cref{app:sec:intro-EPC-ham}), \ie, $u_{\RR, s}^{i}=\frac{1}{\sqrt{N}}\sum_{\qq} u_{\qq, s}^{i} e^{i\qq\cdot (\RR+\rr_{h_s})}$, $\cre{c}{\RR, a} =\frac{1}{\sqrt{N}}\sum_{\kk} \cre{c}{\kk, a} e^{-i\kk(\RR+\rr_{m_a})}$. We consider the $\Gamma_5^+$ bond-stretching mode $u_{\qq,s}^\mu=\frac{1}{\sqrt{2}}(-1)^{s+1}u_{\qq}^\mu$, as defined in \cref{app:eq:phonon_GM5+_basis}.

\paragraph{\textbf{$H_{\text{onsite}}$}. }
For $H_{\text{onsite}}$, we have
\begin{equation}\begin{aligned}
    H_{\text{onsite}} &= \frac{ig_0^b}{\sqrt{2}} \sum_{\RR,s} (-1)^{s+1} u_{s,+}(\RR)N_{s,+}(\RR) + h.c. \\
    &= \frac{ig_0^b}{\sqrt{2N}} \sum_{\RR,s,\qq} (-1)^{s+1} \frac{1}{\sqrt{2}} (u_{\qq,s}^x + i u_{\qq,s}^y) e^{i\qq\cdot(\RR+\rr_{h_s})} N_{s,+}(\RR) + h.c.  \\
    &= \frac{ig_0^b}{2\sqrt{2N}} \sum_{\RR,\qq} (u_{\qq}^x + i u_{\qq}^y) \left( e^{i\qq\cdot(\RR+\rr_{h_1})} N_{1,+}(\RR) + e^{i\qq\cdot(\RR+\rr_{h_2})} N_{2,+}(\RR) \right) + h.c. \\
    &= \frac{ig_0^b}{2\sqrt{2N}} \sum_{\kk,\qq} (u_{\qq}^x + i u_{\qq}^y) \sum_a \left( \cre{c}{\kk+\qq,a} \des{c}{\kk,a} f_{a}(\qq) \right) + h.c.,
    \label{app:eq:EPC_Ham_onsite}
\end{aligned}\end{equation}
where 
\begin{equation}\begin{aligned}
&f_a(\qq) 
=\begin{cases}
    e^{-i\qq\cdot\rr_{m_1}}(e^{i\qq\cdot (\rr_{h_1} -\bm{a}_1)}+e^{i\qq \cdot \rr_{h_2}})
    & a=1 \\
   e^{-i\qq\cdot\rr_{m_2} +\frac{2\pi i}{3}}
   (e^{i\qq\cdot\rr_{h_1}} + e^{i\qq\cdot\rr_{h_2}})
    & a=2 
    \\e^{-i\qq\cdot \rr_{m_3} - \frac{2\pi i}{3}}
    (e^{i\qq\cdot\rr_{h_1}} + e^{i\qq\cdot(\rr_{h_2}-\bm{a}_2)})
    & a=3
    \label{app:eq:EPC_Ham_onsite_faq}
\end{cases}.
\end{aligned}\end{equation}
When $\qq=\bm{0}$, $H_{\text{onsite}}$ reduces to the following 
\begin{equation}\begin{aligned}
 H_{\text{onsite}} &= \frac{i g_0^b}{\sqrt{2N}} \sum_{\kk} (u_{\bm{0}}^x + i u_{\bm{0}}^y)  \left( \cre{c}{\kk,1} \des{c}{\kk,1} + \omega \cre{c}{\kk,2} \des{c}{\kk,2} + \omega^*  \cre{c}{\kk,3} \des{c}{\kk,3} \right) + h.c. \\
\end{aligned}\end{equation}
Explicitly, let $u_{1,2}$ be the bond-stretching mode along $x,y$ directions as defined in \cref{app:eq:phonon_GM5+_basis}, we have
\begin{equation}\begin{aligned}
    (g_{\kk,\Gamma}^{u_1})_{\text{onsite}} &=
    \frac{ig_0^b}{\sqrt{2}} \text{Diag}\left[1, \omega, \omega^* \right] + h.c. = 
    \frac{\sqrt{6}}{2} g_0^b\text{Diag}\left[0, -1, 1\right], \\
    (g_{\kk,\Gamma}^{u_2})_{\text{onsite}} &= -\frac{g_0^b}{\sqrt{2} } \text{Diag}\left[1, \omega, \omega^*\right] + h.c.
    = \sqrt{2}g_0^b \text{Diag}\left[-1, \frac{1}{2}, \frac{1}{2} \right],
    \label{app:eq:onsite_epc_hamiltonian}
\end{aligned}\end{equation}
which is the same as \cref{app:eq:epc_momentum_sp2_bonding}.

\paragraph{\textbf{$H_{\text{NN}}$}. }
For $H_{\text{NN}}$, we have
\begin{equation}\begin{aligned}
    H_{\text{NN}} &= \frac{ig_1^b}{\sqrt{2}}\sum_{\RR,s} (-1)^{s+1} u_{s,+}(\RR) B_{s,+}(\RR) + h.c. \\
    &= \frac{ig_1^b}{\sqrt{2N}} \sum_{\RR,s,\qq} (-1)^{s+1} \frac{1}{\sqrt{2}}(u_{\qq,s}^x + i u_{\qq,s}^y) e^{i\qq\cdot(\RR+\rr_{h_s})} B_{s,+}(\RR) + h.c.  \\
    &= \frac{ig_1^b}{2\sqrt{2N}} \sum_{\RR,\qq} (u_{\qq}^x + i u_{\qq}^y) \left( e^{i\qq\cdot(\RR+\rr_{h_1})} B_{1,+}(\RR) + e^{i\qq\cdot(\RR+\rr_{h_2})} B_{2,+}(\RR) \right) + h.c. \\
    &= \frac{ig_1^b}{2\sqrt{2N}} \sum_{\kk,\qq} (u_{\qq}^x + i u_{\qq}^y) \sum_{a\neq b} \left( \cre{c}{\kk+\qq,a} \des{c}{\kk,b} f_{ab}(\kk, \qq)\right) + h.c.,
\end{aligned}\end{equation}
where 
\begin{equation}\begin{aligned}
&f_{ab}(\kk,\qq) 
=\begin{cases}
    \left( e^{-i\kk\cdot (\rr_{m_2} -\rr_{m_3})+ i\qq \cdot (\rr_{h_1} - \rr_{m_2})} + 
    e^{-i\kk\cdot (\rr_{m_2} -\rr_{m_3} - \bm{a}_1)+ i\qq \cdot (\rr_{h_2} - \rr_{m_2})} \right)
    & a=2, b=3 \\
    \left( e^{i\kk\cdot (\rr_{m_2} -\rr_{m_3})+ i\qq \cdot (\rr_{h_1} - \rr_{m_3})} + 
    e^{i\kk\cdot (\rr_{m_2} -\rr_{m_3} - \bm{a}_1)+ i\qq \cdot (\rr_{h_2} - \rr_{m_3}-\bm{a}_1)} \right)
    & a=3, b=2 \\
   \omega \left( e^{-i\kk\cdot (\rr_{m_3} -\rr_{m_1} - \bm{a}_2) + i\qq \cdot (\rr_{h_1} - \rr_{m_3})} + 
    e^{-i\kk\cdot (\rr_{m_3} +\bm{a}_1 -\rr_{m_1})+ i\qq \cdot (\rr_{h_2} - \rr_{m_3} - \bm{a}_1)} \right)
    & a=3, b=1 \\
    \omega \left( e^{i\kk\cdot (\rr_{m_3} -\rr_{m_1} - \bm{a}_2) + i\qq \cdot (\rr_{h_1} - \rr_{m_1}-\bm{a}_2)} + 
    e^{i\kk\cdot (\rr_{m_3} +\bm{a}_1 -\rr_{m_1})+ i\qq \cdot (\rr_{h_2} - \rr_{m_1})} \right)
    & a=1, b=3 \\
    \omega^* \left( e^{-i\kk\cdot (\rr_{m_1} +\bm{a}_2 - \rr_{m_2})+ i\qq \cdot (\rr_{h_1} - \rr_{m_1} -\bm{a}_2)} + 
    e^{-i\kk\cdot (\rr_{m_1} -\rr_{m_2})+ i\qq \cdot (\rr_{h_2} - \rr_{m_1})} \right)
    & a=1, b=2 \\
    \omega^* \left( e^{i\kk\cdot (\rr_{m_1} +\bm{a}_2 - \rr_{m_2})+ i\qq \cdot (\rr_{h_1} - \rr_{m_2})} + 
    e^{i\kk\cdot (\rr_{m_1} -\rr_{m_2})+ i\qq \cdot (\rr_{h_2} - \rr_{m_2})} \right)
    & a=2, b=1
\end{cases}.
\end{aligned}\end{equation}
When $\qq=\bm{0}$, we have
$f_{23}(\kk)=f_{32}(\kk)=2\cos(\frac{k_1}{2}), 
f_{31}(\kk)=f_{13}(\kk)=2\omega \cos(\frac{k_1+k_2}{2}),
f_{12}(\kk)=f_{21}(\kk)=2\omega^* \cos(\frac{k_2}{2})$, 
and $H_{\text{NN}}$ reduces to the following 
\begin{equation}\begin{aligned}
 H_{\text{NN}} = \frac{i g_1^b}{\sqrt{2N}} \sum_{\kk} (u_{\bm{0}}^x + i u_{\bm{0}}^y)  &\left[ \left(\cre{c}{\kk,2} \des{c}{\kk,3} + \cre{c}{\kk,3} \des{c}{\kk,2} \right)\cos(\frac{k_1}{2}) + \omega \left(\cre{c}{\kk,3} \des{c}{\kk,1} + \cre{c}{\kk,1} \des{c}{\kk,3} \right) \cos(\frac{k_1+k_2}{2}) \right.\\
 &\left. + \omega^* \left( \cre{c}{\kk,1} \des{c}{\kk,2} + \cre{c}{\kk,2} \des{c}{\kk,1} \right) \cos(\frac{k_2}{2}) \right] + h.c.
\end{aligned}\end{equation}
Explicitly, we have
\begin{equation}\begin{aligned}
    (g_{\kk,\Gamma}^{u_1})_{\text{NN}} &=
    \frac{ig_1^b}{\sqrt{2}} \begin{bmatrix}
        0 & \omega^* \cos(\frac{k_2}{2}) & \omega \cos(\frac{k_1+k_2}{2}) \\
        \omega^* \cos(\frac{k_2}{2}) & 0 & \cos(\frac{k_1}{2})\\
        \omega \cos(\frac{k_1+k_2}{2}) & \cos(\frac{k_1}{2}) & 0
    \end{bmatrix} + h.c. 
    = \frac{g_1^b}{\sqrt{2}} \begin{bmatrix}
        0 & \sqrt{3}\cos(\frac{k_2}{2}) &  -\sqrt{3} \cos(\frac{k_1+k_2}{2}) \\
         & 0 & 0 \\
        c.c. & & 0
    \end{bmatrix}, \\
    (g_{\kk,\Gamma}^{u_2})_{\text{NN}} &= 
    -\frac{g_1^b}{\sqrt{2}} \begin{bmatrix}
        0 & \omega^* \cos(\frac{k_2}{2}) & \omega \cos(\frac{k_1+k_2}{2}) \\
        \omega^* \cos(\frac{k_2}{2}) & 0 & \cos(\frac{k_1}{2})\\
        \omega \cos(\frac{k_1+k_2}{2}) & \cos(\frac{k_1}{2}) & 0
    \end{bmatrix} + h.c. 
    = \frac{g_1^b}{\sqrt{2}} \begin{bmatrix}
        0 & \cos(\frac{k_2}{2}) &  \cos(\frac{k_1+k_2}{2}) \\
         & 0 & -2\cos(\frac{k_1}{2})\\
        c.c. & & 0
    \end{bmatrix}.
    \label{app:eq:NN-bond-EPC-ham}
\end{aligned}\end{equation}

\paragraph{\textbf{$H^{(2)}_{\text{NN}}$}. }
For $H^{(2)}_{\text{NN}}$, we have
\begin{equation}\begin{aligned}
    H^{(2)}_{\text{NN}} &= \frac{(g^b_{2r}+ig^b_{2i})}{\sqrt{2}} \sum_{\RR,s} (-1)^{s+1} u_{s,+}(\RR) B^{(2)}_{s,+}(\RR) + h.c. \\
    &= \frac{(g^b_{2r}+ig^b_{2i})}{\sqrt{2N}} \sum_{\RR,s,\qq} (-1)^{s+1} \frac{1}{\sqrt{2}}(u_{\qq,s}^x + i u_{\qq,s}^y) e^{i\qq\cdot(\RR+\rr_{h_s})} B^{(2)}_{s,+}(\RR) + h.c.  \\
    &= \frac{(g^b_{2r}+ig^b_{2i})}{2\sqrt{2N}} \sum_{\RR,\qq} (u_{\qq}^x + i u_{\qq}^y) \left( e^{i\qq\cdot(\RR+\rr_{h_1})} B^{(2)}_{1,+}(\RR) + e^{i\qq\cdot(\RR+\rr_{h_2})} B^{(2)}_{2,+}(\RR) \right) + h.c. \\
    &= \frac{(g^b_{2r}+ig^b_{2i})}{2\sqrt{2N}} \sum_{\kk,\qq} (u_{\qq}^x + i u_{\qq}^y) \sum_{a\neq b} \left( \cre{c}{\kk+\qq,a} \des{c}{\kk,b} f^{(2)}_{ab}(\kk, \qq)\right) + h.c.,
\end{aligned}\end{equation}
where 
\begin{equation}\begin{aligned}
&f^{(2)}_{ab}(\kk,\qq) =\\
&\begin{cases}
    e^{-i\kk\cdot (\rr_{m_2} -\rr_{m_3})} 
    \left[ e^{i\qq \cdot (\rr_{h_1} - \rr_{m_2})} (e^{-i\kk\cdot(-\bm{a}_1)+i\qq\cdot\bm{a}_1} + e^{-i\kk\cdot(-\bm{a}_1)}) + 
    e^{i\qq \cdot (\rr_{h_2} - \rr_{m_2})}
    (e^{i\qq\cdot(-\bm{a}_1)} + 1)
    \right]
    & a=2, b=3 \\
    e^{-i\kk\cdot (\rr_{m_3} -\rr_{m_2})} 
    \left[ e^{i\qq \cdot (\rr_{h_1} - \rr_{m_3})} (e^{-i\kk\cdot(\bm{a}_1)+i\qq\cdot(-\bm{a}_1)} + e^{-i\kk\cdot(\bm{a}_1)}) + 
    e^{i\qq \cdot (\rr_{h_2} - \rr_{m_3})}
    (1 + e^{i\qq\cdot(-\bm{a}_1)})
    \right]
    & a=3, b=2 \\
    \omega e^{-i\kk\cdot (\rr_{m_1} -\rr_{m_2})} 
    \left[ e^{i\qq \cdot (\rr_{h_1} - \rr_{m_1})} (1+ e^{i\qq\cdot(-\bm{a}_2)}) + 
    e^{i\qq \cdot (\rr_{h_2} - \rr_{m_1})}
    (e^{-i\kk\cdot(\bm{a}_2) + i\qq\cdot(-\bm{a}_1)} + e^{-i\kk\cdot \bm{a}_2})
    \right]
    & a=1, b=2 \\
    \omega e^{-i\kk\cdot (\rr_{m_2} -\rr_{m_1})} 
    \left[ e^{i\qq \cdot (\rr_{h_1} - \rr_{m_2})} (e^{i\qq\cdot(-\bm{a}_2)} + 1) + 
    e^{i\qq \cdot (\rr_{h_2} - \rr_{m_2})}
    (e^{-i\kk\cdot(-\bm{a}_2) +i\qq\cdot\bm{a}_2} + e^{-i\kk\cdot(-\bm{a}_2)})
    \right]
    & a=2, b=1 \\
    \omega^* e^{-i\kk\cdot (\rr_{m_3} -\rr_{m_1})} 
    \left[ e^{i\qq \cdot (\rr_{h_1} - \rr_{m_3})} (e^{-i\kk\cdot\bm{a}_1 -i\qq\cdot(\bm{a}_1+\bm{a}_2)} + e^{-i\kk\cdot \bm{a}_1}) + 
    e^{i\qq \cdot (\rr_{h_2} - \rr_{m_3})}
    (e^{i\kk\cdot\bm{a}_2 + i\qq\cdot\bm{a}_2} + e^{i\kk\cdot\bm{a}_2 - i\qq\cdot\bm{a}_1})
    \right]
    & a=3, b=1 \\
    \omega^* e^{-i\kk\cdot (\rr_{m_1} -\rr_{m_3})} 
    \left[ e^{i\qq \cdot (\rr_{h_1} - \rr_{m_1})} (e^{i\kk\cdot\bm{a}_1+i\qq\cdot\bm{a}_1} + e^{i\kk\cdot\bm{a}_1- i\qq\cdot\bm{a}_2}) + 
    e^{i\qq \cdot (\rr_{h_2} - \rr_{m_1})}
    (e^{-i\kk\cdot\bm{a}_2-i\qq\cdot(\bm{a}_1+\bm{a}_2)}+ e^{-i\kk\cdot\bm{a}_2})
    \right]
    & a=1, b=3 
\end{cases}
\label{app:eq:H_NN2_fkq}
\end{aligned}\end{equation}
When $\qq=\bm{0}$, we have
\begin{equation}\begin{aligned}
    f^{(2)}_{23}(\kk) &= f^{(2)}_{32}(\kk)= 4\cos\left(\frac{k_1}{2}\right),\quad
    f^{(2)}_{12}(\kk) = f^{(2)}_{21}(\kk)= 4\omega \cos\left(\frac{k_2}{2}\right),\quad
    f^{(2)}_{31}(\kk) = f^{(2)}_{13}(\kk)= 4\omega^* \cos\left(\frac{k_1+k_2}{2}\right), 
\end{aligned}\end{equation}
and $H^{(2)}_{\text{NN}}$ reduces to the following form
\begin{equation}\begin{aligned}
 H^{(2)}_{\text{NN}} = \frac{2(g^b_{2r}+ig^b_{2i})}{\sqrt{2N}} \sum_{\kk} (u_{\bm{0}}^x + i u_{\bm{0}}^y)  &\left[ \left(\cre{c}{\kk,2} \des{c}{\kk,3} + \cre{c}{\kk,3} \des{c}{\kk,2} \right)\cos(\frac{k_1}{2}) + \omega^* \left(\cre{c}{\kk,3} \des{c}{\kk,1} + \cre{c}{\kk,1} \des{c}{\kk,3} \right) \cos(\frac{k_1+k_2}{2}) \right.\\
 &\left. + \omega \left( \cre{c}{\kk,1} \des{c}{\kk,2} + \cre{c}{\kk,2} \des{c}{\kk,1} \right) \cos(\frac{k_2}{2}) \right] + h.c.
\end{aligned}\end{equation}
Explicitly, we have
\begin{equation}\begin{aligned}
    (g_{\kk,\Gamma}^{u_1})_{\text{NN}}^{(2)} &=
    \sqrt{2}(g^b_{2r}+ig^b_{2i})\begin{bmatrix}
        0 & \omega \cos(\frac{k_2}{2}) & \omega^* \cos(\frac{k_1+k_2}{2}) \\
        \omega \cos(\frac{k_2}{2}) & 0 & \cos(\frac{k_1}{2})\\
        \omega^* \cos(\frac{k_1+k_2}{2}) & \cos(\frac{k_1}{2}) & 0
    \end{bmatrix} + h.c. \\
    &= \sqrt{2} g^b_{2r}
    \begin{bmatrix}
        0 & -\cos(\frac{k_2}{2}) & - \cos(\frac{k_1+k_2}{2}) \\
         & 0 & 2\cos(\frac{k_1}{2})\\
        c.c. & & 0
    \end{bmatrix}
    + \sqrt{2}g_{2i}^b 
    \begin{bmatrix}
        0 & -\sqrt{3}\cos(\frac{k_2}{2}) & \sqrt{3} \cos(\frac{k_1+k_2}{2}) \\
         & 0 & 0 \\
        c.c. & & 0
    \end{bmatrix}, \\
    (g_{\kk,\Gamma}^{u_2})_{\text{NN}}^{(2)} &= 
    i\sqrt{2} (g^b_{2r}+ig^b_{2i}) \begin{bmatrix}
        0 & \omega^* \cos(\frac{k_2}{2}) & \omega \cos(\frac{k_1+k_2}{2}) \\
        \omega^* \cos(\frac{k_2}{2}) & 0 & \cos(\frac{k_1}{2})\\
        \omega \cos(\frac{k_1+k_2}{2}) & \cos(\frac{k_1}{2}) & 0
    \end{bmatrix} + h.c. \\
    &= -\sqrt{2} g^b_{2i}
    \begin{bmatrix}
        0 & -\cos(\frac{k_2}{2}) &  -\cos(\frac{k_1+k_2}{2}) \\
         & 0 & 2\cos(\frac{k_1}{2})\\
        c.c. & & 0
    \end{bmatrix}
    + \sqrt{2}g_{2r}^b 
    \begin{bmatrix}
        0 & -\sqrt{3}\cos(\frac{k_2}{2}) & \sqrt{3} \cos(\frac{k_1+k_2}{2}) \\
         & 0 & 0 \\
        c.c. & & 0
    \end{bmatrix}
    \label{app:eq:NN-bond-EPC-ham2}
\end{aligned}\end{equation}

\subsubsection{EPC in the band basis}\label{app:sec:EPC-band-basis}

Finally, we transform the EPC tensor into the two bands from the $\Gamma_5^+$ mode near the FS in the $sp^2$ bonding states. We show that the $k^2$ decay of FS-averaged \textit{ab initio} EPC in \cref{app:fig:DFT_EPC_band_basis} can be reproduced in the simple analytic EPC model.

\paragraph{\textbf{Onsite EPC only.} } 
We first only consider the onsite EPC term with Hamiltonian \cref{app:eq:onsite_epc_hamiltonian}, and use the two wavefunctions from the NN kagome model expanded near the $\Gamma_5^+$ mode. 

The kagome flatband (FB) wavefunction can be expanded to the second order in momentum $\kk=k(\cos(\theta),\sin(\theta))$ near $\Gamma$, where $k\in[0,\pi]$ is dimensionless in natural lattice units with $a=1$ (see detailed derivation in \cref{app:sec:perturbation_theory}): 
\begin{equation}
    \psi^{FB}_{\kk}\approx \sqrt{\frac{2}{3}}\left\{\left[\cos(\theta), \cos(\theta + \frac{2\pi}{3}), \cos(\theta - \frac{2\pi}{3}) \right] - \frac{k^2}{96}\cos(3\theta)[1,1,1]
    \right\}, 
    \label{app:eq:FB_kp_k2}
\end{equation}
where a normalization factor $\frac{1}{\sqrt{1 +k^4/96(1+\cos 6\theta)}}=1+\mathcal{O}(k^4)$ is omitted. 
The wavefunction of the other dispersive band (DB) from $\Gamma_5^+$ can be obtained similarly: 
\begin{equation}
    \psi^{DB}_{\kk}\approx \sqrt{\frac{2}{3}}\left\{\left[\sin(\theta), \sin(\theta + \frac{2\pi}{3}), \sin(\theta - \frac{2\pi}{3}) \right] - \frac{k^2}{96}\sin(3\theta)[1,1,1]
    \right\}.
    \label{app:eq:DB_kp_k2}
\end{equation}
The onsite EPC Hamiltonian \cref{app:eq:onsite_epc_hamiltonian} can be projected into these two band basis:
\begin{equation}\begin{aligned}
    g^{FB,u_1}_{\kk,\Gamma} &= \frac{g^b_0}{\sqrt{2}}\left(-\sin(2\theta) - \frac{k^2}{24} \sin(\theta)\cos(3\theta)\right),\quad
    g^{FB,u_2}_{\kk,\Gamma} = \frac{g^b_0}{\sqrt{2}} \left(-\cos(2\theta) + \frac{k^2}{48} \left(\cos(2\theta) + \cos(4\theta)\right)\right), \\
    g^{DB,u_1}_{\kk,\Gamma} &= \frac{g^b_0}{\sqrt{2}} \left(\sin(2\theta) + \frac{k^2}{24} \left(\sin(2\theta) + \sin(4\theta)\right)\right),\quad
    g^{DB,u_2}_{\kk,\Gamma} = \frac{g^b_0}{\sqrt{2}} \left(\cos(2\theta) + \frac{k^2}{48} \left(\cos(2\theta) - \cos(4\theta)\right)\right).
\end{aligned}\end{equation}
As a result, the FS average of the onsite EPC term is
\begin{equation}\begin{aligned}
    & \frac{1}{2\pi}\int_{0}^{2\pi} |g_{k\theta,\Gamma}^{FB,u_i}|^2 d\theta =
    \frac{(g_0^b)^2}{48}\left(24-k^2\right) +\mathcal{O}(k^4),\\
    &\frac{1}{2\pi}\int_{0}^{2\pi} |g_{k\theta,\Gamma}^{DB,u_i}|^2 d\theta=
    \frac{(g_0^b)^2}{48}\left(24 + k^2 \right) +\mathcal{O}(k^4).
    \label{app:eq:EPC-FS-avg-onsite-only}
\end{aligned}\end{equation}
It can be seen that although the flatband gives the correct $k^2$ decaying EPC as in the \textit{ab initio} results \cref{app:fig:DFT_EPC_band_basis}, the dispersive band gives the wrong $k^2$ increasing trend (due to a different wavefunction). This indicates that the onsite EPC term alone cannot reproduce the correct \textit{ab initio} results.

\paragraph{\textbf{Onsite and NN bond EPC.} } 
We then further include the NN bond EPC terms $H_{\text{NN}}$ in \cref{app:eq:NN-bond-EPC-ham} and $H^{(2)}_{\text{NN}}$ \cref{app:eq:NN-bond-EPC-ham2}. As the magnitude of $H_{\text{NN}}$ is around 10 times smaller than $H^{(2)}_{\text{NN}}$, we only  consider$H^{(2)}_{\text{NN}}$. 

Using the $\kk\cdot\pp$ expanded wavefunction of $\Gamma_5^+$ mode in \cref{app:eq:FB_kp_k2} and \cref{app:eq:DB_kp_k2}, we obtain
\begin{equation}\begin{aligned}
    g^{FB,u_1}_{\kk,\Gamma} &= \frac{1}{\sqrt{2}}\left[-(g^b_0-4g^b_{2i})\sin(2\theta) + 4g^b_{2r}\cos(2\theta) \right] - 
    \frac{k^2 \cos(3\theta)}{24\sqrt{2}} \left[(g^b_0-4g^b_{2i})\sin(\theta) + 4g^b_{2r} \cos(\theta) \right],\\
    g^{FB,u_2}_{\kk,\Gamma} &= \frac{1}{\sqrt{2}}\left[-(g^b_0+4g^b_{2i})\cos(2\theta) + 4g^b_{2r}\sin(2\theta) \right] + 
    \frac{k^2 \cos(3\theta)}{24\sqrt{2}} \left[(g^b_0+4g^b_{2i})\cos(\theta) + 4g^b_{2r} \sin(\theta) \right], \\
    g^{DB,u_1}_{\kk,\Gamma} &= \frac{1}{\sqrt{2}}\left[(g^b_0-4g^b_{2i})\sin(2\theta) - 4g^b_{2r}\cos(2\theta) \right] \\
    &+\frac{k^2}{48\sqrt{2}} \left[(g^b_0+20g^b_{2i}+ 2(g^b_0-4g^b_{2i})\cos(2\theta))\sin(2\theta)+ 4g^b_{2r}\cos(4\theta) +20g^b_{2r}\cos(2\theta)\right],\\
    g^{DB,u_2}_{\kk,\Gamma} &= \frac{1}{\sqrt{2}}\left[(g^b_0+4g^b_{2i})\cos(2\theta) - 4g^b_{2r}\sin(2\theta) \right] \\
    &+\frac{k^2}{48\sqrt{2}} \left[ (g^b_0-20g^b_{2i})\cos(2\theta) - (g^b_0+4g^b_{2i})\cos(4\theta) - 4g^b_{2r}(-5\sin(2\theta) +\sin(4\theta))\right],
\end{aligned}\end{equation}
As a result, we arrive at
\begin{equation}\begin{aligned}
    \langle |g^{FB}_{\kk}|^2 \rangle = &\frac{1}{2\pi}\int_{0}^{2\pi} \sum_i |g_{k\theta,\Gamma}^{FB,u_i}|^2 d\theta =
    \frac{(g_0^b)^2}{48}\left[24(1 + 16\overline{g}) - (1+16\overline{g})k^2
    \right] +\mathcal{O}(k^4), \\
    \langle |g^{DB}_{\kk}|^2 \rangle = &\frac{1}{2\pi}\int_{0}^{2\pi} \sum_i |g_{k\theta,\Gamma}^{DB,u_i}|^2 d\theta =
    \frac{(g_0^b)^2}{48}\left[24(1 + 16\overline{g}) + (1-80\overline{g})k^2
    \right] +\mathcal{O}(k^4),
    \label{app:eq:EPC-FS-avg-onsite-NN}
\end{aligned}\end{equation}
where $\overline{g}=((g^b_{2r})^2 + (g^b_{2i})^2)/(g_0^b)^2 \approx 0.034$. 
Comparing with \cref{app:eq:EPC-FS-avg-onsite-only} where we considered only the onsite EPC term, here the NN bond EPC terms $\overline{g}$ correct both the zeroth-order and $k^2$-order term coefficients. Plugging in the \textit{ab initio} EPC values in \ch{MgB2} in \cref{app:eq:MgB2_epc_values_sp2}, and add the coefficient $l_B = \sqrt{\hbar/2M_B \omega_{\Gamma_5^+}}\approx 0.053$\SI{}{\angstrom} (where $M_B=10.81$ a.u. and $\omega_{\Gamma_5^+}\approx 70$ meV) that converts \SI{}{eV/\angstrom} to eV, we find 
\begin{equation}\begin{aligned}
     \langle |G^{FB}_{\kk}|^2 \rangle &= l_B^2
    \langle |g^{FB}_{\kk}|^2 \rangle = 0.042 - 0.0018 k^2 \ \text{eV}^2, \\
    \langle |G^{DB}_{\kk}|^2 \rangle &= l_B^2
    \langle |g^{DB}_{\kk}|^2 \rangle = 0.042 - 0.0020 k^2 \ \text{eV}^2,
\end{aligned}\end{equation}
where $k\in[0, \pi]$ is in the natural lattice units with $a=1$. 
Both the flat and dispersive bands reproduce the expected $k^2$ decay of the angle-averaged EPC near $\Gamma$, with the dispersive band decaying more rapidly, consistent with the \textit{ab initio} results in \cref{app:fig:DFT_EPC_band_basis}. In \cref{app:sec:analytic_estimation_lambda}, we fit the \textit{ab initio} data of \cref{app:fig:DFT_EPC_band_basis} using \cref{app:eq:EPC-FS-avg-onsite-NN}, obtaining a qualitative agreement. In the \textit{ab initio} calculations, long-range EPC terms and deviations of the electronic wavefunctions from the NN kagome limit modify the coefficients in \cref{app:eq:EPC-FS-avg-onsite-NN}, but the characteristic $k^2$ decay is preserved. Hence, our simple analytic EPC model correctly captures the essential physics of the $\sigma$ Fermi surface in \ch{MgB2}.

\subsubsection{$k\cdot p$ Hamiltonian and DOS of $sp^2$ bands}\label{app:sec:DOS_analytic}

In the following, we consider the $\kk\cdot\pp$ Hamiltonian of the $\Gamma_5^+$ band in \ch{MgB2} and use it to reproduce the DOS of the $sp^2$ bands. 

We first consider the in-plane $\kk\cdot\pp$ Hamiltonian~\cite{jiang2021k} of the $\Gamma_5^+$ band, which can be obtained by projecting the NN kagome model in the three-bonding-state basis to the eigen basis of $\Gamma_5^+$ IRREP. Recall the NN $s$ orbital kagome model with three orbitals at $(\frac{1}{2},0), (\frac{1}{2},\frac{1}{2}), (0,\frac{1}{2})$ takes the form:
\begin{equation}\begin{aligned}
H_{NN}^{kag}(\bm{k})=2 t_{NN}
\left(
\begin{matrix}
    0 &  \cos(\frac{k_2}{2})  &\cos(\frac{k_1+k_2}{2}) \\
    & 0 & \cos(\frac{k_1}{2})  \\
c.c. &  & 0\\
\end{matrix}
\right).
\label{app:eq:NN_kagome_model}
\end{aligned}\end{equation}
At $\Gamma$, the two degenerate modes that form the $\Gamma_5^+$ IRREPs are
\begin{equation}\begin{aligned}
\psi_1=\frac{1}{\sqrt{3}}[1, \omega,\omega^2], \quad
\psi_2=\frac{1}{\sqrt{3}}[1, \omega^2,\omega], 
\label{app:eq:GM5+_eig_basis}
\end{aligned}\end{equation}
with $\omega=e^{i2\pi/3}$. 
These two eigenmodes have $C_3$ eigenvalues $\omega$ and $\omega^2$, respectively. If we project $H_{NN}^{kag}(\kk)$ onto the $\psi_{1,2}$ basis and expand near $\Gamma$, we obtain
\begin{equation}\begin{aligned}
h^{\kk\cdot\pp}(\kk) &= \frac{t_{NN}}{16}\left[ (k_x^2+k_y^2 - 16) \sigma_0 
+ (k_{+}^2 \sigma_{+} + k_{-}^2 \sigma_{-})\right].
\label{app:eq:kp_GM5+_from_kagome}
\end{aligned}\end{equation}
This $\kk\cdot\pp$ Hamiltonian has eigenmodes
$\psi_{\pm}=\frac{1}{\sqrt{2}}\left[\pm e^{2i\theta}, 1\right]$ with $\theta=\arctan(\frac{k_y}{k_x})$, which gives Berry phase $2\pi$ on a closed loop surrounding $\Gamma$. 
The $2\pi$ Berry phase is protected by $C_{2z}T$ symmetry (equivalently, an Euler-class charge) and cannot be eliminated, but can only be transformed into two Dirac nodes with $\pi$ Berry phase once $C_{3z}$ is broken~\cite{sun2009topological, ahn2019failure, yu2023euler, song2019all, herzog2023hofstadter}. 
To show this, note that the symmetry representation matrix of the NN kagome model in \cref{app:eq:NN_kagome_model}:
\begin{equation}\begin{aligned}
    D^{kag}(C_{2z}T) &= \bm{1}_3 \hat{K}
    ,\quad
    D^{kag}(C_{3z})= \begin{bmatrix}
        0 & 0 & 1\\
        1 & 0 & 0\\
        0 & 1 & 0
    \end{bmatrix}.
\end{aligned}\end{equation}
The corresponding representation matrix in the $\psi_{1,2}$ basis of $\Gamma_5^+$ (defined in \cref{app:eq:GM5+_eig_basis}) is
\begin{equation}
    D^{\psi_{1,2}}(C_{2z}T) = \sigma_x \hat{K}, \quad
    D^{\psi_{1,2}}(C_{3z}) = \text{Diag}[\omega,\omega^2].
\end{equation}
It is more convenient to use a real gauge thanks to $C_{2z}T$ symmetry. To do so, we recombine the $\psi_{1,2}$ basis into
\begin{equation}
    \psi_1^r=\frac{1}{\sqrt{2}}(\psi_1+\psi_2)=\frac{1}{\sqrt{6}}[2,-1,-1],\quad
    \psi_2^r=\frac{1}{\sqrt{2}i}(\psi_1-\psi_2)=\frac{1}{\sqrt{2}}[0,1,-1].
\end{equation}
In $\psi_{1,2}^r$ basis, $C_{2z}T$, $C_3$ representation matrix, and the $\kk\cdot\pp$ Hamiltonian takes the form
\begin{equation}\begin{aligned}
    D^{\psi_{1,2}^r}(C_{2z}T) &= \hat{K}, \quad
    D^{\psi_{1,2}^r}(C_{3z}) = \begin{bmatrix}
        \cos(\frac{2\pi}{3}) & -\sin(\frac{2\pi}{3}) \\
        \sin(\frac{2\pi}{3}) & \cos(\frac{2\pi}{3})
    \end{bmatrix}
    =\exp(i\frac{2\pi}{3}\sigma_y), \\
    h^{\kk\cdot\pp}_{real}(\kk) &= \frac{t_{NN}}{16}\left[ (k_x^2+k_y^2 - 16) \sigma_0 
+ 2k_xk_y\sigma_x + (-k_x^2+k_y^2)\sigma_z
\right].
\end{aligned}\end{equation}
We then add the following $C_3$ breaking term:
\begin{equation}
    H_{C_3break}(\kk) = m_0 \frac{t_{NN}}{16} \sigma_x.
\end{equation}
The eigenvalues of the $C_3$-broken Hamiltonian $h^{\kk\cdot\pp}_{real}(\kk)+H_{C_3break}(\kk)$ are
\begin{equation}
    \epsilon_{\pm}(\kk) = \frac{t_{NN}}{16}\left( (k_x^2+k_y^2-16) \pm  \sqrt{(k_x^2+k_y^2)^2 + 4 k_xk_y m_0 + m_0^2}\right). 
\end{equation}
It can be seen that the 2D degenerate point at $\Gamma$ is gapped, while two Dirac crossings with $\pi$-Berry phase appear at 
\begin{equation}
\kk_{0}^{\pm}=
\left\{
\begin{array}{lr}
\pm \sqrt{\frac{m_0}{2}}[1, -1], & m_0\ge 0,  \\
\pm \sqrt{\frac{|m_0|}{2}}[1, 1], & m_0<0.
\end{array}
\right.
\end{equation}
In summary, with $C_{2z}T=\hat{K}$ (spinless), the Berry phase on any closed loop in 2D is quantized to $n\pi$ (i.e. $0$ or $\pi$ mod $2\pi$). In a gauge where $C_{2z}T$ acts as complex conjugation, the Hamiltonian and its eigenstates can be chosen real, so the $U(1)$ gauge freedom collapses to $\mathbb{Z}_2=\{\pm 1\}$. Consequently, multiplying the wavefunction by a position-dependent phase $e^{in\theta}$ is not an allowed symmetry-respecting gauge transformation, and one cannot change the Berry phase of the $\Gamma_5^+$ node in \cref{app:eq:kp_GM5+_from_kagome} by $2n\pi$.

\paragraph{\textbf{General $\kk\cdot\pp$ Hamiltonian for the $\Gamma_5^+$ IRREP.}}

We then derive the general form of the $\kk\cdot\pp$ Hamiltonian for the $\Gamma_5^+$ IRREP in SG 191. Consider the following representation matrices for $\Gamma_5^+$:
\begin{equation}
    D(C_6) = \begin{bmatrix}
        e^{-i\frac{2\pi}{3}} & 0 \\
        0 & e^{i\frac{2\pi}{3}}
    \end{bmatrix},\quad
    D(C_{2, 110}) = \begin{bmatrix}
        0 & 1 \\
        1 & 0
    \end{bmatrix},\quad
    D(P) = \begin{bmatrix}
        1 & 0 \\
        0 & 1
    \end{bmatrix}.     
\end{equation}
By enforcing the symmetry constraints $D(g)h(\kk)D^{-1}(g)=h(g\kk)$, we obtain the $\kk\cdot\pp$ Hamiltonian for $\Gamma_5^+$ mode up to second order terms:
\begin{equation}\begin{aligned}
h^{\kk\cdot\pp}_{\Gamma_5^+}(\kk) &= E_0- c_1(k_x^2+k_y^2) \sigma_0 
+ c_2 k_z^2 \sigma_0
+ c_3 (k_{+}^2 \sigma_{+} + k_{-}^2 \sigma_{-})
\label{app:eq:kp_ham_GM5+}
\end{aligned}\end{equation}
Odd-order terms are forbidden by the inversion symmetry, \ie, $h^{odd}(\kk)=h^{odd}(-\kk)=-h^{odd}(\kk)=\bm{0}$.  
Note that the $c_3$ term represents the interband coupling, without which the Hamiltonian becomes two degenerate parabolic bands. The dispersion of this Hamiltonian is two quadratic bands:
\begin{equation}\begin{aligned}
E^{\pm}_{\Gamma_5^+}(\kk) &= E_0 + (-c_1 \pm 2c_3) (k_x^2+k_y^2) + c_2 k_z^2.
\label{app:eq:disperion_kp_twoband}
\end{aligned}\end{equation}
In the following, we first compute the DOS for this quadratic $\kk\cdot\pp$ Hamiltonian, which can qualitatively reproduce the \textit{ab initio} DOS in \ch{MgB2} from the $sp^2$ bands. We will further add a 4th-order term to more faithfully reproduce the \textit{ab initio} DOS.

\paragraph{\textbf{DOS of $\kk\cdot\pp$ model}. }
We compute the DOS from the $\kk\cdot\pp$ Hamiltonian. The DOS is defined as 
\begin{equation}\begin{aligned}
D(E) &= 
\sum_{n} \frac{1}{(2\pi)^3} \int d^3k \delta(E-E_n(\kk)) \\
&=\sum_n \frac{1}{(2\pi)^3} \int_{E_n(\kk)=E} \frac{dS}{|\nabla_{\kk} E_n(\kk)|} 
\end{aligned}\end{equation}

\begin{figure}[htbp]
    \centering
    \includegraphics[width=0.8\linewidth]{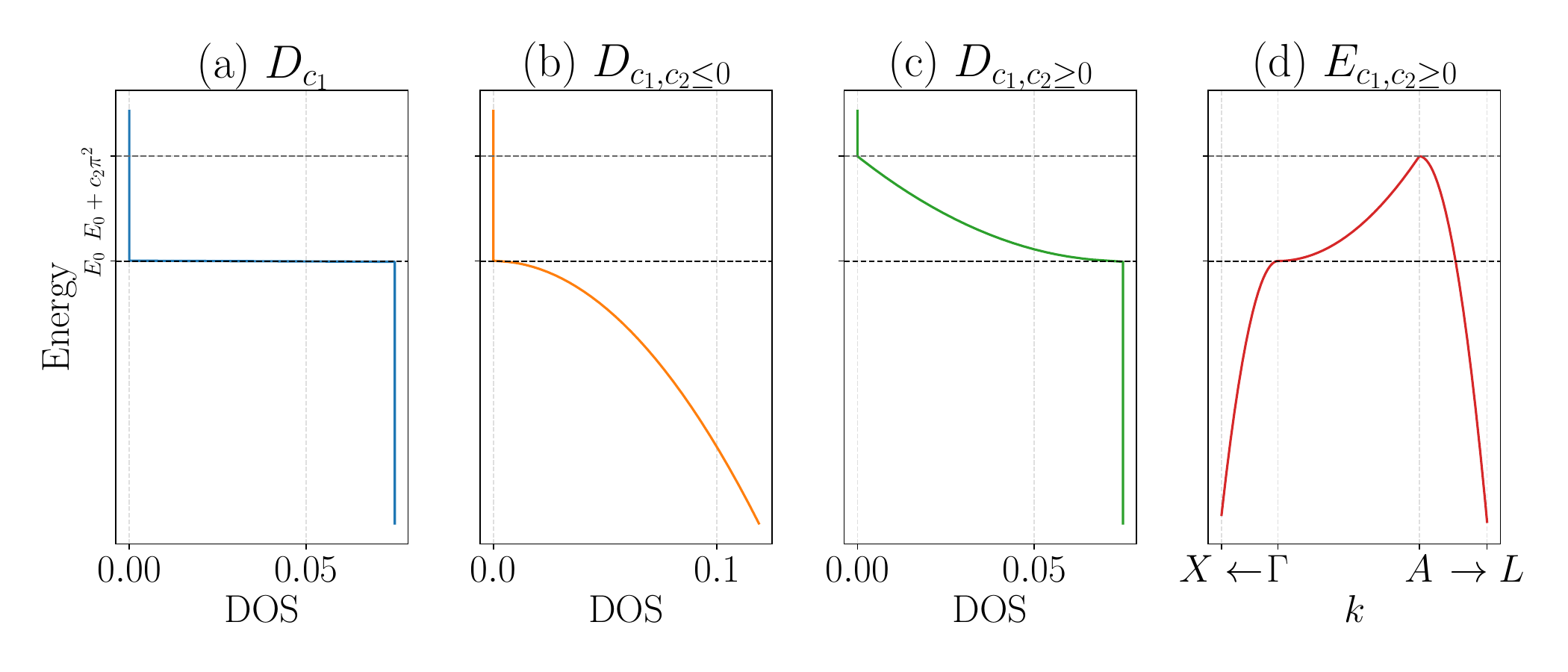}
    \caption{The DOS and dispersion from the $\kk\cdot\pp$ models. (a) DOS of $E(\kk)=E_0 - c_1(k_x^2+k_y^2)$, which is constant for $E\le E_0$. (b) DOS of $E_n(\kk)=E_0 - c_1(k_x^2+k_y^2) + c_2 k_z^2$, with $c_1>0, c_2<0$, where the DOS $\propto \sqrt{E_0-E}$ (see \cref{app:eq:DOS-kp-c1c2-ge0}). (c) Similar to (b) but with $c_1>0, c_2>0$, described by \cref{app:eq:DOS_kp_k^2_with_kz_term}. 
    (d) The dispersion used to compute the DOS in (c), where a saddle point appears at $\Gamma$ at $E_0$. 
    }
    \label{fig:DOS-kp-1}
\end{figure}

We first only include the in-plane $c_1$ term, with dispersion $E_{n}(\kk)=E_0 - c_1(k_x^2+k_y^2)$, which is constant along the $k_z$ direction. With no loss of generality, we assume $c_1>0$. Then $k(E) = \sqrt{\frac{E_0 - E}{c_1}}$, and $|\nabla_{\kk} E_n(\kk)|=2c_1 k=2\sqrt{c_1(E_0-E)}$. In this case, the DOS for each band is:
\begin{equation}\begin{aligned}
D(E) = \frac{1}{(2\pi)^2} \int_{E_n(\kk)=E} \frac{dS}{|\nabla_{\kk} E_n(\kk)|} = \frac{1}{(2\pi)^2} \frac{2\pi \sqrt{\frac{E_0-E}{c_1}}}{2\sqrt{c_1(E_0-E)}} = \frac{1}{4\pi c_1} \Theta(E_0-E),
\end{aligned}\end{equation} 
where $\Theta(E)$ is the Heaviside step function. In this case, the DOS remains constant and is independent of energy except for the jump at $E_0$, as shown in \cref{fig:DOS-kp-1} (a).

We then consider both $c_1$ and $c_2$, with dispersion $E_n(\kk)=E_0 - c_1(k_x^2+k_y^2) + c_2 k_z^2$. We first assume $c_1>0, c_2<0$. This is not the case in \ch{MgB2}, but we include it here for completeness. 
In this case, the DOS for each band has the form
\begin{equation}\begin{aligned}
D(E) &= \frac{1}{(2\pi)^3} \int_{-\infty}^{\infty} dk_z \int_0^{2\pi}d\theta \int_0^{\infty} k_{\parallel} dk_{\parallel} \delta(E-E_0 + c_1 k_{\parallel}^2-c_2 k_z^2) \\
&= \frac{1}{(2\pi)^2} \int_{-\infty}^{\infty} dk_z \int_0^{\infty} k_{\parallel} dk_{\parallel} \delta(\Delta E - c_1 k_{\parallel}^2 + c_2 k_z^2) \quad (\Delta E=E_0 -E)\\
&= \frac{1}{(2\pi)^2} \int_{-\infty}^{\infty} dk_z \frac{1}{2c_1} \Theta(\Delta E + c_2k_z^2) 
\\
&= \frac{1}{(2\pi)^2 \cdot 2c_1} \int_{-\sqrt{\frac{\Delta E}{-c_2}}}^{\sqrt{\frac{\Delta E}{-c_2}}} dk_z \\
&= \frac{\sqrt{E_0 - E}}{4\pi^2c_1\sqrt{-c_2}} \Theta(E_0 - E).
\label{app:eq:DOS-kp-c1c2-ge0}
\end{aligned}\end{equation}
We have used $\int \delta(f(k))dk=\sum_i 1/|f'(k_i)|$, and $\int k \delta(f(k))dk=\sum_i k_i/|f'(k_i)|$, where $k_i$ are the (simple) roots of $f(k)$. 
The $k_z$-dispersion makes the DOS proportional to $\sqrt{E}$, as shown in \cref{fig:DOS-kp-1} (b).

However, in \ch{MgB2}, we have $c_1>0, c_2>0$, \ie, $\Gamma$ becomes a saddle point, which is decreasing in $k_{x,y}$ plane while increasing along $k_z$, as shown in \cref{fig:DOS-kp-1} (d). In this case, the DOS will be divergent in the $\kk\cdot\pp$ model due to the contributions from large $k_z$. To regularize the unphysical divergence, we restrict the $k_z$ integration to $[-\pi, \pi]$. Then the DOS has the formula
\begin{equation}\begin{aligned}
D(E) &= \frac{1}{(2\pi)^2} \int_{-\pi}^{\pi} dk_z \int_0^{\infty} k_{\parallel} dk_{\parallel} \delta(\Delta E + c_1 k_{\parallel}^2-c_2 k_z^2) \\
&= \frac{1}{(2\pi)^2} \int_{-\pi}^{\pi} dk_z \frac{1}{2c_1} \Theta(c_2k_z^2-\Delta E) \\
&=
\begin{cases}
0 & E>E_0 + c_2\pi^2.\\
\frac{1}{4\pi^2c_1} \left(\pi - \sqrt{\frac{E-E_0}{c_2}}\right) & E_0 < E\le E_0 + c_2\pi^2. \\
\frac{1}{4\pi c_1} & E \le E_0.
\end{cases}
\label{app:eq:DOS_kp_k^2_with_kz_term}
\end{aligned}\end{equation}
In this case, the DOS has $\sqrt{E}$ behavior within range $[E_0, E_0+c_2 \pi^2]$, \ie, between $\Gamma$ saddle point at $E_0$ on the $k_z=0$ plane and the band edge at $E_0+c_2 \pi^2$ on the $k_z=\pi$ plane. When the energy is lower than the $\Gamma$ saddle point at $E_0$, the DOS becomes constant, as shown in \cref{fig:DOS-kp-1} (c). We remark that this DOS can only qualitatively reproduce the \textit{ab initio} DOS in \ch{MgB2}. A more faithful DOS can be obtained by including the 4th-order $\kk\cdot\pp$ term, as we show in the following. 

The last $c_3$ term splits the bands into two quadratic branches in \cref{app:eq:disperion_kp_twoband}. In this case, the DOS is simply the sum of the two bands. Note that $D(E)$ needs to be multiplied by 2 to account for spin degeneracy.

\begin{figure}[htbp]
    \centering
    \includegraphics[width=0.8\linewidth]{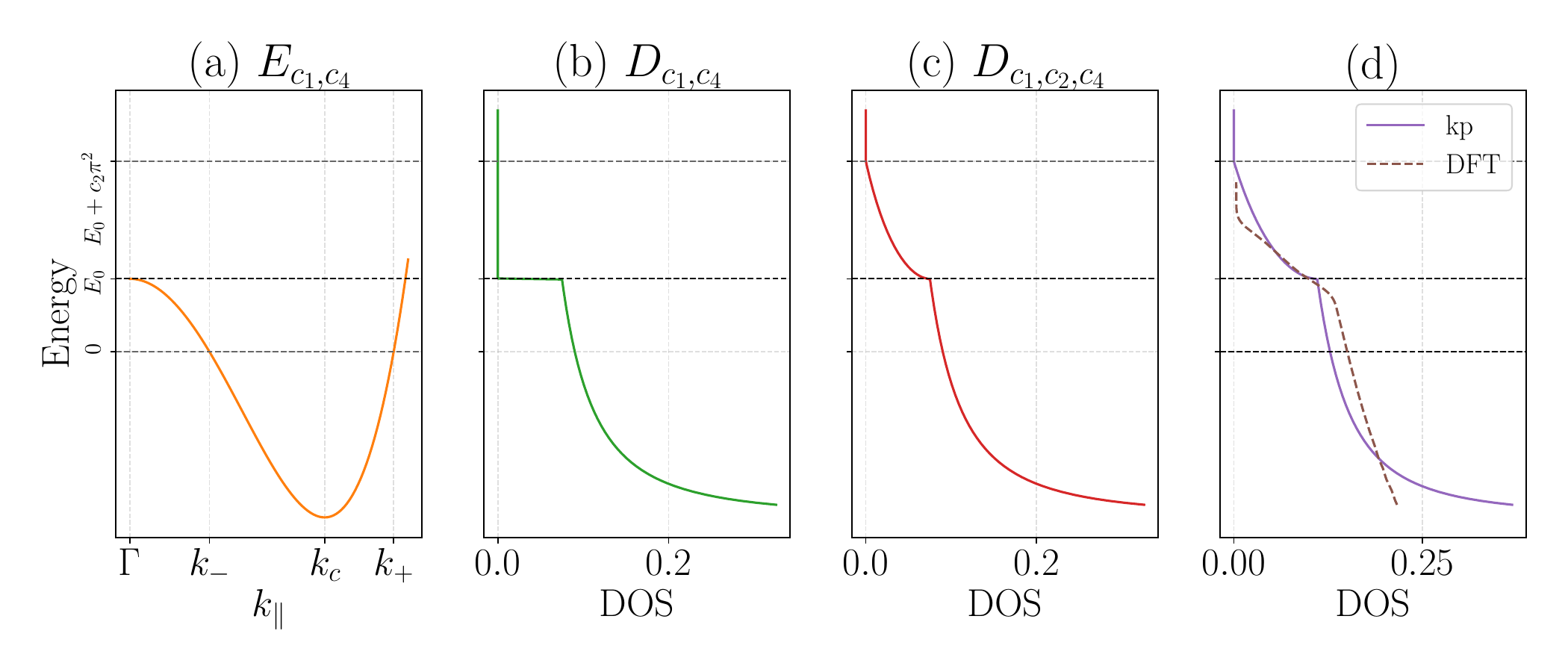}
    \caption{The DOS and dispersion from the $\kk\cdot\pp$ models. (a) The dispersion from $E(\kk)=E_0 - c_1 k_{\parallel}^2 + c_4 k_{\parallel}^4$. For a given energy, we label the two solutions as $k_{\pm}$. Only the $k_-$ branch is kept when evaluating \cref{app:eq:DOS_analytic_4th_order_term}, as the $k_+$ branch is unphysical and only appears in the $\kk\cdot\pp$ model but not in DFT. (b) The DOS from \cref{app:eq:DOS_analytic_4th_order_term}, which is approximately linear for a small energy interval below $E_0$. (c) The DOS from \cref{app:eq:DOS_analytic_4th_order_termP_with_kz}. (d) The comparison of the total DOS from two $\kk\cdot\pp$ bands characterized by parameters in \cref{app:table:fitted_kp_parameters}, and the DOS of $sp^2$ bands in DFT. A reasonably good agreement is observed near $E_f=0$ and $E_0$. }
    \label{app:fig:DOS-kp-2}
\end{figure}

\paragraph{\textbf{Including 4th-order term}. }
In the \textit{ab initio} DOS of \ch{MgB2} shown in \cref{app:fig:DOS_test_convergence}, the DOS becomes linearly increasing when $E<E_0$, where $E_0$ is the energy of the $\Gamma_5^+$ bands at $\Gamma$, which does not agree with the analytic DOS in \cref{app:eq:DOS_kp_k^2_with_kz_term} obtained from the $\kk\cdot\pp$ model. This discrepancy is caused by the higher-order terms in DFT dispersion. To recover the linear DOS for $E<E_0$, we add a fourth-order $\kk\cdot\pp$ term 
\begin{equation}\begin{aligned}
E(\kk) &= E_0 - c_1 k_{\parallel}^2 + c_4 k_{\parallel}^4,
\label{app:eq:kp_4th_order}
\end{aligned}\end{equation}
where $c_1>0, c_4>0$ (see \cref{app:fig:DOS-kp-2} (a) for the dispersion). 
Other fourth-order terms that render the in-plane dispersion anisotropic also exist, but are omitted here for simplicity. 
Note that we require $c_4>0$ so that the DOS increases when $E$ is lower. $c_4$ needs to be small and we only consider small $k_{\parallel}$ near $\Gamma$, otherwise $k_{\parallel}^4$ term will dominate the dispersion. 
For simplicity, we first ignore the $k_z$ dispersion. The DOS is computed as 
\begin{equation}\begin{aligned}
D(E) &= \frac{1}{2\pi} \int_0^{\infty} k_{\parallel}dk_{\parallel} \delta(E-E_0 +c_1 k_{\parallel}^2 - c_4 k_{\parallel}^4) \\
&= \frac{1}{4\pi} \int_0^{\infty} dx \delta(\Delta E -c_1x+c_4x^2) \quad (x=k_{\parallel}^2,\ \Delta E=E_0-E) \\
&= \frac{1}{4\pi\sqrt{c_1^2-4 c_4 (E_0-E)}} \Theta\left(E-(E_0-\frac{c_1^2}{4c_4})\right)\Theta(E_0-E).
\label{app:eq:DOS_analytic_4th_order_term}
\end{aligned}\end{equation}
This DOS is shown in \cref{app:fig:DOS-kp-2} (b). 
When evaluating the last equation in \cref{app:eq:DOS_analytic_4th_order_term}, there are two roots from $\delta(\Delta E -c_1x+c_4x^2)$, \ie, $k^2_{\pm}=\frac{1}{2c_4}\left(c_1\pm \sqrt{c_1^2-4c_4\Delta E}\right)$, and
we only consider $k^2_{-}$ (see \cref{app:fig:DOS-kp-2} (a)). This is because the $k^2_+$ branch corresponds to the dispersion term with large $k_{\parallel}$, which only appears in the $\kk\cdot\pp$ model with the $k_{\parallel}^4$ term but is absent in the realistic band structure. When $c_4(E_0-E)$ is small, we can expand the DOS up to the first order
\begin{equation}\begin{aligned}
D(E) \approx \frac{1}{4\pi c_1}\left(1 + \frac{2c_4}{c_1^2}(E_0-E) \right).
\label{app:eq:DOS_analytic_linear_expansion}
\end{aligned}\end{equation}
Thus, adding a small 4th-order term in the $\kk\cdot\pp$ model leads to a quasi-linear DOS in addition to the constant DOS given by the quadratic term, in agreement with the \textit{ab initio} DOS of \ch{MgB2}.

At last, we add back the (symmetry-allowed lowest-order) $k_z$ term, \ie,
\begin{equation}\begin{aligned}
E(\kk) &= E_0 - c_1 k_{\parallel}^2 + c_2 k_z^2 + c_4 k_{\parallel}^4,
\label{app:eq:kp_4th_order_with_kz}
\end{aligned}\end{equation}
where $c_1,c_2,c_4>0$. 
The corresponding DOS is
\begin{equation}\begin{aligned}
D(E) &= \frac{1}{(2\pi)^2} \int_{-\pi}^{\pi}dk_z \int_0^{\infty} k_{\parallel}dk_{\parallel} \delta(E-E_0 +c_1 k_{\parallel}^2 -c_2k_z^2 - c_4 k_{\parallel}^4) \\
&= \frac{1}{8\pi^2} \int_{-\pi}^{\pi}dk_z  \int_0^{\infty} dx \delta(\Delta E + c_2 k_z^2 -c_1x+c_4x^2) \quad (x=k_{\parallel}^2, \Delta E=E_0-E) \\
&= \frac{1}{8\pi^2} \int_{-\pi}^{\pi}dk_z \frac{1}{\sqrt{c_1^2-4c_4(\Delta E+c_2k_z^2)}} \\ 
&= 
\begin{cases}
0 & E > E_0+c_2\pi^2 \\
\frac{1}{4\pi^2\sqrt{c_1^2-4 c_4 (E_0-E)}} \left(\pi - \sqrt{\frac{E-E_0}{c_2}}\right) & E_0 < E \le E_0+c_2\pi^2 \\
\frac{1}{4\pi\sqrt{c_1^2-4 c_4 (E_0-E)}} & E_0-\frac{c_1^2}{4c_4} <E \le E_0
\end{cases}.
\label{app:eq:DOS_analytic_4th_order_termP_with_kz}
\end{aligned}\end{equation}
The DOS is shown in \cref{app:fig:DOS-kp-2} (c). 
In this case, the DOS has $\sqrt{E}$ behavior in $[E_0,E_0+c_2\pi^2]$, and approximately linear below $E_0$ (note $E_0$ is the energy of the $\Gamma_5^+$ band). 

By fitting to the \textit{ab initio} band structure in \ch{MgB2}, we obtain the $\kk\cdot\pp$ parameters in \cref{app:eq:kp_4th_order_with_kz}, as tabulated in \cref{app:table:fitted_kp_parameters}. The resultant DOS is in agreement with the \textit{ab initio} DOS of \ch{MgB2}, as shown in \cref{app:fig:DOS-kp-2} (d). 

\begin{table}[htbp]
\centering
\begin{tabular}{c|c|c|c|c}
\hline\hline
$\kk\cdot\pp$ parameter/eV & $E_0$ & $c_1$ & $c_4$ & $c_2$ \\ \hline
lower band & 0.43 & 2.24 & 0.30 & 0.07 \\ \hline
upper band & 0.43 & 1.06 & 0.20 & 0.07 \\ \hline\hline
\end{tabular}
\caption{\label{app:table:fitted_kp_parameters} The $\kk\cdot\pp$ parameter in \ch{MgB2} fitted from the DFT band structure. The parameters are defined in \cref{app:eq:kp_4th_order_with_kz}, where $\kk$ is given in direct coordinates and takes values in $[0,2\pi]$. The lower and upper bands correspond to the two $sp^2$ bonding states that contribute to the $\sigma$ Fermi surface.
}
\end{table}

\subsubsection{Analytic estimation of $\lambda$}\label{app:sec:analytic_estimation_lambda}

In the previous two subsections \cref{app:sec:EPC-band-basis} and \cref{app:sec:DOS_analytic}, we obtained the analytic expression for the averaged EPC strength on the FS (\cref{app:eq:EPC-FS-avg-onsite-NN}) and DOS. We now use them to estimate $\lambda$. We restrict the discussion to the 2D case for simplicity, as \ch{MgB2} has weak $k_z$ dispersion. 

Assume the band maximum of the $sp^2$ bonding states at $\Gamma$ has energy $E_0$. We consider the DOS expression from \cref{app:eq:DOS_analytic_4th_order_term}. 
As discussed in \cref{app:eq:EPC-FS-avg-onsite-NN}, the averaged squared EPC strength on the FS can be expressed as a function of Fermi momentum for the two FSs from $\Gamma_5^+$ mode:
\begin{equation}\begin{aligned}
G^2(k_{\parallel}) &= \alpha(\beta - k_{\parallel}^2),
\end{aligned}\end{equation}
where $k_{\parallel}\in[0, \pi]$ is in natural lattice unit with $a=1$, and 
$\alpha, \beta$ are the coefficients determined by the real-space EPC strength of the system. By fitting to the \textit{ab initio} results shown in \cref{app:fig:DFT_EPC_band_basis}, we have 
\begin{equation}\begin{aligned}
&\text{lower band: }\alpha=0.22 \text{ eV}^2, \beta=0.92,\\
&\text{upper band: }\alpha=0.16 \text{ eV}^2, \beta=1.28.
\label{app:eq:fit_alpha_beta_in_G}
\end{aligned}\end{equation}
We note that the fitted parameters deviate from the analytic formula in \cref{app:eq:EPC-FS-avg-onsite-NN}, which was derived considering only the onsite and NN bond EPC and using the simplified electron wavefunction from the NN kagome model.

Consider the $\kk\cdot\pp$ dispersion up to the 4th-order in \cref{app:eq:kp_4th_order} and ignoring $k_z$ dispersion, we express the EPC strength as a function of energy:
\begin{equation}\begin{aligned}
G^2(E) &= \alpha \left[\beta - \frac{c_1-\sqrt{c_1^2 - 4c_4(E_0-E)}}{2c_4}\right]
\end{aligned}\end{equation}

Assume the bond-stretching mode has frequency $\omega_0$. Then we estimate $\lambda$ using \cref{app:eq:lambda_FS_approx} as
\begin{equation}\begin{aligned}
\lambda(E) &= \frac{2}{\omega_0} G^2(E) D(E) \\
&= \frac{\alpha}{2\pi \omega_0} \left(\beta - \frac{c_1-\sqrt{c_1^2 - 4c_4(E_0-E)}}{2c_4}\right) \frac{1}{\sqrt{c_1^2-4 c_4 (E_0-E)}} \\
&= \frac{\alpha}{4\pi \omega_0 c_4} \left(1 + \frac{2\beta c_4-c_1}{\sqrt{c_1^2 - 4c_4(E_0-E)}}\right)
\end{aligned}\end{equation}
where $2$ in the first equation accounts for the spin degeneracy. We then analyze the behavior of $\lambda(E)$:
\begin{itemize}
\item When $E=E_0$, $\lambda(E_0)=\frac{\alpha\beta}{2\pi \omega_0 c_1}$ is finite and independent of 4th-order term $c_4$. 
\item When $E$ is lower and close to $E_0$, we expand $\lambda$ up to the first-order of $E$: 
\begin{equation}\begin{aligned}
\lambda(E) &\approx  \frac{\alpha}{4\pi \omega_0 c_4} \left(1 + \frac{2\beta c_4-c_1}{c_1}(1 + \frac{2c_4}{c_1^2}(E_0-E))\right).
\end{aligned}\end{equation}
Using the fitted values in \cref{app:table:fitted_kp_parameters} and \cref{app:eq:fit_alpha_beta_in_G}, we find that the coefficient $2\beta c_4-c_1$ is negative for both the lower and upper bands from the $sp^2$ FS. Thus $\lambda$ increases when the Fermi level increases, in agreement with the \textit{ab initio} results of \ch{MgB2} in \cref{app:fig:sp2-Tc-rigid-doping-effect}. Physically, this increase in $\lambda$ arises because the EPC strength grows more rapidly than the DOS decreases.

However, the upward trend of $\lambda$ is not universal and depends on the band parameters. For instance, if the fourth-order coefficient $c_4$ is sufficiently large, $\lambda$ can instead decrease as $E_f$ increases. Physically, a larger $c_4$ (coefficient of $k^4$ term) makes the DOS fall more rapidly near the band edge, outpacing the enhancement of the EPC as the Fermi level approaches the degenerate point.

\item When $E_0<E<E_0+c_2\pi^2$ (with $E_0+c_2\pi^2$ the band maximum from the $c_2 k_z^2$ term in \cref{app:eq:kp_4th_order_with_kz}, $\lambda$ decreases and vanishes upon approaching the band maximum.

\end{itemize}

In summary, by including the 4th-order $\kk\cdot\pp$ term and a $k^2$-behaved EPC near $\Gamma$, our analytic estimation of $\lambda$ agrees with the \textit{ab initio} results of \ch{MgB2} shown in \cref{app:fig:sp2-Tc-rigid-doping-effect}.

\subsubsection{Quantum metric of degenerate bands}\label{app:sec:quantum_metric_deg_point}
We briefly discuss the quantum metric for the bands near $E_f$ in \ch{MgB2}. The quantum metric is the real part of the quantum geometric tensor, defined as
\begin{equation}\begin{aligned}
\text{Quantum geometric tensor: }\mathcal{B}_{ij}(\kk) &= 2 \text{Tr}[P(\kk) \partial_i P(\kk) \partial_j P(\kk)], \\
\text{Quantum metric: }
g_{ij}(\kk) &= \text{Re}\mathcal{B}_{ij}(\kk)
=\text{Tr}[\partial_i P(\kk) \partial_j P(\kk)],
\end{aligned}\end{equation}
where $P_{mn}(\kk)=|u_{m\kk}\rangle\langle u_{n\kk}|$ is the projector defined for a chosen set of bands $|u_{m\kk}\rangle$. In practice, it is also useful to define the 2D quantum metric
\begin{equation}\begin{aligned}
g(\kk)=\frac{1}{2}\sum_{i=x,y} \text{Tr}[\partial_i P(\kk) \partial_i P(\kk)]
\end{aligned}\end{equation}

We consider the wavefunction of the NN kagome model defined in \cref{app:eq:kagome_NN_flatband}. The 2D quantum metric is then computed as
\begin{equation}\begin{aligned}
g^{FB}_{\kk} &= 
-\frac{8 \cos ^3\left(\frac{k_x}{2}\right) \cos \left(\frac{\sqrt{3} k_y}{2}\right)+3 \cos (k_x)+\cos \left(\sqrt{3} k_y\right)-12}{8 \left(2 \cos \left(\frac{k_x}{2}\right) \cos \left(\frac{\sqrt{3} k_y}{2}\right)+\cos (k_x)-3\right)^2}.
\end{aligned}\end{equation}
Expansion near $\Gamma$ point leads to
\begin{equation}\begin{aligned}
g^{FB}_{\kk} & \approx \frac{1}{k^2} + \frac{k^2 (5-4 \cos (6 \theta ))}{3840}.
\end{aligned}\end{equation}
For the dispersive band (DB) originating from the $\Gamma_5^+$ mode, we use the $\kk\cdot\pp$ expanded wavefunction near $\Gamma$ in \cref{app:eq:DB_kp_k2}. The 2D quantum metric is computed as
\begin{equation}
    g^{DB}_{\kk} \approx \frac{1}{k^2} + \frac{k^2 (2+ \cos (6 \theta ))}{1536}. 
\end{equation}
It can be seen that the quantum metric is $\frac{1}{k^2}$-divergent for the two $\Gamma_5^+$ bands degenerate at $\Gamma$, which is a generic feature of the quantum metric for (non-accidental) degenerate points.

More generally, consider a 2D degenerate point with $\kk\cdot\pp$ wavefunction $u^{\pm}_{\kk}=\frac{1}{\sqrt{2}}[1,\pm e^{i n\theta}]$, where $n=1$ for a Dirac crossing, and $n=2$ for a quadratic crossing. Since
$\frac{\partial \theta}{\partial k_x}=-\frac{\sin(\theta)}{k},\ \frac{\partial \theta}{\partial k_y}=\frac{\cos(\theta)}{k}$, and $P_{\kk}^{\pm}=\frac{1}{2}\begin{bmatrix}
    1 & \pm e^{-in\theta} \\
    \pm e^{in\theta} & 1
\end{bmatrix}$, we have
\begin{equation}\begin{aligned}
    \partial_{k_x} P^{\pm}_{\kk}= \frac{\partial \theta}{\partial k_x} \frac{\partial P^{\pm}_{\kk}}{\partial \theta} = -\frac{\sin(\theta)}{k} \frac{1}{2}\begin{bmatrix}
        0 & \mp ine^{-in\theta} \\
        \pm in e^{in\theta} & 0
    \end{bmatrix}
    \propto \frac{1}{k}, \\
    \partial_{k_y} P^{\pm}_{\kk}= \frac{\partial \theta}{\partial k_y} \frac{\partial P^{\pm}_{\kk}}{\partial \theta} = \frac{\cos(\theta)}{k} \frac{1}{2}\begin{bmatrix}
        0 & \mp ine^{-in\theta} \\
        \pm in e^{in\theta} & 0
    \end{bmatrix} \propto \frac{1}{k}. 
\end{aligned}\end{equation}
As a result, we have 
\begin{equation}
    g^{\pm}_{\kk} = \frac{n^2}{4k^2}. 
\end{equation}
In \cref{app:sec:EPC_from_GA}, we separate the EPC into energetic and geometric parts using the Gaussian approximation of hoppings, and will evaluate them in \ch{MgB2} in \cref{app:sec:EPC_from_GA_MgB2}. We will show that the geometric part of EPC, containing $\partial_i P_{\kk}$, shows a peak near the $\Gamma_5^+$ degenerate node with approximately $-k^2$ decay and dominates the total EPC.

\subsection{EPC from Gaussian approximation in \ch{MgB2}}\label{app:sec:EPC_from_GA_MgB2}

In this section, we consider the Gaussian approximation (GA) for \ch{MgB2} by using \textit{ab initio} data to extract the EPC tensor and calculate the superconducting properties. Slater-Koster parameters~\cite{slater1954simplified} are introduced to account for the angular dependence of the in-plane $p$ orbitals in \ch{MgB2}.

\subsubsection{Slater-Koster parameterization for $s$ and $p$ orbitals}
Unlike the isotropic $s$ orbital, the $p$ orbitals have angular dependence. The corresponding Gaussian approximation needs to include the angular contributions. To address it, we use the Slater-Koster formalism~\cite{slater1954simplified} to parameterize the hopping integrals by separating the radial and angular parts. Note that the Slater–Koster parameterization satisfies the two-center approximation, in which the hopping depends only on the two orbitals involved. 

\begin{figure}[htbp]
    \centering
    \includegraphics[width=0.7\linewidth]{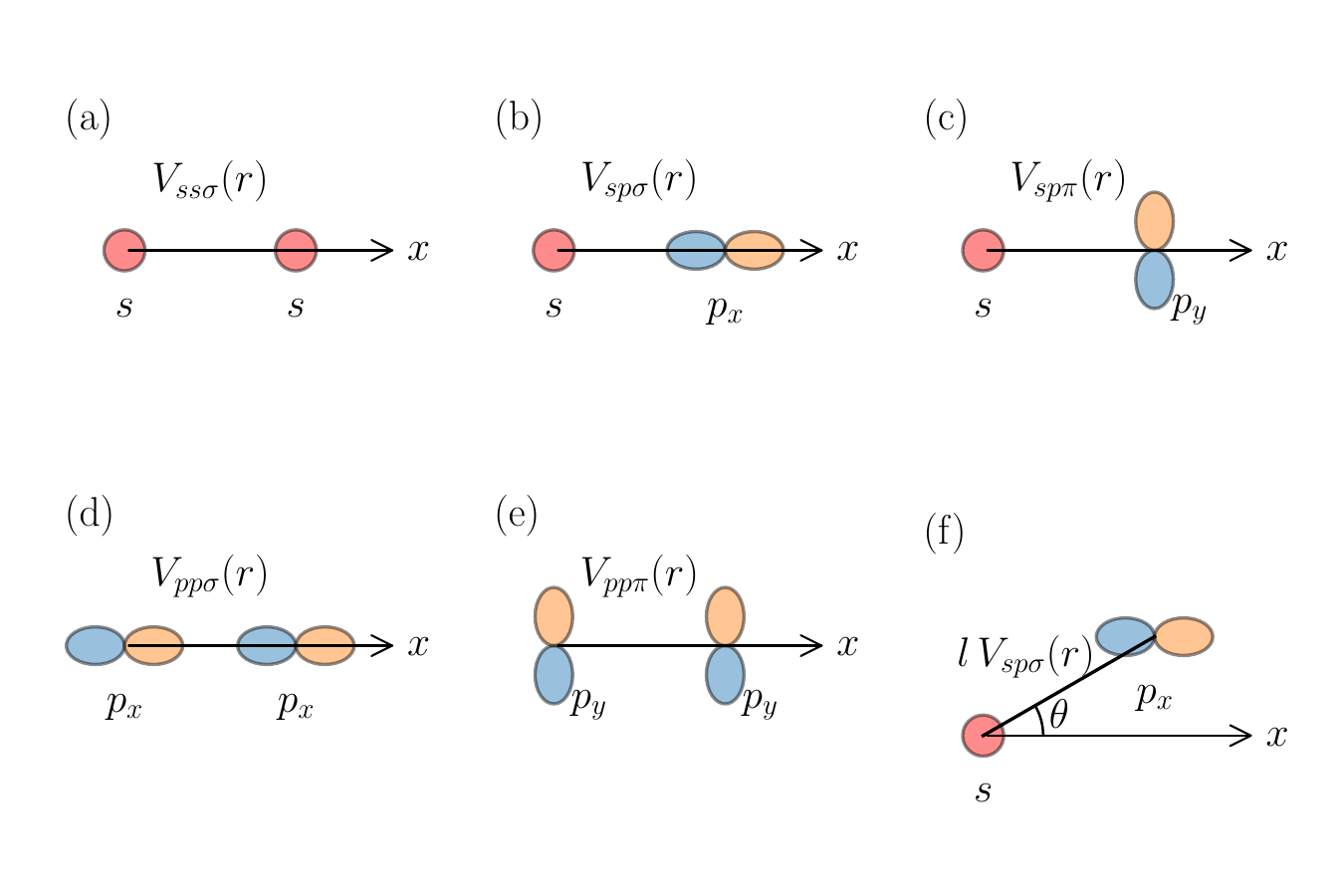}
    \caption{Illustration of hopping integrals in the Slater-Koster (SK) formalism, for (a) $V_{ss\sigma}(r)$, (b) $V_{sp\sigma}(r)$, (c) $V_{sp\pi}(r)$, (d) $V_{pp\sigma}(r)$, (e) $V_{pp\pi}(r)$, and (f) $lV_{sp\sigma}(r)$ with angle $\theta$ and direction cosine $l=\cos(\theta)$. The two orbitals are placed at $\rr_1$ (origin point) and $\rr_2$, with displacement vector $\rr=\rr_2-\rr_1$, with $p_x$ defined w.r.t. the $x$ axis. In (a)-(e), $\rr$ is along the $x$ axis, while in (f), $\rr$ has angle $\theta$ with $x$ axis. 
    }
    \label{fig:SK-orbital-overlap}
\end{figure}

Define the direction cosines for a given vector $\rr=(r_x, r_y, r_z)$:
\begin{equation}\begin{aligned}
l=\frac{r_x}{|\rr|},\quad
m=\frac{r_y}{|\rr|},\quad
n=\frac{r_z}{|\rr|}.
\label{app:eq:direction-cosine}
\end{aligned}\end{equation}
For $s$ and $p$ orbitals, we consider the following radial hopping integrals:
\begin{equation}\begin{aligned}
V_{ss\sigma}(r) &= \langle \psi_{s}(\bm{0}) | H | \psi_s(\rr_x)\rangle, \\
V_{sp\sigma}(r) &= \langle \psi_{s}(\bm{0}) | H | \psi_{p_x}(\rr_x)\rangle, \\
V_{sp\pi}(r) &= \langle \psi_{s}(\bm{0}) | H | \psi_{p_y}(\rr_x)\rangle, \\
V_{pp\sigma}(r) &= \langle \psi_{p_x}(\bm{0}) | H | \psi_{p_x}(\rr_x)\rangle, \\
V_{pp\pi}(r) &= \langle \psi_{p_y}(\bm{0}) | H | \psi_{p_y}(\rr_x)\rangle,
\label{app:eq:hopping_integral_sp}
\end{aligned}\end{equation}
where $\rr_x=(r,0,0)$. The subscript $\sigma$ ($\pi$) denotes the $\sigma$ ($\pi$) bond. Note that $V_{sp\pi}(r)$ is zero when $H$ has a mirror plane that passes the $s$-$p$ bond direction. These hopping integrals are illustrated in \cref{fig:SK-orbital-overlap}.

In the general case, the hopping matrix element $V_{ij\alpha}(\rr)=\langle \psi_{i}(\rr_1) | H | \psi_j(\rr_2)\rangle$ is expressed in terms of direction cosines, as tabulated in
\cref{app:table:SK-parameterization}. 
As an explicit example, we derive the
hopping between $s$ and $p_x$ orbitals, shown in \cref{fig:SK-orbital-overlap}(f). 
Note that in SK parameterization, the hopping matrix elements are evaluated within the two-center approximation (see \cref{app:sec:two-center-approx}), so that the local two-center problem has axial symmetry about the bond direction. In this case, only the component of the $p_x$ orbital projected along $\rr$, namely $\cos \theta p_x$, has a non-vanishing overlap with the $s$ orbital, yielding a hopping $l V_{sp\sigma}(r)$, where $l=\cos\theta$ is the direction cosine of the bond. The orthogonal component $\sin\theta\,p_x$, which is odd under a two-fold rotation along the bond, gives zero contribution.

\begin{table}[htbp]
\centering
\begin{tabular}{c|c|c}
\hline\hline
$i$ & $j$ & $t_{ij}(\mathbf{r})$ \\ \hline
$s$ & $s$ & $V_{ss\sigma}(r)$ \\ \hline
$s$ & $p_x$ & $lV_{sp\sigma}(r)$ \\ \hline
$p_x$ & $p_x$ & $l^2 V_{pp\sigma}(r) + (1 - l^2) V_{pp\pi}(r)$ \\ \hline
$p_x$ & $p_y$ & $lm V_{pp\sigma}(r) - lm V_{pp\pi}(r)$ \\ \hline
$p_x$ & $p_z$ & $ln V_{pp\sigma}(r) - ln V_{pp\pi}(r)$ \\ \hline\hline
\end{tabular}
\caption{The Slater-Koster parameterization for $s$ and $p$ orbitals. $l,m,n$ are direction cosines defined in \cref{app:eq:direction-cosine}. $V_{ijk}(r)$ are radial hopping integrals with no angular dependence.
}
\label{app:table:SK-parameterization}
\end{table}

We then simplify the Slater-Koster parameters based on the geometry of \ch{MgB2}. The two boron atoms are located at two honeycomb sites $\rr_1=\frac{1}{3} \mathbf{a}_1+ \frac{2}{3}\mathbf{a}_2, \rr_2=\frac{2}{3} \mathbf{a}_1+ \frac{1}{3}\mathbf{a}_2$, as shown in \cref{app:fig:crystal_structure}.  
Let $\rr=\rr_2-\rr_1=\frac{1}{3} (\mathbf{a}_1-\mathbf{a}_2)$. 
Since the boron atoms are on the same plane, the direction cosines are simplified into $l=\cos\theta, m=\sin\theta, n=0$, where $\theta=\arccos(\frac{r_x}{|\rr|})=-\frac{\pi}{6}$. 
Consider the hopping between the $s,p_x,p_y$ orbitals of two borons. They have the form
\begin{equation}\begin{aligned}
t_{B_s^1, B_s^2}(\rr) &= V_{ss\sigma}(r),\\ 
t_{B_{s}^1, B_{p_x}^2}(\rr) &= \cos\theta V_{sp\sigma}(r) = -t_{B_{p_x}^1, B_{s}^2}(\rr),\\ 
t_{B_{s}^1, B_{p_y}^2}(\rr) &= \sin\theta V_{sp\sigma}(r) = -t_{B_{p_y}^1, B_{s}^2}(\rr),\\
t_{B_{p_x}^1, B_{p_x}^2}(\rr) &= \cos^2\theta V_{pp\sigma}(r) + \sin^2\theta V_{pp\pi}(r),\\
t_{B_{p_x}^1, B_{p_y}^2}(\rr) &= \cos\theta \sin\theta \left(V_{pp\sigma}(r) - V_{pp\pi}(r) \right) = t_{B_{p_y}^1, B_{p_x}^2}(\rr),\\
t_{B_{p_y}^1, B_{p_y}^2}(\rr) &= \sin^2\theta V_{pp\sigma}(r) + \cos^2\theta V_{pp\pi}(r).
\label{app:eq:SK-hoppings-mgb2}
\end{aligned}\end{equation}

\subsubsection{EPC from Gaussian approximation}

With the Slater-Koster parameterization, we use the Gaussian approximation to extract the real-space EPC. 

To begin with, we assume the radial hopping integrals in the Slater-Koster parameters have the Gaussian form
\begin{equation}
    V_{ij\alpha}(r)=V_{ij\alpha}^0 \exp\left( -\gamma_{ij\alpha} \frac{r^2}{2} \right)
\label{app:eq:gaussian_hop}
\end{equation}
where $i,j$ are orbital indices for $s, p_x, p_y$ orbitals, and $\alpha=\sigma,\pi$ denotes the type of hoppings. 

With the analytical expression of the hopping integrals, we evaluate the EPC generated from them using the Gaussian approximation~\cite{yu2024non}. In the Gaussian approximation, the EPC is generated by taking the derivative of the analytical hoppings:
\begin{equation}
    g^{ij,\mu} (\rr_0) = \left. \frac{\partial}{\partial x_{\mu}} t_{ij}(\rr) \right|_{\rr=\rr_0}
    \label{app:eq:EPC_from_hopping_derivative}
\end{equation}
For the radial hopping integral with the Gaussian form, the corresponding EPC has the analytic expression
\begin{equation}
\begin{aligned}
    \frac{\partial}{\partial r} V_{ij\alpha}(r) &= -\gamma_{ij\alpha} r V_{ij\alpha}(r), \quad 
    \frac{\partial}{\partial x_{\mu}} V_{ij\alpha}(r)
    = -\gamma_{ij\alpha} x_{\mu} V_{ij\alpha}(r).
\end{aligned}
\end{equation}
As the Slater-Koster parameters are written in the polar coordinates, we perform a change of variable from polar to cartesian coordinates in the derivative using $r=\sqrt{x^2+y^2}, \theta=\arctan{\frac{y}{x}}$:
\begin{equation}
    \frac{\partial}{\partial x_i}=   \frac{\partial r}{\partial x_i} \frac{\partial}{\partial r} + \frac{\partial \theta}{\partial x_i} \frac{\partial}{\partial \theta}, 
    \quad \text{where }
    \frac{\partial r}{\partial x_i} = \frac{x_i}{r}, \quad
    \frac{\partial \theta}{\partial x} = -\frac{y}{r^2},\quad  
    \frac{\partial \theta}{\partial y} = \frac{x}{r^2}.
    \label{app:eq:partial_derivative_expansion}
\end{equation}
More compactly, we have 
$\frac{\partial \theta}{\partial x_i} =\epsilon_{ji}\frac{x_j}{r^2}$, where $\epsilon_{12}=1,\epsilon_{21}=-1,\epsilon_{ii}=0$.

Plugging \cref{app:eq:SK-hoppings-mgb2} and \cref{app:eq:partial_derivative_expansion} into \cref{app:eq:EPC_from_hopping_derivative}, we have the expression for EPC:
\begin{equation}
\begin{aligned}
g_{B_s^1, B_s^2, B^2_{x_{\mu}}}(\rr) 
&= -\gamma_{ss\sigma} x_{\mu} t_{B_s^1, B_s^2}(\rr), 
\\
g_{B_s^1, B_{p_x}^2, B^2_{x_{\mu}}}(\rr) 
&=-\left(\gamma_{sp\sigma} x_{\mu} + \epsilon_{\nu\mu} \frac{x_{\nu}}{r^2} \tan\theta \right) t_{B_s^1, B_{p_x}^2}(\rr) 
=-g_{B_{p_x}^1, B_{s}^2, B^2_{x_{\mu}}}(\rr) , \\
g_{B_s^1, B_{p_y}^2, B^2_{x_{\mu}}}(\rr) 
&=-\left(\gamma_{sp\sigma} x_{\mu} - \epsilon_{\nu\mu}\frac{x_{\nu}}{r^2} \cot\theta \right) t_{B_s^1, B_{p_x}^2}(\rr)
=-g_{B_{p_y}^1, B_{s}^2, B^2_{x_{\mu}}}(\rr) , \\
g_{B_{p_x}^1, B_{p_x}^2, B^2_{x_{\mu}}}(\rr) 
&= 
\left(-\gamma_{pp\sigma} x_{\mu}  - \epsilon_{\nu\mu} \frac{2x_{\nu}}{r^2} \tan\theta \right) \cos^2\theta V_{pp\sigma}(r)
+ \left(-\gamma_{pp\pi} x_{\mu}  + \epsilon_{\nu\mu} \frac{2x_{\nu}}{r^2} \cot\theta \right) \sin^2\theta V_{pp\pi}(r) \\
g_{B_{p_x}^1, B_{p_y}^2, B^2_{x_{\mu}}}(\rr) 
&= 
\left(-\gamma_{pp\sigma} x_{\mu}  + \epsilon_{\nu\mu} \frac{x_{\nu}}{r^2} (\cot\theta -\tan\theta) \right) \cos\theta \sin\theta V_{pp\sigma}(r) \\
&- \left(-\gamma_{pp\pi} x_{\mu}  + \epsilon_{\nu\mu} \frac{x_{\nu}}{r^2} (\cot\theta -\tan\theta) \right) \cos\theta\sin\theta V_{pp\pi}(r)
= g_{B_{p_y}^1, B_{p_x}^2, B^2_{x_{\mu}}}(\rr) \\
g_{B_{p_y}^1, B_{p_y}^2, B^2_{x_{\mu}}}(\rr) 
&= 
\left(-\gamma_{pp\sigma} x_{\mu}  + \epsilon_{\nu\mu} \frac{2x_{\nu}}{r^2} \cot\theta  \right) \sin^2\theta V_{pp\sigma}(r)
+ \left(-\gamma_{pp\pi} x_{\mu}  - \epsilon_{\nu\mu} \frac{2x_{\nu}}{r^2} \tan\theta  \right) \cos^2\theta V_{pp\pi}(r)
\end{aligned}
\end{equation}
Note that the GA-derived EPC always obeys the two-center approximation (see \cref{app:sec:two-center-approx}), whereas the DFT-derived EPC does not necessarily satisfy this constraint. 
Numerically, taking $g_{B_s^1, B_{p_x}^2, B^2_{x}}(\rr)$ with $\gamma_{sp\sigma}=$\SI{0.54}{\angstrom}$^{-2}$, $a=$\SI{3.07}{\angstrom}, we find the radial is about 5 times larger than the angular part.

\subsubsection{Momentum-space EPC from GA}

In the GA for $p$ orbitals, the angular dependence of the Slater–Koster (SK) hoppings renders the EPC nontrivial—it is not just a simple spatial derivative of the Hamiltonian but has angular parts. In what follows, we simplify the GA EPC for $p$ orbitals and derive its momentum-space form.

To begin with, we consider two atoms at $\rr_1$ and $\rr_2$ with angle $\theta=\arccos(\frac{r_x}{|\rr|})$, where $\rr=\rr_2-\rr_1$ (see \cref{fig:SK-orbital-overlap}(f)). Each atom has $s, p_x, p_y$ orbitals. 
We first consider the special case where two atoms are aligned on the $x$ axis, i.e., $\rr=(r_x,0,0),\theta=0$ (see \cref{fig:SK-orbital-overlap}(a)). Then the hopping matrix between the $s, p_x, p_y$ orbitals from two atoms reads
\begin{equation}\begin{aligned}
t_0(\rr) &= 
\begin{bmatrix}
    V_{ss\sigma}(r) & V_{sp\sigma}(r) & 0 \\
    -V_{sp\sigma}(r) & V_{pp\sigma}(r) & 0 \\
    0 & 0 & V_{pp\pi}(r)
\end{bmatrix}.
\end{aligned}\end{equation}
Then we consider the general case when $\rr_{1,2}$ are on the same $xy$ plane. The hopping matrix has the form
\begin{equation}\begin{aligned}
t(\rr) &= R(\theta) t_0(\rr) R^\dagger(\theta) \\
&= \begin{bmatrix}
V_{ss\sigma}(r) & \cos\theta V_{sp\sigma}(r) & \sin\theta V_{sp\sigma}(r) \\
-\cos\theta V_{sp\sigma}(r) & \cos^2\theta V_{pp\sigma}(r) + \sin^2\theta V_{pp\pi}(r) & \cos\theta \sin\theta \left(V_{pp\sigma}(r) - V_{pp\pi}(r) \right) \\
-\sin\theta V_{sp\sigma}(r) & \cos\theta \sin\theta \left(V_{pp\sigma}(r) - V_{pp\pi}(r) \right) & \sin^2\theta V_{pp\sigma}(r) + \cos^2\theta V_{pp\pi}(r)
\end{bmatrix}
\\
R(\theta) &= 
\begin{bmatrix}
    1 & 0 & 0 \\
    0 & \cos\theta & -\sin\theta \\
    0 & \sin\theta & \cos\theta
\end{bmatrix}.
\label{app:eq:SK-param-Rtheta-form}
\end{aligned}\end{equation}

We then consider the derivative of the hopping matrix:
\begin{equation}\begin{aligned}
\partial_{x_{\mu}} t(\rr)
&= \left[\partial_{x_{\mu}} R(\theta)\right] t_0(\rr) R^\dag(\theta)
+ R(\theta) \left[\partial_{x_{\mu}} t_0(\rr) \right] R^\dagger(\theta)
+ R(\theta) t_0(\rr) \left[\partial_{x_{\mu}} R^\dagger(\theta)\right].
\end{aligned}\end{equation}
Using the GA in \cref{app:eq:gaussian_hop}, the derivative of $t_0(\rr)$ with respect to $x_\mu$ is
\begin{equation}\begin{aligned}
\partial_{x_{\mu}} t_{0,ij}(\rr) &= 
-\gamma_{ij} x_{\mu} t_{0,ij}(r),
\end{aligned}\end{equation}
where the decaying factor $\gamma_{ij}>0$ depends on the type of the hopping. 
The derivative of the rotation matrix is
\begin{equation}\begin{aligned}
\partial_{x_{\mu}} R(\theta) &=
\sum_{\nu} \epsilon_{\nu\mu} \frac{x_\nu}{r^2} \partial_{\theta} R(\theta) = \sum_{\nu} \epsilon_{\nu\mu} \frac{x_\nu}{r^2} 
\begin{bmatrix}
    0 & 0 & 0 \\
    0 & -\sin\theta & -\cos\theta \\
    0 & \cos\theta & -\sin\theta
\end{bmatrix} 
= \sum_{\nu} \epsilon_{\nu\mu} \frac{x_\nu}{r^2} (-i\tilde{\sigma}_y) R(\theta),
\quad
\tilde{\sigma}_y=
\begin{bmatrix}
    0 & 0\\
    0 & \sigma_y
\end{bmatrix},\\
\partial_{x_{\mu}} R^\dagger(\theta) &= \left(\partial_{x_{\mu}} R(\theta) \right)^\dagger = 
 -\sum_{\nu} \epsilon_{\nu\mu} \frac{x_\nu}{r^2} R^\dag(\theta)(-i\tilde{\sigma}_y).
\end{aligned}\end{equation}
By assuming $\gamma_{pp\sigma}=\gamma_{pp\pi}$, we arrive at 
\begin{equation}\begin{aligned}
\partial_{x_{\mu}} t_{ij}(\rr) &=
-\gamma_{ij} x_{\mu} t_{ij}(\rr) + \sum_{\nu} \epsilon_{\nu\mu} \frac{x_\nu}{r^2}
\left[-i\tilde{\sigma}_y, t(\rr)\right]_{ij}.
\end{aligned}\end{equation}
Following the derivation in \cref{app:sec:EPC_from_GA}, we obtain the real-space EPC tensor
\begin{equation}\begin{aligned}
g^{i j,l\mu}_{\RR_e,\RR_p} 
=& \left\{
-\gamma_{ij} (\RR_e+\rr_j-\rr_i)_{\mu} t_{ij}(\RR_e+\rr_j-\rr_i) + \sum_{\nu} \epsilon_{\nu\mu} \frac{(\RR_e+\rr_j-\rr_i)_\nu}{|\RR_e+\rr_j-\rr_i|^2}
\left[-i\tilde{\sigma}_y, t(\RR_e+\rr_j-\rr_i)\right]_{ij}
\right\}\times \\
& \times(\delta_{j,l}\delta_{\RR_e,\RR_p} - \delta_{i,l}\delta_{\RR_p,\bm{0}}) 
\end{aligned}\end{equation}

We then FT to the momentum space. Consider a set of hoppings with the same distance, \ie, $|\RR_e+\rr_j-\rr_i|\equiv r_{ij}$, for $\RR_e\in\{\RR_e||\RR_e+\rr_j-\rr_i|= r_{ij} \}$. For example, when $\rr_i=(\frac{1}{3},\frac{2}{3}), \rr_j=(\frac{2}{3},\frac{1}{3})$, and $r_{ij}=\frac{a}{\sqrt{3}}$ is the NN distance for two orbitals at honeycomb sites. Then $\RR_e\in\{(0,0), (0,1), (-1,0)\}$, corresponding to three NN honeycomb sites $\rr_j+\RR_e$ in the neighboring unit cells for $\rr_i$. As a result, 
\begin{equation}\begin{aligned}
\sum_{\RR_e \in\{\RR_e||\RR_e+\rr_j-\rr_i|= r_{ij} \}} \left[ \partial_{x_{\mu}} t_{ij}(\RR_e + \rr_j-\rr_i)\right] e^{i\kk\cdot (\RR_e + \rr_j-\rr_i)} &=
i\gamma_{ij} \partial_{k_\mu} t_{ij}(\kk) -\sum_{\nu} \frac{\epsilon_{\nu\mu}}{r_{ij}^2} \left[-i\tilde{\sigma}_y, i\partial_{k_\nu} t(\kk)\right]_{ij},
\label{app:eq:SK-hopping-FT}
\end{aligned}\end{equation}
Note that in the second term, $r_{ij}$ appears as a constant that is independent of $\RR_e$, as we make the assumption that only the $\RR_e$ that satisfies $|\RR_e+\rr_j-\rr_i|\equiv r_{ij}$ is considered. 

Thus
\begin{equation}\begin{aligned}
g_{\kk,\qq}^{i j,l\mu} =& \sum_{\RR_{e},\RR_{p}} g_{\RR_e,\RR_{p}}^{i j,l\mu} e^{i\kk\cdot(\RR_e+\rr_j-\rr_i)+i\qq\cdot(\RR_{p}+\rr_l-\rr_i)} \\
=&
\sum_{\RR_{e},\RR_{p}} 
 \left\{
-\gamma_{ij} (\RR_e+\rr_j-\rr_i)_{\mu} t_{ij}(\RR_e+\rr_j-\rr_i) + \sum_{\nu} \epsilon_{\nu\mu} \frac{(\RR_e+\rr_j-\rr_i)_\nu}{|\RR_e+\rr_j-\rr_i|^2}
\left[-i\tilde{\sigma}_y, t(\RR_e+\rr_j-\rr_i)\right]_{ij}
\right\} \times \\
&\times (\delta_{j,l}\delta_{\RR_e,\RR_p} - \delta_{i,l}\delta_{\RR_p,\bm{0}}) 
e^{i\kk\cdot(\RR_e+\rr_j-\rr_i)+i\qq\cdot(\RR_{p}+\rr_l-\rr_i)} 
\\
=& i\gamma_{ij} \left(
\partial_{k_\mu} t_{ij}(\kk+\qq)\delta_{jl} - \partial_{k_\mu} t_{ij}(\kk)\delta_{il} 
\right) - \sum_{\nu} \frac{\epsilon_{\nu\mu}}{r_{ij}^2} \left\{ \left[\tilde{\sigma}_y,  
\partial_{k_\nu} t(\kk+\qq)\right]_{ij} \delta_{jl} - \left[\tilde{\sigma}_y, \partial_{k_\nu} t(\kk)\right]_{ij}\delta_{il} \right\},
\end{aligned}\end{equation}

We then consider the 6-orbital TB Hamiltonian in \ch{MgB2} from the $s,p_x,p_y$ orbitals of two boron atoms, located at $\rr_A=(\frac{1}{3},\frac{2}{3}),\rr_B=(\frac{2}{3},\frac{1}{3})$ sublattices (same as defined in \cref{app:eq:atomic_coord}). The TB Hamiltonian has the form
\begin{equation}\begin{aligned}
\hat{H} &= 
\begin{bmatrix}
    \hat{h}_{AA} & \hat{h}_{AB} \\
    \hat{h}_{AB}^\dag & \hat{h}_{BB}
\end{bmatrix},
\end{aligned}\end{equation}
where $\hat{h}_{AB}$ contains the NN hopping between two sublattices, and $\hat{h}_{AA/BB}$ contains the onsite energy terms and the NNN hoppings between the same sublattice. For each Hamiltonian block $\hat{h}_{ij}$, we explicitly require that only the sets of hoppings with the same distance are considered, so that \cref{app:eq:SK-hopping-FT} is applicable. 
Longer-range hoppings can be neglected as their values are small in DFT. Both the NN and NNN hoppings can be parameterized using the SK parameters defined in \cref{app:eq:SK-param-Rtheta-form}. Define 
\begin{equation}\begin{aligned}
\Sigma_y=\text{Diag}\left[0,\sigma_y,0,\sigma_y\right].
\end{aligned}\end{equation}
Then the EPC from the 6-orbital TB Hamiltonian in \ch{MgB2} takes the form
\begin{equation}\begin{aligned}
g_{\kk,\qq}^{i j,l\mu} 
=& i\gamma_{ij} \left(
\partial_{k_\mu} h_{ij}(\kk+\qq)\delta_{jl} - \partial_{k_\mu} h_{ij}(\kk)\delta_{il} 
\right) - \sum_{\nu} \frac{\epsilon_{\nu\mu}}{r_{ij}^2} \left\{ \left[\tilde{\Sigma}_y,  
\partial_{k_\nu} h(\kk+\qq)\right]_{ij} \delta_{jl} - \left[\tilde{\Sigma}_y, \partial_{k_\nu} h(\kk)\right]_{ij}\delta_{il} \right\}. \\
=& \left(f_\mu^{ij}(\kk+\qq) \delta_{jl} - f_\mu^{ij}(\kk) \delta_{il} \right) - \sum_{\nu} \frac{\epsilon_{\nu\mu}}{i\gamma_{ij} r_{ij}^2} 
\left\{ \left[\tilde{\sigma}_y,  
f_{\nu}(\kk+\qq)\right]_{ij} \delta_{jl} - \left[\tilde{\sigma}_y, f_{\nu}(\kk)\right]_{ij} \delta_{il} \right\},
\label{app:eq:EPC_GA_spxpy_orbital}
\end{aligned}\end{equation}
where $f^{ij}_{\mu}(\kk)=i\gamma_{ij} \partial_{k_{\mu}} h_{ij}(\kk)$, and $r_{ij}$ is the distance between $ij$ atoms. Note that the NN and NNN hoppings are independent and appear in different matrix blocks of $\hat{H}$ (\ie, $\hat{h}_{AB}$ and $\hat{h}_{AA/BB}$). 
In \cref{app:eq:EPC_GA_spxpy_orbital}, the first term in the EPC is the same as the EPC from GA of $s$ orbitals, as defined in \cref{app:eq:EPC_GA_s_orbital}. The second term, however, arises from the angular dependence in the GA of $p$ orbitals. 
We observe that at $\qq=\bm{0}$, the electron-diagonal part always has zero EPC, \ie, $g_{\kk,\bm{0}}^{ii,l\mu}=\bm{0}$. 

Further separation of the energetic and geometric parts in the EPC is straightforward by defining $f^{\text{E}}_{\mu}(\kk)$ and $f^{\text{geo}}_{\mu}(\kk)$, as discussed in \cref{app:eq:f_E_geo}.

\subsubsection{Fitting GA from \textit{ab initio} data}

The \textit{ab initio} real-space EPC is discussed in \cref{app:sec:EPC_3s_basis}. 
\cref{app:table:epc_DFT_spxpy} tabulates the \textit{ab initio} real-space NN EPC in the home unit cell, obtained from an 8-orbital Wannier model of boron $s,p$ orbitals (two $p_z$ orbitals are omitted). We observe that many terms break the two-center form (see definition in \cref{app:sec:two-center-approx}), which are beyond the GA. There also exist large electron onsite-type EPC terms in \textit{ab initio} which are beyond the two-center approximation, as tabulated in \cref{app:table:epc_DFT_spxpy_onsite} for $(s,p_x,p_y)$ orbital basis.

Although onsite EPC and two-center–breaking terms exist in \ch{MgB2}, we restrict the GA fit to the two-center–preserving contributions, especially the dominant NN bond EPC. The Gaussian parameters are obtained by fitting the \textit{ab initio} NN and NNN hoppings and the first derivatives of NN hoppings (i.e., the EPC) in the $(s,p_x,p_y)$ orbital basis. 

The fitted GA parameters are listed in \cref{app:table:fitted-GA-param-spxpy}, with the corresponding real-space EPC tabulated in \cref{app:table:epc_GA_spxpy}. Compared with the \textit{ab initio} real-space EPC in \cref{app:table:epc_DFT_spxpy}, we observe a good agreement in the two-center preserving terms. After transforming into the $3s$ orbital basis, the dominant NN bond EPC terms (defined in \cref{app:eq:g_sp2_basis_transformation}) also agree with DFT, \ie, $g_{\bm{0},\bm{0}}^{B_{s_1}^1, B_{s_2}^2, B_{x}^2}=7.17$ \SI{}{eV/\angstrom} in GA, and 7.41 \SI{}{eV/\angstrom} in DFT.

\begin{table}[htbp]
\centering
\begin{tabular}{c|c|c|c|c}
\hline\hline
Parameter & $V_{ss\sigma}(r)$ & $V_{sp\sigma}(r)$ & $V_{pp\sigma}(r)$ & $V_{pp\pi}(r)$ \\ \hline
$V^0_{ij\alpha}$/eV & -6.051 &  9.395 &  14.378 &  -6.409 \\\hline
$\gamma_{ij\alpha}$/\AA$^{-2}$ & 0.527 &   0.539 &   0.799 &  0.799 \\ \hline\hline
\end{tabular}
\caption{\label{app:table:fitted-GA-param-spxpy} The fitted Slater-Koster parameters in \ch{MgB2}. They are fitted using the Gaussian form defined in \cref{app:eq:gaussian_hop} to the \textit{ab initio} 8-orbital Wannier model constructed from the Boron $s$ and $p$ orbitals. The fitted onsite energies for the $s$ and $p$ orbitals are -1.186 eV and 3.324 eV, respectively. 
}
\end{table}

\begin{table}[htbp]
\centering
\begin{tabular}{c|c|c|c|c|c|c|c|c|c}
\hline\hline
EPC (eV/\AA) 
 & $(B_s^1, B_s^2)$ & 
 {\color{red}$(B_s^1, B_{p_x}^2)$} & 
 {\color{red}$(B_s^1, B_{p_y}^2)$} & 
 {\color{red}$(B_{p_x}^1, B_{s}^2)$} & 
 $(B_{p_x}^1, B_{p_x}^2)$ & 
 {\color{red}$(B_{p_x}^1, B_{p_y}^2)$} & 
 {\color{red}$(B_{p_y}^1, B_{s}^2)$} & 
 {\color{red}$(B_{p_y}^1, B_{p_x}^2)$} & $(B_{p_y}^1, B_{p_y}^2)$ \\ \hline
 $B_{x}^1$ & 
-2.13 &  2.32 & -2.63 & -2.32 &  1.76 & -3.95 &  2.63 & -3.95 &  1.00 \\ \hline 
$B_{y}^1$ & 
1.23 & -2.63 & -0.73 &  2.63 & -4.31 &  0.38 &  0.73 &  0.38 &  2.71 \\ \hline 
$B_{x}^2$ & 2.13 & -2.32 &  2.63 &  2.32 & -1.76 &  3.95 & -2.63 &  3.95 & -1.00 \\ \hline
$B_{y}^2$ & -1.23 & 2.63 &  0.73 & -2.63 &  4.31 & -0.38 & -0.73 & -0.38 & -2.71 \\
 \hline\hline
\end{tabular}
\caption{\label{app:table:epc_GA_spxpy} The real-space EPC in the $(s, p_x, p_y)$ basis obtained from the Gaussian approximation (GA). The notation is the same as in \cref{app:table:epc_DFT_spxpy}, and only the intra-home-unit-cell NN EPC terms are tabulated. 
We mark in red for EPC columns that break the two-center approximation in \textit{ab initio} (see \cref{app:sec:two-center-approx}), which cannot be well fitted in GA.
}
\end{table}

\begin{figure}[htbp]
    \centering
    \includegraphics[width=1\textwidth]{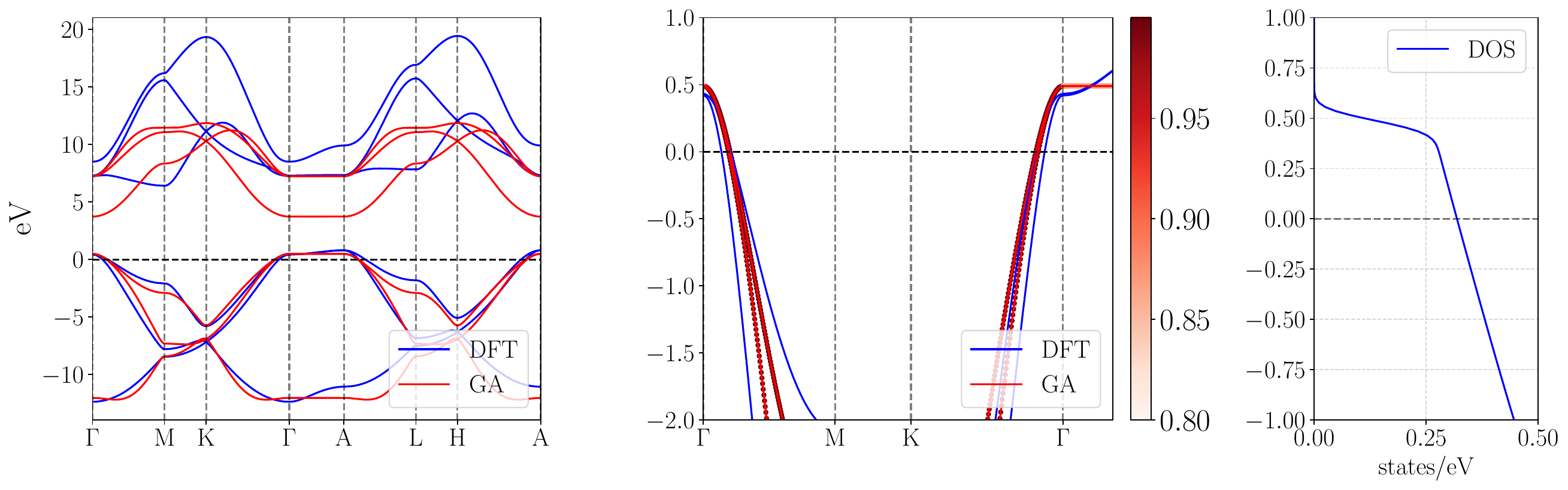}
    \caption{\label{app:fig:GA-band-DOS-3s} The band structure and DOS from Gaussian approximation in \ch{MgB2}. The first plot shows the comparison of DFT and GA dispersion in a large energy scale. The second plot shows the bands near $E_f$, with the colorbar being the overlap between DFT and GA eigenstates. The overlap is larger than 95\% near $E_f$. 
    The last plot is the DOS from GA (spin degree of freedom included), which also shows good agreement with the DOS of the $sp^2$ bands in DFT (see \cref{app:fig:sp2-Tc-DFT-doping-effect}). 
    }
\end{figure}

With the fitted GA parameters, we calculate the corresponding electron band structure and DOS, as shown in \cref{app:fig:GA-band-DOS-3s}. A good agreement with DFT results is observed for both dispersion and wavefunction near $E_f$.

\begin{figure}[htbp]
    \centering
    \includegraphics[width=0.4\textwidth]{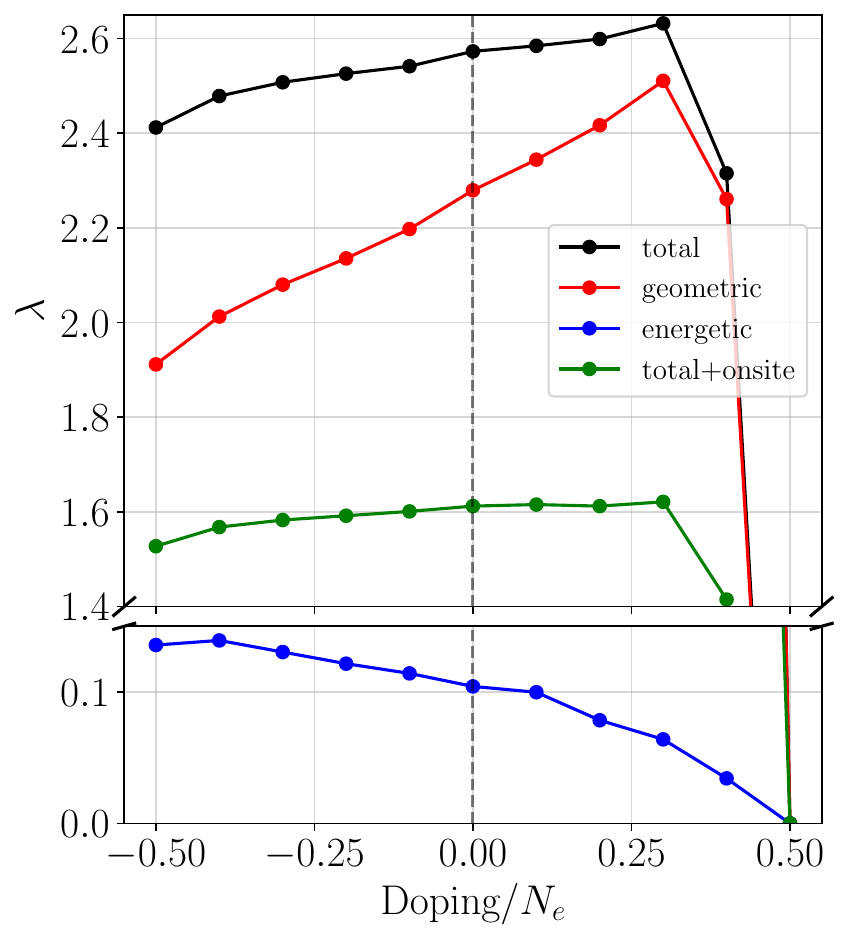}
    \caption{\label{app:fig:GA-lambda-doping-result} Electron–phonon coupling constant $\lambda$ obtained from the Gaussian approximation (GA) as a function of electron doping. We show the total $\lambda$, its geometric ($\lambda^{\text{geo}}$) and energetic ($\lambda^{\text{E}}$) components, and the total after reinstating the onsite EPC terms extracted from DFT (GA + onsite). Adding the onsite terms lowers $\lambda$ from about 2.5 to 1.5. Both the total and geometric contributions rise with light electron doping and then fall, mirroring the fully \textit{ab initio} trend in \cref{app:fig:sp2-Tc-rigid-doping-effect}, whereas the energetic component decreases monotonically, tracking the density of states.
    }
\end{figure}

We next evaluate the superconducting properties using the GA parameters. For the phonons, we retain only the boron in-plane modes, as defined in the simplified phonon model in \cref{app:eq:H-phonon-Bxy}. 
Following Ref.~\cite{yu2024non} and the discussion in \cref{app:sec:EPC_from_GA}, we separate the EPC into geometric and energetic parts (see detailed discussion in \cref{app:sec:EPC_from_GA}): 
\begin{equation}\begin{aligned}
\lambda &= \lambda^{\text{geo}} + \lambda^{\text{E}} + \lambda^{\text{geo-E}}.
\end{aligned}\end{equation}
where the expressions for the three terms are given near \cref{app:eq:lambda_separation}. 
The resultant EPC strength $\lambda$ as a function of the doping level is shown in \cref{app:fig:GA-lambda-doping-result}. 
Both the total EPC $\lambda$ and its geometric part $\lambda^{\text{geo}}$ rise and then fall with electron doping, closely matching the \textit{ab initio} trend in \cref{app:fig:sp2-Tc-rigid-doping-effect}. By contrast, the energetic contribution $\lambda^{\text{E}}$ decreases monotonically, tracking the density of states. At zero doping, we find $\lambda^{\text{geo}}/\lambda=88.6\%$, confirming that the EPC in \ch{MgB2} is dominated by geometric contributions—consistent with the value reported in Ref.~\cite{yu2024non}.

Quantitatively, however, the GA overestimates the absolute coupling: the computed $\lambda$ is roughly three times larger than the \textit{ab initio} value. This discrepancy arises because (i) two-centre–breaking EPC terms present in the \textit{ab initio} Hamiltonian are absent in the GA, and (ii) onsite EPC contributions are not included in the current GA parametrization.

Because the GA parametrization omits the onsite EPC, we restore the onsite EPC terms at their \textit{ab initio} values when evaluating $\lambda$. The updated results are plotted in \cref{app:fig:GA-lambda-doping-result}. Re-introducing the onsite terms lowers $\lambda$ from approximately 2.5 to 1.5, leading to a closer agreement with the \textit{ab initio} value around 0.75. 
The reduction in $\lambda$ after adding the onsite terms is what one expects from the relation derived in \cref{app:eq:epc_3s_bonding_basis}, where the effective EPC in the $sp^{2}$ bonding basis is $g_{\text{eff}}=\sqrt{2}g_0-g_1$, with $g_{0}$ and $g_{1}$ denoting the dominant bond and onsite EPC terms, respectively, in the $3s$ orbital basis. Adding the onsite contribution $g_{1}$ therefore reduces $g_{\text{eff}}$ and hence the total $\lambda$.

\begin{figure}[htbp]
    \centering
    \includegraphics[width=0.7\textwidth]{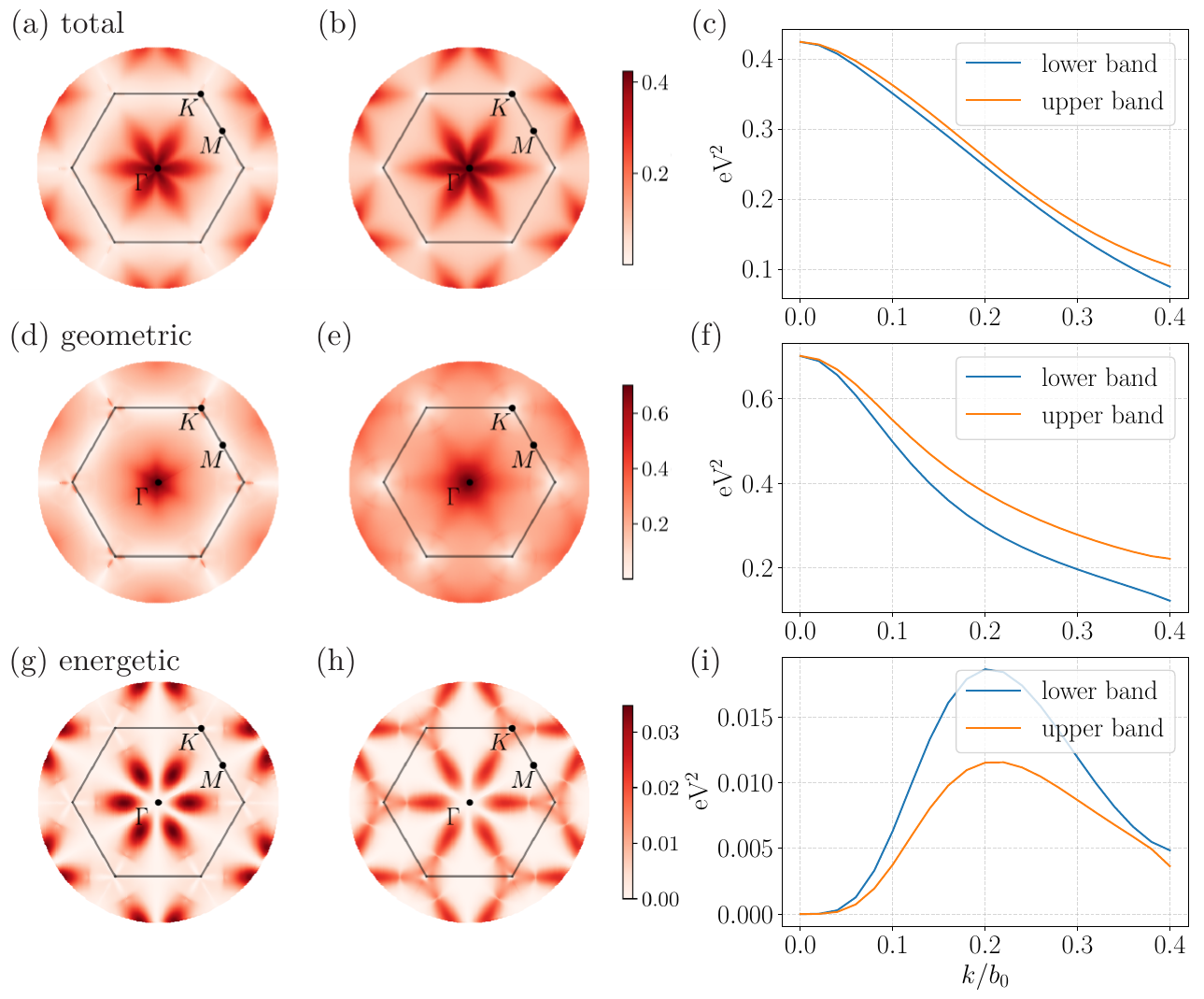}
    \caption{\label{app:fig:GA-epc-band-basis} Electron–phonon coupling (EPC) in the band basis obtained from GA. The three rows show, respectively, the total, geometric, and energetic contributions. In each row, the first two panels plot the band-resolved EPC $G_{\mathbf{k}}$ (defined in \cref{app:eq:Gk_band_basis_sp2_bands}) on the $k_z=0$ plane for phonon momentum $\mathbf{q}=0$, summed over the two bond-stretching modes, for the two $sp^2$ bands that form the $\sigma$ FS. The third panel presents the radial average of $G_{\kk}$ versus the distance $k$ from $\Gamma$. The total GA EPC reproduces the \textit{ab initio} trend in \cref{app:fig:DFT_EPC_band_basis}: $G_{\mathbf{k}}$ peaks at $\Gamma$, remains elevated along the $\Gamma$–K line, and decays approximately as $k^{2}$ away from $\Gamma$. The geometric part shows an even sharper maximum at $\Gamma$, whereas the energetic part is small in magnitude and vanishes at $\Gamma$. 
    }
\end{figure}

In \cref{app:fig:GA-epc-band-basis} we plot the electron–phonon coupling (EPC) in the band basis obtained from the GA, resolved into its total, geometric, and energetic components (see \cref{app:eq:lambda_separation}). The total GA EPC reproduces the \textit{ab initio} trend in \cref{app:fig:DFT_EPC_band_basis}: $G_{\mathbf{k}}$ peaks at $\Gamma$, remains large along the $\Gamma$–K line, and decays approximately as $k^{2}$ away from $\Gamma$. The geometric part shows an even sharper maximum at $\Gamma$, whereas the energetic part is small and vanishes exactly at $\Gamma$. Thus the pronounced peak in the band-basis EPC is almost entirely geometric in origin.

In summary, the GA fit captures both the \textit{ab initio} band structure and the EPC once onsite terms are restored. Decomposition of the coupling reveals that \ch{MgB2} is overwhelmingly dominated by the geometric component, with $\lambda^{\text{geo}}/\lambda \approx 0.9$. In the band basis, the enhancement of EPC $G_{\mathbf{k}}$ at $\Gamma$ is again geometric, while the energetic contribution is exactly zero at $\Gamma$. Because the superconducting $T_c$ increases under light electron doping, precisely as the Fermi level approaches this geometric peak, we conclude that the doping-induced rise in $T_{c}$ is driven by the geometric component of the EPC.

\section{Perturbation theory}\label{app:sec:perturbation_theory}

\subsection{Non-degenerate perturbation}

We first review the non-degenerate perturbation theory to obtain correction terms for both energy and wavefunctions. 
Assume the Hamiltonian is 
\begin{equation}
    H(\lambda) = H_0 + \lambda V,
\end{equation}
where $H_0$ is the unperturbed Hamiltonian, and $V$ is the perturbation term with $\lambda$ being a small parameter. Assume the eigenstates of $H_0$ are known, which serve as the zeroth-order solution of $H$:
\begin{equation}
    H_0|\psi^{(0)}_m\rangle = E^{(0)}_m|\psi^{(0)}_m\rangle, 
\end{equation}
where $m$ denotes different (non-degenerate) energy levels. The zeroth-order wavefunctions $|\psi^{(0)}_m\rangle$ form a complete basis, which can be used to express higher-order correction terms. 
Assume for a given energy level $m$, the energy and wavefunction of $H$ can be expanded in $\lambda$:
\begin{equation}
    E_m(\lambda) = \sum_{n=0}^{\infty} \lambda^n E_m^{(n)},\quad
    |\psi_m(\lambda) \rangle = \sum_{n=0}^{\infty} \lambda^n |\psi_m^{(n)} \rangle. 
\end{equation}
We choose the normalization condition that higher-order correction terms are orthogonal to the zeroth-order term, \ie, 
\begin{equation}
    \langle \psi_m^{(0)}| \psi_m^{(0)}\rangle = 1, \quad
    \langle \psi_m^{(0)}| \psi_m^{(n)}\rangle = 0 \ (n\ge 1). 
\end{equation}
As $\lambda$ is small, we can arrange the eigen equation of $H$ into different orders of $\lambda$:
\begin{equation}
    (H_0 - E_m^{(0)} |\psi_m^{(n)}\rangle = -(V - E_m^{(1)})|\psi^{(n-1)}\rangle - \sum_{l=2}^{n} E_m^{(l)}  |\psi_m^{(n-l)}\rangle. 
    \label{app:eq:RS-rec}
\end{equation}
In the following, we let $|n\rangle=|\psi_n^{(0)}\rangle$ for simplicity. 
Using the recursive equation \cref{app:eq:RS-rec}, we obtain the energy correction terms
\begin{equation}
    E_m^{(n)} = \langle m |V| \psi^{(n-1)}_{m} \rangle,
\end{equation}
with the lowest few order terms being
\begin{equation}
\begin{aligned}
    E_m^{(1)} &= \langle m |V| m \rangle, \\
    E_m^{(2)} &= \langle m |V| \psi^{(1)}_{m} \rangle  = \sum_{l \neq m} \frac{|\langle m |V| l \rangle|^2}{E_m^{(0)} - E_l^{(0)}}, \\
    E_m^{(3)} &= \sum_{k,s \neq m}
    \frac{\langle m|V|k \rangle\langle k|V|s\rangle\langle s|V|m\rangle }{(E^{(0)}_m - E^{(0)}_k)(E^{(0)}_m - E^{(0)}_s)} - \langle m|V|m\rangle \sum_{k\neq m} \frac{|\langle k |V| m\rangle|^2}{(E^{(0)}_m - E^{(0)}_k)^2},
\end{aligned}
\end{equation}
where we have used the first few order wavefunctions $|\psi^{(n)}_{m}\rangle$ with formula given below. 

For wavefunction corrections, let $|\psi^{(n)}_{m}\rangle = \sum_{k \neq m} c_{k}^{(n)} |k \rangle$, where $c_{k}^{(n)}=\langle k |\psi^{(n)}_m\rangle$, with $n\ge 1$, and $c_k^{(0)}=0$. 
Using \cref{app:eq:RS-rec}, we arrive at
\begin{equation}
    c_{k}^{(n)} = \frac{
    \langle k |V| \psi^{(n-1)}_m\rangle - \sum_{l=1}^{n} E^{(l)}_m c^{(n - l)}_k
    }{E^{(0)}_m - E^{(0)}_{k}}, \quad k \neq m. 
\end{equation}
Explicitly, we have
\begin{equation}
\begin{aligned}
    |\psi^{(1)}_m\rangle &= \sum_{k\neq m}\frac{ \langle k |V| m \rangle }{ E^{(0)}_m - E^{(0)}_{k} } |k\rangle , \\
    |\psi^{(2)}_m\rangle &= \sum_{k\neq m}\left[
    \sum_{s\neq m} \frac{\langle k|V|s\rangle\langle s|V|m\rangle }{(E^{(0)}_m - E^{(0)}_k)(E^{(0)}_m - E^{(0)}_s)} - \frac{\langle m |V| m\rangle\langle k|V|m\rangle}{(E^{(0)}_m - E^{(0)}_k)^2}
    \right] |k\rangle.
\end{aligned}
\end{equation}

\subsection{Degenerate perturbation}
When $H_0$ has a degenerate subspace $P_0$ with dimension $d$ and energy $E_0$, the perturbation theory for states in this degenerate subspace needs to be modified, \ie, we need to first use the perturbation $V$ to obtain non-degenerate bases for $P_0$. 

Let $Q_0=1-P_0$ be the complement of $P_0$. A given wavefunction $|\psi_a\rangle$ can be decomposed into 
$|\psi_a\rangle = P_0|\psi_a\rangle + Q_0|\psi_a\rangle = |\psi_{P,a}\rangle + |\psi_{Q,a}\rangle$. We then project the eigen equation into two subspaces:
\begin{equation}\begin{aligned}
    P_0 (H_0 + \lambda V) (|\psi_{P,a}\rangle + |\psi_{Q,a}\rangle) &= E |\psi_{P,a}\rangle, \\
    Q_0 (H_0 + \lambda V) (|\psi_{P,a}\rangle + |\psi_{Q,a}\rangle) &= E |\psi_{Q,a}\rangle.
\end{aligned}
\end{equation}
For the $P_0$ subspace, we have
\begin{equation}
    (E - E_0) |\psi_{P,a}\rangle = \lambda P_0 V \left( |\psi_{P,a}\rangle + |\psi_{Q,a}\rangle \right).
\end{equation}
For the $Q_0$ subspace, we have
\begin{equation}
    \left(E^{(0)}_d - Q_0 H_0 Q_0 \right) |\psi_{Q,a}\rangle = \lambda Q_0 V \left(|\psi_{P,a}\rangle + |\psi_{Q,a}\rangle \right) - \left(E - E_0 \right) |\psi_{Q,a} \rangle.
\end{equation}
Let 
\begin{equation}
    R_0 = \left(E_0 - Q_0 H_0 Q_0 \right)^{-1} = \sum_{\alpha\in Q_0}\frac{|\alpha\rangle\langle \alpha|}{E_0 - E_{\alpha}}, 
\end{equation} 
where $H_0|\alpha\rangle = E_{\alpha}|\alpha\rangle$ is the eigensystem of $H_0$ in the $Q_0$ subspace. We have
\begin{equation}
    |\psi_{Q,a}\rangle = R_0\left[\lambda Q_0 V (|\psi_{P,a}\rangle + |\psi_{Q,a}\rangle) - (E - E_0) |\psi_{Q,a} \rangle \right].
    \label{app:eq:deg_perturb_Q_eq}
\end{equation}

We then diagonalize the perturbation term $V$ within the $P_0$ subspace, \ie, 
\begin{equation}
    P_0 V P_0 |\chi_a\rangle = \epsilon_a |\chi_a\rangle,\quad a=1,\cdots, d. 
    \label{app:eq:deg_perturb_zeroth_eig_eq}
\end{equation}
Note that $|\chi_a\rangle\in P_0$ as $P_0|\chi_a\rangle=|\chi_a\rangle$. As a result, we can use them as the new zeroth-order basis in the $P_0$ subspace, \ie, 
\begin{equation}
    |\psi_{P,a}^{(0)}\rangle = |\chi_a\rangle, \quad \text{with }
    E_a^{(0)} = E_0,\ 
    E_a^{(1)} = \epsilon_a. 
\end{equation}
Note that we assume $\epsilon_a$'s are not degenerate. If not, we need higher-order Hamiltonian terms to disentangle them.

We then separate different orders of eigen energy and eigen states for given state $a$:
\begin{equation}
    E_a = E_0 + \sum_{n\ge 1} \lambda^n E_a^{(n)}, \quad
    |\psi_{P,a}\rangle = \sum_{n \ge 0} \lambda^n |\psi_{P,a}^{(n)} \rangle, \quad
    |\psi_{Q,a}\rangle = \sum_{n \ge 1} \lambda^n |\psi_{Q,a}^{(n)} \rangle,
\end{equation}
with $|\psi_{Q,a}^{(0)} \rangle =0$. Plugging into \cref{app:eq:deg_perturb_Q_eq}, we have the $Q_0$ subspace equation
\begin{equation}
    |\psi_{Q,a}^{(n)}\rangle = R_0\left[ Q_0 V \left(|\psi_{P,a}^{(n-1)}\rangle + |\psi_{Q,a}^{(n-1)} \rangle \right) - \sum_{l=1}^{n-1} E_a^{(l)} |\psi_{Q,a}^{(n-l)}\rangle \right], \quad n\ge 1. 
\end{equation}
The first-order correction term is
\begin{equation}
\begin{aligned}
    |\psi_{Q,a}^{(1)}\rangle &= R_0 QV |\psi_{P,a}^{(0)}\rangle = \sum_{\alpha\in Q_0} \frac{\langle \alpha |V| \chi_a\rangle}{E_0 - E_{\alpha}} |\alpha\rangle. 
    \label{app:eq:deg_perturb_Q1_1st_wfc}
\end{aligned}
\end{equation}

For the $P_0$ subspace, we have
\begin{equation}
    \sum_{l=1}^{n} E^{(l)}_{a} |\psi_{P,a}^{(n-l)}\rangle = P_0 V \left(|\psi_{P,a}^{(n-1)}\rangle + |\psi_{Q,a}^{(n-1)} \rangle \right), \quad n\ge 1. 
\end{equation}
For $n=1$, this gives $P_0VP_0|\chi_a\rangle = E_{a}^{(1)}|\chi_a\rangle$, which is nothing but \cref{app:eq:deg_perturb_zeroth_eig_eq}. 
For $n=2$, we have 
\begin{equation}
    (E^{(1)}_a - P_0 V) |\psi_{P,a}^{(1)}\rangle = P_0V |\psi_{Q,a}^{(1)}\rangle - E^{(2)}_a |\chi_a\rangle. 
    \label{app:eq:P_0_1st_order}
\end{equation}
By using the normalization condition
\begin{equation}
    \langle \chi_a |\psi_{P,a}\rangle = 1 \quad\Rightarrow \quad
    \langle \chi_a |\psi^{(n)}_{P,a}\rangle = 0,\quad (n\ge 1),
\end{equation}
we arrive at the second-order energy correction
\begin{equation}
    E^{(2)}_{a} = \langle \chi_a |V| \psi^{(1)}_{Q,a} \rangle 
    = \sum_{\alpha\in Q_0} \frac{|\langle \alpha |V| \chi_a\rangle|^2 }{E_0 - E_{\alpha}}. 
\end{equation}
The first-order wavefunction correction term in the $P_0$ subspace is obtained by projecting \cref{app:eq:P_0_1st_order} onto $\langle \chi_b|$ for $b\neq a$, \ie, 
$(\epsilon_a - \epsilon_b) \langle \chi_b|\psi^{(1)}_{P,a}\rangle = \langle \chi_b|V|\psi^{(1)}_{Q,a}\rangle$, which gives
\begin{equation}
    |\psi^{(1)}_{P,a}\rangle = \sum_{b\neq a}|\chi_b\rangle \frac{\langle \chi_b|V|\psi^{(1)}_{Q,a}\rangle }{\epsilon_a - \epsilon_b}
    = \sum_{b\neq a}|\chi_b\rangle \sum_{\alpha\in Q_0} \frac{\langle \chi_b|V|\alpha \rangle \langle \alpha|V|\chi_a\rangle }{(\epsilon_a - \epsilon_b)(E_0 - E_{\alpha})}.
    \label{app:eq:deg_perturb_P0_1st_wfc}
\end{equation}
Higher-order correction terms can be obtained similarly.

\subsection{Application to kagome model}

Consider an $s$-orbital kagome model with three orbitals at $(\frac{1}{2},0), (\frac{1}{2},\frac{1}{2}), (0,\frac{1}{2})$ and only the nearest-neighbor (NN) hopping. The Hamiltonian takes the form:
\begin{equation}\begin{aligned}
H_{NN}^{kag}(\kk)=2 t_{0}
\begin{bmatrix}
    0 &  \cos(\frac{k_2}{2})  &\cos(\frac{k_1+k_2}{2}) \\
    & 0 & \cos(\frac{k_1}{2})  \\
c.c. &  & 0\\
\end{bmatrix}.
\end{aligned}\end{equation}
At $\Gamma$, the three eigenvalues are
\begin{equation}
    E_1 = -2 t_{0},\quad
    E_2 = -2 t_{0},\quad
    E_3 = 4 t_{0}, 
\end{equation}
with the corresponding eigen wavefunctions
\begin{equation}\begin{aligned}
|\psi_1\rangle =\frac{1}{\sqrt{3}}[1, \omega,\omega^2]^T, \quad
|\psi_2\rangle =\frac{1}{\sqrt{3}}[1, \omega^2,\omega]^T, 
\quad
|\psi_3\rangle = \frac{1}{\sqrt{3}}[1,1,1]^T.
\end{aligned}\end{equation}
$|\psi_1\rangle$ and $|\psi_2\rangle$ are degenerate and form the $\Gamma_5^+$ IRREPs, with $\omega=e^{i2\pi/3}$. 
These two eigenmodes have $C_3$ eigenvalues $\omega$ and $\omega^2$, respectively.

The flat band in the NN kagome model has a simple analytic wavefunction
\begin{equation}\begin{aligned}
|\psi_{\kk}^{FB} \rangle = \frac{1}{\mathcal{N}_{\kk}}\left[
\sin(\frac{k_1}{2}), -\sin(\frac{k_1+k_2}{2}), \sin(\frac{k_2}{2})
\right]^T,
\label{app:eq:kagome_NN_flatband}
\end{aligned}\end{equation}
where 
$\mathcal{N}_{\kk}=\sqrt{3-\cos (k_1+k_2)-\cos (k_1)-\cos (k_2)}/\sqrt{2}$ 
is the normalization factor. Note that $\psi_{\kk}^{FB}$ in \cref{app:eq:kagome_NN_flatband} is not well-defined at $\kk=\bm{0}$ as $\mathcal{N}_{\kk}=0$ is singular. Instead, one can take $|\psi_{\kk=\bm{0}}^{FB}\rangle =[0,0,1]^T$.

We then use the degenerate perturbation theory to compute the two $\kk\cdot \pp$ wavefunctions originated from $\Gamma_5^+$. Expand 
\begin{equation}
    H^{kag}_{NN}(\kk) = H_0 + H_2 + \mathcal{O}(k^4),\quad
    H_0 = 2t_0 \begin{bmatrix}
        0 & 1 & 1 \\
        1 & 0 & 1 \\
        1 & 1 & 0
    \end{bmatrix},\quad
    H_2 = -\frac{t_0}{4}\begin{bmatrix}
        0 & k_2^2 & (k_1 + k_2)^2 \\
        & 0 & k_1^2 \\
        c.c. & & 0
    \end{bmatrix}. 
\end{equation}
Let $V=H_2$, and $P_0=|\psi_1\rangle\langle\psi_1| + |\psi_2\rangle\langle\psi_2|$, $Q_0=|\psi_3\rangle\langle\psi_3|$. 
After transforming to the polar coordinates, \ie, $k_1=k\cos(\theta),\ k_2=k\cos(\theta-\frac{2\pi}{3})$, we find the following basis for $P_0 VP_0$:
\begin{equation}
\begin{aligned}
    |\psi^{(0)}_{FB}(\theta)\rangle &= \frac{1}{\sqrt{2}}\left(e^{-i\theta}|\psi_1\rangle + e^{i\theta} |\psi_2\rangle \right)  =
    \sqrt{\frac{2}{3}} \left[\cos(\theta), \cos(\theta + \frac{2\pi}{3}), \cos(\theta - \frac{2\pi}{3}) \right]^T, \quad E^{(1)}_{FB} = 0, \\
    |\psi^{(0)}_{DB}(\theta)\rangle &= \frac{1}{\sqrt{2}}\left(e^{-i\theta}|\psi_1\rangle - e^{i\theta} |\psi_2\rangle \right)  
    = \sqrt{\frac{2}{3}} \left[\sin(\theta), \sin(\theta + \frac{2\pi}{3}), \sin(\theta - \frac{2\pi}{3}) \right]^T,\quad 
    E^{(1)}_{DB} = \frac{t_0}{4}. 
\end{aligned}
\end{equation}
To obtain the first-order correction term in the wavefunction, we first note that as $E^{(1)}_{FB} = 0$, the correction terms from the $P_0$ subspace \cref{app:eq:deg_perturb_P0_1st_wfc} are zero. Thus we only need to compute the correction terms from the $Q_0$ subspace using \cref{app:eq:deg_perturb_Q1_1st_wfc}, which gives
\begin{equation}\begin{aligned}
    |\psi^{(1)}_{FB}(\theta)\rangle &= -\frac{\sqrt{2}}{96} k^2 \cos(3\theta) |\psi_3\rangle, \\
    |\psi^{(1)}_{DB}(\theta)\rangle &= -\frac{\sqrt{2}}{96} k^2 \sin(3\theta) |\psi_3\rangle.
\end{aligned}\end{equation}
Consequently, the two $\kk\cdot\pp$ wavefunctions from the $\Gamma_5^+$ doublet expanded to the second order in $k$ near $\Gamma$ is
\begin{equation}\begin{aligned}
    \psi^{FB}_{\kk} &\approx \sqrt{\frac{2}{3}}\left\{\left[\cos(\theta), \cos(\theta + \frac{2\pi}{3}), \cos(\theta - \frac{2\pi}{3}) \right] - \frac{k^2}{96}\cos(3\theta)[1,1,1]
    \right\}, \\
    \psi^{DB}_{\kk} &\approx \sqrt{\frac{2}{3}}\left\{\left[\sin(\theta), \sin(\theta + \frac{2\pi}{3}), \sin(\theta - \frac{2\pi}{3}) \right] - \frac{k^2}{96}\sin(3\theta)[1,1,1]
    \right\}.
\end{aligned}\end{equation}
where a normalization factor $\frac{1}{\sqrt{1 +k^4/96(1+\cos 6\theta)}}=1+\mathcal{O}(k^4)$ is omitted.

\subsection{Schrieffer–Wolff transformation for EPC Hamiltonian}\label{app:sec:SW_transformation}

Schrieffer–Wolff (SW) transformation is a unitary transformation used to obtain effective low-energy Hamiltonians by decoupling weakly coupled subspaces. SW transformation is an operator version of the perturbation theory introduced in previous sections, including the degenerate perturbation. 
Applying the SW transformation to an electron-phonon coupling (EPC) Hamiltonian leads to an effective electron-electron attractive interaction Hamiltonian. 

We start from a general Hamiltonian $\hat{H}=\hat{H}_0+\hat{V}$, where $\hat{V}$ is a small perturbation. Denote the eigen basis of $\hat{H}_0$ as $|m\rangle$ with eigenvalues $E_m$, and assume $\hat{V}$ is off-diagonal in $|m\rangle$, \ie, $\langle m|\hat{V}|m\rangle=0$ (the diagonal part of $\hat{V}$ can be absorbed into $\hat{H}_0$). The SW transformation $e^{\hat{S}}$ gives
\begin{equation}\begin{aligned}
    \hat{H}' &= e^{\hat{S}} \hat{H} e^{-\hat{S}} = \hat{H} + [\hat{S},\hat{H}] + \frac{1}{2}[\hat{S},[\hat{S},\hat{H}]] + \dots \\
    &= \hat{H}_0 + \hat{V} + [\hat{S},\hat{H}_0] + [\hat{S}, \hat{V}] + \frac{1}{2}[\hat{S},[\hat{S},\hat{H}_0]] + \frac{1}{2}[\hat{S},[\hat{S},\hat{V}]] + \dots.
\end{aligned}\end{equation}
We require that
\begin{equation}
    \hat{V} + [\hat{S},\hat{H}_0] = 0
    \label{app:eq:SW_S_eq}
\end{equation}
so that 
\begin{equation}
    \hat{H}'=\hat{H}_0 + \frac{1}{2}[\hat{S}, \hat{V}] + \mathcal{O}(\hat{V}^3). 
\end{equation}
In practice, we need to solve for $\hat{S}$ using \cref{app:eq:SW_S_eq}, and then plug into $[\hat{S},\hat{V}]$ to obtain the effective Hamiltonian $\hat{H}'$. 

We consider a general EPC Hamiltonian in the band basis
\begin{equation}
\begin{aligned}
    \hat{H} &= \hat{H}_e + \hat{H}_p + \hat{H}_g, \\
    \hat{H}_e &= \sum_{\kk,n,\sigma}\epsilon_{\kk,n} \cre{\gamma}{\kk,n\sigma}\des{\gamma}{\kk,n\sigma}, \\
    \hat{H}_p &= \sum_{\qq,s} \omega_{\qq,s} \cre{b}{\qq,s} \des{b}{\qq,s}, \\
    \hat{H}_g &= \frac{1}{\sqrt{N}}\sum_{\kk,\qq,m,n,\sigma,s} \left(G^{mns}_{\kk,\qq} \cre{\gamma}{\kk+\qq, m\sigma} \des{\gamma}{\kk, n\sigma}\des{b}{\qq,s} + h.c.
    \right)
\end{aligned}
\end{equation}
Let $\hat{H}_0=\hat{H}_e + \hat{H}_p$. 
Consider a SW transformation $e^{\hat{S}}$, with
\begin{equation}
    \hat{S} = \hat{L} - \hat{L}^\dagger,\quad
    \hat{L} = \frac{1}{\sqrt{N}}\sum_{\kk,\qq,m,n,\sigma,s} v^{mns}_{\kk,\qq} \cre{\gamma}{\kk+\qq, m\sigma} \des{\gamma}{\kk, n\sigma}\des{b}{\qq,s},
\end{equation}
where $v^{mns}_{\kk,\qq}$ are coefficients that needs to satisfy
\begin{equation}
    [\hat{H}_0, \hat{S}] = \hat{H}_g.
    \label{app:eq:SW_epc_constraint}
\end{equation}
By using $\{\cre{\gamma}{\kk n\sigma}, \des{\gamma}{\kk',m\sigma'}\}=\delta_{\kk,\kk'}\delta_{mn}\delta_{\sigma,\sigma'}$, $[\cre{b}{\qq,s}, \des{b}{\qq',s'}]=-\delta_{\qq,\qq'}\delta_{ss'}$, $[\hat{\gamma}^{(\dagger)}_{\kk,m\sigma}, \hat{b}^{(\dagger)}_{\qq,s}]=0$, $[AB,C]=A[B,C]+[A,C]B$, $[AB,C]=A\{B,C\}-\{A,C\}B$, $[A, BC]=-B\{A,C\} + \{A, B\}C$, and $[\cre{\gamma}{i}\des{\gamma}{j}, \cre{\gamma}{k}\des{\gamma}{l}]=\delta_{jk}\cre{\gamma}{i}\des{\gamma}{l} - \delta_{il}\cre{\gamma}{k}\des{\gamma}{j}$, 
we have 
\begin{equation}
\begin{aligned}
    [\hat{H}_b, \hat{L}] &= -\frac{1}{\sqrt{N}}\sum_{\kk,\qq,m,n,\sigma,s} \omega_{\qq,s} v^{mns}_{\kk,\qq} \cre{\gamma}{\kk+\qq, m\sigma} \des{\gamma}{\kk, n\sigma}\des{b}{\qq,s}, \\
    [\hat{H}_e, \hat{L}] &= \frac{1}{\sqrt{N}}\sum_{\kk,\qq,m,n,\sigma,s} (\epsilon_{\kk+\qq,m} - \epsilon_{\kk,n}) v^{mns}_{\kk,\qq} \cre{\gamma}{\kk+\qq, m\sigma} \des{\gamma}{\kk, n\sigma}\des{b}{\qq,s}, \\
    \Rightarrow [\hat{H}_0, \hat{L}] &= \frac{1}{\sqrt{N}}\sum_{\kk,\qq,m,n,\sigma,s} (\epsilon_{\kk+\qq,m} - \epsilon_{\kk,n} - \omega_{\qq,s}) v^{mns}_{\kk,\qq} \cre{\gamma}{\kk+\qq, m\sigma} \des{\gamma}{\kk, n\sigma}\des{b}{\qq,s}, \\
    \Rightarrow [\hat{H}_0, \hat{S}] &= \frac{1}{\sqrt{N}}\sum_{\kk,\qq,m,n,\sigma,s} (\epsilon_{\kk+\qq,m} - \epsilon_{\kk,n} - \omega_{\qq,s}) v^{mns}_{\kk,\qq} \cre{\gamma}{\kk+\qq, m\sigma} \des{\gamma}{\kk, n\sigma}\des{b}{\qq,s} + h.c.
\end{aligned}
\end{equation}
After plugging into \cref{app:eq:SW_epc_constraint}, we arrive at 
\begin{equation}
    v^{mns}_{\kk,\qq} = \frac{G^{mns}_{\kk,\qq}}{\epsilon_{\kk+\qq,m} - \epsilon_{\kk,n} - \omega_{\qq,s}}. 
\end{equation}
We then compute the effective Hamiltonian after the SW transformation:
\begin{equation}
    \hat{H}'=\hat{H}_0 + \frac{1}{2}[\hat{S}, \hat{H}_g]. 
\end{equation}
Let $\alpha=(\kk,\qq,m,n,\sigma,s)$, $\hat{X}_{\alpha}=\cre{\gamma}{\kk+\qq, m\sigma} \des{\gamma}{\kk, n\sigma}\des{b}{\qq,s}$, and 
$\hat{S}=\frac{1}{\sqrt{N}}\sum_{\alpha} \left(v_{\alpha} \hat{X}_{\alpha} - v^*_{\alpha} \hat{X}^\dagger_{\alpha}\right)$, 
$\hat{H}_g=\frac{1}{\sqrt{N}}\sum_{\beta}\left(g_{\beta} \hat{X}_{\beta} + g_{\beta}^* \hat{X}^{\dagger}_{\beta} \right)$, we have
\begin{equation}
    \frac{1}{2}[\hat{S}, \hat{H}_g] = \frac{1}{N} \sum_{\alpha,\beta} \frac{1}{2}\left(v_{\alpha} g^*_{\beta} + v^*_{\beta} g_{\alpha}\right) [\hat{X}_\alpha, \hat{X}^\dagger_\beta]. 
\end{equation}
We are interested in the electron interaction terms, while other electron self-energy corrections and terms involving phonon operators are omitted. As a result, we have
\begin{equation}
\begin{aligned}
    \frac{1}{2}[\hat{S}, \hat{H}_g] &= 
    \frac{1}{2N} \sum_{\kk,\qq,m,n,\sigma,s} \sum_{\kk',\qq',m',n',\sigma',s'} \left(v^{mns}_{\kk,\qq} G^{m'n's'*}_{\kk',\qq'} + v^{m'n's'*}_{\kk',\qq'} G^{mns}_{\kk,\qq}\right) \left[
    \cre{\gamma}{\kk+\qq,m\sigma}\des{\gamma}{\kk,n\sigma} \des{b}{\qq s}, \cre{\gamma}{\kk',n'\sigma'}\des{\gamma}{\kk'+\qq',m'\sigma'} \cre{b}{\qq' s'}
    \right] \\
    &= \frac{1}{2N} \sum_{\kk,\qq,m,n,\sigma,s} \sum_{\kk',\qq',m',n',\sigma',s'} \left(v^{mns}_{\kk,\qq} G^{m'n's'*}_{\kk',\qq'} + v^{m'n's'*}_{\kk',\qq'} G^{mns}_{\kk,\qq}\right)\left(
    \cre{\gamma}{\kk+\qq,m\sigma}\des{\gamma}{\kk,n\sigma} \cre{\gamma}{\kk',n'\sigma'}\des{\gamma}{\kk'+\qq',m'\sigma'}\left[\des{b}{\qq s},  \cre{b}{\qq' s'}\right] + 
    \right.\\
    &+\left.
    \left[\cre{\gamma}{\kk+\qq,m\sigma}\des{\gamma}{\kk,n\sigma},  \cre{\gamma}{\kk',n'\sigma'}\des{\gamma}{\kk'+\qq',m'\sigma'} \right] \des{b}{\qq s} \cre{b}{\qq' s'}
    \right) 
\end{aligned}
\end{equation}
By ignoring the $\des{b}{\qq s} \cre{b}{\qq' s'}$ and two fermion terms, we obtain
\begin{equation}\begin{aligned}
    \frac{1}{2}[\hat{S}, \hat{H}_g] &\rightarrow
    \frac{1}{2N} \sum_{\kk,\qq,m,n,\sigma,s} \sum_{\kk',m',n',\sigma'} \left(v^{mns}_{\kk,\qq} G^{m'n's*}_{\kk',\qq} + v^{m'n's*}_{\kk',\qq} G^{mns}_{\kk,\qq}\right)
    \cre{\gamma}{\kk+\qq,m\sigma}\des{\gamma}{\kk,n\sigma} \cre{\gamma}{\kk',n'\sigma'}\des{\gamma}{\kk'+\qq,m'\sigma'} \\
    &\rightarrow \frac{1}{2N} \sum_{\kk,\qq,m,n,\sigma,s} \sum_{\kk',m',n',\sigma'} \left(\frac{G^{mns}_{\kk,\qq} G^{m'n's*}_{\kk',\qq}}{\epsilon_{\kk+\qq,m} - \epsilon_{\kk,n} - \omega_{\qq,s}} + \frac{G^{m'n's*}_{\kk',\qq} G^{mns}_{\kk,\qq}}{\epsilon_{\kk'+\qq,m'} - \epsilon_{\kk',n'} - \omega_{\qq,s}} \right)
    \cre{\gamma}{\kk+\qq,m\sigma} \cre{\gamma}{\kk',n'\sigma'} \des{\gamma}{\kk'+\qq,m'\sigma'} \des{\gamma}{\kk,n\sigma} ,
\end{aligned}
\end{equation}
where the second term can be rewritten into
\begin{equation}
\begin{aligned}
    \text{2nd term}=& -\frac{G^{m'n's*}_{\kk',-\qq} G^{mns}_{\kk,-\qq}}{\epsilon_{\kk',n'} - \epsilon_{\kk'-\qq,m'} + \omega_{-\qq,s}} 
    \cre{\gamma}{\kk-\qq,m\sigma} \cre{\gamma}{\kk',n'\sigma'} \des{\gamma}{\kk'-\qq,m'\sigma'} \des{\gamma}{\kk,n\sigma} \quad (\qq\rightarrow -\qq)\\
    &= -\frac{G^{m'n's*}_{\kk+\qq,-\qq} G^{mns}_{\kk'+\qq,-\qq}}{ \epsilon_{\kk+\qq,n'} - \epsilon_{\kk,m'} + \omega_{\qq,s}} 
    \cre{\gamma}{\kk',m\sigma} \cre{\gamma}{\kk+\qq,n'\sigma'} \des{\gamma}{\kk,m'\sigma'} \des{\gamma}{\kk'+\qq,n\sigma} \quad (\kk\rightarrow \kk'+\qq,\kk'\rightarrow \kk+\qq)\\
    &= -\frac{G^{nm s*}_{\kk+\qq,-\qq} G^{n'm's}_{\kk'+\qq,-\qq}}{\epsilon_{\kk+\qq,m} - \epsilon_{\kk,n} + \omega_{\qq,s}} 
    \cre{\gamma}{\kk',n'\sigma'} \cre{\gamma}{\kk+\qq, m\sigma} \des{\gamma}{\kk, n\sigma} \des{\gamma}{\kk'+\qq, m'\sigma'} \quad (\sigma \leftrightarrow \sigma', m\leftrightarrow n', n \leftrightarrow m')\\
    &= -\frac{G^{nm s*}_{\kk+\qq,-\qq} G^{n'm's}_{\kk'+\qq,-\qq}}{\epsilon_{\kk+\qq,m} - \epsilon_{\kk,n} + \omega_{\qq,s}} 
    \cre{\gamma}{\kk+\qq, m\sigma} \cre{\gamma}{\kk',n'\sigma'} \des{\gamma}{\kk'+\qq, m'\sigma'} \des{\gamma}{\kk, n\sigma}.
\end{aligned}
\end{equation}
As a result, we arrive at the effective electron-electron interaction from integrating out the EPC Hamiltonian
\begin{equation}
\begin{aligned}
    \hat{H}_{ee} &= \frac{1}{2N} \sum_{\kk,\qq,m,n,\sigma,s} \sum_{\kk',m',n',\sigma'} \left(\frac{G^{mns}_{\kk,\qq} G^{m'n's*}_{\kk'-\qq,\qq}}{\epsilon_{\kk+\qq,m} - \epsilon_{\kk,n} - \omega_{\qq,s}} 
    - \frac{G^{nm s*}_{\kk+\qq,-\qq} G^{n'm's}_{\kk',-\qq}}{\epsilon_{\kk+\qq,m} - \epsilon_{\kk,n} + \omega_{\qq,s}}  
    \right)
    \cre{\gamma}{\kk+\qq,m\sigma} \cre{\gamma}{\kk'-\qq,n'\sigma'} \des{\gamma}{\kk',m'\sigma'} \des{\gamma}{\kk,n\sigma},
\end{aligned}
\end{equation}
where we further change $\kk'\rightarrow \kk'-\qq$. 
If there is only one relevant band near $E_f$ so that we can suppress the band index, and assume the EPC tensor is $\kk$-independent and satisfies $|G^s_{\qq}|^2=|G^s_{-\qq}|^2$, we arrive at the simplified form
\begin{equation}
\begin{aligned}
    \hat{H}_{ee} &= \frac{1}{N} \sum_{\kk,\kk',\qq,\sigma,\sigma',s} \frac{ \omega_{\qq,s}|G^{s}_{\qq}|^2}{(\epsilon_{\kk+\qq} - \epsilon_{\kk})^2 - \omega_{\qq,s}^2} 
    \cre{\gamma}{\kk+\qq,\sigma} \cre{\gamma}{\kk'-\qq,\sigma'} \des{\gamma}{\kk',\sigma'} \des{\gamma}{\kk,\sigma}.
\end{aligned}
\end{equation}
If we further assume the pairing only happens for electrons with opposite momentum by setting $\kk_1:=\kk+\qq=-(\kk'-\qq)$, $\kk_2:= \kk=-\kk'$, then we have $\qq=\kk_1-\kk_2$, and 
\begin{equation}
\begin{aligned}
    \hat{H}_{ee} &= \frac{1}{N} \sum_{\kk_1,\kk_2,\sigma,\sigma',s} \frac{ \omega_{\qq,s}|G^{s}_{\qq}|^2} {(\epsilon_{\kk_1} - \epsilon_{\kk_2})^2 - \omega_{\qq,s}^2} 
    \cre{\gamma}{\kk_1\sigma} \cre{\gamma}{-\kk_1,\sigma'} \des{\gamma}{-\kk_2, \sigma'} \des{\gamma}{\kk_2,\sigma}.
\end{aligned}
\end{equation}
This interaction is attractive when $|\epsilon_{\kk_1} - \epsilon_{\kk_2}|<\omega_{\qq,s}$.


\begin{thebibliography}{113}%
\makeatletter
\providecommand \@ifxundefined [1]{%
 \@ifx{#1\undefined}
}%
\providecommand \@ifnum [1]{%
 \ifnum #1\expandafter \@firstoftwo
 \else \expandafter \@secondoftwo
 \fi
}%
\providecommand \@ifx [1]{%
 \ifx #1\expandafter \@firstoftwo
 \else \expandafter \@secondoftwo
 \fi
}%
\providecommand \natexlab [1]{#1}%
\providecommand \enquote  [1]{``#1''}%
\providecommand \bibnamefont  [1]{#1}%
\providecommand \bibfnamefont [1]{#1}%
\providecommand \citenamefont [1]{#1}%
\providecommand \href@noop [0]{\@secondoftwo}%
\providecommand \href [0]{\begingroup \@sanitize@url \@href}%
\providecommand \@href[1]{\@@startlink{#1}\@@href}%
\providecommand \@@href[1]{\endgroup#1\@@endlink}%
\providecommand \@sanitize@url [0]{\catcode `\\12\catcode `\$12\catcode `\&12\catcode `\#12\catcode `\^12\catcode `\_12\catcode `\%12\relax}%
\providecommand \@@startlink[1]{}%
\providecommand \@@endlink[0]{}%
\providecommand \url  [0]{\begingroup\@sanitize@url \@url }%
\providecommand \@url [1]{\endgroup\@href {#1}{\urlprefix }}%
\providecommand \urlprefix  [0]{URL }%
\providecommand \Eprint [0]{\href }%
\providecommand \doibase [0]{https://doi.org/}%
\providecommand \selectlanguage [0]{\@gobble}%
\providecommand \bibinfo  [0]{\@secondoftwo}%
\providecommand \bibfield  [0]{\@secondoftwo}%
\providecommand \translation [1]{[#1]}%
\providecommand \BibitemOpen [0]{}%
\providecommand \bibitemStop [0]{}%
\providecommand \bibitemNoStop [0]{.\EOS\space}%
\providecommand \EOS [0]{\spacefactor3000\relax}%
\providecommand \BibitemShut  [1]{\csname bibitem#1\endcsname}%
\let\auto@bib@innerbib\@empty
\bibitem [{\citenamefont {Nagamatsu}\ \emph {et~al.}(2001)\citenamefont {Nagamatsu}, \citenamefont {Nakagawa}, \citenamefont {Muranaka}, \citenamefont {Zenitani},\ and\ \citenamefont {Akimitsu}}]{Nagamatsu2001_MgB2}%
  \BibitemOpen
  \bibfield  {author} {\bibinfo {author} {\bibfnamefont {J.}~\bibnamefont {Nagamatsu}}, \bibinfo {author} {\bibfnamefont {N.}~\bibnamefont {Nakagawa}}, \bibinfo {author} {\bibfnamefont {T.}~\bibnamefont {Muranaka}}, \bibinfo {author} {\bibfnamefont {Y.}~\bibnamefont {Zenitani}},\ and\ \bibinfo {author} {\bibfnamefont {J.}~\bibnamefont {Akimitsu}},\ }\href@noop {} {\bibfield  {journal} {\bibinfo  {journal} {nature}\ }\textbf {\bibinfo {volume} {410}},\ \bibinfo {pages} {63} (\bibinfo {year} {2001})}\BibitemShut {NoStop}%
\bibitem [{\citenamefont {Larbalestier}\ \emph {et~al.}(2001)\citenamefont {Larbalestier}, \citenamefont {Cooley}, \citenamefont {Rikel}, \citenamefont {Polyanskii}, \citenamefont {Jiang}, \citenamefont {Patnaik}, \citenamefont {Cai}, \citenamefont {Feldmann}, \citenamefont {Gurevich}, \citenamefont {Squitieri} \emph {et~al.}}]{larbalestier2001strongly}%
  \BibitemOpen
  \bibfield  {author} {\bibinfo {author} {\bibfnamefont {D.}~\bibnamefont {Larbalestier}}, \bibinfo {author} {\bibfnamefont {L.}~\bibnamefont {Cooley}}, \bibinfo {author} {\bibfnamefont {M.}~\bibnamefont {Rikel}}, \bibinfo {author} {\bibfnamefont {A.}~\bibnamefont {Polyanskii}}, \bibinfo {author} {\bibfnamefont {J.}~\bibnamefont {Jiang}}, \bibinfo {author} {\bibfnamefont {S.}~\bibnamefont {Patnaik}}, \bibinfo {author} {\bibfnamefont {X.}~\bibnamefont {Cai}}, \bibinfo {author} {\bibfnamefont {D.}~\bibnamefont {Feldmann}}, \bibinfo {author} {\bibfnamefont {A.}~\bibnamefont {Gurevich}}, \bibinfo {author} {\bibfnamefont {A.}~\bibnamefont {Squitieri}}, \emph {et~al.},\ }\href@noop {} {\bibfield  {journal} {\bibinfo  {journal} {Nature}\ }\textbf {\bibinfo {volume} {410}},\ \bibinfo {pages} {186} (\bibinfo {year} {2001})}\BibitemShut {NoStop}%
\bibitem [{\citenamefont {Bud'ko}\ \emph {et~al.}(2001)\citenamefont {Bud'ko}, \citenamefont {Lapertot}, \citenamefont {Petrovic}, \citenamefont {Cunningham}, \citenamefont {Anderson},\ and\ \citenamefont {Canfield}}]{bud2001boron}%
  \BibitemOpen
  \bibfield  {author} {\bibinfo {author} {\bibfnamefont {S.~L.}\ \bibnamefont {Bud'ko}}, \bibinfo {author} {\bibfnamefont {G.}~\bibnamefont {Lapertot}}, \bibinfo {author} {\bibfnamefont {C.}~\bibnamefont {Petrovic}}, \bibinfo {author} {\bibfnamefont {C.}~\bibnamefont {Cunningham}}, \bibinfo {author} {\bibfnamefont {N.}~\bibnamefont {Anderson}},\ and\ \bibinfo {author} {\bibfnamefont {P.}~\bibnamefont {Canfield}},\ }\href@noop {} {\bibfield  {journal} {\bibinfo  {journal} {Physical Review Letters}\ }\textbf {\bibinfo {volume} {86}},\ \bibinfo {pages} {1877} (\bibinfo {year} {2001})}\BibitemShut {NoStop}%
\bibitem [{\citenamefont {Yildirim}\ \emph {et~al.}(2001)\citenamefont {Yildirim}, \citenamefont {G{\"u}lseren}, \citenamefont {Lynn}, \citenamefont {Brown}, \citenamefont {Udovic}, \citenamefont {Huang}, \citenamefont {Rogado}, \citenamefont {Regan}, \citenamefont {Hayward}, \citenamefont {Slusky} \emph {et~al.}}]{yildirim2001giant}%
  \BibitemOpen
  \bibfield  {author} {\bibinfo {author} {\bibfnamefont {T.}~\bibnamefont {Yildirim}}, \bibinfo {author} {\bibfnamefont {O.}~\bibnamefont {G{\"u}lseren}}, \bibinfo {author} {\bibfnamefont {J.}~\bibnamefont {Lynn}}, \bibinfo {author} {\bibfnamefont {C.}~\bibnamefont {Brown}}, \bibinfo {author} {\bibfnamefont {T.}~\bibnamefont {Udovic}}, \bibinfo {author} {\bibfnamefont {Q.}~\bibnamefont {Huang}}, \bibinfo {author} {\bibfnamefont {N.}~\bibnamefont {Rogado}}, \bibinfo {author} {\bibfnamefont {K.}~\bibnamefont {Regan}}, \bibinfo {author} {\bibfnamefont {M.}~\bibnamefont {Hayward}}, \bibinfo {author} {\bibfnamefont {J.}~\bibnamefont {Slusky}}, \emph {et~al.},\ }\href@noop {} {\bibfield  {journal} {\bibinfo  {journal} {Physical review letters}\ }\textbf {\bibinfo {volume} {87}},\ \bibinfo {pages} {037001} (\bibinfo {year} {2001})}\BibitemShut {NoStop}%
\bibitem [{\citenamefont {Osborn}\ \emph {et~al.}(2001)\citenamefont {Osborn}, \citenamefont {Goremychkin}, \citenamefont {Kolesnikov},\ and\ \citenamefont {Hinks}}]{osborn2001phonon}%
  \BibitemOpen
  \bibfield  {author} {\bibinfo {author} {\bibfnamefont {R.}~\bibnamefont {Osborn}}, \bibinfo {author} {\bibfnamefont {E.}~\bibnamefont {Goremychkin}}, \bibinfo {author} {\bibfnamefont {A.}~\bibnamefont {Kolesnikov}},\ and\ \bibinfo {author} {\bibfnamefont {D.}~\bibnamefont {Hinks}},\ }\href@noop {} {\bibfield  {journal} {\bibinfo  {journal} {Physical Review Letters}\ }\textbf {\bibinfo {volume} {87}},\ \bibinfo {pages} {017005} (\bibinfo {year} {2001})}\BibitemShut {NoStop}%
\bibitem [{\citenamefont {Lorenz}\ \emph {et~al.}(2001)\citenamefont {Lorenz}, \citenamefont {Meng},\ and\ \citenamefont {Chu}}]{lorenz2001high}%
  \BibitemOpen
  \bibfield  {author} {\bibinfo {author} {\bibfnamefont {B.}~\bibnamefont {Lorenz}}, \bibinfo {author} {\bibfnamefont {R.}~\bibnamefont {Meng}},\ and\ \bibinfo {author} {\bibfnamefont {C.}~\bibnamefont {Chu}},\ }\href@noop {} {\bibfield  {journal} {\bibinfo  {journal} {Physical Review B}\ }\textbf {\bibinfo {volume} {64}},\ \bibinfo {pages} {012507} (\bibinfo {year} {2001})}\BibitemShut {NoStop}%
\bibitem [{\citenamefont {Buzea}\ and\ \citenamefont {Yamashita}(2001)}]{buzea2001review}%
  \BibitemOpen
  \bibfield  {author} {\bibinfo {author} {\bibfnamefont {C.}~\bibnamefont {Buzea}}\ and\ \bibinfo {author} {\bibfnamefont {T.}~\bibnamefont {Yamashita}},\ }\href@noop {} {\bibfield  {journal} {\bibinfo  {journal} {Superconductor Science and Technology}\ }\textbf {\bibinfo {volume} {14}},\ \bibinfo {pages} {R115} (\bibinfo {year} {2001})}\BibitemShut {NoStop}%
\bibitem [{\citenamefont {An}\ and\ \citenamefont {Pickett}(2001{\natexlab{a}})}]{AnPickett2001_MgB2}%
  \BibitemOpen
  \bibfield  {author} {\bibinfo {author} {\bibfnamefont {J.}~\bibnamefont {An}}\ and\ \bibinfo {author} {\bibfnamefont {W.}~\bibnamefont {Pickett}},\ }\href@noop {} {\bibfield  {journal} {\bibinfo  {journal} {Physical Review Letters}\ }\textbf {\bibinfo {volume} {86}},\ \bibinfo {pages} {4366} (\bibinfo {year} {2001}{\natexlab{a}})}\BibitemShut {NoStop}%
\bibitem [{\citenamefont {Kortus}\ \emph {et~al.}(2001)\citenamefont {Kortus}, \citenamefont {Mazin}, \citenamefont {Belashchenko}, \citenamefont {Antropov},\ and\ \citenamefont {Boyer}}]{Kortus2001_MgB2}%
  \BibitemOpen
  \bibfield  {author} {\bibinfo {author} {\bibfnamefont {J.}~\bibnamefont {Kortus}}, \bibinfo {author} {\bibfnamefont {I.}~\bibnamefont {Mazin}}, \bibinfo {author} {\bibfnamefont {K.~D.}\ \bibnamefont {Belashchenko}}, \bibinfo {author} {\bibfnamefont {V.~P.}\ \bibnamefont {Antropov}},\ and\ \bibinfo {author} {\bibfnamefont {L.}~\bibnamefont {Boyer}},\ }\href@noop {} {\bibfield  {journal} {\bibinfo  {journal} {Physical Review Letters}\ }\textbf {\bibinfo {volume} {86}},\ \bibinfo {pages} {4656} (\bibinfo {year} {2001})}\BibitemShut {NoStop}%
\bibitem [{\citenamefont {Liu}\ \emph {et~al.}(2001)\citenamefont {Liu}, \citenamefont {Mazin},\ and\ \citenamefont {Kortus}}]{liu2001beyond}%
  \BibitemOpen
  \bibfield  {author} {\bibinfo {author} {\bibfnamefont {A.~Y.}\ \bibnamefont {Liu}}, \bibinfo {author} {\bibfnamefont {I.~I.}\ \bibnamefont {Mazin}},\ and\ \bibinfo {author} {\bibfnamefont {J.}~\bibnamefont {Kortus}},\ }\href {https://doi.org/10.1103/PhysRevLett.87.087005} {\bibfield  {journal} {\bibinfo  {journal} {Phys. Rev. Lett.}\ }\textbf {\bibinfo {volume} {87}},\ \bibinfo {pages} {087005} (\bibinfo {year} {2001})}\BibitemShut {NoStop}%
\bibitem [{\citenamefont {Kong}\ \emph {et~al.}(2001)\citenamefont {Kong}, \citenamefont {Dolgov}, \citenamefont {Jepsen},\ and\ \citenamefont {Andersen}}]{kong2001electron}%
  \BibitemOpen
  \bibfield  {author} {\bibinfo {author} {\bibfnamefont {Y.}~\bibnamefont {Kong}}, \bibinfo {author} {\bibfnamefont {O.}~\bibnamefont {Dolgov}}, \bibinfo {author} {\bibfnamefont {O.}~\bibnamefont {Jepsen}},\ and\ \bibinfo {author} {\bibfnamefont {O.}~\bibnamefont {Andersen}},\ }\href@noop {} {\bibfield  {journal} {\bibinfo  {journal} {Physical Review B}\ }\textbf {\bibinfo {volume} {64}},\ \bibinfo {pages} {020501} (\bibinfo {year} {2001})}\BibitemShut {NoStop}%
\bibitem [{\citenamefont {Szab{\'o}}\ \emph {et~al.}(2001)\citenamefont {Szab{\'o}}, \citenamefont {Samuely}, \citenamefont {Ka{\v{c}}mar{\v{c}}{\'\i}k}, \citenamefont {Klein}, \citenamefont {Marcus}, \citenamefont {Fruchart}, \citenamefont {Miraglia}, \citenamefont {Marcenat},\ and\ \citenamefont {Jansen}}]{szabo2001evidence}%
  \BibitemOpen
  \bibfield  {author} {\bibinfo {author} {\bibfnamefont {P.}~\bibnamefont {Szab{\'o}}}, \bibinfo {author} {\bibfnamefont {P.}~\bibnamefont {Samuely}}, \bibinfo {author} {\bibfnamefont {J.}~\bibnamefont {Ka{\v{c}}mar{\v{c}}{\'\i}k}}, \bibinfo {author} {\bibfnamefont {T.}~\bibnamefont {Klein}}, \bibinfo {author} {\bibfnamefont {J.}~\bibnamefont {Marcus}}, \bibinfo {author} {\bibfnamefont {D.}~\bibnamefont {Fruchart}}, \bibinfo {author} {\bibfnamefont {S.}~\bibnamefont {Miraglia}}, \bibinfo {author} {\bibfnamefont {C.}~\bibnamefont {Marcenat}},\ and\ \bibinfo {author} {\bibfnamefont {A.}~\bibnamefont {Jansen}},\ }\href@noop {} {\bibfield  {journal} {\bibinfo  {journal} {Physical review letters}\ }\textbf {\bibinfo {volume} {87}},\ \bibinfo {pages} {137005} (\bibinfo {year} {2001})}\BibitemShut {NoStop}%
\bibitem [{\citenamefont {Bouquet}\ \emph {et~al.}(2001)\citenamefont {Bouquet}, \citenamefont {Fisher}, \citenamefont {Phillips}, \citenamefont {Hinks},\ and\ \citenamefont {Jorgensen}}]{bouquet2001specific}%
  \BibitemOpen
  \bibfield  {author} {\bibinfo {author} {\bibfnamefont {F.}~\bibnamefont {Bouquet}}, \bibinfo {author} {\bibfnamefont {R.}~\bibnamefont {Fisher}}, \bibinfo {author} {\bibfnamefont {N.}~\bibnamefont {Phillips}}, \bibinfo {author} {\bibfnamefont {D.}~\bibnamefont {Hinks}},\ and\ \bibinfo {author} {\bibfnamefont {J.}~\bibnamefont {Jorgensen}},\ }\href@noop {} {\bibfield  {journal} {\bibinfo  {journal} {arXiv preprint cond-mat/0104206}\ } (\bibinfo {year} {2001})}\BibitemShut {NoStop}%
\bibitem [{\citenamefont {Giubileo}\ \emph {et~al.}(2001)\citenamefont {Giubileo}, \citenamefont {Roditchev}, \citenamefont {Sacks}, \citenamefont {Lamy}, \citenamefont {Thanh}, \citenamefont {Klein}, \citenamefont {Miraglia}, \citenamefont {Fruchart}, \citenamefont {Marcus},\ and\ \citenamefont {Monod}}]{giubileo2001two}%
  \BibitemOpen
  \bibfield  {author} {\bibinfo {author} {\bibfnamefont {F.}~\bibnamefont {Giubileo}}, \bibinfo {author} {\bibfnamefont {D.}~\bibnamefont {Roditchev}}, \bibinfo {author} {\bibfnamefont {W.}~\bibnamefont {Sacks}}, \bibinfo {author} {\bibfnamefont {R.}~\bibnamefont {Lamy}}, \bibinfo {author} {\bibfnamefont {D.}~\bibnamefont {Thanh}}, \bibinfo {author} {\bibfnamefont {J.}~\bibnamefont {Klein}}, \bibinfo {author} {\bibfnamefont {S.}~\bibnamefont {Miraglia}}, \bibinfo {author} {\bibfnamefont {D.}~\bibnamefont {Fruchart}}, \bibinfo {author} {\bibfnamefont {J.}~\bibnamefont {Marcus}},\ and\ \bibinfo {author} {\bibfnamefont {P.}~\bibnamefont {Monod}},\ }\href@noop {} {\bibfield  {journal} {\bibinfo  {journal} {Physical review letters}\ }\textbf {\bibinfo {volume} {87}},\ \bibinfo {pages} {177008} (\bibinfo {year} {2001})}\BibitemShut {NoStop}%
\bibitem [{\citenamefont {Angst}\ \emph {et~al.}(2002)\citenamefont {Angst}, \citenamefont {Puzniak}, \citenamefont {Wisniewski}, \citenamefont {Jun}, \citenamefont {Kazakov}, \citenamefont {Karpinski}, \citenamefont {Roos},\ and\ \citenamefont {Keller}}]{angst2002temperature}%
  \BibitemOpen
  \bibfield  {author} {\bibinfo {author} {\bibfnamefont {M.}~\bibnamefont {Angst}}, \bibinfo {author} {\bibfnamefont {R.}~\bibnamefont {Puzniak}}, \bibinfo {author} {\bibfnamefont {A.}~\bibnamefont {Wisniewski}}, \bibinfo {author} {\bibfnamefont {J.}~\bibnamefont {Jun}}, \bibinfo {author} {\bibfnamefont {S.}~\bibnamefont {Kazakov}}, \bibinfo {author} {\bibfnamefont {J.}~\bibnamefont {Karpinski}}, \bibinfo {author} {\bibfnamefont {J.}~\bibnamefont {Roos}},\ and\ \bibinfo {author} {\bibfnamefont {H.}~\bibnamefont {Keller}},\ }\href@noop {} {\bibfield  {journal} {\bibinfo  {journal} {Physical review letters}\ }\textbf {\bibinfo {volume} {88}},\ \bibinfo {pages} {167004} (\bibinfo {year} {2002})}\BibitemShut {NoStop}%
\bibitem [{\citenamefont {Souma}\ \emph {et~al.}(2003)\citenamefont {Souma}, \citenamefont {Machida}, \citenamefont {Sato}, \citenamefont {Takahashi}, \citenamefont {Matsui}, \citenamefont {Wang}, \citenamefont {Ding}, \citenamefont {Kaminski}, \citenamefont {Campuzano}, \citenamefont {Sasaki} \emph {et~al.}}]{souma2003origin}%
  \BibitemOpen
  \bibfield  {author} {\bibinfo {author} {\bibfnamefont {S.}~\bibnamefont {Souma}}, \bibinfo {author} {\bibfnamefont {Y.}~\bibnamefont {Machida}}, \bibinfo {author} {\bibfnamefont {T.}~\bibnamefont {Sato}}, \bibinfo {author} {\bibfnamefont {T.}~\bibnamefont {Takahashi}}, \bibinfo {author} {\bibfnamefont {H.}~\bibnamefont {Matsui}}, \bibinfo {author} {\bibfnamefont {S.-C.}\ \bibnamefont {Wang}}, \bibinfo {author} {\bibfnamefont {H.}~\bibnamefont {Ding}}, \bibinfo {author} {\bibfnamefont {A.}~\bibnamefont {Kaminski}}, \bibinfo {author} {\bibfnamefont {J.}~\bibnamefont {Campuzano}}, \bibinfo {author} {\bibfnamefont {S.}~\bibnamefont {Sasaki}}, \emph {et~al.},\ }\href@noop {} {\bibfield  {journal} {\bibinfo  {journal} {Nature}\ }\textbf {\bibinfo {volume} {423}},\ \bibinfo {pages} {65} (\bibinfo {year} {2003})}\BibitemShut {NoStop}%
\bibitem [{\citenamefont {Choi}\ \emph {et~al.}(2002{\natexlab{a}})\citenamefont {Choi}, \citenamefont {Roundy}, \citenamefont {Sun}, \citenamefont {Cohen},\ and\ \citenamefont {Louie}}]{Choi2002_MgB2_twoGap}%
  \BibitemOpen
  \bibfield  {author} {\bibinfo {author} {\bibfnamefont {H.~J.}\ \bibnamefont {Choi}}, \bibinfo {author} {\bibfnamefont {D.}~\bibnamefont {Roundy}}, \bibinfo {author} {\bibfnamefont {H.}~\bibnamefont {Sun}}, \bibinfo {author} {\bibfnamefont {M.~L.}\ \bibnamefont {Cohen}},\ and\ \bibinfo {author} {\bibfnamefont {S.~G.}\ \bibnamefont {Louie}},\ }\href@noop {} {\bibfield  {journal} {\bibinfo  {journal} {Nature}\ }\textbf {\bibinfo {volume} {418}},\ \bibinfo {pages} {758} (\bibinfo {year} {2002}{\natexlab{a}})}\BibitemShut {NoStop}%
\bibitem [{\citenamefont {Choi}\ \emph {et~al.}(2002{\natexlab{b}})\citenamefont {Choi}, \citenamefont {Roundy}, \citenamefont {Sun}, \citenamefont {Cohen},\ and\ \citenamefont {Louie}}]{choi2002first}%
  \BibitemOpen
  \bibfield  {author} {\bibinfo {author} {\bibfnamefont {H.~J.}\ \bibnamefont {Choi}}, \bibinfo {author} {\bibfnamefont {D.}~\bibnamefont {Roundy}}, \bibinfo {author} {\bibfnamefont {H.}~\bibnamefont {Sun}}, \bibinfo {author} {\bibfnamefont {M.~L.}\ \bibnamefont {Cohen}},\ and\ \bibinfo {author} {\bibfnamefont {S.~G.}\ \bibnamefont {Louie}},\ }\href@noop {} {\bibfield  {journal} {\bibinfo  {journal} {Physical Review B}\ }\textbf {\bibinfo {volume} {66}},\ \bibinfo {pages} {020513} (\bibinfo {year} {2002}{\natexlab{b}})}\BibitemShut {NoStop}%
\bibitem [{\citenamefont {Mazin}\ \emph {et~al.}(2002)\citenamefont {Mazin}, \citenamefont {Andersen}, \citenamefont {Jepsen}, \citenamefont {Dolgov}, \citenamefont {Kortus}, \citenamefont {Golubov}, \citenamefont {Kuz’menko},\ and\ \citenamefont {Van Der~Marel}}]{mazin2002superconductivity}%
  \BibitemOpen
  \bibfield  {author} {\bibinfo {author} {\bibfnamefont {I.}~\bibnamefont {Mazin}}, \bibinfo {author} {\bibfnamefont {O.}~\bibnamefont {Andersen}}, \bibinfo {author} {\bibfnamefont {O.}~\bibnamefont {Jepsen}}, \bibinfo {author} {\bibfnamefont {O.}~\bibnamefont {Dolgov}}, \bibinfo {author} {\bibfnamefont {J.}~\bibnamefont {Kortus}}, \bibinfo {author} {\bibfnamefont {A.~A.}\ \bibnamefont {Golubov}}, \bibinfo {author} {\bibfnamefont {.~f.~A.}\ \bibnamefont {Kuz’menko}},\ and\ \bibinfo {author} {\bibfnamefont {D.}~\bibnamefont {Van Der~Marel}},\ }\href@noop {} {\bibfield  {journal} {\bibinfo  {journal} {Physical review letters}\ }\textbf {\bibinfo {volume} {89}},\ \bibinfo {pages} {107002} (\bibinfo {year} {2002})}\BibitemShut {NoStop}%
\bibitem [{\citenamefont {Gurevich}(2003)}]{gurevich2003enhancement}%
  \BibitemOpen
  \bibfield  {author} {\bibinfo {author} {\bibfnamefont {A.}~\bibnamefont {Gurevich}},\ }\href@noop {} {\bibfield  {journal} {\bibinfo  {journal} {Physical Review B}\ }\textbf {\bibinfo {volume} {67}},\ \bibinfo {pages} {184515} (\bibinfo {year} {2003})}\BibitemShut {NoStop}%
\bibitem [{\citenamefont {Braccini}\ \emph {et~al.}(2005)\citenamefont {Braccini}, \citenamefont {Gurevich}, \citenamefont {Giencke}, \citenamefont {Jewell}, \citenamefont {Eom}, \citenamefont {Larbalestier}, \citenamefont {Pogrebnyakov}, \citenamefont {Cui}, \citenamefont {Liu}, \citenamefont {Hu} \emph {et~al.}}]{braccini2005high}%
  \BibitemOpen
  \bibfield  {author} {\bibinfo {author} {\bibfnamefont {V.}~\bibnamefont {Braccini}}, \bibinfo {author} {\bibfnamefont {A.}~\bibnamefont {Gurevich}}, \bibinfo {author} {\bibfnamefont {J.}~\bibnamefont {Giencke}}, \bibinfo {author} {\bibfnamefont {M.}~\bibnamefont {Jewell}}, \bibinfo {author} {\bibfnamefont {C.}~\bibnamefont {Eom}}, \bibinfo {author} {\bibfnamefont {D.}~\bibnamefont {Larbalestier}}, \bibinfo {author} {\bibfnamefont {A.}~\bibnamefont {Pogrebnyakov}}, \bibinfo {author} {\bibfnamefont {Y.}~\bibnamefont {Cui}}, \bibinfo {author} {\bibfnamefont {B.}~\bibnamefont {Liu}}, \bibinfo {author} {\bibfnamefont {Y.}~\bibnamefont {Hu}}, \emph {et~al.},\ }\href@noop {} {\bibfield  {journal} {\bibinfo  {journal} {Physical Review B—Condensed Matter and Materials Physics}\ }\textbf {\bibinfo {volume} {71}},\ \bibinfo {pages} {012504} (\bibinfo {year} {2005})}\BibitemShut {NoStop}%
\bibitem [{\citenamefont {Choi}\ \emph {et~al.}(2009)\citenamefont {Choi}, \citenamefont {Louie},\ and\ \citenamefont {Cohen}}]{choi2009anisotropic}%
  \BibitemOpen
  \bibfield  {author} {\bibinfo {author} {\bibfnamefont {H.~J.}\ \bibnamefont {Choi}}, \bibinfo {author} {\bibfnamefont {S.~G.}\ \bibnamefont {Louie}},\ and\ \bibinfo {author} {\bibfnamefont {M.~L.}\ \bibnamefont {Cohen}},\ }\href@noop {} {\bibfield  {journal} {\bibinfo  {journal} {Physical Review B—Condensed Matter and Materials Physics}\ }\textbf {\bibinfo {volume} {79}},\ \bibinfo {pages} {094518} (\bibinfo {year} {2009})}\BibitemShut {NoStop}%
\bibitem [{\citenamefont {Dou}\ \emph {et~al.}(2002)\citenamefont {Dou}, \citenamefont {Soltanian}, \citenamefont {Horvat}, \citenamefont {Wang}, \citenamefont {Zhou}, \citenamefont {Ionescu}, \citenamefont {Liu}, \citenamefont {Munroe},\ and\ \citenamefont {Tomsic}}]{dou2002enhancement}%
  \BibitemOpen
  \bibfield  {author} {\bibinfo {author} {\bibfnamefont {S.~X.}\ \bibnamefont {Dou}}, \bibinfo {author} {\bibfnamefont {S.}~\bibnamefont {Soltanian}}, \bibinfo {author} {\bibfnamefont {J.}~\bibnamefont {Horvat}}, \bibinfo {author} {\bibfnamefont {X.}~\bibnamefont {Wang}}, \bibinfo {author} {\bibfnamefont {S.}~\bibnamefont {Zhou}}, \bibinfo {author} {\bibfnamefont {M.}~\bibnamefont {Ionescu}}, \bibinfo {author} {\bibfnamefont {H.-K.}\ \bibnamefont {Liu}}, \bibinfo {author} {\bibfnamefont {P.}~\bibnamefont {Munroe}},\ and\ \bibinfo {author} {\bibfnamefont {M.}~\bibnamefont {Tomsic}},\ }\href@noop {} {\bibfield  {journal} {\bibinfo  {journal} {Applied Physics Letters}\ }\textbf {\bibinfo {volume} {81}},\ \bibinfo {pages} {3419} (\bibinfo {year} {2002})}\BibitemShut {NoStop}%
\bibitem [{\citenamefont {Slusky}\ \emph {et~al.}(2001)\citenamefont {Slusky}, \citenamefont {Rogado}, \citenamefont {Regan}, \citenamefont {Hayward}, \citenamefont {Khalifah}, \citenamefont {He}, \citenamefont {Inumaru}, \citenamefont {Loureiro}, \citenamefont {Haas}, \citenamefont {Zandbergen} \emph {et~al.}}]{Slusky2001_AlDoping}%
  \BibitemOpen
  \bibfield  {author} {\bibinfo {author} {\bibfnamefont {J.}~\bibnamefont {Slusky}}, \bibinfo {author} {\bibfnamefont {N.}~\bibnamefont {Rogado}}, \bibinfo {author} {\bibfnamefont {K.}~\bibnamefont {Regan}}, \bibinfo {author} {\bibfnamefont {M.}~\bibnamefont {Hayward}}, \bibinfo {author} {\bibfnamefont {P.}~\bibnamefont {Khalifah}}, \bibinfo {author} {\bibfnamefont {T.}~\bibnamefont {He}}, \bibinfo {author} {\bibfnamefont {K.}~\bibnamefont {Inumaru}}, \bibinfo {author} {\bibfnamefont {S.}~\bibnamefont {Loureiro}}, \bibinfo {author} {\bibfnamefont {M.}~\bibnamefont {Haas}}, \bibinfo {author} {\bibfnamefont {H.}~\bibnamefont {Zandbergen}}, \emph {et~al.},\ }\href@noop {} {\bibfield  {journal} {\bibinfo  {journal} {Nature}\ }\textbf {\bibinfo {volume} {410}},\ \bibinfo {pages} {343} (\bibinfo {year} {2001})}\BibitemShut {NoStop}%
\bibitem [{\citenamefont {Lee}\ \emph {et~al.}(2003)\citenamefont {Lee}, \citenamefont {Masui}, \citenamefont {Yamamoto}, \citenamefont {Uchiyama},\ and\ \citenamefont {Tajima}}]{lee2003carbon}%
  \BibitemOpen
  \bibfield  {author} {\bibinfo {author} {\bibfnamefont {S.}~\bibnamefont {Lee}}, \bibinfo {author} {\bibfnamefont {T.}~\bibnamefont {Masui}}, \bibinfo {author} {\bibfnamefont {A.}~\bibnamefont {Yamamoto}}, \bibinfo {author} {\bibfnamefont {H.}~\bibnamefont {Uchiyama}},\ and\ \bibinfo {author} {\bibfnamefont {S.}~\bibnamefont {Tajima}},\ }\href@noop {} {\bibfield  {journal} {\bibinfo  {journal} {Physica C: Superconductivity}\ }\textbf {\bibinfo {volume} {397}},\ \bibinfo {pages} {7} (\bibinfo {year} {2003})}\BibitemShut {NoStop}%
\bibitem [{\citenamefont {Kazakov}\ \emph {et~al.}(2005)\citenamefont {Kazakov}, \citenamefont {Puzniak}, \citenamefont {Rogacki}, \citenamefont {Mironov}, \citenamefont {Zhigadlo}, \citenamefont {Jun}, \citenamefont {Soltmann}, \citenamefont {Batlogg},\ and\ \citenamefont {Karpinski}}]{kazakov2005carbon}%
  \BibitemOpen
  \bibfield  {author} {\bibinfo {author} {\bibfnamefont {S.}~\bibnamefont {Kazakov}}, \bibinfo {author} {\bibfnamefont {R.}~\bibnamefont {Puzniak}}, \bibinfo {author} {\bibfnamefont {K.}~\bibnamefont {Rogacki}}, \bibinfo {author} {\bibfnamefont {A.}~\bibnamefont {Mironov}}, \bibinfo {author} {\bibfnamefont {N.}~\bibnamefont {Zhigadlo}}, \bibinfo {author} {\bibfnamefont {J.}~\bibnamefont {Jun}}, \bibinfo {author} {\bibfnamefont {C.}~\bibnamefont {Soltmann}}, \bibinfo {author} {\bibfnamefont {B.}~\bibnamefont {Batlogg}},\ and\ \bibinfo {author} {\bibfnamefont {J.}~\bibnamefont {Karpinski}},\ }\href@noop {} {\bibfield  {journal} {\bibinfo  {journal} {Physical Review B—Condensed Matter and Materials Physics}\ }\textbf {\bibinfo {volume} {71}},\ \bibinfo {pages} {024533} (\bibinfo {year} {2005})}\BibitemShut {NoStop}%
\bibitem [{\citenamefont {Pogrebnyakov}\ \emph {et~al.}(2004)\citenamefont {Pogrebnyakov}, \citenamefont {Redwing}, \citenamefont {Raghavan}, \citenamefont {Vaithyanathan}, \citenamefont {Schlom}, \citenamefont {Xu}, \citenamefont {Li}, \citenamefont {Tenne}, \citenamefont {Soukiassian}, \citenamefont {Xi} \emph {et~al.}}]{pogrebnyakov2004enhancement}%
  \BibitemOpen
  \bibfield  {author} {\bibinfo {author} {\bibfnamefont {A.}~\bibnamefont {Pogrebnyakov}}, \bibinfo {author} {\bibfnamefont {J.}~\bibnamefont {Redwing}}, \bibinfo {author} {\bibfnamefont {S.}~\bibnamefont {Raghavan}}, \bibinfo {author} {\bibfnamefont {V.}~\bibnamefont {Vaithyanathan}}, \bibinfo {author} {\bibfnamefont {D.}~\bibnamefont {Schlom}}, \bibinfo {author} {\bibfnamefont {S.}~\bibnamefont {Xu}}, \bibinfo {author} {\bibfnamefont {Q.}~\bibnamefont {Li}}, \bibinfo {author} {\bibfnamefont {D.}~\bibnamefont {Tenne}}, \bibinfo {author} {\bibfnamefont {A.}~\bibnamefont {Soukiassian}}, \bibinfo {author} {\bibfnamefont {X.}~\bibnamefont {Xi}}, \emph {et~al.},\ }\href@noop {} {\bibfield  {journal} {\bibinfo  {journal} {Physical review letters}\ }\textbf {\bibinfo {volume} {93}},\ \bibinfo {pages} {147006} (\bibinfo {year} {2004})}\BibitemShut {NoStop}%
\bibitem [{\citenamefont {Xi}\ \emph {et~al.}(2007)\citenamefont {Xi}, \citenamefont {Pogrebnyakov}, \citenamefont {Xu}, \citenamefont {Chen}, \citenamefont {Cui}, \citenamefont {Maertz}, \citenamefont {Zhuang}, \citenamefont {Li}, \citenamefont {Lamborn}, \citenamefont {Redwing} \emph {et~al.}}]{xi2007mgb2}%
  \BibitemOpen
  \bibfield  {author} {\bibinfo {author} {\bibfnamefont {X.}~\bibnamefont {Xi}}, \bibinfo {author} {\bibfnamefont {A.}~\bibnamefont {Pogrebnyakov}}, \bibinfo {author} {\bibfnamefont {S.}~\bibnamefont {Xu}}, \bibinfo {author} {\bibfnamefont {K.}~\bibnamefont {Chen}}, \bibinfo {author} {\bibfnamefont {Y.}~\bibnamefont {Cui}}, \bibinfo {author} {\bibfnamefont {E.}~\bibnamefont {Maertz}}, \bibinfo {author} {\bibfnamefont {C.}~\bibnamefont {Zhuang}}, \bibinfo {author} {\bibfnamefont {Q.}~\bibnamefont {Li}}, \bibinfo {author} {\bibfnamefont {D.}~\bibnamefont {Lamborn}}, \bibinfo {author} {\bibfnamefont {J.~M.}\ \bibnamefont {Redwing}}, \emph {et~al.},\ }\href@noop {} {\bibfield  {journal} {\bibinfo  {journal} {Physica C: Superconductivity}\ }\textbf {\bibinfo {volume} {456}},\ \bibinfo {pages} {22} (\bibinfo {year} {2007})}\BibitemShut {NoStop}%
\bibitem [{\citenamefont {Zheng}\ and\ \citenamefont {Zhu}(2006)}]{zheng2006searching}%
  \BibitemOpen
  \bibfield  {author} {\bibinfo {author} {\bibfnamefont {J.-C.}\ \bibnamefont {Zheng}}\ and\ \bibinfo {author} {\bibfnamefont {Y.}~\bibnamefont {Zhu}},\ }\href@noop {} {\bibfield  {journal} {\bibinfo  {journal} {Physical Review B—Condensed Matter and Materials Physics}\ }\textbf {\bibinfo {volume} {73}},\ \bibinfo {pages} {024509} (\bibinfo {year} {2006})}\BibitemShut {NoStop}%
\bibitem [{\citenamefont {Bekaert}\ \emph {et~al.}(2017)\citenamefont {Bekaert}, \citenamefont {Aperis}, \citenamefont {Partoens}, \citenamefont {Oppeneer},\ and\ \citenamefont {Milo\ifmmode \check{s}\else \v{s}\fi{}evi\ifmmode~\acute{c}\else \'{c}\fi{}}}]{bekaert2017evolution}%
  \BibitemOpen
  \bibfield  {author} {\bibinfo {author} {\bibfnamefont {J.}~\bibnamefont {Bekaert}}, \bibinfo {author} {\bibfnamefont {A.}~\bibnamefont {Aperis}}, \bibinfo {author} {\bibfnamefont {B.}~\bibnamefont {Partoens}}, \bibinfo {author} {\bibfnamefont {P.~M.}\ \bibnamefont {Oppeneer}},\ and\ \bibinfo {author} {\bibfnamefont {M.~V.}\ \bibnamefont {Milo\ifmmode \check{s}\else \v{s}\fi{}evi\ifmmode~\acute{c}\else \'{c}\fi{}}},\ }\href {https://doi.org/10.1103/PhysRevB.96.094510} {\bibfield  {journal} {\bibinfo  {journal} {Phys. Rev. B}\ }\textbf {\bibinfo {volume} {96}},\ \bibinfo {pages} {094510} (\bibinfo {year} {2017})}\BibitemShut {NoStop}%
\bibitem [{\citenamefont {Zhang}\ and\ \citenamefont {Zhang}(2011)}]{zhang2011effect}%
  \BibitemOpen
  \bibfield  {author} {\bibinfo {author} {\bibfnamefont {C.}~\bibnamefont {Zhang}}\ and\ \bibinfo {author} {\bibfnamefont {X.}~\bibnamefont {Zhang}},\ }\href@noop {} {\bibfield  {journal} {\bibinfo  {journal} {Computational Materials Science}\ }\textbf {\bibinfo {volume} {50}},\ \bibinfo {pages} {1097} (\bibinfo {year} {2011})}\BibitemShut {NoStop}%
\bibitem [{\citenamefont {Johansson}\ \emph {et~al.}(2022)\citenamefont {Johansson}, \citenamefont {Tasn{\'a}di}, \citenamefont {Ektarawong}, \citenamefont {Rosen},\ and\ \citenamefont {Alling}}]{johansson2022effect}%
  \BibitemOpen
  \bibfield  {author} {\bibinfo {author} {\bibfnamefont {E.}~\bibnamefont {Johansson}}, \bibinfo {author} {\bibfnamefont {F.}~\bibnamefont {Tasn{\'a}di}}, \bibinfo {author} {\bibfnamefont {A.}~\bibnamefont {Ektarawong}}, \bibinfo {author} {\bibfnamefont {J.}~\bibnamefont {Rosen}},\ and\ \bibinfo {author} {\bibfnamefont {B.}~\bibnamefont {Alling}},\ }\href@noop {} {\bibfield  {journal} {\bibinfo  {journal} {Journal of Applied Physics}\ }\textbf {\bibinfo {volume} {131}} (\bibinfo {year} {2022})}\BibitemShut {NoStop}%
\bibitem [{\citenamefont {Resta}(2011)}]{resta2011insulating}%
  \BibitemOpen
  \bibfield  {author} {\bibinfo {author} {\bibfnamefont {R.}~\bibnamefont {Resta}},\ }\href@noop {} {\bibfield  {journal} {\bibinfo  {journal} {The European Physical Journal B}\ }\textbf {\bibinfo {volume} {79}},\ \bibinfo {pages} {121} (\bibinfo {year} {2011})}\BibitemShut {NoStop}%
\bibitem [{\citenamefont {Provost}\ and\ \citenamefont {Vallee}(1980)}]{provost1980riemannian}%
  \BibitemOpen
  \bibfield  {author} {\bibinfo {author} {\bibfnamefont {J.}~\bibnamefont {Provost}}\ and\ \bibinfo {author} {\bibfnamefont {G.}~\bibnamefont {Vallee}},\ }\href@noop {} {\bibfield  {journal} {\bibinfo  {journal} {Communications in Mathematical Physics}\ }\textbf {\bibinfo {volume} {76}},\ \bibinfo {pages} {289} (\bibinfo {year} {1980})}\BibitemShut {NoStop}%
\bibitem [{\citenamefont {Peotta}\ and\ \citenamefont {T{\"o}rm{\"a}}(2015)}]{PeottaTorma2015_superfluidWeight}%
  \BibitemOpen
  \bibfield  {author} {\bibinfo {author} {\bibfnamefont {S.}~\bibnamefont {Peotta}}\ and\ \bibinfo {author} {\bibfnamefont {P.}~\bibnamefont {T{\"o}rm{\"a}}},\ }\href@noop {} {\bibfield  {journal} {\bibinfo  {journal} {Nature communications}\ }\textbf {\bibinfo {volume} {6}},\ \bibinfo {pages} {8944} (\bibinfo {year} {2015})}\BibitemShut {NoStop}%
\bibitem [{\citenamefont {Julku}\ \emph {et~al.}(2016)\citenamefont {Julku}, \citenamefont {Peotta}, \citenamefont {Vanhala}, \citenamefont {Kim},\ and\ \citenamefont {T\"orm\"a}}]{Julku2016_quantumMetric}%
  \BibitemOpen
  \bibfield  {author} {\bibinfo {author} {\bibfnamefont {A.}~\bibnamefont {Julku}}, \bibinfo {author} {\bibfnamefont {S.}~\bibnamefont {Peotta}}, \bibinfo {author} {\bibfnamefont {T.~I.}\ \bibnamefont {Vanhala}}, \bibinfo {author} {\bibfnamefont {D.-H.}\ \bibnamefont {Kim}},\ and\ \bibinfo {author} {\bibfnamefont {P.}~\bibnamefont {T\"orm\"a}},\ }\href {https://doi.org/10.1103/PhysRevLett.117.045303} {\bibfield  {journal} {\bibinfo  {journal} {Phys. Rev. Lett.}\ }\textbf {\bibinfo {volume} {117}},\ \bibinfo {pages} {045303} (\bibinfo {year} {2016})}\BibitemShut {NoStop}%
\bibitem [{\citenamefont {Peotta}\ \emph {et~al.}(2023)\citenamefont {Peotta}, \citenamefont {Huhtinen},\ and\ \citenamefont {T{\"o}rm{\"a}}}]{peotta2308quantum}%
  \BibitemOpen
  \bibfield  {author} {\bibinfo {author} {\bibfnamefont {S.}~\bibnamefont {Peotta}}, \bibinfo {author} {\bibfnamefont {K.-E.}\ \bibnamefont {Huhtinen}},\ and\ \bibinfo {author} {\bibfnamefont {P.}~\bibnamefont {T{\"o}rm{\"a}}},\ }\href@noop {} {\bibfield  {journal} {\bibinfo  {journal} {arXiv preprint arXiv:2308.08248}\ } (\bibinfo {year} {2023})}\BibitemShut {NoStop}%
\bibitem [{\citenamefont {T{\"o}rm{\"a}}\ \emph {et~al.}(2022)\citenamefont {T{\"o}rm{\"a}}, \citenamefont {Peotta},\ and\ \citenamefont {Bernevig}}]{torma2022superconductivity}%
  \BibitemOpen
  \bibfield  {author} {\bibinfo {author} {\bibfnamefont {P.}~\bibnamefont {T{\"o}rm{\"a}}}, \bibinfo {author} {\bibfnamefont {S.}~\bibnamefont {Peotta}},\ and\ \bibinfo {author} {\bibfnamefont {B.~A.}\ \bibnamefont {Bernevig}},\ }\href@noop {} {\bibfield  {journal} {\bibinfo  {journal} {Nature Reviews Physics}\ }\textbf {\bibinfo {volume} {4}},\ \bibinfo {pages} {528} (\bibinfo {year} {2022})}\BibitemShut {NoStop}%
\bibitem [{\citenamefont {T{\"o}rm{\"a}}(2023)}]{torma2023essay}%
  \BibitemOpen
  \bibfield  {author} {\bibinfo {author} {\bibfnamefont {P.}~\bibnamefont {T{\"o}rm{\"a}}},\ }\href@noop {} {\bibfield  {journal} {\bibinfo  {journal} {Physical Review Letters}\ }\textbf {\bibinfo {volume} {131}},\ \bibinfo {pages} {240001} (\bibinfo {year} {2023})}\BibitemShut {NoStop}%
\bibitem [{\citenamefont {Yu}\ \emph {et~al.}(2025)\citenamefont {Yu}, \citenamefont {Bernevig}, \citenamefont {Queiroz}, \citenamefont {Rossi}, \citenamefont {T{\"o}rm{\"a}},\ and\ \citenamefont {Yang}}]{yu2025quantum}%
  \BibitemOpen
  \bibfield  {author} {\bibinfo {author} {\bibfnamefont {J.}~\bibnamefont {Yu}}, \bibinfo {author} {\bibfnamefont {B.~A.}\ \bibnamefont {Bernevig}}, \bibinfo {author} {\bibfnamefont {R.}~\bibnamefont {Queiroz}}, \bibinfo {author} {\bibfnamefont {E.}~\bibnamefont {Rossi}}, \bibinfo {author} {\bibfnamefont {P.}~\bibnamefont {T{\"o}rm{\"a}}},\ and\ \bibinfo {author} {\bibfnamefont {B.-J.}\ \bibnamefont {Yang}},\ }\href@noop {} {\bibfield  {journal} {\bibinfo  {journal} {npj Quantum Materials}\ }\textbf {\bibinfo {volume} {10}},\ \bibinfo {pages} {101} (\bibinfo {year} {2025})}\BibitemShut {NoStop}%
\bibitem [{\citenamefont {Liu}\ \emph {et~al.}(2025)\citenamefont {Liu}, \citenamefont {Qiang}, \citenamefont {Lu},\ and\ \citenamefont {Xie}}]{liu2025quantum}%
  \BibitemOpen
  \bibfield  {author} {\bibinfo {author} {\bibfnamefont {T.}~\bibnamefont {Liu}}, \bibinfo {author} {\bibfnamefont {X.-B.}\ \bibnamefont {Qiang}}, \bibinfo {author} {\bibfnamefont {H.-Z.}\ \bibnamefont {Lu}},\ and\ \bibinfo {author} {\bibfnamefont {X.}~\bibnamefont {Xie}},\ }\href@noop {} {\bibfield  {journal} {\bibinfo  {journal} {National Science Review}\ }\textbf {\bibinfo {volume} {12}},\ \bibinfo {pages} {nwae334} (\bibinfo {year} {2025})}\BibitemShut {NoStop}%
\bibitem [{\citenamefont {Verma}\ \emph {et~al.}(2026)\citenamefont {Verma}, \citenamefont {Moll}, \citenamefont {Holder},\ and\ \citenamefont {Queiroz}}]{verma2025quantum}%
  \BibitemOpen
  \bibfield  {author} {\bibinfo {author} {\bibfnamefont {N.}~\bibnamefont {Verma}}, \bibinfo {author} {\bibfnamefont {P.~J.}\ \bibnamefont {Moll}}, \bibinfo {author} {\bibfnamefont {T.}~\bibnamefont {Holder}},\ and\ \bibinfo {author} {\bibfnamefont {R.}~\bibnamefont {Queiroz}},\ }\href@noop {} {\bibfield  {journal} {\bibinfo  {journal} {Nature Reviews Physics}\ ,\ \bibinfo {pages} {1}} (\bibinfo {year} {2026})}\BibitemShut {NoStop}%
\bibitem [{\citenamefont {Yu}\ \emph {et~al.}(2024)\citenamefont {Yu}, \citenamefont {Ciccarino}, \citenamefont {Bianco}, \citenamefont {Errea}, \citenamefont {Narang},\ and\ \citenamefont {Bernevig}}]{yu2024non}%
  \BibitemOpen
  \bibfield  {author} {\bibinfo {author} {\bibfnamefont {J.}~\bibnamefont {Yu}}, \bibinfo {author} {\bibfnamefont {C.~J.}\ \bibnamefont {Ciccarino}}, \bibinfo {author} {\bibfnamefont {R.}~\bibnamefont {Bianco}}, \bibinfo {author} {\bibfnamefont {I.}~\bibnamefont {Errea}}, \bibinfo {author} {\bibfnamefont {P.}~\bibnamefont {Narang}},\ and\ \bibinfo {author} {\bibfnamefont {B.~A.}\ \bibnamefont {Bernevig}},\ }\href@noop {} {\bibfield  {journal} {\bibinfo  {journal} {Nature Physics}\ ,\ \bibinfo {pages} {1}} (\bibinfo {year} {2024})}\BibitemShut {NoStop}%
\bibitem [{\citenamefont {Xu}\ \emph {et~al.}(2021)\citenamefont {Xu}, \citenamefont {Elcoro}, \citenamefont {Li}, \citenamefont {Song}, \citenamefont {Regnault}, \citenamefont {Yang}, \citenamefont {Sun}, \citenamefont {Parkin}, \citenamefont {Felser},\ and\ \citenamefont {Bernevig}}]{xu2021three}%
  \BibitemOpen
  \bibfield  {author} {\bibinfo {author} {\bibfnamefont {Y.}~\bibnamefont {Xu}}, \bibinfo {author} {\bibfnamefont {L.}~\bibnamefont {Elcoro}}, \bibinfo {author} {\bibfnamefont {G.}~\bibnamefont {Li}}, \bibinfo {author} {\bibfnamefont {Z.-D.}\ \bibnamefont {Song}}, \bibinfo {author} {\bibfnamefont {N.}~\bibnamefont {Regnault}}, \bibinfo {author} {\bibfnamefont {Q.}~\bibnamefont {Yang}}, \bibinfo {author} {\bibfnamefont {Y.}~\bibnamefont {Sun}}, \bibinfo {author} {\bibfnamefont {S.}~\bibnamefont {Parkin}}, \bibinfo {author} {\bibfnamefont {C.}~\bibnamefont {Felser}},\ and\ \bibinfo {author} {\bibfnamefont {B.~A.}\ \bibnamefont {Bernevig}},\ }\href@noop {} {\bibfield  {journal} {\bibinfo  {journal} {arXiv preprint arXiv:2111.02433}\ } (\bibinfo {year} {2021})}\BibitemShut {NoStop}%
\bibitem [{\citenamefont {Xu}\ \emph {et~al.}(2024{\natexlab{a}})\citenamefont {Xu}, \citenamefont {Elcoro}, \citenamefont {Song}, \citenamefont {Vergniory}, \citenamefont {Felser}, \citenamefont {Parkin}, \citenamefont {Regnault}, \citenamefont {Ma{\~n}es},\ and\ \citenamefont {Bernevig}}]{xu2024filling}%
  \BibitemOpen
  \bibfield  {author} {\bibinfo {author} {\bibfnamefont {Y.}~\bibnamefont {Xu}}, \bibinfo {author} {\bibfnamefont {L.}~\bibnamefont {Elcoro}}, \bibinfo {author} {\bibfnamefont {Z.-D.}\ \bibnamefont {Song}}, \bibinfo {author} {\bibfnamefont {M.}~\bibnamefont {Vergniory}}, \bibinfo {author} {\bibfnamefont {C.}~\bibnamefont {Felser}}, \bibinfo {author} {\bibfnamefont {S.~S.}\ \bibnamefont {Parkin}}, \bibinfo {author} {\bibfnamefont {N.}~\bibnamefont {Regnault}}, \bibinfo {author} {\bibfnamefont {J.~L.}\ \bibnamefont {Ma{\~n}es}},\ and\ \bibinfo {author} {\bibfnamefont {B.~A.}\ \bibnamefont {Bernevig}},\ }\href@noop {} {\bibfield  {journal} {\bibinfo  {journal} {Physical Review B}\ }\textbf {\bibinfo {volume} {109}},\ \bibinfo {pages} {165139} (\bibinfo {year} {2024}{\natexlab{a}})}\BibitemShut {NoStop}%
\bibitem [{\citenamefont {Yao}\ \emph {et~al.}(2007)\citenamefont {Yao}, \citenamefont {Ye}, \citenamefont {Qi}, \citenamefont {Zhang},\ and\ \citenamefont {Fang}}]{yao2007spin}%
  \BibitemOpen
  \bibfield  {author} {\bibinfo {author} {\bibfnamefont {Y.}~\bibnamefont {Yao}}, \bibinfo {author} {\bibfnamefont {F.}~\bibnamefont {Ye}}, \bibinfo {author} {\bibfnamefont {X.-L.}\ \bibnamefont {Qi}}, \bibinfo {author} {\bibfnamefont {S.-C.}\ \bibnamefont {Zhang}},\ and\ \bibinfo {author} {\bibfnamefont {Z.}~\bibnamefont {Fang}},\ }\href@noop {} {\bibfield  {journal} {\bibinfo  {journal} {Physical Review B—Condensed Matter and Materials Physics}\ }\textbf {\bibinfo {volume} {75}},\ \bibinfo {pages} {041401} (\bibinfo {year} {2007})}\BibitemShut {NoStop}%
\bibitem [{\citenamefont {Abergel}\ \emph {et~al.}(2010)\citenamefont {Abergel}, \citenamefont {Apalkov}, \citenamefont {Berashevich}, \citenamefont {Ziegler},\ and\ \citenamefont {Chakraborty}}]{abergel2010properties}%
  \BibitemOpen
  \bibfield  {author} {\bibinfo {author} {\bibfnamefont {D.}~\bibnamefont {Abergel}}, \bibinfo {author} {\bibfnamefont {V.}~\bibnamefont {Apalkov}}, \bibinfo {author} {\bibfnamefont {J.}~\bibnamefont {Berashevich}}, \bibinfo {author} {\bibfnamefont {K.}~\bibnamefont {Ziegler}},\ and\ \bibinfo {author} {\bibfnamefont {T.}~\bibnamefont {Chakraborty}},\ }\href@noop {} {\bibfield  {journal} {\bibinfo  {journal} {Advances in Physics}\ }\textbf {\bibinfo {volume} {59}},\ \bibinfo {pages} {261} (\bibinfo {year} {2010})}\BibitemShut {NoStop}%
\bibitem [{\citenamefont {Aroyo}\ \emph {et~al.}(2011)\citenamefont {Aroyo}, \citenamefont {Perez-Mato}, \citenamefont {Orobengoa}, \citenamefont {Tasci}, \citenamefont {de~la Flor},\ and\ \citenamefont {Kirov}}]{aroyo2011crystallography}%
  \BibitemOpen
  \bibfield  {author} {\bibinfo {author} {\bibfnamefont {M.~I.}\ \bibnamefont {Aroyo}}, \bibinfo {author} {\bibfnamefont {J.~M.}\ \bibnamefont {Perez-Mato}}, \bibinfo {author} {\bibfnamefont {D.}~\bibnamefont {Orobengoa}}, \bibinfo {author} {\bibfnamefont {E.}~\bibnamefont {Tasci}}, \bibinfo {author} {\bibfnamefont {G.}~\bibnamefont {de~la Flor}},\ and\ \bibinfo {author} {\bibfnamefont {A.}~\bibnamefont {Kirov}},\ }\href@noop {} {\bibfield  {journal} {\bibinfo  {journal} {Bulg. Chem. Commun}\ }\textbf {\bibinfo {volume} {43}},\ \bibinfo {pages} {183} (\bibinfo {year} {2011})}\BibitemShut {NoStop}%
\bibitem [{\citenamefont {Aroyo}\ \emph {et~al.}(2006{\natexlab{a}})\citenamefont {Aroyo}, \citenamefont {Perez-Mato}, \citenamefont {Capillas}, \citenamefont {Kroumova}, \citenamefont {Ivantchev}, \citenamefont {Madariaga}, \citenamefont {Kirov},\ and\ \citenamefont {Wondratschek}}]{aroyo2006bilbao1}%
  \BibitemOpen
  \bibfield  {author} {\bibinfo {author} {\bibfnamefont {M.~I.}\ \bibnamefont {Aroyo}}, \bibinfo {author} {\bibfnamefont {J.~M.}\ \bibnamefont {Perez-Mato}}, \bibinfo {author} {\bibfnamefont {C.}~\bibnamefont {Capillas}}, \bibinfo {author} {\bibfnamefont {E.}~\bibnamefont {Kroumova}}, \bibinfo {author} {\bibfnamefont {S.}~\bibnamefont {Ivantchev}}, \bibinfo {author} {\bibfnamefont {G.}~\bibnamefont {Madariaga}}, \bibinfo {author} {\bibfnamefont {A.}~\bibnamefont {Kirov}},\ and\ \bibinfo {author} {\bibfnamefont {H.}~\bibnamefont {Wondratschek}},\ }\href@noop {} {\bibfield  {journal} {\bibinfo  {journal} {Zeitschrift f{\"u}r Kristallographie-Crystalline Materials}\ }\textbf {\bibinfo {volume} {221}},\ \bibinfo {pages} {15} (\bibinfo {year} {2006}{\natexlab{a}})}\BibitemShut {NoStop}%
\bibitem [{\citenamefont {Aroyo}\ \emph {et~al.}(2006{\natexlab{b}})\citenamefont {Aroyo}, \citenamefont {Kirov}, \citenamefont {Capillas}, \citenamefont {Perez-Mato},\ and\ \citenamefont {Wondratschek}}]{aroyo2006bilbao2}%
  \BibitemOpen
  \bibfield  {author} {\bibinfo {author} {\bibfnamefont {M.~I.}\ \bibnamefont {Aroyo}}, \bibinfo {author} {\bibfnamefont {A.}~\bibnamefont {Kirov}}, \bibinfo {author} {\bibfnamefont {C.}~\bibnamefont {Capillas}}, \bibinfo {author} {\bibfnamefont {J.}~\bibnamefont {Perez-Mato}},\ and\ \bibinfo {author} {\bibfnamefont {H.}~\bibnamefont {Wondratschek}},\ }\href@noop {} {\bibfield  {journal} {\bibinfo  {journal} {Acta Crystallographica Section A: Foundations of Crystallography}\ }\textbf {\bibinfo {volume} {62}},\ \bibinfo {pages} {115} (\bibinfo {year} {2006}{\natexlab{b}})}\BibitemShut {NoStop}%
\bibitem [{\citenamefont {Gao}\ \emph {et~al.}(2022)\citenamefont {Gao}, \citenamefont {Qian}, \citenamefont {Jia}, \citenamefont {Guo}, \citenamefont {Fang}, \citenamefont {Liu}, \citenamefont {Weng},\ and\ \citenamefont {Wang}}]{gao2022unconventional}%
  \BibitemOpen
  \bibfield  {author} {\bibinfo {author} {\bibfnamefont {J.}~\bibnamefont {Gao}}, \bibinfo {author} {\bibfnamefont {Y.}~\bibnamefont {Qian}}, \bibinfo {author} {\bibfnamefont {H.}~\bibnamefont {Jia}}, \bibinfo {author} {\bibfnamefont {Z.}~\bibnamefont {Guo}}, \bibinfo {author} {\bibfnamefont {Z.}~\bibnamefont {Fang}}, \bibinfo {author} {\bibfnamefont {M.}~\bibnamefont {Liu}}, \bibinfo {author} {\bibfnamefont {H.}~\bibnamefont {Weng}},\ and\ \bibinfo {author} {\bibfnamefont {Z.}~\bibnamefont {Wang}},\ }\href@noop {} {\bibfield  {journal} {\bibinfo  {journal} {Science bulletin}\ }\textbf {\bibinfo {volume} {67}},\ \bibinfo {pages} {598} (\bibinfo {year} {2022})}\BibitemShut {NoStop}%
\bibitem [{\citenamefont {Yang}\ \emph {et~al.}(2024)\citenamefont {Yang}, \citenamefont {Sheng}, \citenamefont {Guo}, \citenamefont {Zhang}, \citenamefont {Wu}, \citenamefont {Weng}, \citenamefont {Fang},\ and\ \citenamefont {Wang}}]{yang2024superconductivity}%
  \BibitemOpen
  \bibfield  {author} {\bibinfo {author} {\bibfnamefont {Z.}~\bibnamefont {Yang}}, \bibinfo {author} {\bibfnamefont {H.}~\bibnamefont {Sheng}}, \bibinfo {author} {\bibfnamefont {Z.}~\bibnamefont {Guo}}, \bibinfo {author} {\bibfnamefont {R.}~\bibnamefont {Zhang}}, \bibinfo {author} {\bibfnamefont {Q.}~\bibnamefont {Wu}}, \bibinfo {author} {\bibfnamefont {H.}~\bibnamefont {Weng}}, \bibinfo {author} {\bibfnamefont {Z.}~\bibnamefont {Fang}},\ and\ \bibinfo {author} {\bibfnamefont {Z.}~\bibnamefont {Wang}},\ }\href@noop {} {\bibfield  {journal} {\bibinfo  {journal} {npj Computational Materials}\ }\textbf {\bibinfo {volume} {10}},\ \bibinfo {pages} {25} (\bibinfo {year} {2024})}\BibitemShut {NoStop}%
\bibitem [{\citenamefont {Wang}\ \emph {et~al.}(2022)\citenamefont {Wang}, \citenamefont {Jiang}, \citenamefont {Liu}, \citenamefont {Zhang}, \citenamefont {Li}, \citenamefont {Liu}, \citenamefont {Sun}, \citenamefont {Weng},\ and\ \citenamefont {Chen}}]{wang2022two}%
  \BibitemOpen
  \bibfield  {author} {\bibinfo {author} {\bibfnamefont {L.}~\bibnamefont {Wang}}, \bibinfo {author} {\bibfnamefont {Y.}~\bibnamefont {Jiang}}, \bibinfo {author} {\bibfnamefont {J.}~\bibnamefont {Liu}}, \bibinfo {author} {\bibfnamefont {S.}~\bibnamefont {Zhang}}, \bibinfo {author} {\bibfnamefont {J.}~\bibnamefont {Li}}, \bibinfo {author} {\bibfnamefont {P.}~\bibnamefont {Liu}}, \bibinfo {author} {\bibfnamefont {Y.}~\bibnamefont {Sun}}, \bibinfo {author} {\bibfnamefont {H.}~\bibnamefont {Weng}},\ and\ \bibinfo {author} {\bibfnamefont {X.-Q.}\ \bibnamefont {Chen}},\ }\href@noop {} {\bibfield  {journal} {\bibinfo  {journal} {Physical Review B}\ }\textbf {\bibinfo {volume} {106}},\ \bibinfo {pages} {155144} (\bibinfo {year} {2022})}\BibitemShut {NoStop}%
\bibitem [{\citenamefont {C{\u{a}}lug{\u{a}}ru}\ \emph {et~al.}(2025)\citenamefont {C{\u{a}}lug{\u{a}}ru}, \citenamefont {Jiang}, \citenamefont {Guo}, \citenamefont {Sajan}, \citenamefont {Wang}, \citenamefont {Hu}, \citenamefont {Yu}, \citenamefont {Bernevig}, \citenamefont {de~Juan},\ and\ \citenamefont {Ugeda}}]{cualuguaru2025probing}%
  \BibitemOpen
  \bibfield  {author} {\bibinfo {author} {\bibfnamefont {D.}~\bibnamefont {C{\u{a}}lug{\u{a}}ru}}, \bibinfo {author} {\bibfnamefont {Y.}~\bibnamefont {Jiang}}, \bibinfo {author} {\bibfnamefont {H.}~\bibnamefont {Guo}}, \bibinfo {author} {\bibfnamefont {S.}~\bibnamefont {Sajan}}, \bibinfo {author} {\bibfnamefont {Y.}~\bibnamefont {Wang}}, \bibinfo {author} {\bibfnamefont {H.}~\bibnamefont {Hu}}, \bibinfo {author} {\bibfnamefont {J.}~\bibnamefont {Yu}}, \bibinfo {author} {\bibfnamefont {B.~A.}\ \bibnamefont {Bernevig}}, \bibinfo {author} {\bibfnamefont {F.}~\bibnamefont {de~Juan}},\ and\ \bibinfo {author} {\bibfnamefont {M.~M.}\ \bibnamefont {Ugeda}},\ }\href@noop {} {\bibfield  {journal} {\bibinfo  {journal} {arXiv preprint arXiv:2501.09063}\ } (\bibinfo {year} {2025})}\BibitemShut {NoStop}%
\bibitem [{\citenamefont {Holbrook}\ \emph {et~al.}(2024)\citenamefont {Holbrook}, \citenamefont {Ingham}, \citenamefont {Kaplan}, \citenamefont {Holtzman}, \citenamefont {Bierman}, \citenamefont {Olson}, \citenamefont {Nashabeh}, \citenamefont {Liu}, \citenamefont {Zhu}, \citenamefont {Rhodes} \emph {et~al.}}]{holbrook2024real}%
  \BibitemOpen
  \bibfield  {author} {\bibinfo {author} {\bibfnamefont {M.}~\bibnamefont {Holbrook}}, \bibinfo {author} {\bibfnamefont {J.}~\bibnamefont {Ingham}}, \bibinfo {author} {\bibfnamefont {D.}~\bibnamefont {Kaplan}}, \bibinfo {author} {\bibfnamefont {L.}~\bibnamefont {Holtzman}}, \bibinfo {author} {\bibfnamefont {B.}~\bibnamefont {Bierman}}, \bibinfo {author} {\bibfnamefont {N.}~\bibnamefont {Olson}}, \bibinfo {author} {\bibfnamefont {L.}~\bibnamefont {Nashabeh}}, \bibinfo {author} {\bibfnamefont {S.}~\bibnamefont {Liu}}, \bibinfo {author} {\bibfnamefont {X.}~\bibnamefont {Zhu}}, \bibinfo {author} {\bibfnamefont {D.}~\bibnamefont {Rhodes}}, \emph {et~al.},\ }\href@noop {} {\bibfield  {journal} {\bibinfo  {journal} {arXiv preprint arXiv:2412.02813}\ } (\bibinfo {year} {2024})}\BibitemShut {NoStop}%
\bibitem [{\citenamefont {Hiorth}\ \emph {et~al.}(2026)\citenamefont {Hiorth}, \citenamefont {Gutierrez-Amigo}, \citenamefont {Cavignac}, \citenamefont {Haule}, \citenamefont {Marques},\ and\ \citenamefont {T{\"o}rm{\"a}}}]{hiorth2026ab}%
  \BibitemOpen
  \bibfield  {author} {\bibinfo {author} {\bibfnamefont {K.~H.}\ \bibnamefont {Hiorth}}, \bibinfo {author} {\bibfnamefont {M.}~\bibnamefont {Gutierrez-Amigo}}, \bibinfo {author} {\bibfnamefont {T.}~\bibnamefont {Cavignac}}, \bibinfo {author} {\bibfnamefont {K.}~\bibnamefont {Haule}}, \bibinfo {author} {\bibfnamefont {M.~A.~L.}\ \bibnamefont {Marques}},\ and\ \bibinfo {author} {\bibfnamefont {P.}~\bibnamefont {T{\"o}rm{\"a}}},\ }\href {https://doi.org/10.1103/2xrg-6fy6} {\bibfield  {journal} {\bibinfo  {journal} {Phys. Rev. B}\ }\textbf {\bibinfo {volume} {113}},\ \bibinfo {pages} {224509} (\bibinfo {year} {2026})}\BibitemShut {NoStop}%
\bibitem [{\citenamefont {Liang}\ \emph {et~al.}(2017)\citenamefont {Liang}, \citenamefont {Vanhala}, \citenamefont {Peotta}, \citenamefont {Siro}, \citenamefont {Harju},\ and\ \citenamefont {T\"orm\"a}}]{Liang2017}%
  \BibitemOpen
  \bibfield  {author} {\bibinfo {author} {\bibfnamefont {L.}~\bibnamefont {Liang}}, \bibinfo {author} {\bibfnamefont {T.~I.}\ \bibnamefont {Vanhala}}, \bibinfo {author} {\bibfnamefont {S.}~\bibnamefont {Peotta}}, \bibinfo {author} {\bibfnamefont {T.}~\bibnamefont {Siro}}, \bibinfo {author} {\bibfnamefont {A.}~\bibnamefont {Harju}},\ and\ \bibinfo {author} {\bibfnamefont {P.}~\bibnamefont {T\"orm\"a}},\ }\href {https://doi.org/10.1103/PhysRevB.95.024515} {\bibfield  {journal} {\bibinfo  {journal} {Phys. Rev. B}\ }\textbf {\bibinfo {volume} {95}},\ \bibinfo {pages} {024515} (\bibinfo {year} {2017})}\BibitemShut {NoStop}%
\bibitem [{\citenamefont {Huhtinen}\ \emph {et~al.}(2022)\citenamefont {Huhtinen}, \citenamefont {Herzog-Arbeitman}, \citenamefont {Chew}, \citenamefont {Bernevig},\ and\ \citenamefont {T\"orm\"a}}]{Huhtinen2022}%
  \BibitemOpen
  \bibfield  {author} {\bibinfo {author} {\bibfnamefont {K.-E.}\ \bibnamefont {Huhtinen}}, \bibinfo {author} {\bibfnamefont {J.}~\bibnamefont {Herzog-Arbeitman}}, \bibinfo {author} {\bibfnamefont {A.}~\bibnamefont {Chew}}, \bibinfo {author} {\bibfnamefont {B.~A.}\ \bibnamefont {Bernevig}},\ and\ \bibinfo {author} {\bibfnamefont {P.}~\bibnamefont {T\"orm\"a}},\ }\href {https://doi.org/10.1103/PhysRevB.106.014518} {\bibfield  {journal} {\bibinfo  {journal} {Phys. Rev. B}\ }\textbf {\bibinfo {volume} {106}},\ \bibinfo {pages} {014518} (\bibinfo {year} {2022})}\BibitemShut {NoStop}%
\bibitem [{\citenamefont {Pires}\ \emph {et~al.}(2026)\citenamefont {Pires}, \citenamefont {{da Silva}}, \citenamefont {Gao}, \citenamefont {Hiorth}, \citenamefont {Cerqueira}, \citenamefont {Cavignac}, \citenamefont {{De Breuck}}, \citenamefont {Wang}, \citenamefont {Đorđe Dangić}, \citenamefont {Fang}, \citenamefont {Sanna}, \citenamefont {Cui}, \citenamefont {Errea}, \citenamefont {Törmä},\ and\ \citenamefont {Marques}}]{Pires2026}%
  \BibitemOpen
  \bibfield  {author} {\bibinfo {author} {\bibfnamefont {P.~R.}\ \bibnamefont {Pires}}, \bibinfo {author} {\bibfnamefont {T.~H.}\ \bibnamefont {{da Silva}}}, \bibinfo {author} {\bibfnamefont {K.}~\bibnamefont {Gao}}, \bibinfo {author} {\bibfnamefont {K.~H.}\ \bibnamefont {Hiorth}}, \bibinfo {author} {\bibfnamefont {T.~F.}\ \bibnamefont {Cerqueira}}, \bibinfo {author} {\bibfnamefont {T.}~\bibnamefont {Cavignac}}, \bibinfo {author} {\bibfnamefont {P.-P.}\ \bibnamefont {{De Breuck}}}, \bibinfo {author} {\bibfnamefont {H.-C.}\ \bibnamefont {Wang}}, \bibinfo {author} {\bibnamefont {Đorđe Dangić}}, \bibinfo {author} {\bibfnamefont {Y.-W.}\ \bibnamefont {Fang}}, \bibinfo {author} {\bibfnamefont {A.}~\bibnamefont {Sanna}}, \bibinfo {author} {\bibfnamefont {W.}~\bibnamefont {Cui}}, \bibinfo {author} {\bibfnamefont {I.}~\bibnamefont {Errea}}, \bibinfo {author} {\bibfnamefont {P.}~\bibnamefont {Törmä}},\ and\ \bibinfo {author} {\bibfnamefont {M.~A.}\ \bibnamefont {Marques}},\ }\href
  {https://doi.org/https://doi.org/10.1016/j.commt.2026.100052} {\bibfield  {journal} {\bibinfo  {journal} {Computational Materials Today}\ }\textbf {\bibinfo {volume} {10}},\ \bibinfo {pages} {100052} (\bibinfo {year} {2026})}\BibitemShut {NoStop}%
\bibitem [{\citenamefont {Giustino}(2017)}]{giustino2017electron}%
  \BibitemOpen
  \bibfield  {author} {\bibinfo {author} {\bibfnamefont {F.}~\bibnamefont {Giustino}},\ }\href@noop {} {\bibfield  {journal} {\bibinfo  {journal} {Reviews of Modern Physics}\ }\textbf {\bibinfo {volume} {89}},\ \bibinfo {pages} {015003} (\bibinfo {year} {2017})}\BibitemShut {NoStop}%
\bibitem [{\citenamefont {Haoyu}\ \emph {et~al.}(2026)\citenamefont {Haoyu}, \citenamefont {Zenan}, \citenamefont {B.~Andrei},\ and\ \citenamefont {et~al}}]{EPCpaper}%
  \BibitemOpen
  \bibfield  {author} {\bibinfo {author} {\bibfnamefont {H.}~\bibnamefont {Haoyu}}, \bibinfo {author} {\bibfnamefont {D.}~\bibnamefont {Zenan}}, \bibinfo {author} {\bibfnamefont {B.}~\bibnamefont {B.~Andrei}},\ and\ \bibinfo {author} {\bibnamefont {et~al}},\ }\href@noop {} {\bibfield  {journal} {\bibinfo  {journal} {In preparation}\ } (\bibinfo {year} {2026})}\BibitemShut {NoStop}%
\bibitem [{\citenamefont {Giustino}\ \emph {et~al.}(2007)\citenamefont {Giustino}, \citenamefont {Cohen},\ and\ \citenamefont {Louie}}]{giustino2007electron}%
  \BibitemOpen
  \bibfield  {author} {\bibinfo {author} {\bibfnamefont {F.}~\bibnamefont {Giustino}}, \bibinfo {author} {\bibfnamefont {M.~L.}\ \bibnamefont {Cohen}},\ and\ \bibinfo {author} {\bibfnamefont {S.~G.}\ \bibnamefont {Louie}},\ }\href@noop {} {\bibfield  {journal} {\bibinfo  {journal} {Physical Review B—Condensed Matter and Materials Physics}\ }\textbf {\bibinfo {volume} {76}},\ \bibinfo {pages} {165108} (\bibinfo {year} {2007})}\BibitemShut {NoStop}%
\bibitem [{\citenamefont {Lee}\ \emph {et~al.}(2023)\citenamefont {Lee}, \citenamefont {Ponc{\'e}}, \citenamefont {Bushick}, \citenamefont {Hajinazar}, \citenamefont {Lafuente-Bartolome}, \citenamefont {Leveillee}, \citenamefont {Lian}, \citenamefont {Lihm}, \citenamefont {Macheda}, \citenamefont {Mori} \emph {et~al.}}]{lee2023electron}%
  \BibitemOpen
  \bibfield  {author} {\bibinfo {author} {\bibfnamefont {H.}~\bibnamefont {Lee}}, \bibinfo {author} {\bibfnamefont {S.}~\bibnamefont {Ponc{\'e}}}, \bibinfo {author} {\bibfnamefont {K.}~\bibnamefont {Bushick}}, \bibinfo {author} {\bibfnamefont {S.}~\bibnamefont {Hajinazar}}, \bibinfo {author} {\bibfnamefont {J.}~\bibnamefont {Lafuente-Bartolome}}, \bibinfo {author} {\bibfnamefont {J.}~\bibnamefont {Leveillee}}, \bibinfo {author} {\bibfnamefont {C.}~\bibnamefont {Lian}}, \bibinfo {author} {\bibfnamefont {J.-M.}\ \bibnamefont {Lihm}}, \bibinfo {author} {\bibfnamefont {F.}~\bibnamefont {Macheda}}, \bibinfo {author} {\bibfnamefont {H.}~\bibnamefont {Mori}}, \emph {et~al.},\ }\href@noop {} {\bibfield  {journal} {\bibinfo  {journal} {npj Computational Materials}\ }\textbf {\bibinfo {volume} {9}},\ \bibinfo {pages} {156} (\bibinfo {year} {2023})}\BibitemShut {NoStop}%
\bibitem [{\citenamefont {Ponc{\'e}}\ \emph {et~al.}(2016)\citenamefont {Ponc{\'e}}, \citenamefont {Margine}, \citenamefont {Verdi},\ and\ \citenamefont {Giustino}}]{ponce2016epw}%
  \BibitemOpen
  \bibfield  {author} {\bibinfo {author} {\bibfnamefont {S.}~\bibnamefont {Ponc{\'e}}}, \bibinfo {author} {\bibfnamefont {E.~R.}\ \bibnamefont {Margine}}, \bibinfo {author} {\bibfnamefont {C.}~\bibnamefont {Verdi}},\ and\ \bibinfo {author} {\bibfnamefont {F.}~\bibnamefont {Giustino}},\ }\href@noop {} {\bibfield  {journal} {\bibinfo  {journal} {Computer Physics Communications}\ }\textbf {\bibinfo {volume} {209}},\ \bibinfo {pages} {116} (\bibinfo {year} {2016})}\BibitemShut {NoStop}%
\bibitem [{\citenamefont {Margine}\ and\ \citenamefont {Giustino}(2013)}]{margine2013anisotropic}%
  \BibitemOpen
  \bibfield  {author} {\bibinfo {author} {\bibfnamefont {E.~R.}\ \bibnamefont {Margine}}\ and\ \bibinfo {author} {\bibfnamefont {F.}~\bibnamefont {Giustino}},\ }\href@noop {} {\bibfield  {journal} {\bibinfo  {journal} {Physical Review B—Condensed Matter and Materials Physics}\ }\textbf {\bibinfo {volume} {87}},\ \bibinfo {pages} {024505} (\bibinfo {year} {2013})}\BibitemShut {NoStop}%
\bibitem [{\citenamefont {Zhou}\ \emph {et~al.}(2021)\citenamefont {Zhou}, \citenamefont {Park}, \citenamefont {Lu}, \citenamefont {Maliyov}, \citenamefont {Tong},\ and\ \citenamefont {Bernardi}}]{zhou2021perturbo}%
  \BibitemOpen
  \bibfield  {author} {\bibinfo {author} {\bibfnamefont {J.-J.}\ \bibnamefont {Zhou}}, \bibinfo {author} {\bibfnamefont {J.}~\bibnamefont {Park}}, \bibinfo {author} {\bibfnamefont {I.-T.}\ \bibnamefont {Lu}}, \bibinfo {author} {\bibfnamefont {I.}~\bibnamefont {Maliyov}}, \bibinfo {author} {\bibfnamefont {X.}~\bibnamefont {Tong}},\ and\ \bibinfo {author} {\bibfnamefont {M.}~\bibnamefont {Bernardi}},\ }\href@noop {} {\bibfield  {journal} {\bibinfo  {journal} {Computer Physics Communications}\ }\textbf {\bibinfo {volume} {264}},\ \bibinfo {pages} {107970} (\bibinfo {year} {2021})}\BibitemShut {NoStop}%
\bibitem [{\citenamefont {Luo}\ \emph {et~al.}(2024)\citenamefont {Luo}, \citenamefont {Desai}, \citenamefont {Chang}, \citenamefont {Park},\ and\ \citenamefont {Bernardi}}]{luo2024data}%
  \BibitemOpen
  \bibfield  {author} {\bibinfo {author} {\bibfnamefont {Y.}~\bibnamefont {Luo}}, \bibinfo {author} {\bibfnamefont {D.}~\bibnamefont {Desai}}, \bibinfo {author} {\bibfnamefont {B.~K.}\ \bibnamefont {Chang}}, \bibinfo {author} {\bibfnamefont {J.}~\bibnamefont {Park}},\ and\ \bibinfo {author} {\bibfnamefont {M.}~\bibnamefont {Bernardi}},\ }\href@noop {} {\bibfield  {journal} {\bibinfo  {journal} {Physical Review X}\ }\textbf {\bibinfo {volume} {14}},\ \bibinfo {pages} {021023} (\bibinfo {year} {2024})}\BibitemShut {NoStop}%
\bibitem [{\citenamefont {McMillan}(1968)}]{mcmillan1968transition}%
  \BibitemOpen
  \bibfield  {author} {\bibinfo {author} {\bibfnamefont {W.}~\bibnamefont {McMillan}},\ }\href@noop {} {\bibfield  {journal} {\bibinfo  {journal} {Physical Review}\ }\textbf {\bibinfo {volume} {167}},\ \bibinfo {pages} {331} (\bibinfo {year} {1968})}\BibitemShut {NoStop}%
\bibitem [{\citenamefont {Hopfield}(1969)}]{hopfield1969angular}%
  \BibitemOpen
  \bibfield  {author} {\bibinfo {author} {\bibfnamefont {J.}~\bibnamefont {Hopfield}},\ }\href@noop {} {\bibfield  {journal} {\bibinfo  {journal} {Physical Review}\ }\textbf {\bibinfo {volume} {186}},\ \bibinfo {pages} {443} (\bibinfo {year} {1969})}\BibitemShut {NoStop}%
\bibitem [{\citenamefont {Hu}\ \emph {et~al.}(2025)\citenamefont {Hu}, \citenamefont {Jiang}, \citenamefont {C{\u{a}}lug{\u{a}}ru}, \citenamefont {Feng}, \citenamefont {Subires}, \citenamefont {Vergniory}, \citenamefont {Felser}, \citenamefont {Blanco-Canosa},\ and\ \citenamefont {Bernevig}}]{hu2023kagome}%
  \BibitemOpen
  \bibfield  {author} {\bibinfo {author} {\bibfnamefont {H.}~\bibnamefont {Hu}}, \bibinfo {author} {\bibfnamefont {Y.}~\bibnamefont {Jiang}}, \bibinfo {author} {\bibfnamefont {D.}~\bibnamefont {C{\u{a}}lug{\u{a}}ru}}, \bibinfo {author} {\bibfnamefont {X.}~\bibnamefont {Feng}}, \bibinfo {author} {\bibfnamefont {D.}~\bibnamefont {Subires}}, \bibinfo {author} {\bibfnamefont {M.~G.}\ \bibnamefont {Vergniory}}, \bibinfo {author} {\bibfnamefont {C.}~\bibnamefont {Felser}}, \bibinfo {author} {\bibfnamefont {S.}~\bibnamefont {Blanco-Canosa}},\ and\ \bibinfo {author} {\bibfnamefont {B.~A.}\ \bibnamefont {Bernevig}},\ }\href@noop {} {\bibfield  {journal} {\bibinfo  {journal} {Physical Review B}\ }\textbf {\bibinfo {volume} {111}},\ \bibinfo {pages} {054113} (\bibinfo {year} {2025})}\BibitemShut {NoStop}%
\bibitem [{\citenamefont {An}\ and\ \citenamefont {Pickett}(2001{\natexlab{b}})}]{an2001superconductivity}%
  \BibitemOpen
  \bibfield  {author} {\bibinfo {author} {\bibfnamefont {J.}~\bibnamefont {An}}\ and\ \bibinfo {author} {\bibfnamefont {W.}~\bibnamefont {Pickett}},\ }\href@noop {} {\bibfield  {journal} {\bibinfo  {journal} {Physical Review Letters}\ }\textbf {\bibinfo {volume} {86}},\ \bibinfo {pages} {4366} (\bibinfo {year} {2001}{\natexlab{b}})}\BibitemShut {NoStop}%
\bibitem [{\citenamefont {Mitra}(1969)}]{mitra1969electron}%
  \BibitemOpen
  \bibfield  {author} {\bibinfo {author} {\bibfnamefont {T.}~\bibnamefont {Mitra}},\ }\href@noop {} {\bibfield  {journal} {\bibinfo  {journal} {Journal of Physics C: Solid State Physics}\ }\textbf {\bibinfo {volume} {2}},\ \bibinfo {pages} {52} (\bibinfo {year} {1969})}\BibitemShut {NoStop}%
\bibitem [{\citenamefont {Baroni}\ \emph {et~al.}(2001)\citenamefont {Baroni}, \citenamefont {De~Gironcoli}, \citenamefont {Dal~Corso},\ and\ \citenamefont {Giannozzi}}]{baroni2001phonons}%
  \BibitemOpen
  \bibfield  {author} {\bibinfo {author} {\bibfnamefont {S.}~\bibnamefont {Baroni}}, \bibinfo {author} {\bibfnamefont {S.}~\bibnamefont {De~Gironcoli}}, \bibinfo {author} {\bibfnamefont {A.}~\bibnamefont {Dal~Corso}},\ and\ \bibinfo {author} {\bibfnamefont {P.}~\bibnamefont {Giannozzi}},\ }\href@noop {} {\bibfield  {journal} {\bibinfo  {journal} {Reviews of modern Physics}\ }\textbf {\bibinfo {volume} {73}},\ \bibinfo {pages} {515} (\bibinfo {year} {2001})}\BibitemShut {NoStop}%
\bibitem [{\citenamefont {Li}\ \emph {et~al.}(2024)\citenamefont {Li}, \citenamefont {Tang}, \citenamefont {Fu}, \citenamefont {Dong}, \citenamefont {Zou}, \citenamefont {Gong}, \citenamefont {Duan},\ and\ \citenamefont {Xu}}]{li2024deep}%
  \BibitemOpen
  \bibfield  {author} {\bibinfo {author} {\bibfnamefont {H.}~\bibnamefont {Li}}, \bibinfo {author} {\bibfnamefont {Z.}~\bibnamefont {Tang}}, \bibinfo {author} {\bibfnamefont {J.}~\bibnamefont {Fu}}, \bibinfo {author} {\bibfnamefont {W.-H.}\ \bibnamefont {Dong}}, \bibinfo {author} {\bibfnamefont {N.}~\bibnamefont {Zou}}, \bibinfo {author} {\bibfnamefont {X.}~\bibnamefont {Gong}}, \bibinfo {author} {\bibfnamefont {W.}~\bibnamefont {Duan}},\ and\ \bibinfo {author} {\bibfnamefont {Y.}~\bibnamefont {Xu}},\ }\href@noop {} {\bibfield  {journal} {\bibinfo  {journal} {Physical Review Letters}\ }\textbf {\bibinfo {volume} {132}},\ \bibinfo {pages} {096401} (\bibinfo {year} {2024})}\BibitemShut {NoStop}%
\bibitem [{\citenamefont {Kresse}\ and\ \citenamefont {Furthm{\"u}ller}(1996{\natexlab{a}})}]{kresse1996efficiency}%
  \BibitemOpen
  \bibfield  {author} {\bibinfo {author} {\bibfnamefont {G.}~\bibnamefont {Kresse}}\ and\ \bibinfo {author} {\bibfnamefont {J.}~\bibnamefont {Furthm{\"u}ller}},\ }\href@noop {} {\bibfield  {journal} {\bibinfo  {journal} {Computational materials science}\ }\textbf {\bibinfo {volume} {6}},\ \bibinfo {pages} {15} (\bibinfo {year} {1996}{\natexlab{a}})}\BibitemShut {NoStop}%
\bibitem [{\citenamefont {Kresse}\ and\ \citenamefont {Hafner}(1993{\natexlab{a}})}]{kresse1993ab1}%
  \BibitemOpen
  \bibfield  {author} {\bibinfo {author} {\bibfnamefont {G.}~\bibnamefont {Kresse}}\ and\ \bibinfo {author} {\bibfnamefont {J.}~\bibnamefont {Hafner}},\ }\href@noop {} {\bibfield  {journal} {\bibinfo  {journal} {Physical Review B}\ }\textbf {\bibinfo {volume} {48}},\ \bibinfo {pages} {13115} (\bibinfo {year} {1993}{\natexlab{a}})}\BibitemShut {NoStop}%
\bibitem [{\citenamefont {Kresse}\ and\ \citenamefont {Hafner}(1993{\natexlab{b}})}]{kresse1993ab2}%
  \BibitemOpen
  \bibfield  {author} {\bibinfo {author} {\bibfnamefont {G.}~\bibnamefont {Kresse}}\ and\ \bibinfo {author} {\bibfnamefont {J.}~\bibnamefont {Hafner}},\ }\href@noop {} {\bibfield  {journal} {\bibinfo  {journal} {Physical review B}\ }\textbf {\bibinfo {volume} {47}},\ \bibinfo {pages} {558} (\bibinfo {year} {1993}{\natexlab{b}})}\BibitemShut {NoStop}%
\bibitem [{\citenamefont {Kresse}\ and\ \citenamefont {Hafner}(1994)}]{kresse1994ab}%
  \BibitemOpen
  \bibfield  {author} {\bibinfo {author} {\bibfnamefont {G.}~\bibnamefont {Kresse}}\ and\ \bibinfo {author} {\bibfnamefont {J.}~\bibnamefont {Hafner}},\ }\href@noop {} {\bibfield  {journal} {\bibinfo  {journal} {Physical Review B}\ }\textbf {\bibinfo {volume} {49}},\ \bibinfo {pages} {14251} (\bibinfo {year} {1994})}\BibitemShut {NoStop}%
\bibitem [{\citenamefont {Kresse}\ and\ \citenamefont {Furthm{\"u}ller}(1996{\natexlab{b}})}]{kresse1996efficient}%
  \BibitemOpen
  \bibfield  {author} {\bibinfo {author} {\bibfnamefont {G.}~\bibnamefont {Kresse}}\ and\ \bibinfo {author} {\bibfnamefont {J.}~\bibnamefont {Furthm{\"u}ller}},\ }\href@noop {} {\bibfield  {journal} {\bibinfo  {journal} {Physical review B}\ }\textbf {\bibinfo {volume} {54}},\ \bibinfo {pages} {11169} (\bibinfo {year} {1996}{\natexlab{b}})}\BibitemShut {NoStop}%
\bibitem [{\citenamefont {Perdew}\ \emph {et~al.}(1996)\citenamefont {Perdew}, \citenamefont {Burke},\ and\ \citenamefont {Ernzerhof}}]{perdew1996generalized}%
  \BibitemOpen
  \bibfield  {author} {\bibinfo {author} {\bibfnamefont {J.~P.}\ \bibnamefont {Perdew}}, \bibinfo {author} {\bibfnamefont {K.}~\bibnamefont {Burke}},\ and\ \bibinfo {author} {\bibfnamefont {M.}~\bibnamefont {Ernzerhof}},\ }\href@noop {} {\bibfield  {journal} {\bibinfo  {journal} {Physical review letters}\ }\textbf {\bibinfo {volume} {77}},\ \bibinfo {pages} {3865} (\bibinfo {year} {1996})}\BibitemShut {NoStop}%
\bibitem [{\citenamefont {Marzari}\ and\ \citenamefont {Vanderbilt}(1997)}]{marzari1997maximally}%
  \BibitemOpen
  \bibfield  {author} {\bibinfo {author} {\bibfnamefont {N.}~\bibnamefont {Marzari}}\ and\ \bibinfo {author} {\bibfnamefont {D.}~\bibnamefont {Vanderbilt}},\ }\href@noop {} {\bibfield  {journal} {\bibinfo  {journal} {Physical review B}\ }\textbf {\bibinfo {volume} {56}},\ \bibinfo {pages} {12847} (\bibinfo {year} {1997})}\BibitemShut {NoStop}%
\bibitem [{\citenamefont {Souza}\ \emph {et~al.}(2001)\citenamefont {Souza}, \citenamefont {Marzari},\ and\ \citenamefont {Vanderbilt}}]{souza2001maximally}%
  \BibitemOpen
  \bibfield  {author} {\bibinfo {author} {\bibfnamefont {I.}~\bibnamefont {Souza}}, \bibinfo {author} {\bibfnamefont {N.}~\bibnamefont {Marzari}},\ and\ \bibinfo {author} {\bibfnamefont {D.}~\bibnamefont {Vanderbilt}},\ }\href@noop {} {\bibfield  {journal} {\bibinfo  {journal} {Physical Review B}\ }\textbf {\bibinfo {volume} {65}},\ \bibinfo {pages} {035109} (\bibinfo {year} {2001})}\BibitemShut {NoStop}%
\bibitem [{\citenamefont {Marzari}\ \emph {et~al.}(2012)\citenamefont {Marzari}, \citenamefont {Mostofi}, \citenamefont {Yates}, \citenamefont {Souza},\ and\ \citenamefont {Vanderbilt}}]{marzari2012maximally}%
  \BibitemOpen
  \bibfield  {author} {\bibinfo {author} {\bibfnamefont {N.}~\bibnamefont {Marzari}}, \bibinfo {author} {\bibfnamefont {A.~A.}\ \bibnamefont {Mostofi}}, \bibinfo {author} {\bibfnamefont {J.~R.}\ \bibnamefont {Yates}}, \bibinfo {author} {\bibfnamefont {I.}~\bibnamefont {Souza}},\ and\ \bibinfo {author} {\bibfnamefont {D.}~\bibnamefont {Vanderbilt}},\ }\href@noop {} {\bibfield  {journal} {\bibinfo  {journal} {Reviews of Modern Physics}\ }\textbf {\bibinfo {volume} {84}},\ \bibinfo {pages} {1419} (\bibinfo {year} {2012})}\BibitemShut {NoStop}%
\bibitem [{\citenamefont {Pizzi}\ \emph {et~al.}(2020)\citenamefont {Pizzi}, \citenamefont {Vitale}, \citenamefont {Arita}, \citenamefont {Bl{\"u}gel}, \citenamefont {Freimuth}, \citenamefont {G{\'e}ranton}, \citenamefont {Gibertini}, \citenamefont {Gresch}, \citenamefont {Johnson}, \citenamefont {Koretsune} \emph {et~al.}}]{pizzi2020wannier90}%
  \BibitemOpen
  \bibfield  {author} {\bibinfo {author} {\bibfnamefont {G.}~\bibnamefont {Pizzi}}, \bibinfo {author} {\bibfnamefont {V.}~\bibnamefont {Vitale}}, \bibinfo {author} {\bibfnamefont {R.}~\bibnamefont {Arita}}, \bibinfo {author} {\bibfnamefont {S.}~\bibnamefont {Bl{\"u}gel}}, \bibinfo {author} {\bibfnamefont {F.}~\bibnamefont {Freimuth}}, \bibinfo {author} {\bibfnamefont {G.}~\bibnamefont {G{\'e}ranton}}, \bibinfo {author} {\bibfnamefont {M.}~\bibnamefont {Gibertini}}, \bibinfo {author} {\bibfnamefont {D.}~\bibnamefont {Gresch}}, \bibinfo {author} {\bibfnamefont {C.}~\bibnamefont {Johnson}}, \bibinfo {author} {\bibfnamefont {T.}~\bibnamefont {Koretsune}}, \emph {et~al.},\ }\href@noop {} {\bibfield  {journal} {\bibinfo  {journal} {Journal of Physics: Condensed Matter}\ }\textbf {\bibinfo {volume} {32}},\ \bibinfo {pages} {165902} (\bibinfo {year} {2020})}\BibitemShut {NoStop}%
\bibitem [{\citenamefont {Giannozzi}\ \emph {et~al.}(2009)\citenamefont {Giannozzi}, \citenamefont {Baroni}, \citenamefont {Bonini}, \citenamefont {Calandra}, \citenamefont {Car}, \citenamefont {Cavazzoni}, \citenamefont {Ceresoli}, \citenamefont {Chiarotti}, \citenamefont {Cococcioni}, \citenamefont {Dabo} \emph {et~al.}}]{giannozzi2009quantum}%
  \BibitemOpen
  \bibfield  {author} {\bibinfo {author} {\bibfnamefont {P.}~\bibnamefont {Giannozzi}}, \bibinfo {author} {\bibfnamefont {S.}~\bibnamefont {Baroni}}, \bibinfo {author} {\bibfnamefont {N.}~\bibnamefont {Bonini}}, \bibinfo {author} {\bibfnamefont {M.}~\bibnamefont {Calandra}}, \bibinfo {author} {\bibfnamefont {R.}~\bibnamefont {Car}}, \bibinfo {author} {\bibfnamefont {C.}~\bibnamefont {Cavazzoni}}, \bibinfo {author} {\bibfnamefont {D.}~\bibnamefont {Ceresoli}}, \bibinfo {author} {\bibfnamefont {G.~L.}\ \bibnamefont {Chiarotti}}, \bibinfo {author} {\bibfnamefont {M.}~\bibnamefont {Cococcioni}}, \bibinfo {author} {\bibfnamefont {I.}~\bibnamefont {Dabo}}, \emph {et~al.},\ }\href@noop {} {\bibfield  {journal} {\bibinfo  {journal} {Journal of physics: Condensed matter}\ }\textbf {\bibinfo {volume} {21}},\ \bibinfo {pages} {395502} (\bibinfo {year} {2009})}\BibitemShut {NoStop}%
\bibitem [{\citenamefont {Giannozzi}\ \emph {et~al.}(2017)\citenamefont {Giannozzi}, \citenamefont {Andreussi}, \citenamefont {Brumme}, \citenamefont {Bunau}, \citenamefont {Nardelli}, \citenamefont {Calandra}, \citenamefont {Car}, \citenamefont {Cavazzoni}, \citenamefont {Ceresoli}, \citenamefont {Cococcioni} \emph {et~al.}}]{giannozzi2017advanced}%
  \BibitemOpen
  \bibfield  {author} {\bibinfo {author} {\bibfnamefont {P.}~\bibnamefont {Giannozzi}}, \bibinfo {author} {\bibfnamefont {O.}~\bibnamefont {Andreussi}}, \bibinfo {author} {\bibfnamefont {T.}~\bibnamefont {Brumme}}, \bibinfo {author} {\bibfnamefont {O.}~\bibnamefont {Bunau}}, \bibinfo {author} {\bibfnamefont {M.~B.}\ \bibnamefont {Nardelli}}, \bibinfo {author} {\bibfnamefont {M.}~\bibnamefont {Calandra}}, \bibinfo {author} {\bibfnamefont {R.}~\bibnamefont {Car}}, \bibinfo {author} {\bibfnamefont {C.}~\bibnamefont {Cavazzoni}}, \bibinfo {author} {\bibfnamefont {D.}~\bibnamefont {Ceresoli}}, \bibinfo {author} {\bibfnamefont {M.}~\bibnamefont {Cococcioni}}, \emph {et~al.},\ }\href@noop {} {\bibfield  {journal} {\bibinfo  {journal} {Journal of physics: Condensed matter}\ }\textbf {\bibinfo {volume} {29}},\ \bibinfo {pages} {465901} (\bibinfo {year} {2017})}\BibitemShut {NoStop}%
\bibitem [{\citenamefont {Giannozzi}\ \emph {et~al.}(2020)\citenamefont {Giannozzi}, \citenamefont {Baseggio}, \citenamefont {Bonf{\`a}}, \citenamefont {Brunato}, \citenamefont {Car}, \citenamefont {Carnimeo}, \citenamefont {Cavazzoni}, \citenamefont {De~Gironcoli}, \citenamefont {Delugas}, \citenamefont {Ferrari~Ruffino} \emph {et~al.}}]{giannozzi2020quantum}%
  \BibitemOpen
  \bibfield  {author} {\bibinfo {author} {\bibfnamefont {P.}~\bibnamefont {Giannozzi}}, \bibinfo {author} {\bibfnamefont {O.}~\bibnamefont {Baseggio}}, \bibinfo {author} {\bibfnamefont {P.}~\bibnamefont {Bonf{\`a}}}, \bibinfo {author} {\bibfnamefont {D.}~\bibnamefont {Brunato}}, \bibinfo {author} {\bibfnamefont {R.}~\bibnamefont {Car}}, \bibinfo {author} {\bibfnamefont {I.}~\bibnamefont {Carnimeo}}, \bibinfo {author} {\bibfnamefont {C.}~\bibnamefont {Cavazzoni}}, \bibinfo {author} {\bibfnamefont {S.}~\bibnamefont {De~Gironcoli}}, \bibinfo {author} {\bibfnamefont {P.}~\bibnamefont {Delugas}}, \bibinfo {author} {\bibfnamefont {F.}~\bibnamefont {Ferrari~Ruffino}}, \emph {et~al.},\ }\href@noop {} {\bibfield  {journal} {\bibinfo  {journal} {The Journal of chemical physics}\ }\textbf {\bibinfo {volume} {152}} (\bibinfo {year} {2020})}\BibitemShut {NoStop}%
\bibitem [{\citenamefont {Noffsinger}\ \emph {et~al.}(2010)\citenamefont {Noffsinger}, \citenamefont {Giustino}, \citenamefont {Malone}, \citenamefont {Park}, \citenamefont {Louie},\ and\ \citenamefont {Cohen}}]{noffsinger2010epw}%
  \BibitemOpen
  \bibfield  {author} {\bibinfo {author} {\bibfnamefont {J.}~\bibnamefont {Noffsinger}}, \bibinfo {author} {\bibfnamefont {F.}~\bibnamefont {Giustino}}, \bibinfo {author} {\bibfnamefont {B.~D.}\ \bibnamefont {Malone}}, \bibinfo {author} {\bibfnamefont {C.-H.}\ \bibnamefont {Park}}, \bibinfo {author} {\bibfnamefont {S.~G.}\ \bibnamefont {Louie}},\ and\ \bibinfo {author} {\bibfnamefont {M.~L.}\ \bibnamefont {Cohen}},\ }\href@noop {} {\bibfield  {journal} {\bibinfo  {journal} {Computer Physics Communications}\ }\textbf {\bibinfo {volume} {181}},\ \bibinfo {pages} {2140} (\bibinfo {year} {2010})}\BibitemShut {NoStop}%
\bibitem [{\citenamefont {Perdew}\ \emph {et~al.}(2008)\citenamefont {Perdew}, \citenamefont {Ruzsinszky}, \citenamefont {Csonka}, \citenamefont {Vydrov}, \citenamefont {Scuseria}, \citenamefont {Constantin}, \citenamefont {Zhou},\ and\ \citenamefont {Burke}}]{perdew2008restoring}%
  \BibitemOpen
  \bibfield  {author} {\bibinfo {author} {\bibfnamefont {J.~P.}\ \bibnamefont {Perdew}}, \bibinfo {author} {\bibfnamefont {A.}~\bibnamefont {Ruzsinszky}}, \bibinfo {author} {\bibfnamefont {G.~I.}\ \bibnamefont {Csonka}}, \bibinfo {author} {\bibfnamefont {O.~A.}\ \bibnamefont {Vydrov}}, \bibinfo {author} {\bibfnamefont {G.~E.}\ \bibnamefont {Scuseria}}, \bibinfo {author} {\bibfnamefont {L.~A.}\ \bibnamefont {Constantin}}, \bibinfo {author} {\bibfnamefont {X.}~\bibnamefont {Zhou}},\ and\ \bibinfo {author} {\bibfnamefont {K.}~\bibnamefont {Burke}},\ }\href@noop {} {\bibfield  {journal} {\bibinfo  {journal} {Physical review letters}\ }\textbf {\bibinfo {volume} {100}},\ \bibinfo {pages} {136406} (\bibinfo {year} {2008})}\BibitemShut {NoStop}%
\bibitem [{\citenamefont {Van~Setten}\ \emph {et~al.}(2018)\citenamefont {Van~Setten}, \citenamefont {Giantomassi}, \citenamefont {Bousquet}, \citenamefont {Verstraete}, \citenamefont {Hamann}, \citenamefont {Gonze},\ and\ \citenamefont {Rignanese}}]{van2018pseudodojo}%
  \BibitemOpen
  \bibfield  {author} {\bibinfo {author} {\bibfnamefont {M.~J.}\ \bibnamefont {Van~Setten}}, \bibinfo {author} {\bibfnamefont {M.}~\bibnamefont {Giantomassi}}, \bibinfo {author} {\bibfnamefont {E.}~\bibnamefont {Bousquet}}, \bibinfo {author} {\bibfnamefont {M.~J.}\ \bibnamefont {Verstraete}}, \bibinfo {author} {\bibfnamefont {D.~R.}\ \bibnamefont {Hamann}}, \bibinfo {author} {\bibfnamefont {X.}~\bibnamefont {Gonze}},\ and\ \bibinfo {author} {\bibfnamefont {G.-M.}\ \bibnamefont {Rignanese}},\ }\href@noop {} {\bibfield  {journal} {\bibinfo  {journal} {Computer Physics Communications}\ }\textbf {\bibinfo {volume} {226}},\ \bibinfo {pages} {39} (\bibinfo {year} {2018})}\BibitemShut {NoStop}%
\bibitem [{\citenamefont {Agrestini}\ \emph {et~al.}(2004)\citenamefont {Agrestini}, \citenamefont {Metallo}, \citenamefont {Filippi}, \citenamefont {Campi}, \citenamefont {Sanipoli}, \citenamefont {De~Negri}, \citenamefont {Giovannini}, \citenamefont {Saccone}, \citenamefont {Latini},\ and\ \citenamefont {Bianconi}}]{agrestini2004sc}%
  \BibitemOpen
  \bibfield  {author} {\bibinfo {author} {\bibfnamefont {S.}~\bibnamefont {Agrestini}}, \bibinfo {author} {\bibfnamefont {C.}~\bibnamefont {Metallo}}, \bibinfo {author} {\bibfnamefont {M.}~\bibnamefont {Filippi}}, \bibinfo {author} {\bibfnamefont {G.}~\bibnamefont {Campi}}, \bibinfo {author} {\bibfnamefont {C.}~\bibnamefont {Sanipoli}}, \bibinfo {author} {\bibfnamefont {S.}~\bibnamefont {De~Negri}}, \bibinfo {author} {\bibfnamefont {M.}~\bibnamefont {Giovannini}}, \bibinfo {author} {\bibfnamefont {A.}~\bibnamefont {Saccone}}, \bibinfo {author} {\bibfnamefont {A.}~\bibnamefont {Latini}},\ and\ \bibinfo {author} {\bibfnamefont {A.}~\bibnamefont {Bianconi}},\ }\href@noop {} {\bibfield  {journal} {\bibinfo  {journal} {Journal of Physics and Chemistry of Solids}\ }\textbf {\bibinfo {volume} {65}},\ \bibinfo {pages} {1479} (\bibinfo {year} {2004})}\BibitemShut {NoStop}%
\bibitem [{\citenamefont {Bradlyn}\ \emph {et~al.}(2017)\citenamefont {Bradlyn}, \citenamefont {Elcoro}, \citenamefont {Cano}, \citenamefont {Vergniory}, \citenamefont {Wang}, \citenamefont {Felser}, \citenamefont {Aroyo},\ and\ \citenamefont {Bernevig}}]{bradlyn2017topological}%
  \BibitemOpen
  \bibfield  {author} {\bibinfo {author} {\bibfnamefont {B.}~\bibnamefont {Bradlyn}}, \bibinfo {author} {\bibfnamefont {L.}~\bibnamefont {Elcoro}}, \bibinfo {author} {\bibfnamefont {J.}~\bibnamefont {Cano}}, \bibinfo {author} {\bibfnamefont {M.~G.}\ \bibnamefont {Vergniory}}, \bibinfo {author} {\bibfnamefont {Z.}~\bibnamefont {Wang}}, \bibinfo {author} {\bibfnamefont {C.}~\bibnamefont {Felser}}, \bibinfo {author} {\bibfnamefont {M.~I.}\ \bibnamefont {Aroyo}},\ and\ \bibinfo {author} {\bibfnamefont {B.~A.}\ \bibnamefont {Bernevig}},\ }\href@noop {} {\bibfield  {journal} {\bibinfo  {journal} {Nature}\ }\textbf {\bibinfo {volume} {547}},\ \bibinfo {pages} {298} (\bibinfo {year} {2017})}\BibitemShut {NoStop}%
\bibitem [{\citenamefont {Cano}\ \emph {et~al.}(2018)\citenamefont {Cano}, \citenamefont {Bradlyn}, \citenamefont {Wang}, \citenamefont {Elcoro}, \citenamefont {Vergniory}, \citenamefont {Felser}, \citenamefont {Aroyo},\ and\ \citenamefont {Bernevig}}]{cano2018building}%
  \BibitemOpen
  \bibfield  {author} {\bibinfo {author} {\bibfnamefont {J.}~\bibnamefont {Cano}}, \bibinfo {author} {\bibfnamefont {B.}~\bibnamefont {Bradlyn}}, \bibinfo {author} {\bibfnamefont {Z.}~\bibnamefont {Wang}}, \bibinfo {author} {\bibfnamefont {L.}~\bibnamefont {Elcoro}}, \bibinfo {author} {\bibfnamefont {M.~G.}\ \bibnamefont {Vergniory}}, \bibinfo {author} {\bibfnamefont {C.}~\bibnamefont {Felser}}, \bibinfo {author} {\bibfnamefont {M.~I.}\ \bibnamefont {Aroyo}},\ and\ \bibinfo {author} {\bibfnamefont {B.~A.}\ \bibnamefont {Bernevig}},\ }\href@noop {} {\bibfield  {journal} {\bibinfo  {journal} {Physical Review B}\ }\textbf {\bibinfo {volume} {97}},\ \bibinfo {pages} {035139} (\bibinfo {year} {2018})}\BibitemShut {NoStop}%
\bibitem [{\citenamefont {Elcoro}\ \emph {et~al.}(2021)\citenamefont {Elcoro}, \citenamefont {Wieder}, \citenamefont {Song}, \citenamefont {Xu}, \citenamefont {Bradlyn},\ and\ \citenamefont {Bernevig}}]{elcoro2021magnetic}%
  \BibitemOpen
  \bibfield  {author} {\bibinfo {author} {\bibfnamefont {L.}~\bibnamefont {Elcoro}}, \bibinfo {author} {\bibfnamefont {B.~J.}\ \bibnamefont {Wieder}}, \bibinfo {author} {\bibfnamefont {Z.}~\bibnamefont {Song}}, \bibinfo {author} {\bibfnamefont {Y.}~\bibnamefont {Xu}}, \bibinfo {author} {\bibfnamefont {B.}~\bibnamefont {Bradlyn}},\ and\ \bibinfo {author} {\bibfnamefont {B.~A.}\ \bibnamefont {Bernevig}},\ }\href@noop {} {\bibfield  {journal} {\bibinfo  {journal} {Nature communications}\ }\textbf {\bibinfo {volume} {12}},\ \bibinfo {pages} {5965} (\bibinfo {year} {2021})}\BibitemShut {NoStop}%
\bibitem [{\citenamefont {Griffiths}\ and\ \citenamefont {Schroeter}(2018)}]{griffiths2018introduction}%
  \BibitemOpen
  \bibfield  {author} {\bibinfo {author} {\bibfnamefont {D.~J.}\ \bibnamefont {Griffiths}}\ and\ \bibinfo {author} {\bibfnamefont {D.~F.}\ \bibnamefont {Schroeter}},\ }\href@noop {} {\emph {\bibinfo {title} {Introduction to quantum mechanics}}}\ (\bibinfo  {publisher} {Cambridge university press},\ \bibinfo {year} {2018})\BibitemShut {NoStop}%
\bibitem [{\citenamefont {Slater}(1930)}]{PhysRev.36.57}%
  \BibitemOpen
  \bibfield  {author} {\bibinfo {author} {\bibfnamefont {J.~C.}\ \bibnamefont {Slater}},\ }\href {https://doi.org/10.1103/PhysRev.36.57} {\bibfield  {journal} {\bibinfo  {journal} {Phys. Rev.}\ }\textbf {\bibinfo {volume} {36}},\ \bibinfo {pages} {57} (\bibinfo {year} {1930})}\BibitemShut {NoStop}%
\bibitem [{\citenamefont {Zak}(1980)}]{zak1980symmetry}%
  \BibitemOpen
  \bibfield  {author} {\bibinfo {author} {\bibfnamefont {J.}~\bibnamefont {Zak}},\ }\href@noop {} {\bibfield  {journal} {\bibinfo  {journal} {Physical Review Letters}\ }\textbf {\bibinfo {volume} {45}},\ \bibinfo {pages} {1025} (\bibinfo {year} {1980})}\BibitemShut {NoStop}%
\bibitem [{\citenamefont {Zak}(1981)}]{zak1981band}%
  \BibitemOpen
  \bibfield  {author} {\bibinfo {author} {\bibfnamefont {J.}~\bibnamefont {Zak}},\ }\href@noop {} {\bibfield  {journal} {\bibinfo  {journal} {Physical Review B}\ }\textbf {\bibinfo {volume} {23}},\ \bibinfo {pages} {2824} (\bibinfo {year} {1981})}\BibitemShut {NoStop}%
\bibitem [{\citenamefont {Miao}\ \emph {et~al.}(2015)\citenamefont {Miao}, \citenamefont {Zhong-Yi},\ and\ \citenamefont {Tao}}]{miao2015finding}%
  \BibitemOpen
  \bibfield  {author} {\bibinfo {author} {\bibfnamefont {G.}~\bibnamefont {Miao}}, \bibinfo {author} {\bibfnamefont {L.}~\bibnamefont {Zhong-Yi}},\ and\ \bibinfo {author} {\bibfnamefont {X.}~\bibnamefont {Tao}},\ }\href@noop {} {\bibfield  {journal} {\bibinfo  {journal} {Physics}\ }\textbf {\bibinfo {volume} {44}},\ \bibinfo {pages} {421} (\bibinfo {year} {2015})}\BibitemShut {NoStop}%
\bibitem [{\citenamefont {Jiang}\ \emph {et~al.}(2025)\citenamefont {Jiang}, \citenamefont {Hu}, \citenamefont {C{\u{a}}lug{\u{a}}ru}, \citenamefont {Felser}, \citenamefont {Blanco-Canosa}, \citenamefont {Weng}, \citenamefont {Xu},\ and\ \citenamefont {Bernevig}}]{jiang2025fege}%
  \BibitemOpen
  \bibfield  {author} {\bibinfo {author} {\bibfnamefont {Y.}~\bibnamefont {Jiang}}, \bibinfo {author} {\bibfnamefont {H.}~\bibnamefont {Hu}}, \bibinfo {author} {\bibfnamefont {D.}~\bibnamefont {C{\u{a}}lug{\u{a}}ru}}, \bibinfo {author} {\bibfnamefont {C.}~\bibnamefont {Felser}}, \bibinfo {author} {\bibfnamefont {S.}~\bibnamefont {Blanco-Canosa}}, \bibinfo {author} {\bibfnamefont {H.}~\bibnamefont {Weng}}, \bibinfo {author} {\bibfnamefont {Y.}~\bibnamefont {Xu}},\ and\ \bibinfo {author} {\bibfnamefont {B.~A.}\ \bibnamefont {Bernevig}},\ }\href@noop {} {\bibfield  {journal} {\bibinfo  {journal} {Physical Review B}\ }\textbf {\bibinfo {volume} {111}},\ \bibinfo {pages} {125163} (\bibinfo {year} {2025})}\BibitemShut {NoStop}%
\bibitem [{\citenamefont {et~al}(2026)}]{PaperMassSeparation}%
  \BibitemOpen
  \bibfield  {author} {\bibinfo {author} {\bibfnamefont {Y.~J.}\ \bibnamefont {et~al}},\ }\href@noop {} {\bibfield  {journal} {\bibinfo  {journal} {In preparation}\ } (\bibinfo {year} {2026})}\BibitemShut {NoStop}%
\bibitem [{\citenamefont {Xu}\ \emph {et~al.}(2024{\natexlab{b}})\citenamefont {Xu}, \citenamefont {Vergniory}, \citenamefont {Ma}, \citenamefont {Ma{\~n}es}, \citenamefont {Song}, \citenamefont {Bernevig}, \citenamefont {Regnault},\ and\ \citenamefont {Elcoro}}]{xu2024catalog}%
  \BibitemOpen
  \bibfield  {author} {\bibinfo {author} {\bibfnamefont {Y.}~\bibnamefont {Xu}}, \bibinfo {author} {\bibfnamefont {M.}~\bibnamefont {Vergniory}}, \bibinfo {author} {\bibfnamefont {D.-S.}\ \bibnamefont {Ma}}, \bibinfo {author} {\bibfnamefont {J.~L.}\ \bibnamefont {Ma{\~n}es}}, \bibinfo {author} {\bibfnamefont {Z.-D.}\ \bibnamefont {Song}}, \bibinfo {author} {\bibfnamefont {B.~A.}\ \bibnamefont {Bernevig}}, \bibinfo {author} {\bibfnamefont {N.}~\bibnamefont {Regnault}},\ and\ \bibinfo {author} {\bibfnamefont {L.}~\bibnamefont {Elcoro}},\ }\href@noop {} {\bibfield  {journal} {\bibinfo  {journal} {Science}\ }\textbf {\bibinfo {volume} {384}},\ \bibinfo {pages} {eadf8458} (\bibinfo {year} {2024}{\natexlab{b}})}\BibitemShut {NoStop}%
\bibitem [{\citenamefont {Loudon}\ \emph {et~al.}(2015)\citenamefont {Loudon}, \citenamefont {Yazdi}, \citenamefont {Kasama}, \citenamefont {Zhigadlo},\ and\ \citenamefont {Karpinski}}]{loudon_measurement_2015}%
  \BibitemOpen
  \bibfield  {author} {\bibinfo {author} {\bibfnamefont {J.~C.}\ \bibnamefont {Loudon}}, \bibinfo {author} {\bibfnamefont {S.}~\bibnamefont {Yazdi}}, \bibinfo {author} {\bibfnamefont {T.}~\bibnamefont {Kasama}}, \bibinfo {author} {\bibfnamefont {N.~D.}\ \bibnamefont {Zhigadlo}},\ and\ \bibinfo {author} {\bibfnamefont {J.}~\bibnamefont {Karpinski}},\ }\bibfield  {journal} {\bibinfo  {journal} {Physical Review B}\ }\textbf {\bibinfo {volume} {91}},\ \href {https://doi.org/10.1103/physrevb.91.054505} {10.1103/physrevb.91.054505} (\bibinfo {year} {2015}),\ \bibinfo {note} {publisher: American Physical Society (APS)}\BibitemShut {NoStop}%
\bibitem [{\citenamefont {Eltsev}\ \emph {et~al.}(2002)\citenamefont {Eltsev}, \citenamefont {Lee}, \citenamefont {Nakao}, \citenamefont {Chikumoto}, \citenamefont {Tajima}, \citenamefont {Koshizuka},\ and\ \citenamefont {Murakami}}]{eltsev_anisotropic_2002}%
  \BibitemOpen
  \bibfield  {author} {\bibinfo {author} {\bibfnamefont {Y.}~\bibnamefont {Eltsev}}, \bibinfo {author} {\bibfnamefont {S.}~\bibnamefont {Lee}}, \bibinfo {author} {\bibfnamefont {K.}~\bibnamefont {Nakao}}, \bibinfo {author} {\bibfnamefont {N.}~\bibnamefont {Chikumoto}}, \bibinfo {author} {\bibfnamefont {S.}~\bibnamefont {Tajima}}, \bibinfo {author} {\bibfnamefont {N.}~\bibnamefont {Koshizuka}},\ and\ \bibinfo {author} {\bibfnamefont {M.}~\bibnamefont {Murakami}},\ }\href {https://doi.org/10.1103/PhysRevB.65.140501} {\bibfield  {journal} {\bibinfo  {journal} {Physical Review B}\ }\textbf {\bibinfo {volume} {65}},\ \bibinfo {pages} {140501} (\bibinfo {year} {2002})}\BibitemShut {NoStop}%
\bibitem [{\citenamefont {Klein}\ \emph {et~al.}(2006)\citenamefont {Klein}, \citenamefont {Lyard}, \citenamefont {Marcus}, \citenamefont {Holanova},\ and\ \citenamefont {Marcenat}}]{klein_magnetic_2006}%
  \BibitemOpen
  \bibfield  {author} {\bibinfo {author} {\bibfnamefont {T.}~\bibnamefont {Klein}}, \bibinfo {author} {\bibfnamefont {L.}~\bibnamefont {Lyard}}, \bibinfo {author} {\bibfnamefont {J.}~\bibnamefont {Marcus}}, \bibinfo {author} {\bibfnamefont {Z.}~\bibnamefont {Holanova}},\ and\ \bibinfo {author} {\bibfnamefont {C.}~\bibnamefont {Marcenat}},\ }\href {https://doi.org/10.1103/PhysRevB.73.184513} {\bibfield  {journal} {\bibinfo  {journal} {Physical Review B}\ }\textbf {\bibinfo {volume} {73}},\ \bibinfo {pages} {184513} (\bibinfo {year} {2006})}\BibitemShut {NoStop}%
\bibitem [{\citenamefont {Tan}\ \emph {et~al.}(2015)\citenamefont {Tan}, \citenamefont {Wolak}, \citenamefont {Acharya}, \citenamefont {Krick}, \citenamefont {Lang}, \citenamefont {Sloppy}, \citenamefont {Taheri}, \citenamefont {Civale}, \citenamefont {Chen},\ and\ \citenamefont {Xi}}]{tan_enhancement_2015}%
  \BibitemOpen
  \bibfield  {author} {\bibinfo {author} {\bibfnamefont {T.}~\bibnamefont {Tan}}, \bibinfo {author} {\bibfnamefont {M.~A.}\ \bibnamefont {Wolak}}, \bibinfo {author} {\bibfnamefont {N.}~\bibnamefont {Acharya}}, \bibinfo {author} {\bibfnamefont {A.}~\bibnamefont {Krick}}, \bibinfo {author} {\bibfnamefont {A.~C.}\ \bibnamefont {Lang}}, \bibinfo {author} {\bibfnamefont {J.}~\bibnamefont {Sloppy}}, \bibinfo {author} {\bibfnamefont {M.~L.}\ \bibnamefont {Taheri}}, \bibinfo {author} {\bibfnamefont {L.}~\bibnamefont {Civale}}, \bibinfo {author} {\bibfnamefont {K.}~\bibnamefont {Chen}},\ and\ \bibinfo {author} {\bibfnamefont {X.~X.}\ \bibnamefont {Xi}},\ }\href {https://doi.org/10.1063/1.4916696} {\bibfield  {journal} {\bibinfo  {journal} {APL Materials}\ }\textbf {\bibinfo {volume} {3}},\ \bibinfo {pages} {041101} (\bibinfo {year} {2015})}\BibitemShut {NoStop}%
\bibitem [{\citenamefont {Jiang}\ \emph {et~al.}(2021)\citenamefont {Jiang}, \citenamefont {Fang},\ and\ \citenamefont {Fang}}]{jiang2021k}%
  \BibitemOpen
  \bibfield  {author} {\bibinfo {author} {\bibfnamefont {Y.}~\bibnamefont {Jiang}}, \bibinfo {author} {\bibfnamefont {Z.}~\bibnamefont {Fang}},\ and\ \bibinfo {author} {\bibfnamefont {C.}~\bibnamefont {Fang}},\ }\href@noop {} {\bibfield  {journal} {\bibinfo  {journal} {Chinese Physics Letters}\ }\textbf {\bibinfo {volume} {38}},\ \bibinfo {pages} {077104} (\bibinfo {year} {2021})}\BibitemShut {NoStop}%
\bibitem [{\citenamefont {Sun}\ \emph {et~al.}(2009)\citenamefont {Sun}, \citenamefont {Yao}, \citenamefont {Fradkin},\ and\ \citenamefont {Kivelson}}]{sun2009topological}%
  \BibitemOpen
  \bibfield  {author} {\bibinfo {author} {\bibfnamefont {K.}~\bibnamefont {Sun}}, \bibinfo {author} {\bibfnamefont {H.}~\bibnamefont {Yao}}, \bibinfo {author} {\bibfnamefont {E.}~\bibnamefont {Fradkin}},\ and\ \bibinfo {author} {\bibfnamefont {S.~A.}\ \bibnamefont {Kivelson}},\ }\href@noop {} {\bibfield  {journal} {\bibinfo  {journal} {Physical review letters}\ }\textbf {\bibinfo {volume} {103}},\ \bibinfo {pages} {046811} (\bibinfo {year} {2009})}\BibitemShut {NoStop}%
\bibitem [{\citenamefont {Ahn}\ \emph {et~al.}(2019)\citenamefont {Ahn}, \citenamefont {Park},\ and\ \citenamefont {Yang}}]{ahn2019failure}%
  \BibitemOpen
  \bibfield  {author} {\bibinfo {author} {\bibfnamefont {J.}~\bibnamefont {Ahn}}, \bibinfo {author} {\bibfnamefont {S.}~\bibnamefont {Park}},\ and\ \bibinfo {author} {\bibfnamefont {B.-J.}\ \bibnamefont {Yang}},\ }\href@noop {} {\bibfield  {journal} {\bibinfo  {journal} {Physical Review X}\ }\textbf {\bibinfo {volume} {9}},\ \bibinfo {pages} {021013} (\bibinfo {year} {2019})}\BibitemShut {NoStop}%
\bibitem [{\citenamefont {Yu}\ \emph {et~al.}(2023)\citenamefont {Yu}, \citenamefont {Xie}, \citenamefont {Wu},\ and\ \citenamefont {Das~Sarma}}]{yu2023euler}%
  \BibitemOpen
  \bibfield  {author} {\bibinfo {author} {\bibfnamefont {J.}~\bibnamefont {Yu}}, \bibinfo {author} {\bibfnamefont {M.}~\bibnamefont {Xie}}, \bibinfo {author} {\bibfnamefont {F.}~\bibnamefont {Wu}},\ and\ \bibinfo {author} {\bibfnamefont {S.}~\bibnamefont {Das~Sarma}},\ }\href@noop {} {\bibfield  {journal} {\bibinfo  {journal} {Physical Review B}\ }\textbf {\bibinfo {volume} {107}},\ \bibinfo {pages} {L201106} (\bibinfo {year} {2023})}\BibitemShut {NoStop}%
\bibitem [{\citenamefont {Song}\ \emph {et~al.}(2019)\citenamefont {Song}, \citenamefont {Wang}, \citenamefont {Shi}, \citenamefont {Li}, \citenamefont {Fang},\ and\ \citenamefont {Bernevig}}]{song2019all}%
  \BibitemOpen
  \bibfield  {author} {\bibinfo {author} {\bibfnamefont {Z.}~\bibnamefont {Song}}, \bibinfo {author} {\bibfnamefont {Z.}~\bibnamefont {Wang}}, \bibinfo {author} {\bibfnamefont {W.}~\bibnamefont {Shi}}, \bibinfo {author} {\bibfnamefont {G.}~\bibnamefont {Li}}, \bibinfo {author} {\bibfnamefont {C.}~\bibnamefont {Fang}},\ and\ \bibinfo {author} {\bibfnamefont {B.~A.}\ \bibnamefont {Bernevig}},\ }\href@noop {} {\bibfield  {journal} {\bibinfo  {journal} {Physical review letters}\ }\textbf {\bibinfo {volume} {123}},\ \bibinfo {pages} {036401} (\bibinfo {year} {2019})}\BibitemShut {NoStop}%
\bibitem [{\citenamefont {Herzog-Arbeitman}\ \emph {et~al.}(2023)\citenamefont {Herzog-Arbeitman}, \citenamefont {Song}, \citenamefont {Elcoro},\ and\ \citenamefont {Bernevig}}]{herzog2023hofstadter}%
  \BibitemOpen
  \bibfield  {author} {\bibinfo {author} {\bibfnamefont {J.}~\bibnamefont {Herzog-Arbeitman}}, \bibinfo {author} {\bibfnamefont {Z.-D.}\ \bibnamefont {Song}}, \bibinfo {author} {\bibfnamefont {L.}~\bibnamefont {Elcoro}},\ and\ \bibinfo {author} {\bibfnamefont {B.~A.}\ \bibnamefont {Bernevig}},\ }\href@noop {} {\bibfield  {journal} {\bibinfo  {journal} {Physical review letters}\ }\textbf {\bibinfo {volume} {130}},\ \bibinfo {pages} {236601} (\bibinfo {year} {2023})}\BibitemShut {NoStop}%
\bibitem [{\citenamefont {Slater}\ and\ \citenamefont {Koster}(1954)}]{slater1954simplified}%
  \BibitemOpen
  \bibfield  {author} {\bibinfo {author} {\bibfnamefont {J.~C.}\ \bibnamefont {Slater}}\ and\ \bibinfo {author} {\bibfnamefont {G.~F.}\ \bibnamefont {Koster}},\ }\href@noop {} {\bibfield  {journal} {\bibinfo  {journal} {Physical review}\ }\textbf {\bibinfo {volume} {94}},\ \bibinfo {pages} {1498} (\bibinfo {year} {1954})}\BibitemShut {NoStop}%
\end{thebibliography}
\end{document}